\documentclass[a4paper,fleqn,usenatbib]{mnras}

\usepackage{newtxtext,newtxmath}
\usepackage[T1]{fontenc}
\usepackage{ae,aecompl}

\usepackage{graphicx}	% Including figure files
\usepackage{amsmath}	% Advanced maths commands
\usepackage{multicol}        % Multi-column entries in tables
\usepackage{longtable}
\usepackage{pdflscape}	% Landscape pages
\usepackage{hyperref}
\usepackage{deluxetable}
\usepackage[caption=false]{subfig}

\title[LOSS photometry of 70 stripped SNe]{The Lick Observatory Supernova Search follow-up program: photometry data release of 70 stripped-envelope supernovae}

\author[W. Zheng et al.]
{WeiKang Zheng,$^{1}$\thanks{E-mail: weikang@berkeley.edu; benjamin\_stahl@berkeley.edu; tdejaeger@berkeley.edu; afilippenko@berkeley.edu; shanqinwang@gxu.edu.cn}
Benjamin E. Stahl,$^{1,2}$ 
Thomas de Jaeger,$^{1,3}$ 
Alexei V. Filippenko,$^{1,4}$
\newauthor
Shan-Qin Wang,$^{5}$
Wen-Pei Gan,$^{5}$
Thomas G. Brink,$^{1}$
Ivan Altunin,$^{1}$
\newauthor
Raphael Baer-Way,$^{1}$
Andrew Bigley,$^{1}$
Kyle Blanchard,$^{1}$
Peter K. Blanchard,$^{6}$
James Bradley,$^{1}$
\newauthor
Samantha K. Cargill,$^{1}$
Chadwick Casper,$^{1}$
Teagan Chapman,$^{1}$
Vidhi Chander,$^{1}$
\newauthor
Sanyum Channa,$^{2,7}$
Byung Yun Choi,$^{1}$
Nick Choksi,$^{1}$
Matthew Chu,$^{2}$
Kelsey I. Clubb,$^{1}$
\newauthor
Daniel P. Cohen,$^{8}$
Paul A. Dalba,$^{9,10}$ 
Asia deGraw,$^{1}$
Maxime de Kouchkovsky,$^{1}$
\newauthor
Michael Ellison,$^{1}$
Edward Falcon,$^{1}$
Ori D. Fox,$^{11}$
Kiera Fuller,$^{1}$
Mohan Ganeshalingam,$^{12}$
\newauthor
Nachiket Girish,$^{2}$
Carolina Gould,$^{1}$
Goni Halevi,$^{1,13}$
Andrew Halle,$^{1}$
Kevin T. Hayakawa,$^{8}$
\newauthor
Romain Hardy,$^{8}$
Julia Hestenes,$^{1}$
Andrew M. Hoffman,$^{1}$
Michael Hyland,$^{1}$
\newauthor
Benjamin T. Jeffers,$^{1}$
Connor Jennings,$^{1}$
Michael T. Kandrashoff,$^{1}$
Anthony Khodanian,$^{1}$
\newauthor
Minkyu Kim,$^{1}$
Haejung Kim,$^{1}$
Michelle E. Kislak,$^{1,14}$
Daniel Krishnan,$^{1}$
\newauthor
Sahana Kumar,$^{1,15}$
Snehaa Ganesh Kumar,$^{1}$
Joel Leja,$^{16,17,18}$
Erin J. Leonard,$^{1,19}$
Gary Z. Li,$^{20}$
\newauthor
Weidong Li,$^{1}$\thanks{Deceased 2011 December 12}
Ji-Shun Lian,$^{5}$
Evelyn Liu,$^{1}$
Thomas B. Lowe,$^{3}$
Philip Lu,$^{21}$
Emily Ma,$^{1}$
\newauthor
Michelle N. Mason,$^{22}$
Michael May,$^{1}$
Kyle McAllister,$^{1}$
Emma McGinness,$^{1}$
\newauthor
Shaunak Modak,$^{2,13}$
Jeffrey Molloy,$^{1}$
Yukei S. Murakami,$^{23}$
Omnarayani Nayak,$^{11}$  %aka Isha Nayak
\newauthor
Derek Perera,$^{1}$
Kenia Pina,$^{1}$
Druv Punjabi,$^{1}$
Andrew Rikhter,$^{24}$
Timothy W. Ross,$^{1}$
\newauthor
Jackson Sipple,$^{1}$
Costas Soler,$^{1}$
Samantha Stegman,$^{1,25}$
Haynes Stephens,$^{1}$
James Sunseri,$^{1,2}$
\newauthor
Kevin Tang,$^{1}$
Stephen Taylor,$^{1}$
Patrick Thrasher,$^{1}$
Schuyler D. Van Dyk,$^{26}$
Xiang-Gao Wang,$^{5}$
\newauthor
Jeremy Wayland,$^{1}$
Andrew Wilkins,$^{1}$
Abel Yagubyan,$^{1}$
Heechan Yuk,$^{27}$
Sameen Yunus,$^{1}$
\newauthor
and Keto D. Zhang$^{1}$
\\
\\
$^{1}$Department of Astronomy, University of California, Berkeley, CA 94720-3411, USA\\
$^{2}$Department of Physics, University of California, Berkeley, CA 94720-7300, USA\\
$^{3}$Institute for Astronomy, University of Hawaii, 2680 Woodlawn Drive, Honolulu, HI 96822, USA\\
$^{4}$Miller Institute for Basic Research in Science, University of California, Berkeley, CA 94720, USA\\
$^{5}$Guangxi Key Laboratory for Relativistic Astrophysics, School of Physical Science and Technology, Guangxi University, Nanning 530004, China\\
$^{6}$Center for Interdisciplinary Exploration and Research in Astrophysics (CIERA) and Department of Physics and Astronomy,\\
      ~~Northwestern University, 1800 Sherman Ave., Evanston, IL 60201, USA\\
$^{7}$Department of Physics, Stanford University, Stanford, CA 94305, USA\\
$^{8}$Department of Physics and Astronomy, University of California, Los Angeles, CA 90095, USA\\
$^{9}$Department of Astronomy and Astrophysics, University of California Santa Cruz, 1156 High St., Santa Cruz, CA 95064, USA\\
$^{10}$Department of Earth and Planetary Sciences, University of California Riverside, 900 University Ave., Riverside, CA 92521, USA\\
$^{11}$Space Telescope Science Institute, 3700 San Martin Drive, Baltimore, MD 21218, USA\\
$^{12}$Lawrence Berkeley National Laboratory, 1 Cyclotron Rd, Berkeley, CA 94720, USA\\
$^{13}$ Department of Astrophysical Sciences, Princeton University, 4 Ivy Lane, Princeton, NJ 08544, USA\\
$^{14}$Netflix, Inc., 100 Winchester Cir, Los Gatos, CA 95032, USA\\
$^{15}$Department of Physics, Florida State University, Tallahassee, FL 32306, USA\\
$^{16}$Department of Astronomy \& Astrophysics, The Pennsylvania State University, University Park, PA 16802, USA\\
$^{17}$Institute for Computational \& Data Sciences, The Pennsylvania State University, University Park, PA, USA\\
$^{18}$Institute for Gravitation and the Cosmos, The Pennsylvania State University, University Park, PA 16802, USA\\
$^{19}$Jet Propulsion Laboratory, California Institute of Technology, Pasadena, CA 91109, USA\\
$^{20}$The Aerospace Corporation, 2310 E. El Segundo Blvd., El Segundo, CA, 90245, USA\\
$^{21}$Center for Theoretical Physics, Department of Physics and Astronomy, Seoul National University, Seoul 08826, Korea\\
$^{22}$Department of Physics and Astronomy, University of Wyoming, 1000 E. University, Dept. 3905, Laramie, WY 82071, USA\\
$^{23}$Department of Physics and Astronomy, Johns Hopkins University, Baltimore, MD 21218, USA\\
$^{24}$Department of Physics, University of California San Diego, La Jolla, CA 92093, USA\\
$^{25}$Argonne National Laboratory, 9700 S. Cass Avenue, Lemont, IL 60439, USA\\
$^{26}$Caltech/Spitzer Science Center, Caltech/IPAC, Mailcode 100-22, Pasadena, CA 91125, USA\\
$^{27}$Department of Physics and Astronomy, University of Oklahoma, 440 W. Brooks St., Norman, OK 73019, USA\\
}

\date{Accepted XXX. Received YYY; in original form ZZZ}

\pubyear{2021}

\begin{document}

\label{firstpage}
\pagerange{\pageref{firstpage}--\pageref{lastpage}}
\maketitle

\begin{abstract}
We present $BVRI$ and unfiltered ({\it Clear}) light curves of 70 stripped-envelope supernovae (SESNe), observed between 2003 and 2020, from
the Lick Observatory Supernova Search (LOSS) follow-up program.
Our SESN sample consists of 19 spectroscopically normal SNe~Ib, two peculiar SNe~Ib, six SNe~Ibn, 14 normal SNe~Ic,
one peculiar SN~Ic, ten SNe~Ic-BL, 15 SNe~IIb, one ambiguous SN~IIb/Ib/c, and two superluminous SNe.
Our follow-up photometry has (on a per-SN basis) a mean coverage of 81 photometric points (median of 58 points) and a mean cadence of 3.6\,d (median of 1.2\,d).
From our full sample, a subset of 38 SNe have pre-maximum coverage in at least one passband, allowing for the peak brightness of each SN in this subset to be quantitatively determined.
We describe our data collection and processing techniques, with emphasis toward our automated photometry pipeline, from which we derive publicly available data products to enable and encourage further study by the community.
Using these data products, we derive host-galaxy extinction values through the empirical colour evolution relationship and, for the first time, produce accurate rise-time measurements for a large sample of SESNe in both optical and infrared passbands.
By modeling multiband light curves, we find that SNe~Ic tend to have lower ejecta masses and lower ejecta velocities than SNe~Ib and IIb, but higher $^{56}$Ni masses.
\end{abstract}


%%%%%%%%%%%%%%%%%%%%%%%%%%%%%%%%%%%%%%%%%%%%%%%%%%

%%%%%%%%%%%%%%%%% BODY OF PAPER %%%%%%%%%%%%%%%%%%

\section{Introduction}
\label{sec:introduction}

It is well established that massive stars (i.e., those having $M \gtrsim 8$\,M$_{\sun}$) have short lives that end in catastrophic explosions
known as core-collapse supernovae (CCSNe). Among CCSNe, those whose spectra show features of hydrogen
are classified as Type II SNe (see, e.g., \citealt{filippenko97,gal17} for reviews
of SN classification). In contrast, hydrogen-poor CCSNe are classified as Type
Ib or Ic, depending on whether their optical spectra contain obvious helium features \citep{matheson01}.
The progenitor stars of hydrogen-poor CCSNe have their outer envelopes stripped away before explosion by
strong winds during the Wolf-Rayet phase \citep[e.g.,][]{conti75,smith06,gal14},
By interaction with a binary companion \citep[e.g.,][]{podsiadlowski92,sana12,eldridge13}, or some combination of these two modes.
If the envelope-stripping process is highly efficient, the helium shell is also removed
before explosion, leading to the differentiation between SNe~Ib (He-rich) and SNe~Ic (He-poor).
Hydrogen-poor CCSNe are generally referred to as stripped-envelope supernovae (SESNe).

SESNe are found to be observationally heterogeneous. For example, in some cases, the stripping
process is incomplete and thus the envelope is left with some fraction of hydrogen.
These SNe typically show H lines at early times that rapidly disappear
after maximum light \citep{filippenko88,filippenko93}, and their spectra resemble SNe~Ib at late times.
A small subset of SN~Ib-like events show evidence of interaction with dense circumstellar material (CSM);
having relatively narrow spectral emission lines \citep[e.g.,][]{pastorello07,foley07,hosseinzadeh17b}, these
objects have been dubbed SNe~Ibn (``n'' for ``narrow lines'').
In addition, a subset of SNe~Ic characterised by the presence broad spectral lines which indicate extremely
high ejecta velocities ($\gtrsim 15,000$\,km\,s$^{-1}$) are designated as SNe~Ic-BL \citep[e.g.,][]{modjaz14}.
Objects within this subclass have been found to be associated with long-duration gamma-ray bursts
\citep[e.g.,][]{woosley06}.
Recently, a new class, SNe~Icn, has been proposed by \cite{gal21} based on the prototype
SN~2019hgp, followed by SNe~2021csp \citep{perley21,fraser21} and 2021ckj \citep{pastorello21}. The early-time spectra of these objects
are dominated by narrow lines with profiles similar to those seen in SNe~Ibn, but originating from
carbon and oxygen rather than He.
In any case, all of the aforementioned SN classifications (Ib, IIb, Ibn, Ic, Ic-BL, and Icn) are,
to some extent, related to the envelope of their progenitor star being stripped. Accordingly, we consider
all of them to be SESNe in the analysis presented herein.

Owing to the efforts of various SN surveys spanning the globe, the study of SESNe with large
light-curve samples has proliferated.
\cite{li11} presented a set of roughly two dozen unfiltered SESN light curves within $\sim 60$\,Mpc.
In the same year, \citet{drout11} presented $V$- and $R$-band light curves of 25 SESNe from the Palomar 60 inch telescope.
\cite{bianco14} published multiband light curves of 64 SESNe obtained by the Harvard-Smithsonian Center
for Astrophysics (CfA) SN group, and
\cite{taddia15} presented expanded sets of multiband light curves of 20 SESNe from the
Sloan Digital Sky Survey (SDSS) SN Survey II.
In addition, \cite{stritzinger18a} published 34 SESN light curves from the first phase 
of the Carnegie Supernova Project (CSP-I). Despite these impressive efforts, the state of
large-scale photometric studies of SESNe substantially lags that of other SNe
(e.g., SNe~Ia, which are routinely studied photometrically at the hundreds-of-objects scale).

Over the past two decades, our Lick Observatory Supernova Search (LOSS; \citealt{filippenko01}) program
has invested considerably in both discovering and monitoring all kinds of SNe, including SESNe.
Large light-curve samples have already been published by \cite{ganeshalingam10} and \cite{stahl19} for SNe~Ia,
and by \cite{dejaeger19} for SNe~II. In this paper, we release the light curves of 70 SESNe observed
by LOSS since 2003. 
In the remainder of the paper, we describe the sample (Sec.~\ref{sec:data}) and our data-reduction strategies (Sec.~\ref{sec:datareduction})
before presenting an analysis of the light curves (Sec.~\ref{sec:results}) and offering our conclusions (Sec.~\ref{sec:conclusion}).

\section{Data Sample}
\label{sec:data}

The Berkeley SESN sample consists of 70 objects observed between 2003 and 2020.
Two main telescopes were used for follow-up observations: (i)
the fully robotic 0.76\,m Katzman Automatic Imaging Telescope (KAIT; \citealt{filippenko01}),
and (ii) the 1\,m Anna Nickel telescope,
both located at Lick Observatory on Mount Hamilton, near San Jose, CA, USA.
Most SESNe in our sample were observed in multiple optical passbands ($B$, $V$, $R$, $I$), and some have additional {\it Clear}-band (unfiltered) data.\footnote{A small fraction are covered with only {\it Clear}-band observations.} 
For a large fraction of SESNe in our sample, spectra were also obtained by our group using multiple facilities.
A detailed analysis and release of the LOSS spectra of SESNe was published by 
\cite{shivvers19}, so the present paper focuses exclusively on our photometric observations.

\begin{figure}
        \includegraphics[width=1.0\columnwidth]{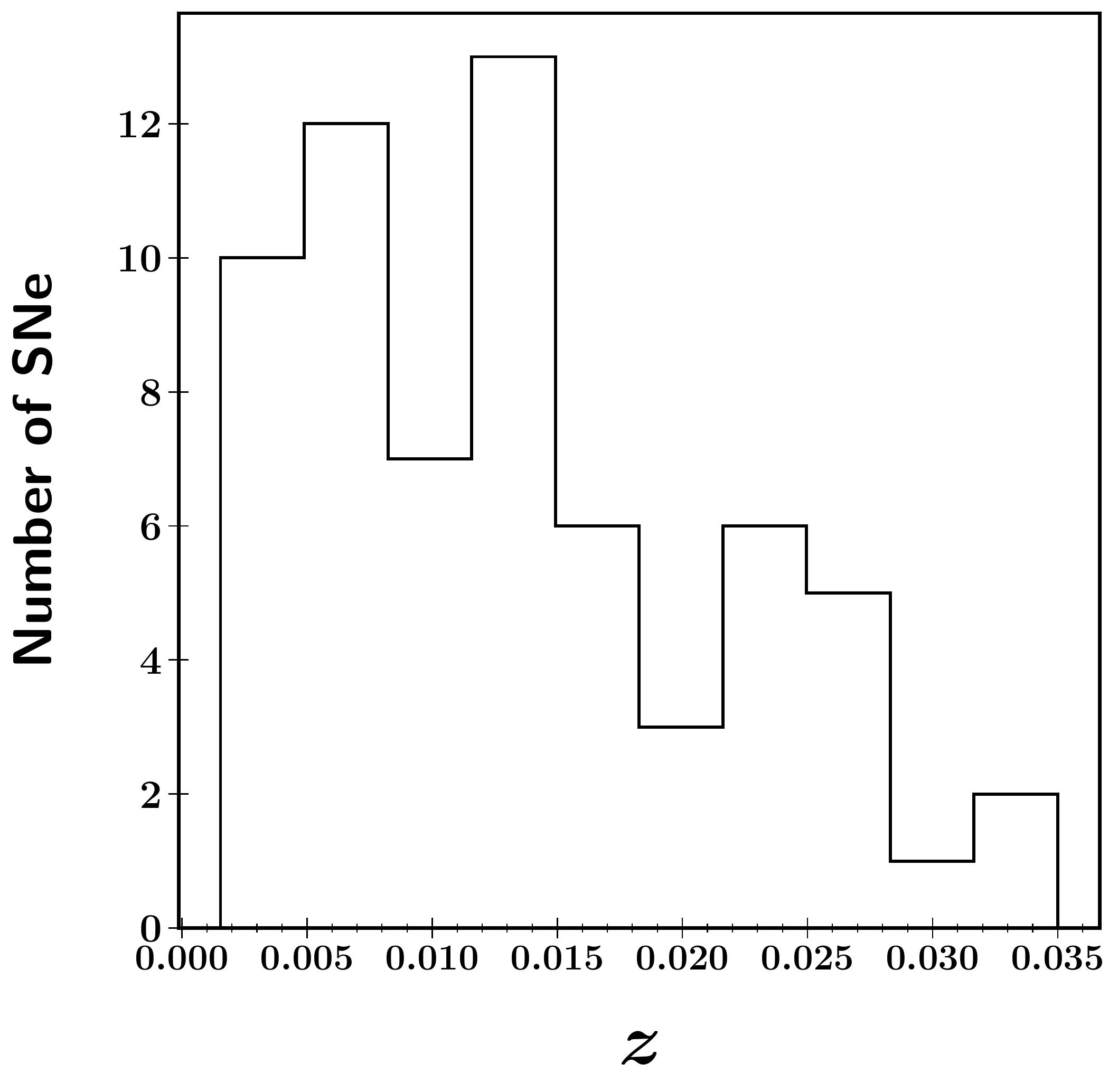}
        \caption{The redshift distribution of the 65 (out of 70 in total) SESNe in our sample
                 which have reliable heliocentric velocity measurements from their host galaxies.
                 The average redshift value is 0.0136 with a standard deviation of 0.0082, and
                 40 SESNe have $z > 0.01$.}
\label{Fig_z_hist}
\end{figure}

Table~\ref{SESNsamplebasicinformation} provides the basic information for each SN in our sample, including its spectroscopic classification, host-galaxy name, distance,
recession velocity, and Galactic extinction \citep{schlafly11}. The listed classifications are adopted from the
Transient Name Server\footnote{\url{www.wis-tns.org}} (TNS), but where they conflict with those given by \cite{shivvers19}, we adopt
the latter owing to the more sophisticated and systematic approach that resulted in their determinations.
The single exception to this paradigm is SN~2008fz, which was classified as an SN~Ic by \cite{shivvers19} but for which we adopt the classification of
superluminous SN~IIn (i.e., SLSN-IIn) from \cite{drake10}.
In summary, our SESN sample consists of 19 spectroscopically normal SNe~Ib, two peculiar SNe~Ib, six SNe~Ibn, 14 normal SNe~Ic, one peculiar SN~Ic, ten SNe~Ic-BL, 15 SNe~IIb, one ambiguous SN~IIb/Ib/c, and two SLSNe.

The host-galaxy distances reported in Table~\ref{SESNsamplebasicinformation} were obtained from the NASA/IPAC Extragalactic Database of
redshift-independent distances (NED-D\footnote{\url{http://ned.ipac.caltech.edu/Library/Distances/}}) when available, and otherwise calculated using a standard cosmological model with H$_0$ = 70\,km\,s$^{-1}$\,Mpc$^{-1}$,
$\Omega_M = 0.30$, and $\Omega_{\Lambda} = 0.70$.
Figure~\ref{Fig_z_hist} shows the redshift distribution of the 65 SESNe in our sample which have reliable
host-galaxy heliocentric velocity measurements.
The redshifts range from 0.0015 (SN~2011dh) to 0.0350 (SN~2009er), with an
average value of 0.0136 and a standard deviation of 0.0082. 40 SESNe have $z > 0.01$.
For the remaining five SESNe without host heliocentric velocity measurements, we adopt
redshifts from various literature sources (see the footnote of Table~\ref{SESNsamplebasicinformation} for more details).

\section{Data Reduction}
\label{sec:datareduction}

All photometric data published herein were obtained with the Lick KAIT and 1\,m Nickel telescopes.
Over the past two decades, the KAIT CCD and filter have been updated several times, resulting in
four different CCD/filter combinations, which we refer as KAIT1-4 (see \citealt[][]{ganeshalingam10,stahl19}
for more details).
In its current configuration (KAIT4), KAIT is equipped with a Finger Lakes Instrument camera 
with $512 \times 512$ pixels covering a $6\overset{\prime}{.}7 \times 6\overset{\prime}{.}7$ field of view.
The 1\,m Nickel is a human-operated telescope, but it can be remotely operated from the UC Berkeley campus.
It is equipped with a thinned, Loral, $2048 \times 2048$~pixel CCD (binned by a factor of two along both axes
to reduce the readout time), and has a $6\overset{\prime}{.}3 \times 6\overset{\prime}{.}3$ field of view. 
The filter set on the Nickel telescope was updated once, in March 2009; we refer to the configuration
before and after this change as Nickel1 and Nickel2, respectively (again, see \citealt[][]{ganeshalingam10,stahl19} for more details).

A novel automated photometry pipeline\footnote{\url{https://github.com/benstahl92/LOSSPhotPypeline}}
was developed by \cite{stahl19} to process the significant volume of SN observations produced by LOSS 
consistently and accurately, while at the same time requiring minimal human intervention.
Although we defer the details to \cite{stahl19}, we briefly summarise the main procedures here.
All images are first treated to remove the bias level, and are then flat-fielded before being
astrometrically calibrated using code provided by astrometry.net\footnote{\url{https://astrometry.net}}
\citep{lang10}. Where necessary, image subtraction is applied so as to remove host-galaxy contamination, with
the template images being sourced on dark nights using the Nickel telescope after the SNe have faded
beyond detection (generally $> 6$ months after discovery).
Point-spread-function (PSF) photometry is obtained using DAOPHOT \citep{stetson87}
from the IDL Astronomy User’s Library\footnote{\url{http://idlastro.gsfc.nasa.gov/}}.
Nearby stars are chosen from the
Pan-STARRS1\footnote{\url{http://archive.stsci.edu/panstarrs/search.php}} catalogue for calibration.
%or the SDSS\footnote\url{{https://www.sdss.org}} catalog if the Pan-STARRS1 is not available.
Their magnitudes are first transformed into the Landolt system \citep{landolt83,landolt92}
using the empirical prescription (Eq.~6) presented by \cite{tonry12},
and then transformed to the appropriate KAIT/Nickel natural system (i.e., KAIT1-4 or Nickel1-2 as appropriate
based on the equipment configuration on the date of observation).
All apparent magnitudes are measured in the natural system,
and the final results are then transformed to the standard system (see Eqs. 1a--1d of \citealt{stahl19})
using local calibrators and the appropriate colour terms as given by \cite{ganeshalingam10} and \cite{stahl19}.
Note that when transforming from the natural system back to the standard system, there are additional errors associated with the transformation that are not accounted for owing to differences between the spectral energy distributions of SNe and the reference stars \citep[e.g.,][]{stritzinger05}.

\begin{landscape}
\begin{deluxetable}{lllllllrrrrc}
 \tabcolsep 0.4mm
 \tablewidth{0pt}
 \tablecaption{SESN sample}
  \tablehead{\colhead{SN} & \colhead{Type} & \colhead{RA (h,m,s)} & \colhead{Dec.($^\circ,',''$} & \colhead{Discovery Date} & \colhead{Host galaxy} & \colhead{$A_V$} & \colhead{Distance} & \colhead{Error} & \colhead{$V_{{\rm helio}}^{1}$ (host)} & \colhead{Error} & \colhead{Subtraction$^2$?}\\
             \colhead{} & \colhead{} & \colhead{(J2000)} & \colhead{(J2000)} & \colhead{(UT)} & \colhead{} & \colhead{(MW)} & \colhead{(Mpc)} & \colhead{(Mpc)} & \colhead{(km\,s$^{-1}$)} & \colhead{(km\,s$^{-1}$)} & \colhead{}
}
\startdata
\hline
                 2003gk &        Ib &  23:01:42.989 &     +02:16:08.69 &  2003-07-01 &                   NGC7460 & 0.24 &  48.2 &    28.0 &  3,192 &   7 &   Y \\
                 2006el &       IIb &   22:47:38.50 &     +39:52:27.59 &  2006-08-25 &                 UGC 12188 & 0.31 &    ---&      ---&  5,115 &  14 &   Y \\
                 2006ep &        Ib &   00:41:24.88 &     +25:29:46.72 &  2006-08-30 &                   NGC 214 & 0.10 &  51.1 &    12.3 &  4,537 &   4 &   N \\
                 2006jc &       Ibn &   09:17:20.78 &     +41:54:32.69 &  2006-10-09 &                  UGC 4904 & 0.06 &    ---&      ---&  1,670 &   4 &   Y \\
                 2006lc &       Ibn &   22:44:24.45 &   $-$00:09:53.89 &  2006-10-21 &                  NGC 7364 & 0.18 &  70.6 &    22.1 &  4,865 &   5 &   Y \\
                  2007C &        Ib &   13:08:49.30 &   $-$06:47:01.00 &  2007-01-07 &                  NGC 4981 & 0.12 &  22.7 &     3.1 &  1,680 &   4 &   Y \\
                  2007D &     Ic-BL &   03:18:38.71 &     +37:36:26.39 &  2007-01-09 &                  UGC 2653 & 0.92 & 103.8 &     5.6 &  6,939 &   5 &   Y \\
                 2007ag &        Ib &   10:01:35.99 &     +21:36:42.01 &  2007-03-07 &                  UGC 5392 & 0.08 & 114.8 &     1.7 &  6,209 &   4 &   Y \\
                 2007cl &        Ic &   17:48:21.19 &     +54:09:05.18 &  2007-05-23 &                  NGC 6479 & 0.12 &    ---&      ---&  6,650 &  43 &   Y \\
                 2007kj &        Ib &   00:01:19.58 &     +13:06:30.60 &  2007-10-02 &                  NGC 7803 & 0.22 &    ---&      ---&  5,366 &   6 &   Y \\
                 2007ru &     Ic-BL &   23:07:23.14 &     +43:35:33.68 &  2007-11-27 &                 UGC 12381 & 0.71 &    ---&      ---&  4,636 &   6 &   N \\
                 2007rw &       IIb &   12:38:03.64 &   $-$02:15:40.10 &  2007-11-29 &                  UGC 7798 & 0.09 &    ---&      ---&  2,568 &   5 &   Y \\
                 2007rz &        Ic &   04:31:10.84 &     +07:37:51.49 &  2007-12-08 &                  NGC 1590 & 0.55 &    ---&      ---&  3,897 &   7 &   Y \\
                 2007uy &    Ib-pec &   09:09:35.35 &     +33:07:08.90 &  2007-12-31 &                  NGC 2770 & 0.06 &  28.7 &     4.2 &  1,947 &   2 &   Y \\
                 2008aq &       IIb &   12:50:30.42 &   $-$10:52:01.42 &  2008-02-27 &             MCG -02-33-20 & 0.12 &  32.0 &     3.0 &  2,390 &   5 &   N \\
                 2008cw &       IIb &   16:32:38.27 &     +41:27:33.19 &  2008-06-01 &  SDSS J163238.15+412730.8 & 0.02 &    ---&      ---&  9,726 &  25 &   Y \\
                 2008dq &        Ic &   16:06:03.11 &     +55:25:37.42 &  2008-06-25 &                 UGC 10214 & 0.03 &    ---&      ---&  9,401 &  15 &   Y \\
                 2008eb &        Ib &   18:11:52.17 &     +14:58:50.59 &  2008-07-07 &                  NGC 6574 & 0.48 &  33.7 &     7.2 &  2,282 &   5 &   Y \\
                 2008ew &        Ic &   16:58:28.92 &     +20:02:38.00 &  2008-08-10 &                   IC 1236 & 0.22 &  38.0 &      ---&  6,030 &   5 &   Y \\
                 2008fi &       IIb &   01:53:23.17 &     +29:21:28.40 &  2008-08-26 &  SDSS J015322.95+292131.2 & 0.17 &    ---&      ---&    --- & --- &   N \\
                 2008fz &  SLSN-IIn &   23:16:16.60 &     +11:42:47.48 &  2008-09-22 &                     Anon. & 0.12 &    ---&      ---&    --- & --- &   N \\
                 2008gj &        Ic &   22:36:28.57 &     +21:37:55.31 &  2008-10-19 &                  NGC 7321 & 0.13 &  92.0 &    10.9 &  7,145 &   5 &   Y \\
                  2009C &       IIb &   23:13:42.84 &     +49:40:47.21 &  2009-01-02 &                 UGC 12433 & 0.79 &    ---&      ---&  6,985 &  32 &   Y \\
                  2009K &       IIb &   04:36:36.77 &   $-$00:08:35.59 &  2009-01-14 &                  NGC 1620 & 0.16 &  40.2 &     4.7 &  3,512 &   1 &   Y \\
                  2009Z &       IIb &   14:01:53.61 &   $-$01:20:30.19 &  2009-02-02 &  SDSS J140153.80-012035.5 & 0.13 &    ---&      ---&  7,534 &   3 &   N \\
                 2009er &    Ib-pec &   15:39:29.84 &     +24:26:05.32 &  2009-05-22 &  SDSS J153930.49+242614.8 & 0.12 &    ---&      ---& 10,492 &  67 &   N \\
                 2009gk &       IIb &   21:44:27.28 &     +14:53:57.30 &  2009-06-23 &                 UGC 11803 & 0.24 &    ---&      ---&  7,946 &  34 &   Y \\
                 2009hy &        Ic &   22:16:27.02 &     +16:28:13.01 &  2009-08-02 &                  NGC 7244 & 0.14 &    ---&      ---&  7,564 &   7 &   Y \\
                 2009jf &        Ib &   23:04:52.98 &     +12:19:59.48 &  2009-09-27 &                  NGC 7479 & 0.31 &  28.3 &     6.1 &  2,381 &   1 &   Y \\
                 2010cn &       IIb &   11:04:06.57 &     +04:49:58.69 &  2010-05-04 &  SDSS J110406.40+044955.5 & 0.13 &    ---&      ---&  7,795 & --- &   Y \\
                 2010gd &        Ic &   17:57:40.98 &     +27:49:48.11 &  2010-07-08 &                 UGC 11064 & 0.16 & 107.1 &    21.6 &  7,043 &  10 &   Y \\
                 2010hy &    SLSN-I &   18:59:32.89 &     +19:24:25.88 &  2010-09-04 &                     Anon. & 1.45 &    ---&      ---&    --- & --- &   N \\
                 2011dh &       IIb &   13:30:05.12 &     +47:10:10.81 &  2011-06-01 &                  NGC 5194 & 0.10 &   7.2 &     2.1 &    463 &   3 &   Y \\
                 2011fu &       IIb &   02:08:21.41 &     +41:29:12.30 &  2011-09-21 &                  UGC 1626 & 0.21 &    ---&      ---&  5,543 &  11 &   Y \\
                 2011gd &        Ib &   16:34:25.67 &     +21:32:28.39 &  2011-08-28 &                  NGC 6186 & 0.13 &    ---&      ---&  2,937 &  29 &   Y \\
                 2012aa &        Ic &   14:52:33.48 &   $-$03:31:54.01 &  2012-01-29 &                     Anon. & 0.28 &    ---&      ---&    --- & --- &   Y \\
                 2012ap &     Ic-BL &   05:00:13.72 &   $-$03:20:51.22 &  2012-02-10 &                  NGC 1729 & 0.14 &  39.3 &     3.3 &  3,632 &   4 &   N \\
                 2012au &        Ib &   12:54:52.18 &   $-$10:14:50.21 &  2012-03-14 &                  NGC 4790 & 0.13 &  22.9 &     2.8 &  1,344 &   5 &   Y \\
                 2012fh &  IIb$/$Ib$/$c &   10:43:34.05 &  +24:53:29.00 &  2012-10-18 &                  NGC 3344 & 0.09 &  11.9 &     6.2 &    580 &   1 &   Y \\
                 2013dk &        Ic &   12:01:52.72 &   $-$18:52:18.30 &  2013-06-22 &                  NGC 4038 & 0.13 &  21.1 &     3.9 &  1,642 &  12 &   Y \\
                  2014C &        Ib &   22:37:05.60 &     +34:24:31.90 &  2014-01-05 &                  NGC 7331 & 0.25 &  13.4 &     2.7 &    816 &   1 &   Y \\
                  2014L &        Ic &   12:18:48.68 &     +14:24:43.49 &  2014-01-26 &                  NGC 4254 & 0.11 &  15.2 &     2.0 &  2,407 &   3 &   Y \\
                 2014as &     Ic-BL &   14:00:54.49 &     +40:58:59.59 &  2014-04-18 &                  NGC 5410 & 0.04 &    ---&      ---&  3,738 &  26 &   Y \\
                 2014cp &     Ic-BL &   02:25:30.46 &   $-$25:37:37.99 &  2014-06-23 &            ESO 479- G 001 & 0.05 &  45.1 &     7.2 &  4,846 &   3 &   N \\
                 2014ds &       IIb &   08:11:16.45 &     +25:10:47.39 &  2014-10-11 &                  NGC 2536 & 0.12 &    ---&      ---&  4,118 &  17 &   Y \\
                 2014eh &        Ic &   20:25:03.86 &   $-$24:49:13.30 &  2014-11-03 &                  NGC 6907 & 0.17 &  32.5 &     6.2 &  3,182 &   4 &   N \\
                 2014ei &        Ib &   05:03:16.39 &   $-$02:56:11.00 &  2014-11-05 &             MCG -01-13-50 & 0.18 &  57.9 &     3.5 &  4,329 &   4 &   Y \\
                  2015G &       Ibn &   20:37:25.58 &     +66:07:11.50 &  2015-03-23 &                  NGC 6951 & 1.02 &  23.1 &     3.5 &  1,424 &   1 &   Y \\
                  2015K &        Ic &   23:35:52.26 &     +23:36:52.09 &  2015-04-25 &                  NGC 7712 & 0.15 &  49.0 &     3.2 &  3,053 &   2 &   N \\
                  2015Q &        Ib &  11:47:35.081 &     +55:58:14.70 &  2015-06-17 &                  NGC 3888 & 0.03 &  39.9 &     1.7 &  2,408 &  11 &   Y \\
                  2015U &       Ibn &   07:28:53.87 &     +33:49:10.60 &  2015-02-13 &                  NGC 2388 & 0.16 &  60.9 &     2.3 &  4,134 &   5 &   Y \\
                  2015Y &        Ib &   09:02:37.87 &     +25:56:04.20 &  2015-04-11 &                  NGC 2735 & 0.11 &  51.1 &     9.0 &  2,450 &   5 &   Y \\
                 2015ap &        Ib &   02:05:13.32 &     +06:06:08.39 &  2015-09-08 &                    IC1776 & 0.12 &    ---&      ---&  3,410 &   5 &   Y \\
                  2016G &     Ic-BL &   03:03:57.74 &     +43:24:03.50 &  2016-01-09 &                  NGC 1171 & 0.43 &  26.6 &     6.2 &  2,742 &   6 &   Y \\
                  2016P &     Ic-BL &   13:57:31.10 &     +06:05:51.00 &  2016-01-19 &                  NGC 5374 & 0.07 &  68.7 &    10.8 &  4,382 &   7 &   Y \\
                2016ajo &        Ib &   18:44:12.49 &     +24:09:29.70 &  2016-02-20 &                 UGC 11344 & 0.34 &  58.5 &     4.2 &  3,836 &   4 &   Y \\
                2016bau &        Ib &   11:20:59.02 &     +53:10:25.60 &  2016-03-13 &                  NGC 3631 & 0.04 &  10.3 &     5.4 &  1,156 &   1 &   N \\
                2016coi &     Ic-BL &   21:59:04.14 &     +18:11:10.46 &  2016-05-27 &                 UGC 11868 & 0.23 &  17.2 &      ---&  1,093 &   5 &   Y \\
                2016gcm &        Ic &   21:04:55.22 &     +65:42:29.30 &  2016-09-08 &                 PGC166705 & 1.33 &    ---&      ---&  7,263 &  50 &   Y \\
                2016gkg &       IIb &   01:34:14.46 &   $-$29:26:25.00 &  2016-09-20 &                   NGC 613 & 0.05 &  20.9 &     5.7 &  1,481 &   5 &   N \\
                2016gqv &    Ic-pec &   04:02:48.53 &     +01:58:15.60 &  2016-09-28 &                 UGC 02936 & 1.23 &  42.8 &     5.1 &  3,813 &   7 &   Y \\
                2016iyc &       IIb &   22:09:14.28 &     +21:31:17.51 &  2016-12-18 &                 UGC 11924 & 0.21 &    ---&      ---&  3,803 &   5 &   Y \\
                2017ein &        Ic &   11:52:53.25 &     +44:07:26.20 &  2017-05-25 &                   NGC3938 & 0.06 &  12.7 &     7.8 &    809 &   4 &   N \\
                2017iro &        Ib &   14:06:23.11 &     +50:43:20.20 &  2017-11-30 &                  NGC 5480 & 0.05 &  24.2 &     6.0 &  1,856 &   5 &   Y \\
                2018cow &     Ic-BL &   16:16:00.22 &     +22:16:04.83 &  2018-06-16 &              CGCG 137-068 & 0.24 &    ---&      ---&  4,241 &  39 &   N \\
                 2018ie &     Ic-BL &   10:54:01.06 &   $-$16:01:21.40 &  2018-01-18 &                  NGC 3456 & 0.19 &  47.5 &     8.6 &  4,267 &   7 &   N \\
                2019wep &       Ibn &  11:04:37.033 &     +45:58:38.95 &  2019-12-07 &                 UGC 06136 & 0.03 &    ---&      ---&  7,521 &  11 &   N \\
                2020nxt &       Ibn &  22:37:36.235 &     +35:00:07.68 &  2020-07-03 &  SDSS J223736.60+350007.4 & 0.21 &    ---&      ---&    --- & --- &   Y \\
 MOTJ120451.50+265946.6 &        Ib &   12:04:51.50 &     +26:59:46.60 &  2014-10-28 &                  NGC 4808 & 0.07 &  18.4 &     2.0 &    567 &   4 &   N \\
              iPTF13bvn &        Ib &  15:00:00.152 &     +01:52:53.17 &  2013-06-16 &                   NGC5806 & 0.14 &  24.7 &     3.2 &  1,359 &   5 &   Y \\
\enddata
\tablenotetext{1}{For the five SNe without host heliocentric velocity measurements, redshifts are adopted as follows:
$z_{\rm SN~2008fi} = 0.02600$ \citep{shivvers19},
$z_{\rm SN~2008fz} = 0.133$   \citep{drake10},
$z_{\rm SN~2010hy} = 0.19010$ \citep{shivvers19},
$z_{\rm SN~2012aa} = 0.07990$ \citep{shivvers19}, and
$z_{\rm SN~2020nxt} \approx 0.02$ \citep{srivastav20}.}
\tablenotetext{2}{Image subtraction is applied to remove host-galaxy contamination; see text for details.}
\label{SESNsamplebasicinformation}
\end{deluxetable}
\end{landscape}

\newpage
\begin{figure*}
        \includegraphics[width=2.0\columnwidth]{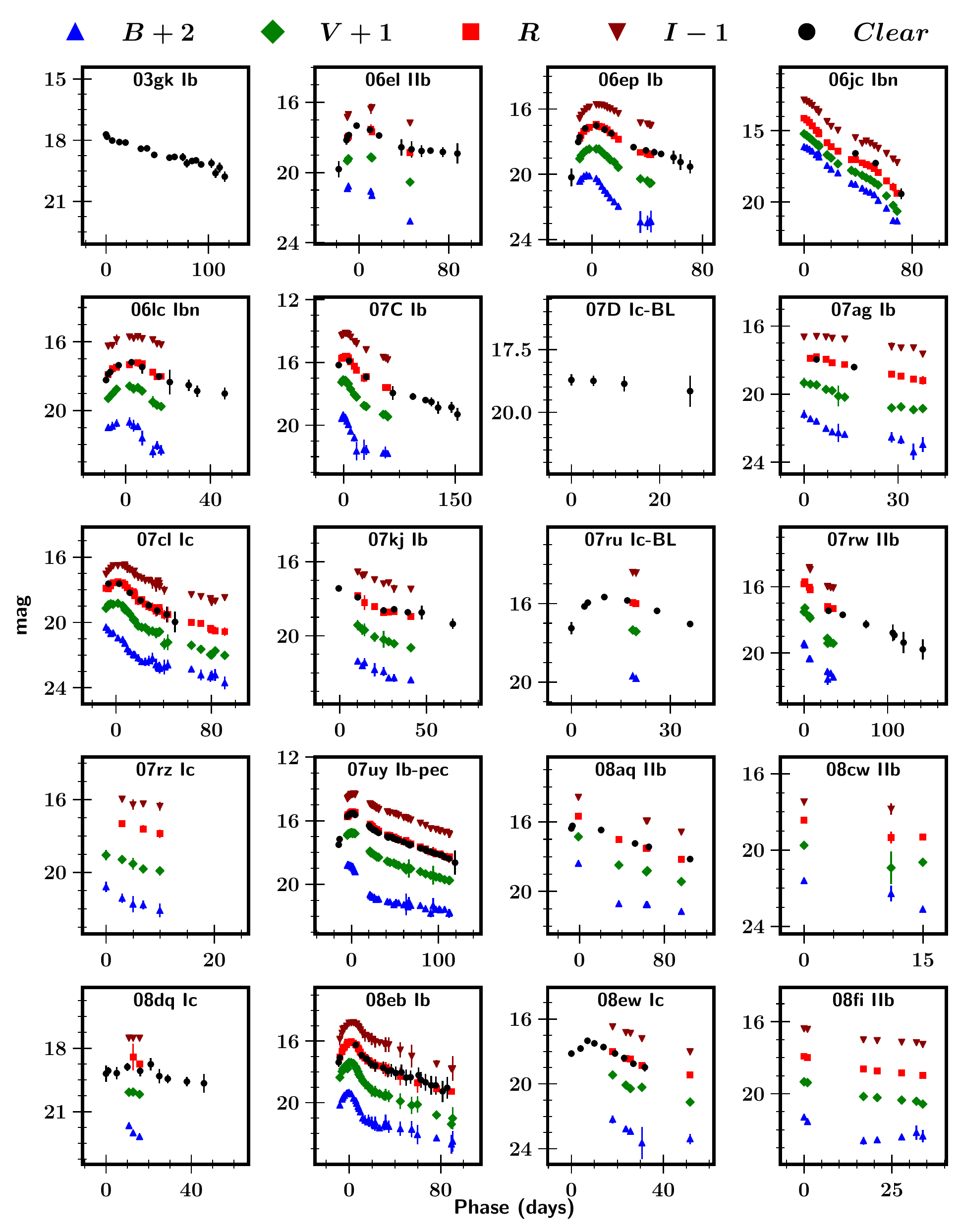}
        \caption{Apparent-magnitude light curves of the SESNe in our sample in the standard system (except for SN~2016P, which has data only in the natural system
                 as shown in Fig. \ref{lc_natural}), without any corrections for extinction.
                 Blue triangles are magnitudes in $B$, green diamonds are $V$, red squares are $R$,
                 black circles are {\it Clear}, and dark-red inverted triangles are $I$.
                 All dates have been shifted relative to the time of maximum $V$-band brightness
                 if determined, and relative to the time of the first epoch otherwise.
                 In each panel, the IAU name and the type are given.
                 }
\label{lc_standard}
\end{figure*}
\begin{figure*}
\ContinuedFloat
        \includegraphics[width=2.0\columnwidth]{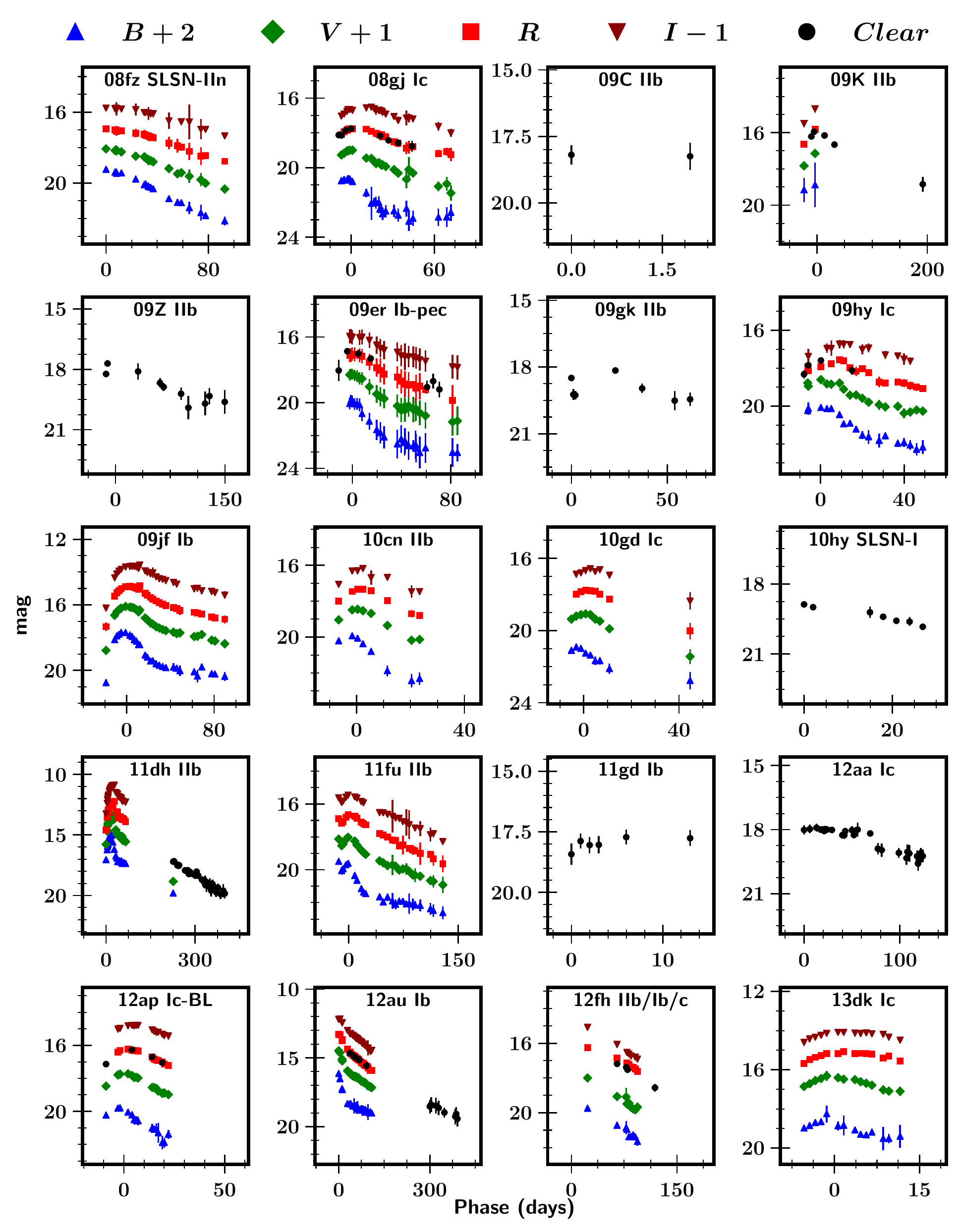}
        \caption{Continued.}
\end{figure*}
\begin{figure*}
\ContinuedFloat
        \includegraphics[width=2.0\columnwidth]{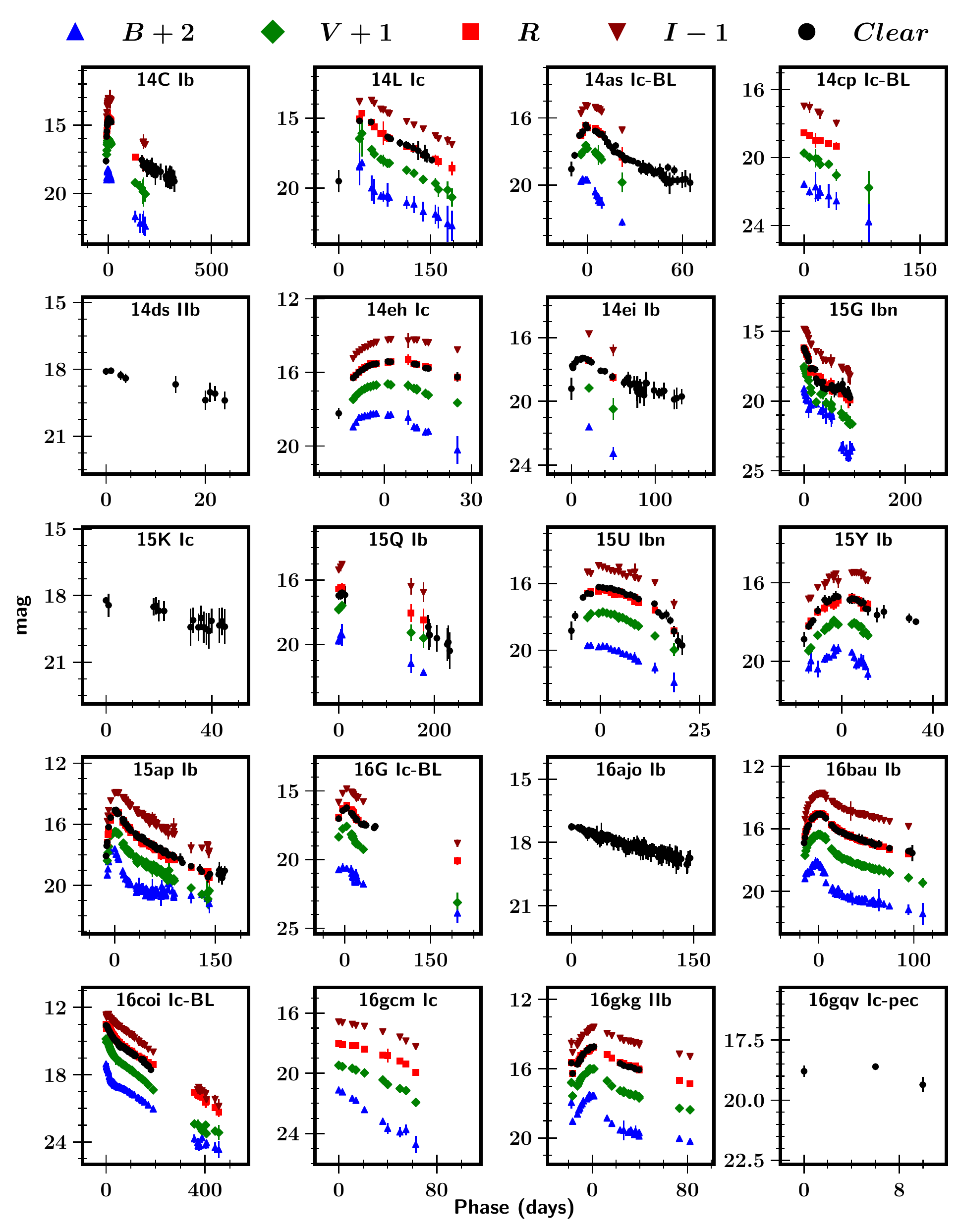}
        \caption{Continued.}
\end{figure*}
\begin{figure*}
\ContinuedFloat
        \includegraphics[width=2.0\columnwidth]{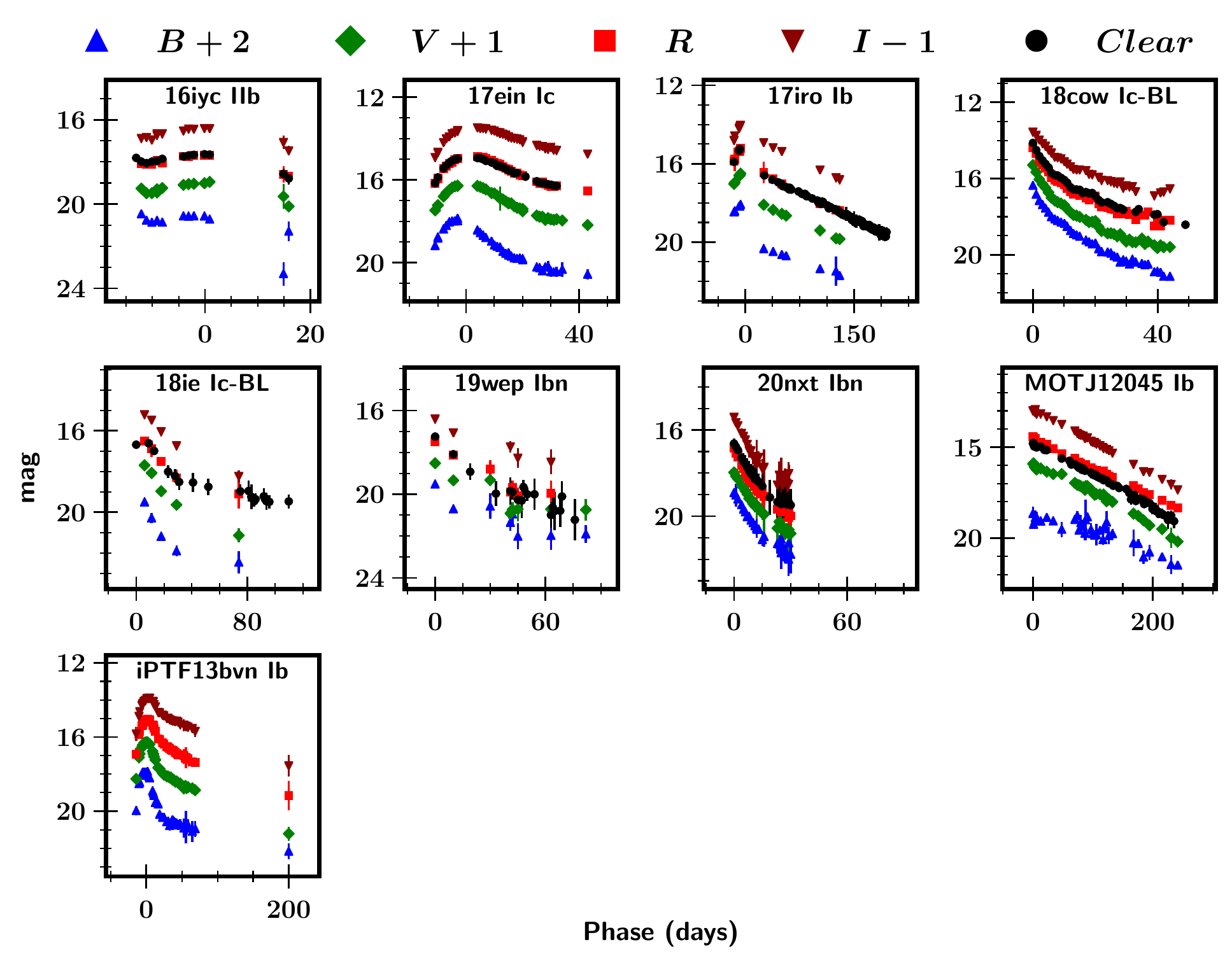}
        \caption{Continued.}
\end{figure*}

\section{Results}
\label{sec:results}

\subsection{Photometry data release}

We provide our final photometry in Tables \ref{lightcurvedatainstandardsystem} (standard system) and \ref{lightcurvedatainnaturalsystem}
(natural system) for all 70 SESNe in our sample.
On average, each SN has 81 observations (median of 58)
%with a standard deviation of 81 points,
and 22 photometric points (median of 13 points) per filter, 
at a cadence of 3.6\,d (median of 1.2\,d). SN~2016coi has the best coverage (434 points), followed by SN~2015ap (351).
The total number of distinct photometric observations published in this work is 5682.
Of the 70 SNe in our sample, 38 have pre-maximum coverage and thus have their peak brightness measured in at least one band.
Note that although several SNe in our sample have had their photometry previously published in individual papers, the magnitudes presented
herein supersede these earlier measurements because (i) for some fields requiring image subtraction we have obtained new, higher-quality templates, 
and (ii) better calibration sources are now available that were not used previously.
Moreover, our processing is now significantly more systematic and self-consistent owing to our use of the battle-tested
\texttt{LOSSPhotPypeline} \citep{stahl19,dejaeger19,stahl20}.

Note that in this release, we do not include the systematic uncertainty of 0.03\,mag
in $BVRI$ that was determined and discussed in detail by \cite{stahl19}.
This amount of systematic uncertainty was estimated by investigating many factors that may contribute to the error, including evolution of colour terms, evolution of atmospheric terms, configurations between different telescopes, and galaxy-subtraction procedures. None of these factors contributed uncertainty over 0.03~mag, consistent with the estimate of \cite{ganeshalingam10}.
Though not included in our photometry tables (Table~\ref{lightcurvedatainstandardsystem} and \ref{lightcurvedatainnaturalsystem})
or light-curve figures (e.g., Figure~\ref{lc_standard} and Figure~\ref{lc_natural}), 
this uncertainty must be accounted for when combining our dataset with others.
Alongside the recent LOSS photometry release for SNe~Ia \citep{stahl19} and SNe~II \citep{dejaeger19},
we aim for our SESN photometry to be used and further analysed by the astronomical community.

\subsection{Light curves}
\label{LightCurves}

Figure~\ref{lc_standard} shows the apparent-magnitude light curves of all SESNe from our sample in the standard Landolt system without any extinction
corrections applied. Note that we also include {\it Clear}-band light curves where available. Although unfiltered and thus nonstandard, it is 
most similar to the $R$ band \citep{li03}. The temporal axes are all in the observer frame and shifted such that times are measured
relative to the times of maximum $V$-band brightness as determined by fitting the near-maximum data with low-order Legendre polynomials.
In the 37 cases where no maximum could be found via this method, the temporal axes are shifted relative to the time of the first observation.
Such fitting was also applied to other bands if the data --- after being supplemented with corresponding observations from
\cite{drout11}, \cite{bianco14}, or \cite{stritzinger18a} --- had sufficient near-maximum coverage.

\subsection{Colour evolution and host extinction}

\begin{figure*}
        \includegraphics[width=1.07\columnwidth]{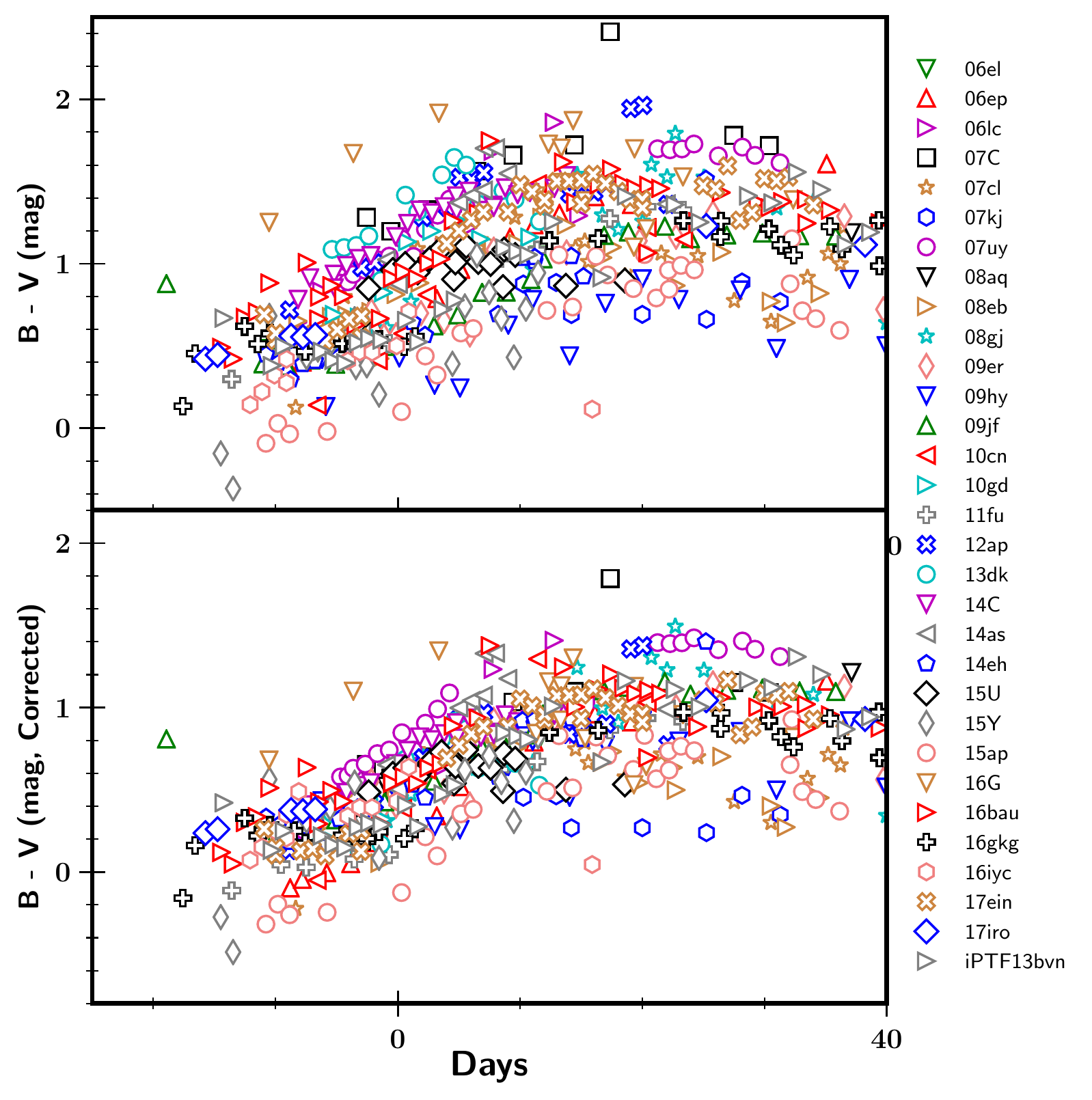}
        \includegraphics[width=0.93\columnwidth]{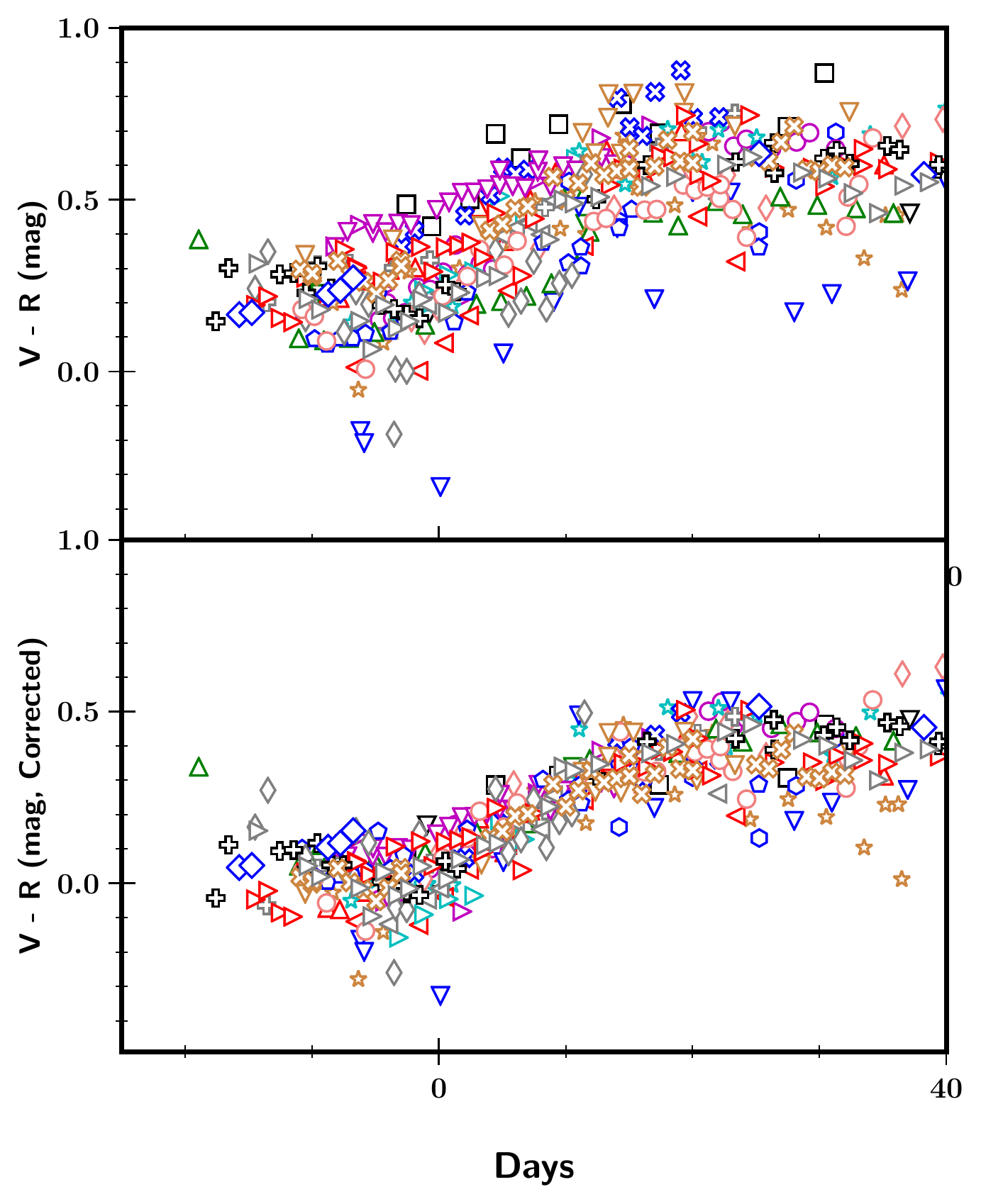}
        \caption{Left: ($B - V$) colour evolution of the SESNe in our sample having simultaneous
observations in both bands, after correction for Galactic extinction (top), and with further
correction for host extinction (bottom) using the empirical colour method (see text for more details).
Right: similar to the left panel, but for the ($V - R$) colour.
}
\label{ColorEvolution}
\end{figure*}

Figure~\ref{ColorEvolution} shows the ($B - V$) (top left) and ($V - R$) (top right) colour evolution of the SESNe in our sample which
have the requisite observations in both bands, after correcting for Milky Way (MW) Galactic extinction \citep{schlafly11} but not host-galaxy extinction. Individually, we find that
the colours trend blue at very early times (e.g., $\sim 10$\,d before $V$-band peak), and then become progressively more red until
$\sim 20$\,d after $V$-band maximum.
As a whole, however, the colours we observe span a wide range at nearly every epoch, in part because we have not (yet) performed
any corrections for host extinction. As SESNe often reside in dusty star-forming regions \citep[e.g.,][]{vandyk96,kelly08},
it is likely that the extinction due to the galactic hosts of our SNe is generally the dominant component of the total line-of-sight extinction.

One useful way to estimate host extinction is to exploit its relationship with the equivalent width of Na~I~D absorption as measured from high-resolution spectra \citep[e.g.,][]{poznanski12,stritzinger18b}. Unfortunately, such data are difficult to obtain, and low-resolution spectra, though easier to procure, are usually not of sufficient quality for such measurements \citep{poznanski11}.

\begin{figure}
        \includegraphics[width=1.0\columnwidth]{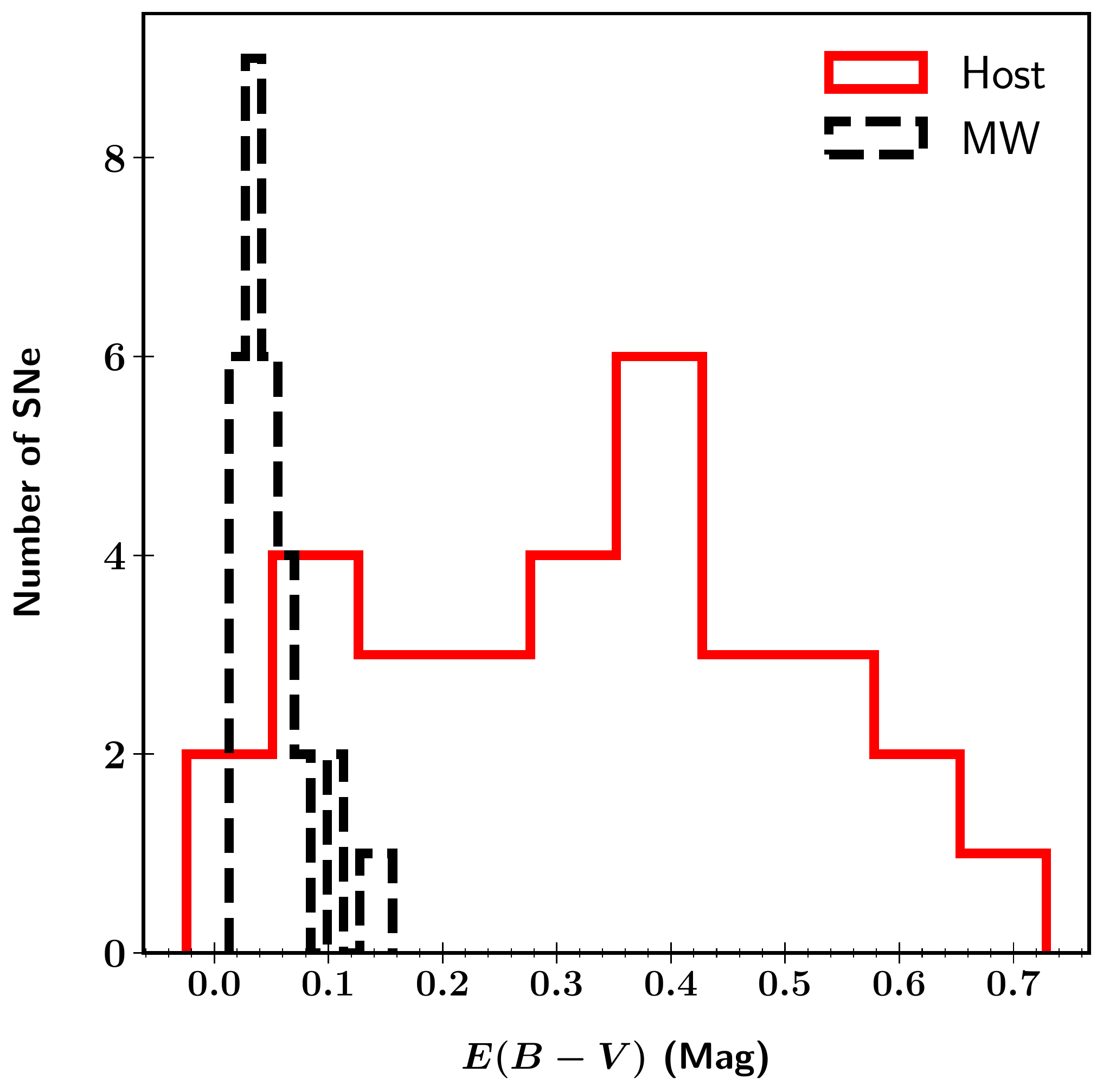}
        \caption{Distribution of host-galaxy extinction values for the 31 SESNe in our sample (solid red), revealing a mean $E(B - V)$ value of
                 0.32\,mag with a standard deviation of 0.19\,mag.
                 %--- significantly higher than the MW extinction from the same sample (dashed black)
                 %with a mean $E(B - V)$ value of\,0.05 mag and a standard deviation of\,0.03 mag.
                 The corresponding MW extinction is shown in dashed black.
                 }
\label{HostExtinction}
\end{figure}

Distinct from spectral proxies using Na~I~D absorption, \cite{drout11} found that the ($V - R$) colour evolution of SESNe has a small dispersion
at $\sim 10$\,d after $V$-band peak, and thus can be exploited as a useful diagnostic for estimating host-galaxy
extinction \citep{drout11,taddia15,stritzinger18b}. We therefore follow the approach of \cite{drout11} to estimate the
extinction induced by the host galaxies of the SNe in our sample. To do this, we measure the ($V - R$) colour at 10\,d after $V$-band maximum,
and apply extra extinction (assumed to be from the host) so that the ($V - R$) colour reaches 0.26\,mag --- the mean ($V - R$) colour 
found by \cite{drout11} after all corrections.
We successfully applied this method to 31 SESNe in our sample with available data in both $V$ and $R$ at 10\,d after $V$-band maximum.
We note that a few SESNe in our measured sample have negative implied host-extinction values
by doing this, but such values are sufficiently small to be consistent with no host extinction.

Figure~\ref{HostExtinction} shows a histogram of the estimated $E(B - V)$ values for the 31 SESNe in our sample (also listed in Tab.~\ref{hostextinctionandpeakmag})
assuming a \cite{cardelli89} reddening law with $R_V = 3.1$. It clearly shows that
SESNe usually suffer moderately-high extinction from their host galaxies, with a mean $E(B - V)$ value of 0.32\,mag and standard deviation of 0.19\,mag ---
consistent with the value found by other works using different SESN samples \citep{drout11,taddia15,stritzinger18b}.
This mean host extinction is 
%significantly higher (i.e., by a factor of 6) than the Galactic (MW) extinction (see also \citealt{prentice16}) from the same sample
%($E(B - V) \approx 0.05$\,mag with a standard deviation of 0.03\,mag; shown in the same figure), and also 
higher than recently studied samples of
SNe~Ia with mean $E(B - V) \approx 0.11$\,mag \citep{burns11,stahl19} and SNe~II with mean $A_V \approx 0.9$\,mag ($E(B - V) \approx 0.29$; \citealt{smartt09a}).

After correcting for host extinction as described above, we plot the extinction-corrected ($B - V$) (bottom left) and ($V - R$) (bottom right) colour curves
in Figure~\ref{ColorEvolution}. The scatter in ($V - R$) colours is significantly decreased at 10\,d after $V$-band maximum,
consistent with the results of \citet{drout11}, \citet{taddia15}, and \citet{stritzinger18b}.
However, although its scatter is modestly reduced, the ($B - V$) colour curve is by no means ``tight'' like we observe for ($V - R$).
This is not surprising because the host extinction is estimated from the ($V - R$) colour, not the ($B - V$) colour as proposed by \cite{drout11}.
This indicates that the ($B - V$) colour may not be as good a proxy as the ($V - R$) colour, or that the host extinction is more complicated,
e.g., the hosts of SESNe are known to be with a range of $R_V$ \citep[e.g.,][]{stritzinger18b}, while adopting a constant $R_V = 3.1$ to derive $E(V-R)$ would lead to errors in $E(B-V)$ and more scatter.
In either case, a better method for estimating the host-galaxy extinction for SESNe would be a valuable contribution to the field.

\subsection{Absolute light curve and peak magnitude}

\begin{landscape}
\begin{deluxetable}{llcccccccccccc}
 \tabcolsep 0.4mm
 \tablewidth{0pt}
 \tablecaption{Light-Curve Properties}
%  \tablehead{\colhead{SN} & \colhead{type} & \colhead{$z$} & \colhead{$E(B - V)$ (host)} & \colhead{$M_{\rm max}(R)$} &\colhead{$m_{\rm max}(U)$} & \colhead{$m_{\rm max}(B)$} & \colhead{$m_{\rm max}(g)$} & \colhead{$m_{\rm max}(V)$} & \colhead{${m_{\rm max}(R)$}$} & \colhead{$m_{\rm max}(r)$} & \colhead{$m_{\rm max}(clear)$} & \colhead{$m_{\rm max}(I)$} & \colhead{$m_{\rm max}(i)$}}
  \tablehead{\colhead{SN} & \colhead{Type} & \colhead{$z$} & \colhead{$E(B - V)_{{\rm host}}$} & \colhead{$M_{{\rm max}}(R)$} & \colhead{$m_{{\rm max}}(U)$} & \colhead{$m_{{\rm max}}(B)$} & \colhead{$m_{{\rm max}}(g)$} & \colhead{$m_{{\rm max}}(V)$} & \colhead{$m_{{\rm max}}(R)$} & \colhead{$m_{{\rm max}}(r)$} & \colhead{$m_{{\rm max}}({\rm clear})$} & \colhead{$m_{{\rm max}}(I)$} & \colhead{$m_{{\rm max}}(i)$} }
\startdata
%\hline
    2006el &       IIb &  0.017062 &     0.25 &     $-$17.88 &   --- & 18.28 &   --- & 17.60 & 17.33 &   --- &   --- &   --- & 17.44 \\
    2006ep &        Ib &  0.015134 &     0.45 &     $-$17.73 & 18.57 & 17.97 & 17.65 & 17.39 & 16.98 & 17.19 &   --- & 16.72 & 17.05 \\
    2006lc &       Ibn &  0.016228 &     0.45 &     $-$18.26 & 19.79 & 18.51 & 18.09 & 17.67 & 17.23 & 17.32 &   --- & 16.71 & 17.22 \\
     2007C &        Ib &  0.005604 &     0.62 &     $-$17.82 &   --- & 17.28 &   --- & 16.13 & 15.58 & 15.76 &   --- & 15.13 & 15.50 \\
    2007cl &        Ic &  0.022182 &     0.35 &     $-$18.27 &   --- &   --- &   --- & 17.85 & 17.60 & 17.48 &   --- & 17.42 & 17.69 \\
    2007kj &        Ib &  0.017899 &     0.42 &          --- & 18.30 & 17.92 & 17.74 & 17.66 &   --- & 17.56 &   --- &   --- & 17.62 \\
    2007ru &     Ic-BL &  0.015464 &      --- &          --- &   --- &   --- &   --- &   --- &   --- &   --- & 15.63 &   --- &   --- \\
    2007uy &    Ib-pec &  0.006494 &     0.30 &     $-$17.62 &   --- & 16.72 &   --- & 15.76 & 15.46 & 15.57 &   --- & 15.30 & 15.63 \\
    2008aq &       IIb &  0.007972 &  $-$0.02 &          --- &   --- & 16.40 &   --- & 15.84 &   --- &   --- &   --- &   --- & 15.95 \\
    2008eb &        Ib &  0.007612 &     0.37 &     $-$17.89 &   --- & 17.36 &   --- & 16.39 & 16.03 &   --- &   --- & 15.76 &   --- \\
    2008ew &        Ic &  0.020114 &      --- &          --- &   --- &   --- &   --- &   --- &   --- &   --- & 17.51 &   --- &   --- \\
    2008gj &        Ic &  0.023833 &     0.30 &     $-$17.94 &   --- & 18.64 &   --- & 18.01 & 17.71 &   --- &   --- & 17.47 &   --- \\
     2009K &       IIb &  0.011715 &      --- &          --- & 17.46 & 16.64 & 16.36 & 16.08 &   --- & 15.83 &   --- &   --- &   --- \\
     2009Z &       IIb &  0.025131 &      --- &          --- & 18.36 & 17.75 & 17.48 & 17.25 &   --- & 17.16 &   --- &   --- & 17.16 \\
    2009er &    Ib-pec &  0.034998 &     0.16 &     $-$19.38 &   --- & 17.92 &   --- & 17.24 & 17.05 & 17.17 &   --- &   --- & 17.25 \\
    2009gk &       IIb &  0.026505 &      --- &          --- &   --- &   --- &   --- &   --- &   --- &   --- & 18.05 &   --- &   --- \\
    2009hy &        Ic &  0.025231 &  $-$0.01 &     $-$17.63 &   --- & 18.02 &   --- & 17.68 & 17.66 &   --- &   --- & 17.77 &   --- \\
    2009jf &        Ib &  0.007942 &     0.07 &     $-$17.81 &   --- & 15.66 &   --- & 15.07 & 14.86 & 14.99 &   --- & 14.63 & 15.00 \\
    2010cn &       IIb &  0.026001 &     0.19 &     $-$18.51 &   --- & 17.90 &   --- & 17.46 & 17.33 &   --- &   --- & 17.19 &   --- \\
    2010gd &        Ic &  0.023493 &     0.50 &     $-$18.75 &   --- & 18.91 &   --- & 18.09 & 17.76 &   --- &   --- & 17.62 &   --- \\
    2011fu &       IIb &  0.018489 &     0.41 &     $-$19.02 &   --- & 17.66 &   --- & 17.06 & 16.67 &   --- &   --- & 16.49 &   --- \\
    2012ap &     Ic-BL &  0.012115 &     0.59 &     $-$18.29 &   --- & 17.79 &   --- & 16.71 & 16.23 &   --- &   --- & 15.78 &   --- \\
    2013dk &        Ic &  0.005477 &     0.73 &     $-$18.38 &   --- & 16.65 &   --- & 15.37 & 15.12 &   --- &   --- & 15.07 &   --- \\
     2014C &        Ib &  0.002722 &     0.50 &     $-$17.55 &   --- & 16.25 &   --- & 15.05 & 14.52 &   --- & 14.57 & 14.07 &   --- \\
     2014L &        Ic &  0.008029 &      --- &          --- &   --- &   --- &   --- &   --- &   --- &   --- &   --- &   --- &   --- \\
    2014as &     Ic-BL &  0.012469 &     0.37 &     $-$18.09 &   --- & 17.63 &   --- & 16.78 & 16.51 &   --- & 16.47 & 16.26 &   --- \\
    2014eh &        Ic &  0.010614 &     0.11 &     $-$17.57 &   --- & 16.23 &   --- & 15.63 & 15.40 &   --- & 15.41 & 15.17 &   --- \\
    2014ei &        Ib &  0.014440 &      --- &          --- &   --- &   --- &   --- &   --- &   --- &   --- & 17.30 &   --- &   --- \\
     2015U &       Ibn &  0.013790 &     0.36 &     $-$18.51 &   --- & 17.73 &   --- & 16.74 & 16.43 &   --- & 16.23 & 16.01 &   --- \\
     2015Y &        Ib &  0.008172 &     0.12 &     $-$17.14 &   --- & 17.27 &   --- & 16.84 & 16.78 &   --- & 16.66 & 16.41 &   --- \\
    2015ap &        Ib &  0.011375 &     0.22 &     $-$18.92 &   --- & 15.53 &   --- & 15.45 & 15.18 &   --- & 15.09 & 14.89 &   --- \\
     2016G &     Ic-BL &  0.009146 &     0.57 &     $-$17.79 &   --- & 18.58 &   --- & 16.56 & 16.07 &   --- & 16.26 & 15.86 &   --- \\
   2016bau &        Ib &  0.003856 &     0.37 &     $-$15.99 &   --- & 16.20 &   --- & 15.36 & 15.02 &   --- & 15.05 & 14.75 &   --- \\
   2016gkg &       IIb &  0.004940 &     0.29 &     $-$17.56 &   --- & 15.58 &   --- & 14.99 & 14.79 &   --- & 14.78 & 14.68 &   --- \\
   2016iyc &       IIb &  0.012685 &     0.07 &     $-$16.36 &   --- & 18.55 &   --- & 17.99 & 17.68 &   --- & 17.65 & 17.41 &   --- \\
   2017ein &        Ic &  0.002699 &     0.43 &     $-$16.79 &   --- & 15.91 &   --- & 15.19 & 14.82 &   --- & 14.87 & 14.45 &   --- \\
   2017iro &        Ib &  0.006191 &     0.18 &     $-$17.54 &   --- & 16.07 &   --- & 15.39 & 14.87 &   --- & 14.98 & 14.79 &   --- \\
 iPTF13bvn &        Ib &  0.004533 &     0.25 &     $-$17.65 &   --- & 15.86 &   --- & 15.26 & 15.03 &   --- &   --- & 14.90 &   ---
\enddata
%\tablenotetext{a}{In units of \kms\,d$^{-1}$.}
\label{hostextinctionandpeakmag}
\end{deluxetable}
\end{landscape}

\begin{figure*}
        \includegraphics[width=1.9\columnwidth]{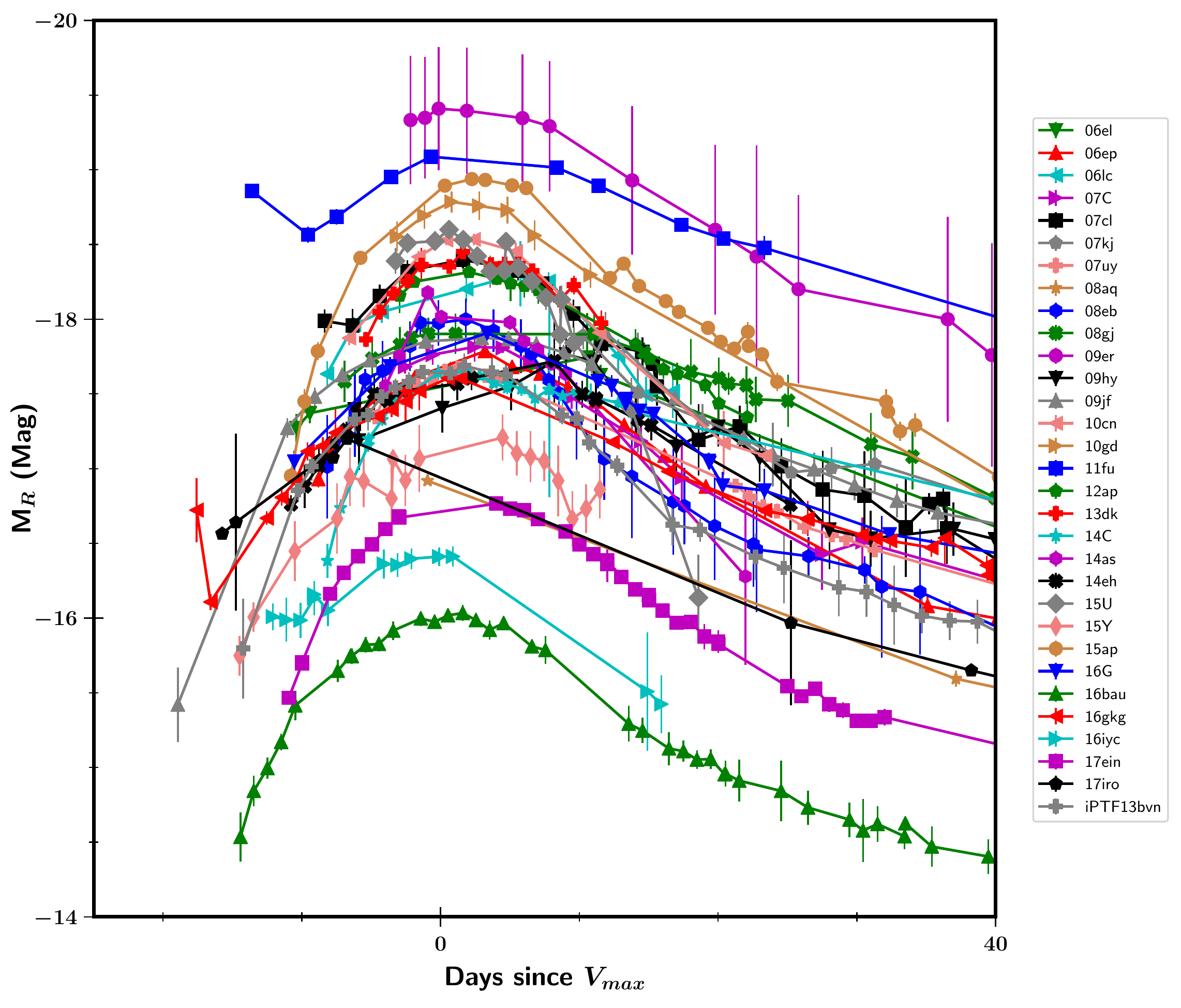}
        \caption{Absolute $R$-band light curves of the 31 SESNe in our sample that can be fully extinction corrected,
          only photometric error, not distance or host-extinction error, was included in this plot.
SESNe show smooth light-curve shapes, with $\sim 10$--20\,d rise times, followed by a slow post-maximum decay.
A few Type IIb SESNe show a decline dip at very early times, which can be attributed to the shock-breakout cooling tail.
}
\label{AbsoluteRBandLC}
\end{figure*}

Figure~\ref{AbsoluteRBandLC} illustrates the absolute $R$-band light curves of the 31 SESNe in our sample for which we are able 
to fully correct for extinction (i.e., correct for both MW and host-galaxy effects).
Overall, the SESNe show smooth light-curve shapes with $\sim 10$--20\,d rise times before maximum brightness
(see Sec.~\ref{sec:firstlighttimeandrisetime} for more details),
followed by a slow decay. A few Type IIb SESNe (e.g., SNe~2011fu, 2016iyc, 2016gkg) exhibit a decline dip at very early times
($\sim 15$\,d before maximum) before rising. The early-time dips for these SNe~IIb can
be attributed to the shock-breakout cooling tail.

\begin{figure}
        \includegraphics[width=1.0\columnwidth]{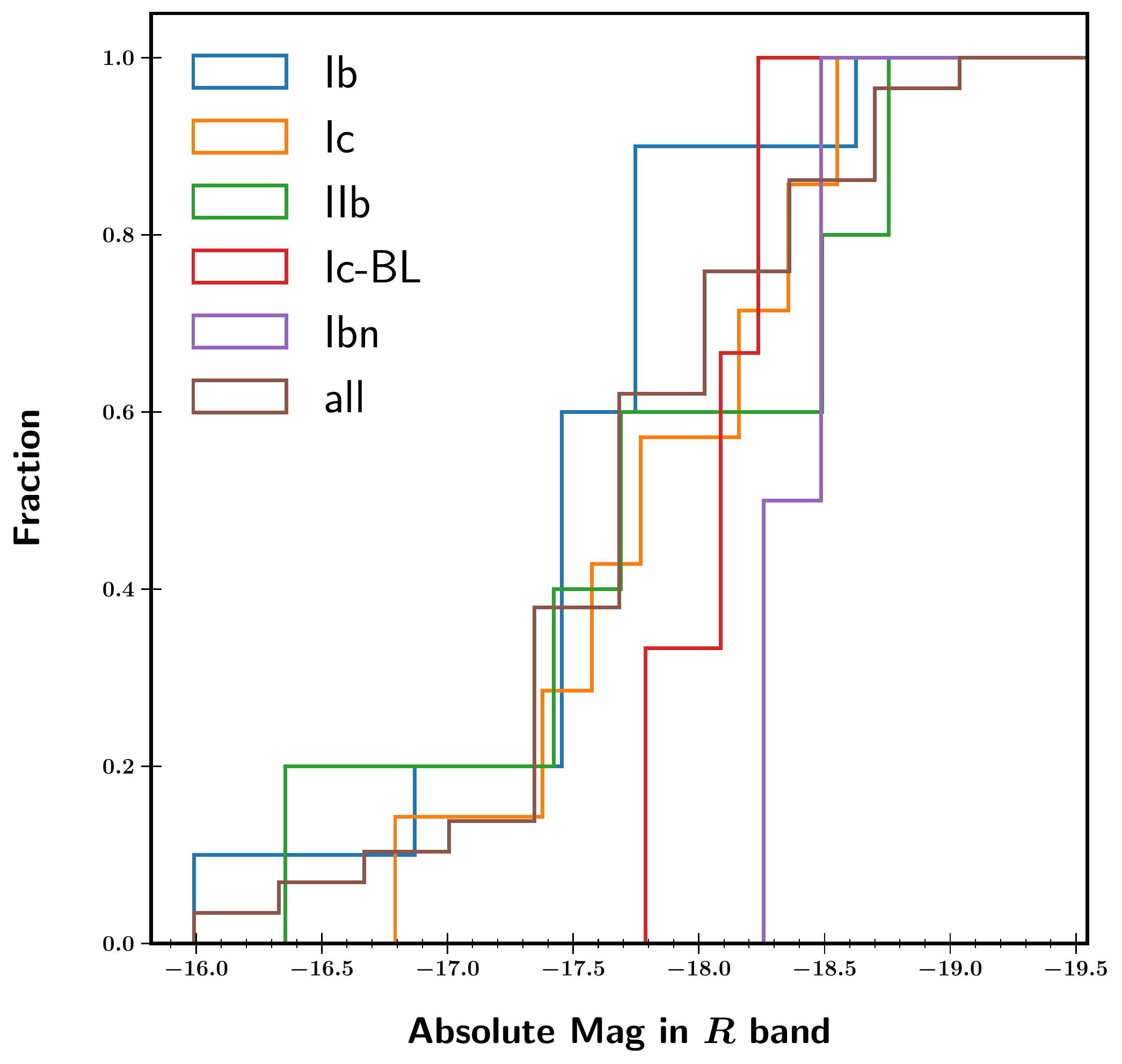}
        \caption{Cumulative distribution of the absolute $R$-band peak magnitudes of the 31 SESNe in our extinction-corrected sample.
SESNe show a wide range from $-16$\,mag to brighter than $-19$\,mag. SNe~Ic-BL and SNe~Ibn appear on average to be more luminous than
both SNe~Ib and SNe~Ic, and SNe~Ic seem to be slightly brighter than SNe~Ib.}
\label{HistRBandAbsoluteMag}
\end{figure}

The peak $R$-band absolute magnitude of SESNe spans a wide range from $-16$\,mag to over $-19$\,mag. In order to compare the peak magnitudes
between different subgroups, we plot the cumulative distribution of $R$-band absolute magnitude (after extinction correction) in
Figure~\ref{HistRBandAbsoluteMag}.
We find that our whole sample (29 SNe with both $R$ peak and host-extinction measurements) has an average mean peak $R$ absolute magnitude of $-17.9 \pm 0.7$\,mag.
We also calculate the mean peak brightness for each subgroup and find a value for SNe~Ib of $-17.6 \pm 0.7$\,mag (ten SNe),
$-17.9 \pm 0.6$\,mag for seven SNe~Ic, and $-17.9 \pm 1.0$\,mag for five SNe~IIb.
With admittedly small samples of three SNe~Ic-BL and two SNe~Ibn, we found $-18.1 \pm 0.3$\,mag and $-18.4 \pm 0.2$\,mag, respectively ---
brighter than the other subgroups (SNe~Ib, Ic, IIb) and consistent with the conclusion reported by \cite{taddia15} that SNe~Ic-BL are more
luminous than both SNe~Ib and SNe~Ic, and that SNe~Ic appear slightly brighter than SNe~Ib.
%Similar conclusion was also reported by \cite{taddia18} showing that SNe Ic appears slightly brighter than SNe Ib.
Our reported mean magnitudes are consistent with the results from \cite{drout11}, who found $-17.9 \pm 0.9$\,mag for SNe~Ib and
$-18.5 \pm 0.8$\,mag for SNe~Ic in the $R$ band. \cite{taddia18} also reported similar trends, but in the $r$ band, with 
$-17.22 \pm 0.60$\,mag, $-17.66 \pm 0.21$\,mag, and $-17.45 \pm 0.54$\, for SNe~Ib, Ic, and IIb, respectively (see their Table 5).

\subsection{First-light time and rise time}
\label{sec:firstlighttimeandrisetime}

\begin{figure}
        \includegraphics[width=1.0\columnwidth]{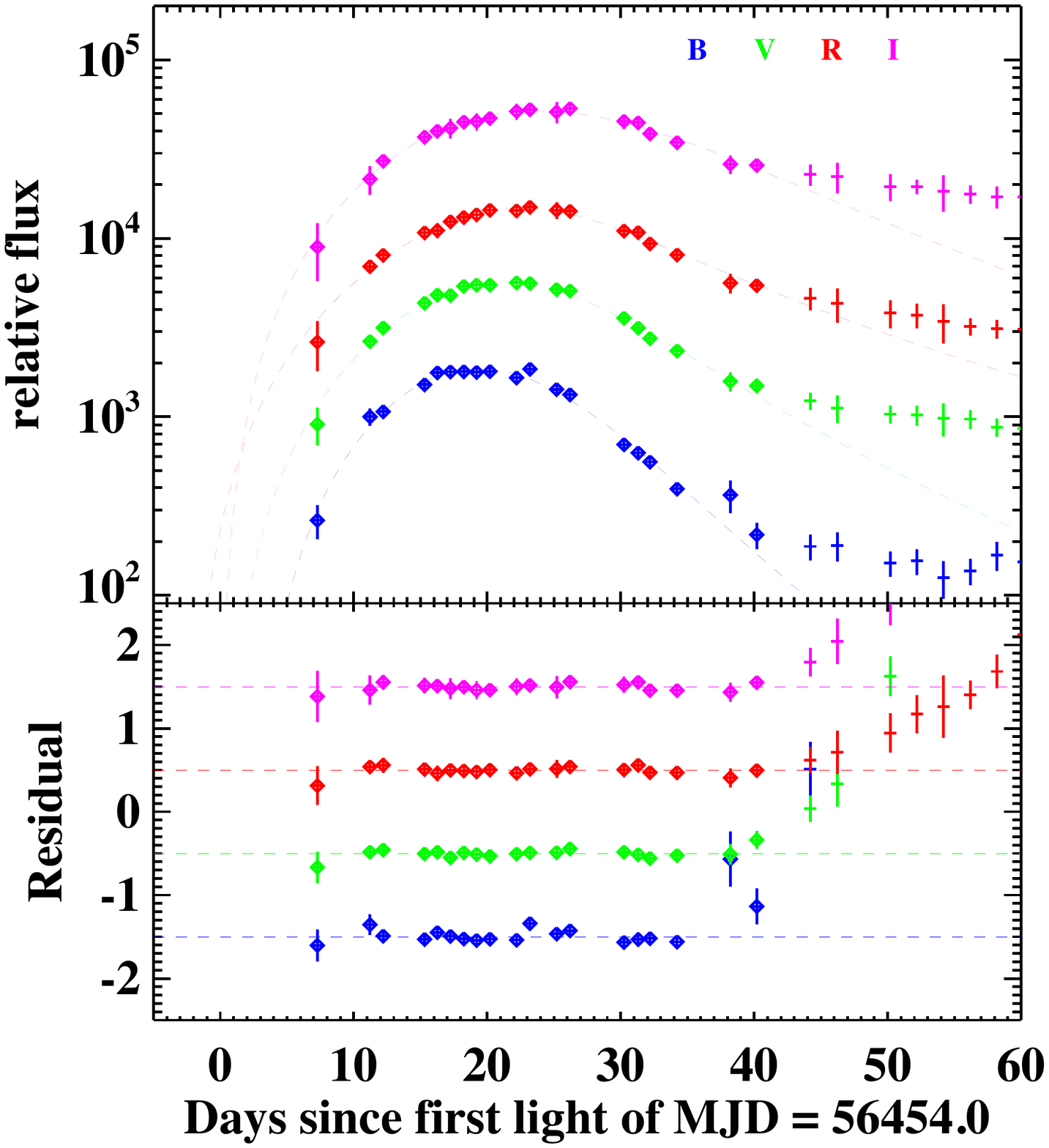}
\caption{An example of multiband light-curve fitting to the well-observed SN~iPTF13bvn
using the function (see text) proposed by Zheng \& Filippenko (2017), which was originally used for fitting SN~Ia light curves.
``Diamond" data points are included in the fitting while ``cross" data points are not.
This example demonstrates that the same function can also be used for fitting SESN light curves.
}
\label{LCfitsniPTF13bvn}
\end{figure}

With progressively more SESNe discovered and observed at very early phases, it has become possible to measure the true first-light time
by fitting the observed light curve. Overall, the light-curve shape of SESNe resembles that of SNe~Ia.
Several empirical functions have been proposed to fit SESN light curves. For example, \cite{taddia15} used a phenomenological model
--- which was first employed by \cite{bazin11} for fitting SNe~Ia --- to fit their SESN light curves. \cite{taddia18} also
proposed a three-component function that has proven to work well.
Motivated by the former, we adopt a function proposed by \cite{zheng17b} for SN~Ia light-curve fitting \citep{zheng17c} to fit the 
SESN light curves in our sample. The function is given as
\begin{equation}
L = A'\left(\frac{t-t_0}{t_b}\right)^{\alpha_{r}} \Big{[} 1 + \left(\frac{t-t_0}{t_b}\right)^{s({\alpha}_{d})}\Big{]}^{-2/s},
\end{equation}
where $A'$ is a scaling constant, $t_0$ is the first-light time, $t_b$ is the break time,
${\alpha}_r$ and ${\alpha}_d$ are the two power-law indices before and after the break, and
$s$ is a smoothing parameter.
We have found that this function can provide satisfactory fits to the SESN light curves in our sample. Figure~\ref{LCfitsniPTF13bvn}
presents an example of this function fit to the well-observed SN~Ib iPTF13bvn.
This method directly takes the first-light time $t_0$ as a parameter in the fitting.
Following the same procedure as \cite{zheng17c}, we fit for each filter with good-quality data.
We finally adopt the mean value of the first-light time from the
fitting if there are more than two measurements in different filters.

\begin{landscape}
\begin{deluxetable}{llccccccccccccccc}
 \tabcolsep 0.4mm
 \tablewidth{0pt}
 \tablecaption{Rise-time information}
  \tablehead{\colhead{SN} & \colhead{Type} & \colhead{$z$} & \colhead{$t_0$ (MJD)} & \colhead{$t_{r}(u)$} & \colhead{$t_{r}(B)$} & \colhead{$t_{r}(g)$} & \colhead{$t_{r}(V)$} & \colhead{$t_{r}(R)$} & \colhead{$t_{r}(r)$} & \colhead{$t_{r}(Clear)$} & \colhead{$t_{r}(I)$} & \colhead{$t_{r}(i)$} & \colhead{$t_{r}(Y)$} & \colhead{$t_{r}(J)$} & \colhead{$t_{r}(H)$} & \colhead{$t_{r}(K)$} }
\startdata
\hline
     2006el &     IIb & 0.017062 &  53,958.0$\pm$1.2 &         ---    &  24.5$\pm$3.8 &         ---    &  25.0$\pm$2.5 &  25.0$\pm$5.6 &         ---    &        ---    &        ---    &   28.5$\pm$1.5 &        ---    &        ---    &        ---    &        ---    \\
     2006ep &      Ib & 0.015134 &  53,964.7$\pm$2.4 &   17.7$\pm$3.0 &  19.8$\pm$2.4 &   20.4$\pm$3.0 &  22.2$\pm$2.6 &  24.3$\pm$3.0 &   24.0$\pm$2.8 &        ---    &  25.8$\pm$2.5 &   25.8$\pm$2.4 &  28.7$\pm$2.4 &  30.2$\pm$2.4 &  30.4$\pm$2.4 &        ---    \\
     2006lc &     Ibn & 0.016228 &  54,022.0$\pm$0.9 &   17.0$\pm$3.6 &  18.3$\pm$2.1 &   18.3$\pm$3.6 &  19.1$\pm$2.1 &  21.8$\pm$1.6 &   20.5$\pm$2.0 &        ---    &  22.7$\pm$1.1 &   21.4$\pm$1.1 &  25.2$\pm$1.6 &  26.2$\pm$1.6 &  26.4$\pm$1.6 &        ---    \\
     2007ru &   Ic-BL & 0.015464 &  54,426.5$\pm$2.1 &         ---    &        ---    &         ---    &        ---    &        ---    &         ---    &  13.8$\pm$3.2 &        ---    &         ---    &        ---    &        ---    &        ---    &        ---    \\
     2007uy &  Ib-pec & 0.006494 &  54,458.0$\pm$1.0 &         ---    &  19.2$\pm$1.1 &         ---    &  23.0$\pm$1.1 &  25.3$\pm$1.1 &   25.5$\pm$1.1 &        ---    &  26.1$\pm$1.1 &   26.1$\pm$1.1 &        ---    &  31.9$\pm$1.0 &  33.1$\pm$1.0 &  34.2$\pm$1.0 \\
     2008eb &      Ib & 0.007612 &  54,645.4$\pm$0.9 &         ---    &  17.0$\pm$1.0 &         ---    &  18.0$\pm$1.1 &  18.7$\pm$1.1 &         ---    &        ---    &  19.7$\pm$1.1 &         ---    &        ---    &        ---    &        ---    &        ---    \\
      2009K &     IIb & 0.011715 &  54,838.3$\pm$1.6 &   26.4$\pm$3.2 &  28.2$\pm$1.9 &   28.9$\pm$1.9 &  31.0$\pm$1.7 &        ---    &   32.9$\pm$1.7 &        ---    &        ---    &         ---    &  38.3$\pm$1.6 &  37.9$\pm$1.6 &  41.9$\pm$1.6 &        ---    \\
     2009jf &      Ib & 0.007942 &  55,097.8$\pm$1.1 &         ---    &  21.0$\pm$1.5 &         ---    &  23.3$\pm$1.3 &  25.3$\pm$1.8 &   25.5$\pm$1.5 &        ---    &  26.5$\pm$1.3 &   26.8$\pm$1.8 &        ---    &  31.1$\pm$1.1 &  32.3$\pm$1.1 &  35.0$\pm$1.1 \\
  iPTF13bvn &      Ib & 0.004533 &  56,454.0$\pm$1.4 &         ---    &  19.5$\pm$1.5 &         ---    &  21.4$\pm$1.6 &  22.8$\pm$1.5 &         ---    &        ---    &  24.0$\pm$1.6 &         ---    &        ---    &        ---    &        ---    &        ---    \\
      2014C &      Ib & 0.002722 &  56,656.6$\pm$0.8 &         ---    &  11.1$\pm$0.9 &         ---    &  13.7$\pm$0.9 &  14.8$\pm$0.9 &         ---    &  14.4$\pm$1.4 &  15.9$\pm$0.9 &         ---    &        ---    &        ---    &        ---    &        ---    \\
     2014as &   Ic-BL & 0.012469 &  56,758.9$\pm$1.1 &         ---    &  10.5$\pm$1.3 &         ---    &  12.3$\pm$1.2 &  13.3$\pm$2.1 &         ---    &  13.1$\pm$2.1 &  13.5$\pm$2.1 &         ---    &        ---    &        ---    &        ---    &        ---    \\
     2014eh &      Ic & 0.010614 &  56,955.1$\pm$1.0 &         ---    &  16.9$\pm$1.7 &         ---    &  19.6$\pm$1.7 &  21.9$\pm$2.3 &         ---    &  22.0$\pm$2.9 &  24.4$\pm$2.3 &         ---    &        ---    &        ---    &        ---    &        ---    \\
     2014ei &      Ib & 0.014440 &  56,960.5$\pm$2.3 &         ---    &        ---    &         ---    &        ---    &        ---    &         ---    &  17.9$\pm$2.4 &        ---    &         ---    &        ---    &        ---    &        ---    &        ---    \\
      2015U &     Ibn & 0.013790 &  57,061.0$\pm$0.9 &         ---    &   7.6$\pm$1.1 &         ---    &   9.5$\pm$1.0 &   9.7$\pm$1.0 &         ---    &   9.9$\pm$1.0 &   9.8$\pm$1.0 &         ---    &        ---    &        ---    &        ---    &        ---    \\
      2015Y &      Ib & 0.008172 &  57,113.6$\pm$2.2 &         ---    &  23.3$\pm$3.0 &         ---    &  23.9$\pm$3.0 &  25.5$\pm$3.0 &         ---    &  23.9$\pm$2.8 &  26.6$\pm$3.0 &         ---    &        ---    &        ---    &        ---    &        ---    \\
     2015ap &      Ib & 0.011375 &  57,270.1$\pm$0.9 &         ---    &  12.2$\pm$2.4 &         ---    &  14.9$\pm$2.4 &  16.6$\pm$1.1 &         ---    &  15.3$\pm$1.0 &  18.0$\pm$1.0 &         ---    &        ---    &        ---    &        ---    &        ---    \\
    2016bau &      Ib & 0.003856 &  57,452.4$\pm$1.0 &         ---    &  22.1$\pm$1.2 &         ---    &  25.3$\pm$1.1 &  25.5$\pm$1.1 &         ---    &  25.7$\pm$1.1 &  27.0$\pm$1.1 &         ---    &        ---    &        ---    &        ---    &        ---    \\
    2016gkg &     IIb & 0.004940 &  57,644.4$\pm$0.8 &         ---    &  24.9$\pm$1.1 &         ---    &  26.4$\pm$1.1 &  27.3$\pm$0.9 &         ---    &  28.1$\pm$4.0 &  28.3$\pm$4.0 &         ---    &        ---    &        ---    &        ---    &        ---    \\
    2017ein &      Ic & 0.002699 &  57,895.7$\pm$0.8 &         ---    &  14.8$\pm$2.6 &         ---    &  17.5$\pm$2.6 &  19.0$\pm$2.6 &         ---    &  18.1$\pm$2.6 &  20.1$\pm$2.6 &         ---    &        ---    &        ---    &        ---    &        ---    \\
\hline
\multicolumn{17}{c} {Literature sample.} \\
\hline
     2004dk &      Ib & 0.005247 &  53,213.9$\pm$0.7 &         ---    &        ---    &         ---    &  24.7$\pm$1.9 &  27.7$\pm$1.9 &         ---    &        ---    &        ---    &         ---    &        ---    &        ---    &        ---    &        ---    \\
     2004ex &     IIb & 0.017549 &  53,281.7$\pm$1.0 &   22.9$\pm$1.7 &  23.9$\pm$1.7 &   24.3$\pm$1.7 &  25.4$\pm$1.1 &        ---    &   26.5$\pm$1.1 &        ---    &        ---    &   28.4$\pm$1.1 &  32.1$\pm$1.0 &  27.5$\pm$1.0 &  30.3$\pm$1.0 &        ---    \\
     2004fe &      Ic & 0.017896 &  53,302.6$\pm$1.2 &   11.9$\pm$1.3 &  13.3$\pm$1.6 &   13.9$\pm$2.1 &  15.1$\pm$1.6 &  16.7$\pm$1.7 &   16.6$\pm$1.4 &        ---    &        ---    &   18.0$\pm$1.3 &        ---    &        ---    &        ---    &        ---    \\
     2004gv &      Ib & 0.019927 &  53,343.9$\pm$0.9 &   19.2$\pm$1.1 &  20.6$\pm$1.1 &   20.6$\pm$1.1 &  21.9$\pm$1.6 &        ---    &   23.7$\pm$1.6 &        ---    &        ---    &   25.7$\pm$1.4 &  26.9$\pm$0.9 &  26.8$\pm$0.9 &  32.1$\pm$0.9 &        ---    \\
     2005hg &      Ib & 0.021308 &  53,659.4$\pm$3.2 &   19.9$\pm$3.2 &  21.5$\pm$3.2 &         ---    &  23.1$\pm$3.2 &  25.0$\pm$3.4 &   25.1$\pm$3.2 &        ---    &        ---    &   27.3$\pm$3.2 &        ---    &  32.3$\pm$3.2 &  33.6$\pm$3.2 &  35.8$\pm$3.2 \\
      2006T &     IIb & 0.008092 &  53,762.7$\pm$1.1 &   15.5$\pm$1.5 &  16.3$\pm$1.2 &   16.6$\pm$1.5 &  17.7$\pm$1.3 &        ---    &   18.5$\pm$1.3 &        ---    &        ---    &   19.8$\pm$1.3 &  20.7$\pm$1.1 &  21.6$\pm$1.1 &  23.8$\pm$2.0 &        ---    \\
      2007Y &      Ib & 0.004637 &  54,142.4$\pm$2.0 &   18.9$\pm$2.1 &  19.9$\pm$2.1 &   20.5$\pm$2.1 &  21.9$\pm$2.1 &        ---    &   23.3$\pm$2.1 &        ---    &        ---    &   24.1$\pm$2.3 &  23.3$\pm$2.0 &  25.2$\pm$2.0 &  24.5$\pm$2.3 &        ---    \\
      2008D &      Ib & 0.006494 &  54,468.9$\pm$0.9 &         ---    &  22.2$\pm$3.7 &         ---    &  23.9$\pm$3.3 &        ---    &   24.7$\pm$3.7 &        ---    &        ---    &   25.5$\pm$1.0 &        ---    &        ---    &        ---    &        ---    \\
     2009bb &      Ic & 0.009877 &  54,909.6$\pm$1.0 &    9.1$\pm$1.4 &  10.3$\pm$1.4 &   11.5$\pm$2.7 &  12.9$\pm$2.7 &        ---    &   13.9$\pm$2.7 &        ---    &        ---    &   14.9$\pm$2.7 &  20.2$\pm$1.0 &  18.5$\pm$1.0 &  17.5$\pm$1.0 &        ---    \\
     2009iz &      Ib & 0.014196 &  55,077.3$\pm$1.1 &         ---    &  28.2$\pm$1.2 &         ---    &  31.0$\pm$2.8 &        ---    &   33.1$\pm$2.8 &        ---    &        ---    &   36.3$\pm$1.3 &        ---    &  42.9$\pm$1.1 &  47.1$\pm$1.1 &  45.3$\pm$1.1 \\
\enddata
%\tablenotetext{a}{In units of \kms\,d$^{-1}$.}
\label{risetimeinallband}
\end{deluxetable}
\end{landscape}

Using the estimated first-light time along with the peak time derived above (see Sec.~\ref{LightCurves}), one can measure
the rise time after correcting for the redshift. Table~\ref{risetimeinallband} gives the first-light time, peak time, and rise time for
the SESNe in our sample. The rise times for different bands are calculated separately when different peak times are available.
In addition, we collect the infrared peak time in the $Y$, $J$, $H$, and $K_s$ filters for those SNe that are also presented by
\cite{bianco14} or \cite{taddia18}, in order to derive the rise time in infrared bands. Ten additional SESNe are added to
the rise-time sample for this analysis;
they are taken from the samples published by \cite{drout11}, \cite{bianco14}, \cite{stritzinger18a}, and \cite{taddia18},
and are listed at the bottom of Table~\ref{risetimeinallband}.

\begin{figure*}
        \includegraphics[width=1.8\columnwidth]{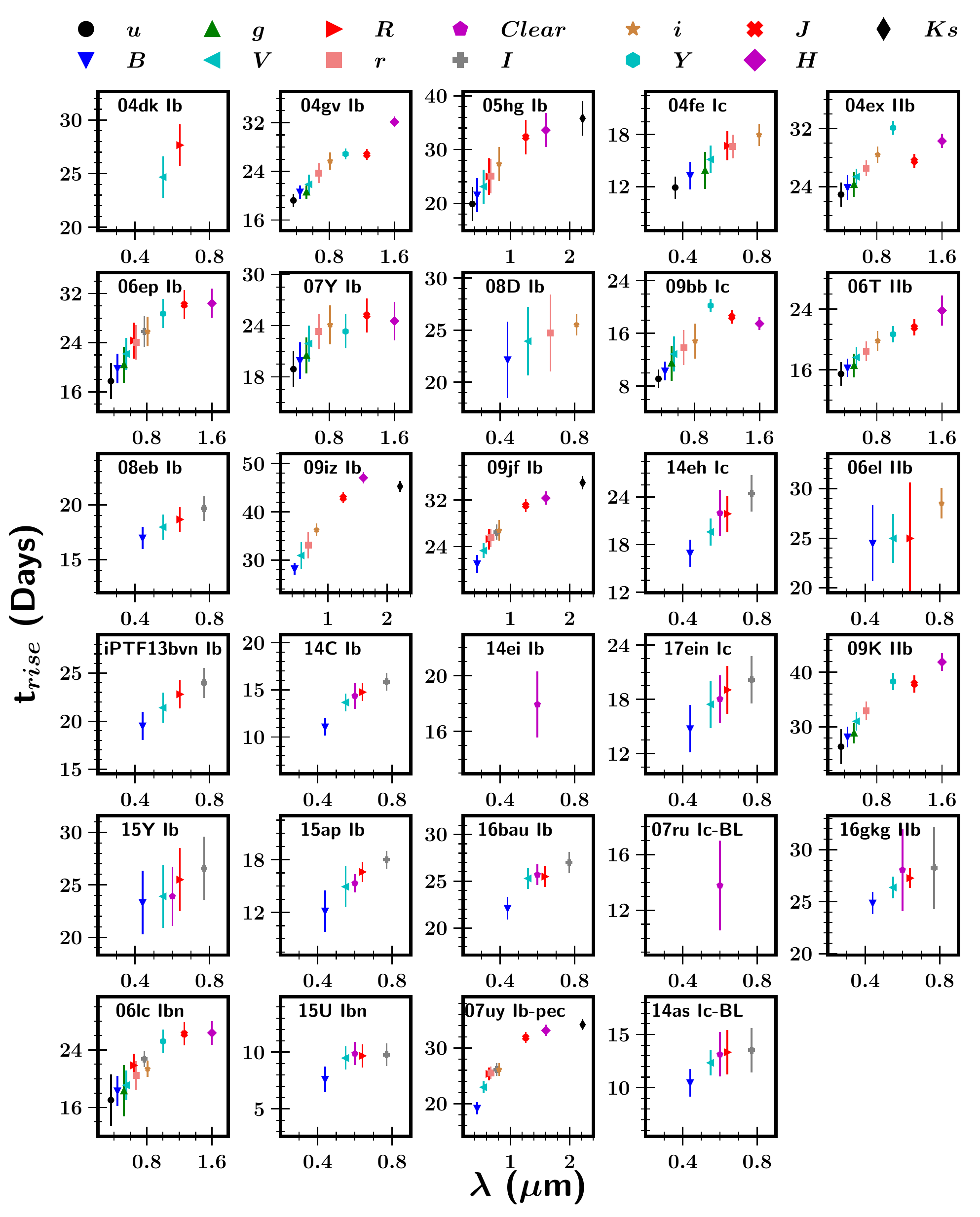}
        \caption{Rise time derived from 29 SESNe as a function of the effective wavelengths for different bands,
colour coded as shown in the top legend. The rise time is generally longer in redder filters than it is in bluer filters.
The rise time in infrared bands is typically a factor of $\sim 1.5$ longer than in blue bands ($U$ or $B$).
}
\label{RiseTimeForEachBand}
\end{figure*}

Figure~\ref{RiseTimeForEachBand} displays the rise time as a function of the effective wavelengths for different bands
using all available fitting results for our sample. This figure is similar to Figure 10 of \cite{taddia15}
and Figure 3 of \cite{taddia18}. However, note that \cite{taddia15} use the explosion time --- defined as the average between
the epochs of last nondetection and first detection --- instead of the first-light time, and \cite{taddia18} use the offset
of peak time in different filters relative to the $r$-band peak time in their Figure 10.
Consequently, we claim that the first-light times and rise times presented herein are the first true measurements of such for a large sample of SESNe.
Also for the first time, we measure accurate rise times of a large sample of SESNe in infrared bands. These measurements are
important for understanding the explosion properties of SESNe (see Sec.~\ref{LCfittingandmodeling}).

\begin{figure*}
        \includegraphics[width=1.8\columnwidth]{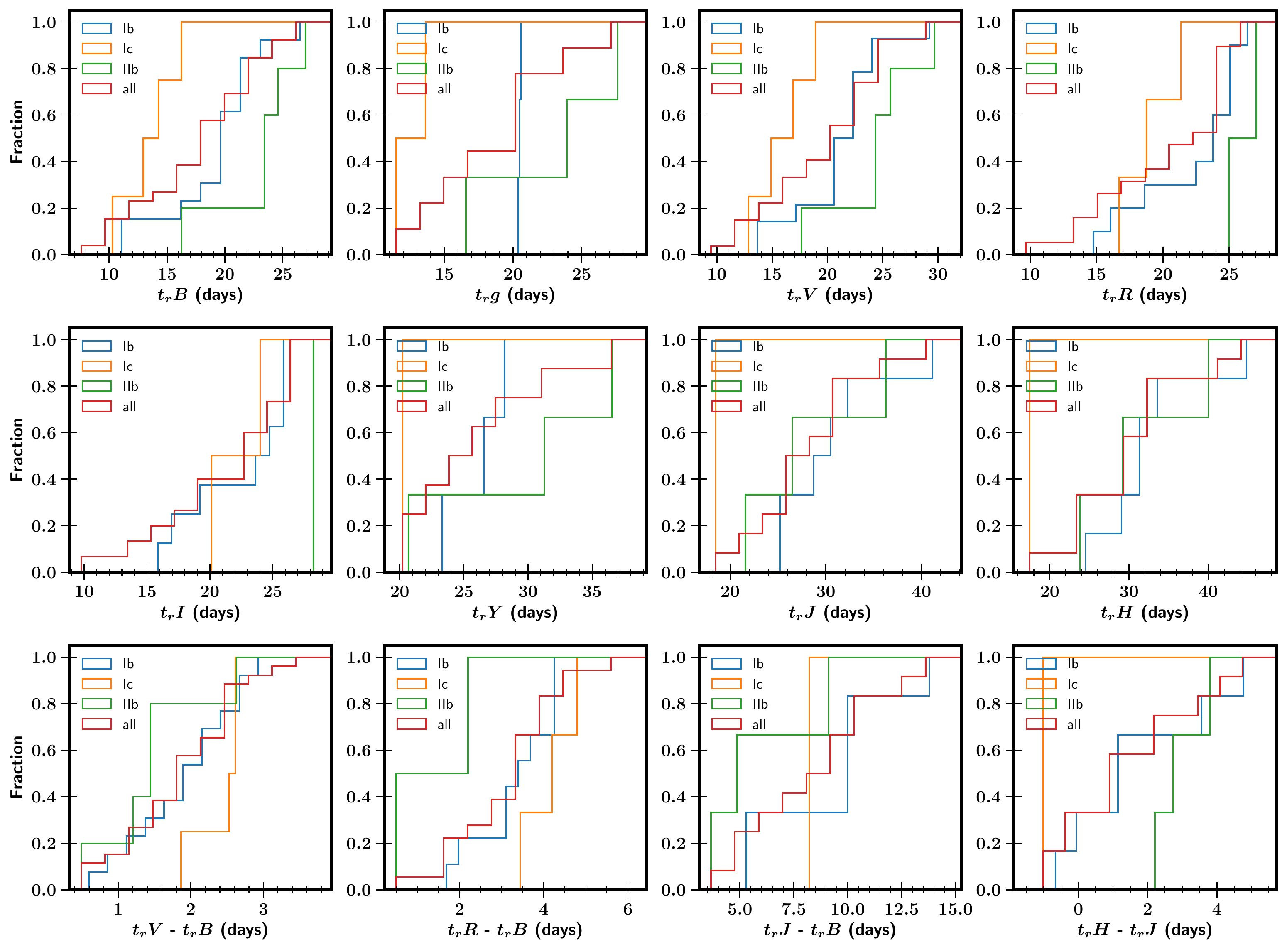}
        \caption{Top two rows: Cumulative distribution of rise time in eight different filters.
K-S tests show significant differences between SNe~Ib and SNe~Ic, and also between SNe~IIb and SNe~Ib.
The average rise times for SNe~IIb, Ib, and Ic are 23.5, 19.9, and 13.8\,d (respectively) in the $B$ band.
For almost all bands, SNe~IIb have the longest rise time, while SNe~Ic have the shortest rise time.
Bottom row: The difference of rise time in the selected two bands.
SNe~Ic take more time for the redder band to reach peak after $B$-band peak compared to SNe~IIb, opposite to
the rise-time relation.
}
\label{RiseTimeCumulativeDistributionForEachBand}
\end{figure*}

As can be seen in Figure~\ref{RiseTimeForEachBand}, the rise time is generally longer in red filters than it is in blue filters,
confirming the similar result found by \citet{bianco14}, \citet{taddia15}, and \citet{taddia18}. The rise time in infrared bands is typically a factor of $\sim 1.5$ 
longer than in blue bands ($U$ or $B$). The rise times in different subtypes of SNe also show differences, which we visualise via the
cumulative distribution of the rise time for each subgroup consisting of SNe~Ib, Ic, and IIb in
Figure~\ref{RiseTimeCumulativeDistributionForEachBand}.
A Kolmogorov-Smirnov (K-S) test comparing SNe~Ib and SNe~Ic in the
$B$ band gives a $p$-value of 0.013, showing a significant difference between the two populations, and an analogous
K-S test between SNe~IIb and SNe~Ib in the $B$ band gives a $p$-value of 0.025 --- also showing a significant difference between the two populations.
We find that the average rise time for SNe~IIb, Ib, and Ic are 23.5, 19.9, and 13.8\,d (respectively) in the $B$ band, and 
26.1, 22.6, and 19.2\,d (respectively) in the $R$ band.
It is obvious that for almost all bands, SNe~IIb have the longest rise times, while
SNe~Ic have the shortest, consistent with the findings of \citet{valenti11}, \citet{taddia15}, and \citet{taddia18}.
We also notice that the different rise times between each SN subtype are less clear in infrared bands compared to optical bands,
though we caution that the infrared sample is much smaller.

The bottom row in Figure~\ref{RiseTimeCumulativeDistributionForEachBand} shows the difference between rise times in the selected two bands.
Typically, SNe~Ic take more time for redder bands to reach peak after $B$-band peak compared with SNe~IIb, opposite to
the aforementioned rise-time relations.

\subsection{Light-curve fitting and modeling}
\label{LCfittingandmodeling}

\begin{table*}
%\centering
\tabletypesize{\scriptsize}
\caption{The definitions, units, and prior ranges of the parameters of the fitting models \label{tab:parameters}.}
\begin{tabular}{cccc}
\hline
\hline
        \colhead{Parameter}             & \colhead{Definition}                                 &  \colhead{Unit}          &     \colhead{Posterior}  \\
\hline
\multicolumn{4}{c} {$^{56}$Ni model} \\
\hline
        $M_{\rm ej}$                    & ejecta mass                                      &   M$_\odot$               &    $[0.1, 50]$          \\
        $v_9$                           & ejecta velocity                                  &   $10^9$\,cm\,s$^{-1}$      &    $[0.1, 5.0 (10.0)]^{a}$  \\
        $M_{\rm Ni}$                    & $^{56}$Ni mass                                   &   M$_\odot$               &    $[0.001, 2.0 (20.0)]^{a}$  \\
        $\kappa_{\rm \gamma, Ni}$       & gamma-ray opacity of $^{56}$Ni-cascade-decay photons &   cm$^2$\,g$^{-1}$          &    $[0.027, 10^4]^{c}$   \\
        $T_{\rm f}$                     & temperature floor of the photosphere             &   10$^3$\,K                &    $[1000, 10,000] $    \\
        $t_{\rm shift}^{b}$         & explosion time relative to the first data        &   days                    &    $[-20, 0]$          \\
\hline
\multicolumn{4}{c} {Cooling model with three additional parameters compared to the $^{56}$Ni model.} \\
\hline
        $M_{\rm e}$                     & envelope mass                                    &   M$_\odot$               &    $[0.01, 30]$          \\
        $R_{\rm e,12}$                  & envelope radius                                  &   $10^{12}$\,cm            &    $[10, 3000]$          \\
        $E_{\rm e,50}$                  & energy passed into the envelope from SN core     &   $10^{50}$\,erg s$^{-1}$  &    $[10^{-5}, 10^3]^{c}$     \\
\hline
\multicolumn{4}{c} {Magnetar model with three different parameters compared to the $^{56}$Ni model, but dropped $M_{\rm Ni}$ and $\kappa_{\rm \gamma, Ni}$.} \\
\hline
        $P_{\rm 0}$                     & initial period of the magnetar                   &   ms                      &    $[0.8, 50]$          \\
        $B_{\rm p,14}$                  & magnetic field strength of the magnetar          &   $10^{14}$\,G             &    $[0.1, 100]$          \\
        $\kappa_{\rm \gamma, mag}$      & gamma-ray opacity of magnetar photons            &   cm$^2$\,\,g$^{-1}$          &    $[0.01, 10^4]^{c}$  \\
\hline\hline
\noalign{\smallskip}
\end{tabular}
\tablenotetext{a}{For four luminous or superluminous SNe (SNe~2008fz, 2010hy, 2012aa, and 2018cow; see Table \ref{table:LC_Param_Mag}), the upper limits of prior of $v_9$ and $M_{\rm Ni}$ are set to be 10 and 20.0, respectively.}
\tablenotetext{b}{For the SNe whose explosion date ($t_0$) had been inferred (see Table \ref{risetimeinallband}), the $t_{\rm shift}$ parameter was set to be fixed.}
\tablenotetext{c}{Parameter was distributed in log space.}
\end{table*}

\begin{figure*}
        \includegraphics[width=2.0\columnwidth]{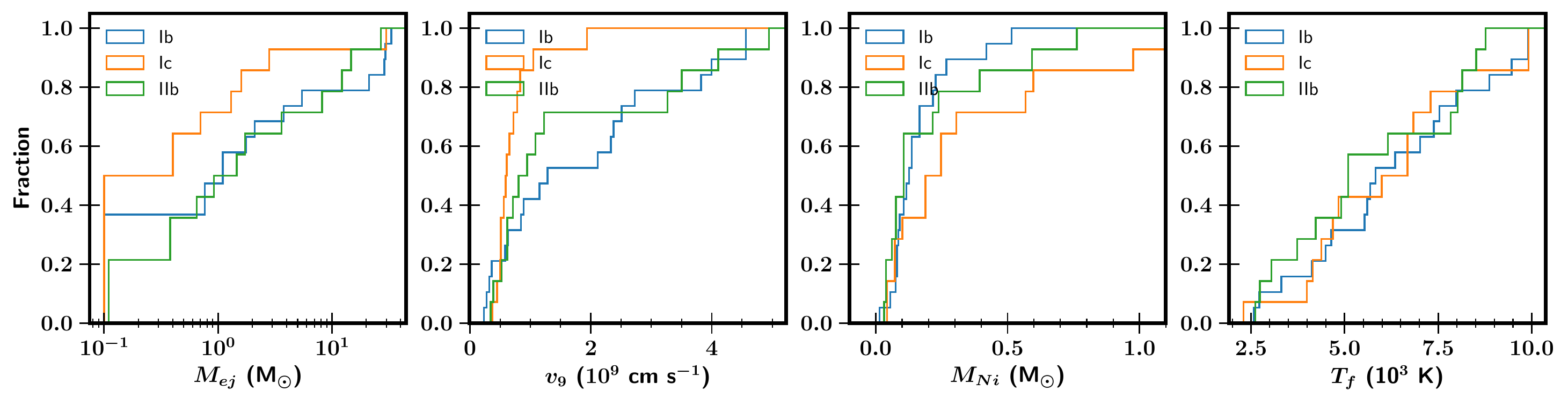}
        \caption{Cumulative distributions of the four parameters ($M_{\rm ej}$, $v_9$, $M_{\rm Ni}$, and $T_{\rm f}$)
                 derived from model fitting. Compared to SNe~Ib and SNe~IIb, SNe~Ic tend to have higher ejecta masses and
                 also higher ejecta velocities, on average, but lower $^{56}$Ni mass. The temperature floor of the photosphere
                 distribution shows no significant difference between different subtypes.
}
\label{FourParsCumulativeDistributionFromModelFitting}
\end{figure*}

To further study the physical properties of the SESNe in our sample, we model
the multiband light curves using the $^{56}$Ni model subject to the following assumptions:
(i) the bolometric luminosities of the photospheres of the SNe are powered
by $^{56}$Ni cascade decay \citep{arnett82,chatzopoulos12,wang15}, (ii) the spectral energy distributions
of the SNe can be described by the blackbody or ultraviolet-absorbed blackbody function
(see \citealt{nicholl17}, and references therein), and (iii)
the velocities of the SN photospheres are constant at early times, and the radii of the photospheres 
are determined by the bolometric luminosities and the temperature at the late epochs when the 
temperature no longer changes (see, e.g., Eqs. 8 and 9 of \citealt{nicholl17}). 
We do not consider the case of interactions with dense circumstellar material, which usually make
the light curves flattened at late times and are more likely related to Type IIn SNe,
while the light curves in our sample do not show enough evidence for such cases.
The definitions, units, and prior ranges of the free parameters of the $^{56}$Ni model are given in
%Like some other studies for SESNe, the opacity of the ejecta is set to be 0.07\,cm$^2$\,g$^{-1}$.
Table~\ref{tab:parameters}, where the prior is uniformly sampled either linearly or in log space over the range.
We employ a Markov Chain Monte Carlo (MCMC) method via the \texttt{emcee} Python package
\citep{foreman13} to get the best-fit parameters and $3\sigma$ confidence ranges of the fitted parameters. 

The $^{56}$Ni model-fitting results for all SESNe in our sample are shown in Figure~\ref{modelfittinglcallsesne}, except SN~2009C for which
there is not enough data to meaningfully constrain the fitting. The best-fitted parameter values are given in Table~\ref{table:LC_Param_Ni},
where we also list in parentheses the median values for comparison.
As one can see, for most of the SNe, though the $B$-band fitting deviates more than other filters for some cases,
our $^{56}$Ni model can fit the general observed light curves with reasonable physical parameters.

However, for a group of SNe with double peaks or rebrightening after initial fading (including SNe~2011fu, 2015Y, 2016gkg, and 2016iyc),
our simple $^{56}$Ni model fails to adequately fit the early-time initial decay. In these cases, we therefore adopt an additional cooling component
and refit the light curve. The new cooling plus $^{56}$Ni model contains three additional parameters:
(i) the mass of the extended envelope ($M_{\rm e}$), (ii) the radius of the extended envelope ($R_{\rm e,12}$),
and (iii) the energy passed into the extended envelope ($E_{\rm e,50}$) from the SN core \citep{piro21}.
The cooling plus $^{56}$Ni model can better fit the light curves as shown in Figure~\ref{modelfittinglcof4coolingmodelsesne},
and the new fitting results are given in Table~\ref{table:LC_Param_Cooling}.

For another group of four SNe (including SNe~2008fz, 2010hy, 2012aa, and 2018cow), the $^{56}$Ni model results suggest
extremely high $^{56}$Ni masses, some even higher than the ejecta mass, which is clearly unphysical.
For these four luminous or superluminous SNe, we therefore
adopt the magnetar model \citep{kasen10b,woosley10,chatzopoulos12,wang15,dai16} to refit the light curves.
Three new parameters are included compared to the $^{56}$Ni model (also listed in Table~\ref{table:LC_Param_Ni}):
the initial period ($P_{\rm 0}$), the magnetic field strength of the magnetar ($B_{\rm p,14}$),
and the gamma-ray opacity of magnetar photons ($\kappa_{\rm \gamma, mag}$).
The magnetar model fitting is shown in Figure~\ref{modelfittinglcof4magnitarmodelsesne}.
and the results are given in Table~\ref{table:LC_Param_Mag}.
Compared to the $^{56}$Ni model, the magnetar model provides comparable fitting results for the light curves,
but with more reasonable physical parameters, indicating that a small fraction of SESNe may be powered by
central magnetars\footnote{Note that we cannot exclude the possibility of a hidden (fainter) magnetar in other SESNe,
but those SESNe do not require the fainter magnetar in the model fitting.}.

To summarise, Figure~\ref{FourParsCumulativeDistributionFromModelFitting} shows the cumulative distribution of the four parameters
($M_{\rm ej}$, $v_9$, $M_{\rm Ni}$, and $T_{\rm f}$) from the model fitting.
We find that SNe~Ic tend to have lower ejecta masses and also slower ejecta velocities, on average, compared to SNe~Ib and SNe~IIb.
On the other hand, the $^{56}$Ni mass of SNe~Ic tends to be higher than in SNe~Ib and SNe~IIb, consistent with the findings
Of \cite{prentice16} (see Table 10 in the paper). \cite{anderson19} reached a similar conclusion (see Table 1 in their paper) that
SNe~Ic have higher $^{56}$Ni mass than SNe~IIb, though their estimate of the $^{56}$Ni mass for SNe~Ib is comparable to that of SNe~Ic.
The temperature floor of the photosphere distribution shows no significant difference between different subtypes.

%\section{Discussion}
%\label{sec:discussion}

\section{Conclusion}
\label{sec:conclusion}

In this paper, we have presented multiband ($BVRI$, along with some {\it Clear}) light curves of a large sample of SESNe observed by
the KAIT and Nickel telescopes at Lick Observatory under the LOSS follow-up program from 2003 through 2020. Our data are processed in a homogeneous fashion,
and here we publicly release all derived data products to the supernova community. Our main results are as follows.

(i) We significantly enlarge the SESN sample by adding 70 SESNe observed by LOSS.

(ii) We confirm that SESNe usually suffer moderately high extinction from their host galaxies. Quantitatively, we find
a mean $E(B - V)$ value of 0.32\,mag, substantially higher than the MW extinction, and also higher than that of SN~Ia and SN~II samples.

(iii) The peak $R$-band absolute magnitude of SESNe shows a wide range from $-16$\,mag to brighter than $-19$\,mag.
SNe~Ic-BL are more luminous than both SNe~Ib and SNe~Ic, and SNe~Ic appear to be slightly brighter than SNe~Ib.

(iv) SESNe exhibit smooth light-curve shapes with a $\sim 10$--20\,d rising phase before reaching maximum brightness, followed by
a slow decay. A few SNe~IIb show a decline dip at very early times before rising again,
which can be attributed to the shock-breakout cooling tail.

(v) For the first time, we derive reliable, robust measurements of the rise times for a large sample of SESNe
in both optical and infrared bands. Our results show that SESNe rise faster in blue bands than in red bands.

(vi) Helium-poor SNe (SNe~Ic) rise to maximum faster than helium-rich SNe (SNe~Ib and IIb).
Average rise times for SNe~IIb, Ib, and Ic are 23.5, 19.9, and 13.8\,d (respectively) in the $B$ band, and
26.1, 22.6, and 19.2\,d (respectively) in the $R$ band.
K-S tests show significant differences between normal SNe~Ib and SNe~Ic, and also between SNe~IIb and SNe~Ib.

(vii) SNe~Ic tend to have lower ejecta masses and also slower ejecta velocities, on average, compared to SNe~Ib and SNe~IIb,
but with higher $^{56}$Ni mass.

\section*{Acknowledgements}

We thank Jenifer Rene Gross and Alessondra Springmann for their effort in taking Lick/Nickel data. We are grateful to the staff at Lick Observatory for their assistance with the Nickel telescope and KAIT. KAIT and its ongoing operation were made possible by donations from Sun Microsystems, Inc., the Hewlett-Packard Company, Auto Scope Corporation, Lick Observatory, the National Science Foundation (NSF), the University of California, the Sylvia \& Jim Katzman Foundation, and the TABASGO Foundation.
Research at Lick Observatory is partially supported by a generous gift from Google. 

Support for A.V.F.'s supernova group has been provided by the NSF, 
Marc J. Staley (whose fellowship partly funded B.E.S. whilst contributing to the work presented herein as a graduate student), the Richard and Rhoda Goldman Fund, the TABASGO Foundation, 
Gary and Cynthia Bengier (who provided financial support for T.deJ. via the Bengier Postdoctoral fellowship), the Christopher R. Redlich Fund, and the UC Berkeley Miller
Institute for Basic Research in Science (in which A.V.F. was a Miller Senior Fellow at the time of this research). 
In addition, we greatly appreciate contributions from 
numerous individuals, including
% Robert and Marina Arnott % no, they don't want to be acknowledged
Charles Baxter and Jinee Tao,
George and Sharon Bensch % confirmed 9/17/19
Greg and Patty Bernstein, % confirmed 2/19/22
Firmin Berta,
Jack Bertges, % confirmed 2/20/22
Marc and Cristina Bensadoun,
Greg and Patty Bernstein, % confirmed 1/5/22
Frank and Roberta Bliss,
Ann and Gordon Brown, % confirmed 12/30/21
Eliza Brown and Hal Candee,
Kathy Burck and Gilbert Montoya,
Alan and Jane Chew,
Christopher Cook, % confirmed 1/6/22
% Lawrence Cool % He prefers to be anonymous
David and Linda Cornfield,
Michael Danylchuk,
Robert Davenport, % confirmed 1/5/22
Jim and Hildy DeFrisco,
Alli and Byron Deeter, % confirmed
Tim and Melissa Draper, % confirmed 12/30/22
William and Phyllis Draper,
Luke Ellis and Laura Sawczuk,
Jim Erbs and Shan Atkins,    
Alan Eustace and Kathy Kwan,
Art and Cindy Folker, % confirmed 12/30/21, sort of; asked again on 2/19/22
Peter and Robin Frazier, % confirmed 9/17/19
David Friedberg,             
Harvey Glasser,              
Charles and Gretchen Gooding,
Alan Gould and Diane Tokugawa,
Richard Gregor, % confirmed 1/4/22
Thomas and Dana Grogan,
Timothy and Judi Hachman % still need to ask him
Michael and Virginia Halloran, % confirmed 1/15/22
Gregory Hirsch and Kathy Long, % confirmed 1/5/22
Alan and Gladys Hoefer,
Jerry and Patti Hume, % confirmed 1/10/22, sort of; asked again 2/19/22
Charles and Patricia Hunt, 
Stephen and Catherine Imbler,
Adam and Rita Kablanian,
Heidi Gerster Kikawada, % confirmed 2/19/22
Roger and Jody Lawler,
Arthur and Rita Levinson, % confirmed 2/19/22
Jesse Levinson, % confirmed 12/31/21
Kenneth and Gloria Levy,
Greg Losito and Ronnie Bayduza, % confirmed 12/31/21
Walter and Karen Loewenstern, % confirmed 1/1/22
Peter Maier, 
DuBose and Nancy Montgomery,
Rand Morimoto and Ana Henderson,
Sunil Nagaraj and Mary Katherine Stimmler, 
Peter and Kristan Norvig,   
James and Marie O'Brient,  
Emilie and Doug Ogden,   
Paul and Sandra Otellini,
Margaret Renn, % confirmed 1/10/22
Robina Riccitiello, % confirmed 2/19/22
% Walter and Thelma Ritchie % No, don't want to be acknowledged; 2/24/22
Leslie Roberts, % confirmed 2/20/22
Jeanne and Sanford Robertson,
Paul Robinson, % confirmed 2/19/22
% Cat Rondeau, % prefers not, 12/30/21
Eric Rudney, % confirmed 2/19/22
Sissy Sailors and Red Conger, % confirmed 9/17/19
Geraldine and David Sandor, % confirmed 2/21/22
Tom and Cathy Saxton, % confirmed 3/2/22
Stanley and Miriam Schiffman,
Thomas and Alison Schneider,
Ajay Shah and Lata Krishnan, 
Alex and Irina Shubat,     
Silicon Valley Community Foundation,
Bruce and Debby Smith,  % confirmed 2/19/22
Mary-Lou Smulders and Nicholas Hodson,
Hans Spiller,
Alan and Janet Stanford,
Richard and Shari Stegman, % confirmed 12/31/22
Hugh Stuart Center Charitable Trust,
Toby Stuart, % confirmed (sort of) 1/6/22; asked again 2/19/22
% Eric and Stephanie Tilenius % No, don't want to be; divorced; 1/2/22
Gerald and Virginia Weiss, % confirmed 2/20/22
Clark and Sharon Winslow,
Ron and Geri Wohl, % confirmed 2/21/22
Weldon and Ruth Wood,
David and Angie Yancey,
Tom Zdeblick, % confirmed 2/19/22
and many others.
X.G.W. is supported by the National Natural Science Foundation of China (NSFC grant 11673006) and the Guangxi Science Foundation (grants 2016GXNSFFA380006 and 2017AD22006).

%We thank (mostly U.C. Berkeley undergraduate students)
%Carmen Anderson,
%James Bradley,
%Stanley Browne,
%Ian Crossfield,
%Griffin Foster,
%Don Gavel,
%Mark Gleed,
%Christopher Griffith,
%Anthony Khodanian,
%Laura Langland,
%Paul Lynam,
%Kyle McAllister,
%Alekzandir Morton,
%Daniel Perley,
%Tyler Pritchard,
%Andrew Rikhter,
%Jackson Sipple,
%Costas Soler,
%Stephen Taylor,
%Jeremy Wayland,
%and Dustin Winslow,
%for their effort in taking Lick/Nickel data.

This research has made use of the NASA/IPAC Extragalactic Database (NED), which is operated by the Jet Propulsion Laboratory, California Institute of Technology, under contract with NASA. The Pan-STARRS1 Surveys (PS1) and the PS1 public science archive have been made possible through contributions by the Institute for Astronomy, the University of Hawaii, the Pan-STARRS Project Office, the Max-Planck Society and its participating institutes, the Max Planck Institute for Astronomy, Heidelberg and the Max Planck Institute for Extraterrestrial Physics, Garching, The Johns Hopkins University, Durham University, the University of Edinburgh, the Queen's University Belfast, the Harvard-Smithsonian Center for Astrophysics, the Las Cumbres Observatory Global Telescope Network Incorporated, the National Central University of Taiwan, the Space Telescope Science Institute, NASA under grant NNX08AR22G issued through the Planetary Science Division of the NASA Science Mission Directorate, the National Science Foundation grant AST-1238877, the University of Maryland, Eotvos Lorand University (ELTE), the Los Alamos National Laboratory, and the Gordon and Betty Moore Foundation. Funding for the Sloan Digital Sky Survey (SDSS) has been provided by the Alfred P. Sloan Foundation, the Participating Institutions, NASA, the NSF, the U.S. Department of Energy, the Japanese Monbukagakusho, and the Max Planck Society. The SDSS Web site is \url{http://www.sdss.org/}. The SDSS is managed by the Astrophysical Research Consortium (ARC) for the Participating Institutions. The Participating Institutions are The University of Chicago, Fermilab, the Institute for Advanced Study, the Japan Participation Group, The Johns Hopkins University, Los Alamos National Laboratory, the Max-Planck-Institute for Astronomy (MPIA), the Max-Planck-Institute for Astrophysics (MPA), New Mexico State University, University of Pittsburgh, Princeton University, the United States Naval Observatory, and the University of Washington.

%%%%%%%%%%%%%%%%%%%%%%%%%%%%%%%%%%%%%%%%%%%%%%%%%%

\section*{Availability of data}
The data underlying this article are available in the article and in its online supplementary material.



\bibliographystyle{mnras}
\bibliography{references}

%%%%%%%%%%%%%%%%%%%%%%%%%%%%%%%%%%%%%%%%%%%%%%%%%%

%%%%%%%%%%%%%%%%% APPENDICES %%%%%%%%%%%%%%%%%%%%%

\appendix

\section{Online Supplementary}
%\section{Additional light-curve information}
%\subsection{Light curves in the natural system}
\begin{figure*}
        \includegraphics[width=2.0\columnwidth]{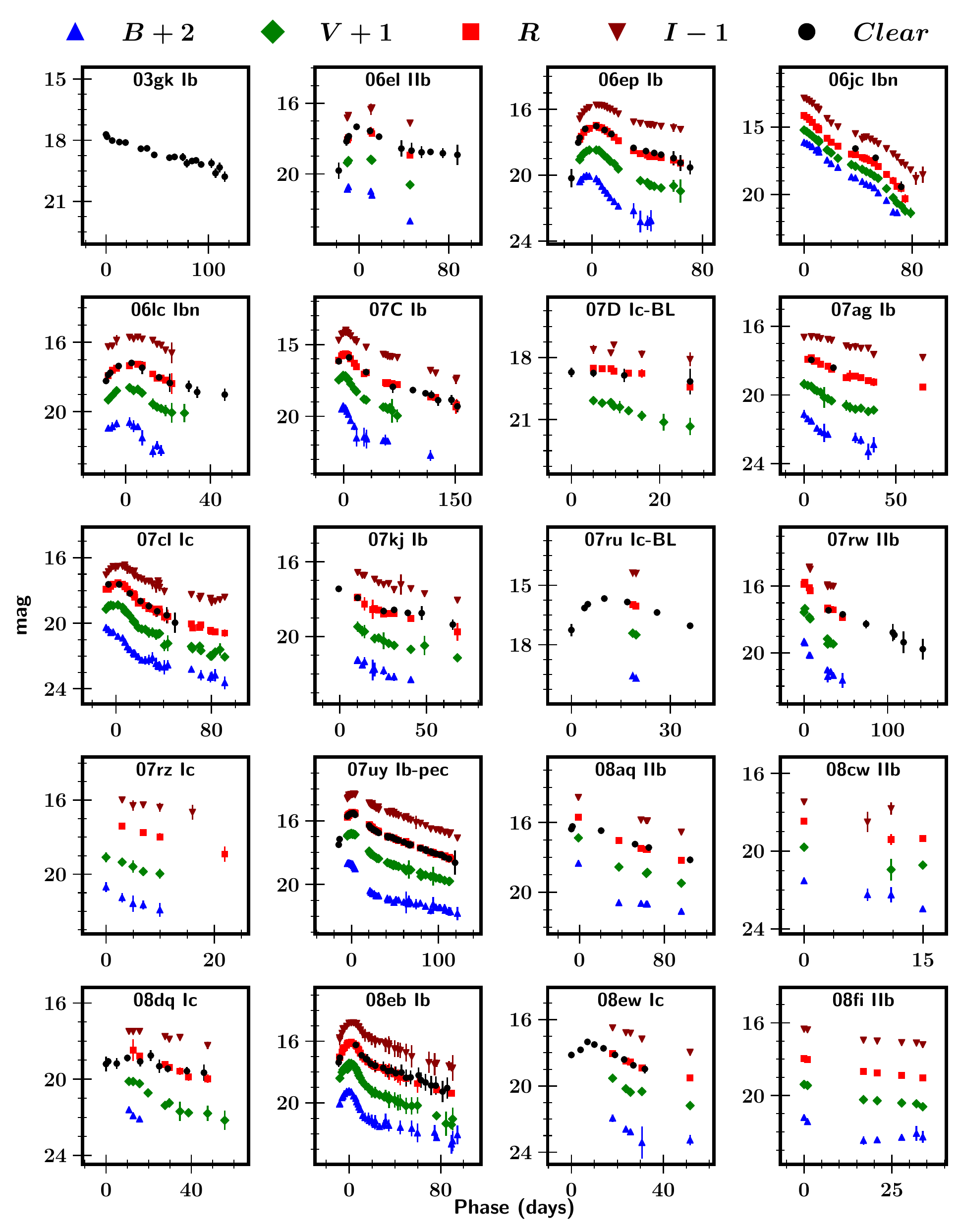}
        \caption{Same as Figure \ref{lc_standard}, but in the natural system.
                 }
\label{lc_natural}
\end{figure*}
\begin{figure*}
\ContinuedFloat
        \includegraphics[width=2.0\columnwidth]{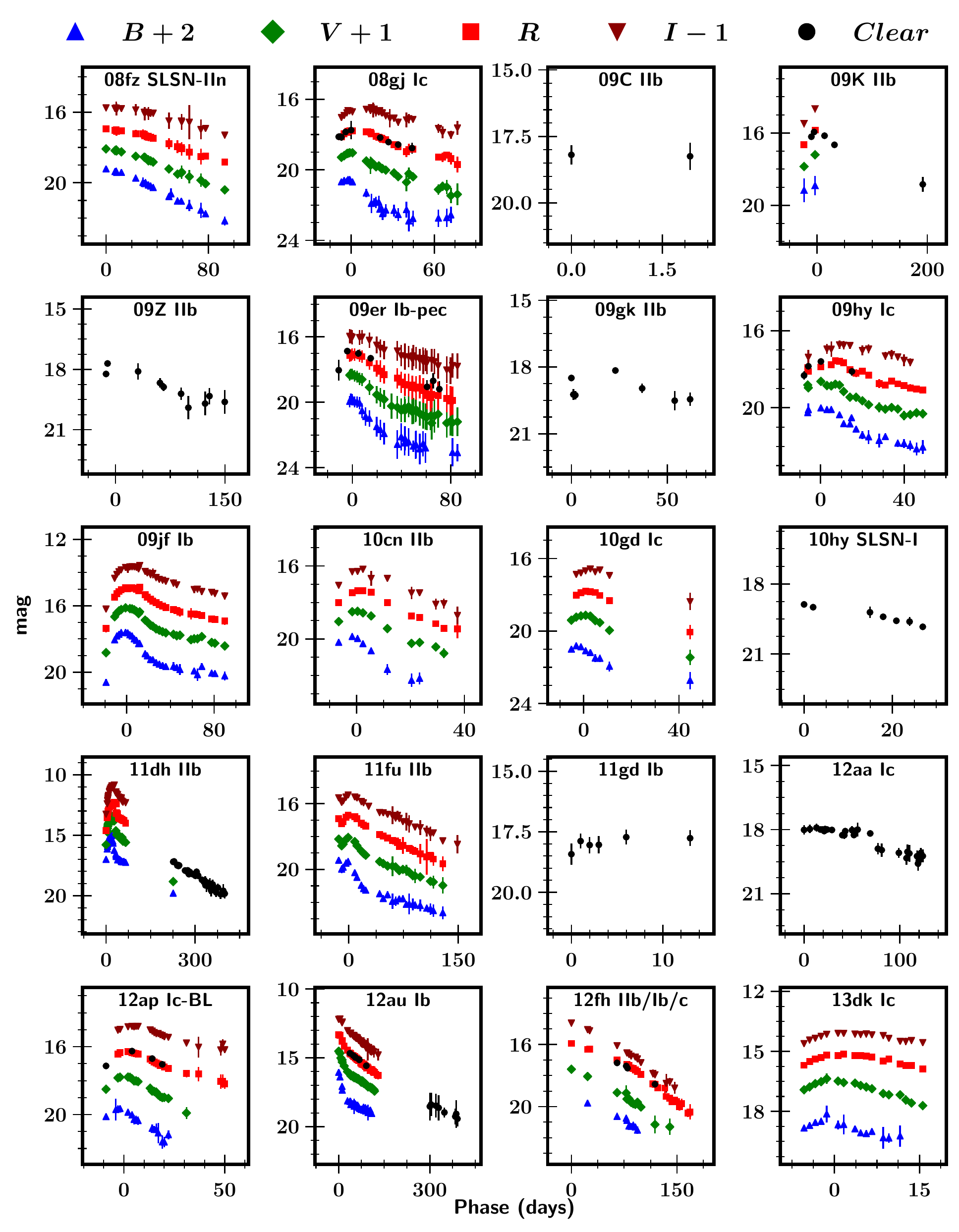}
        \caption{Continued.}
\end{figure*}
\begin{figure*}
\ContinuedFloat
        \includegraphics[width=2.0\columnwidth]{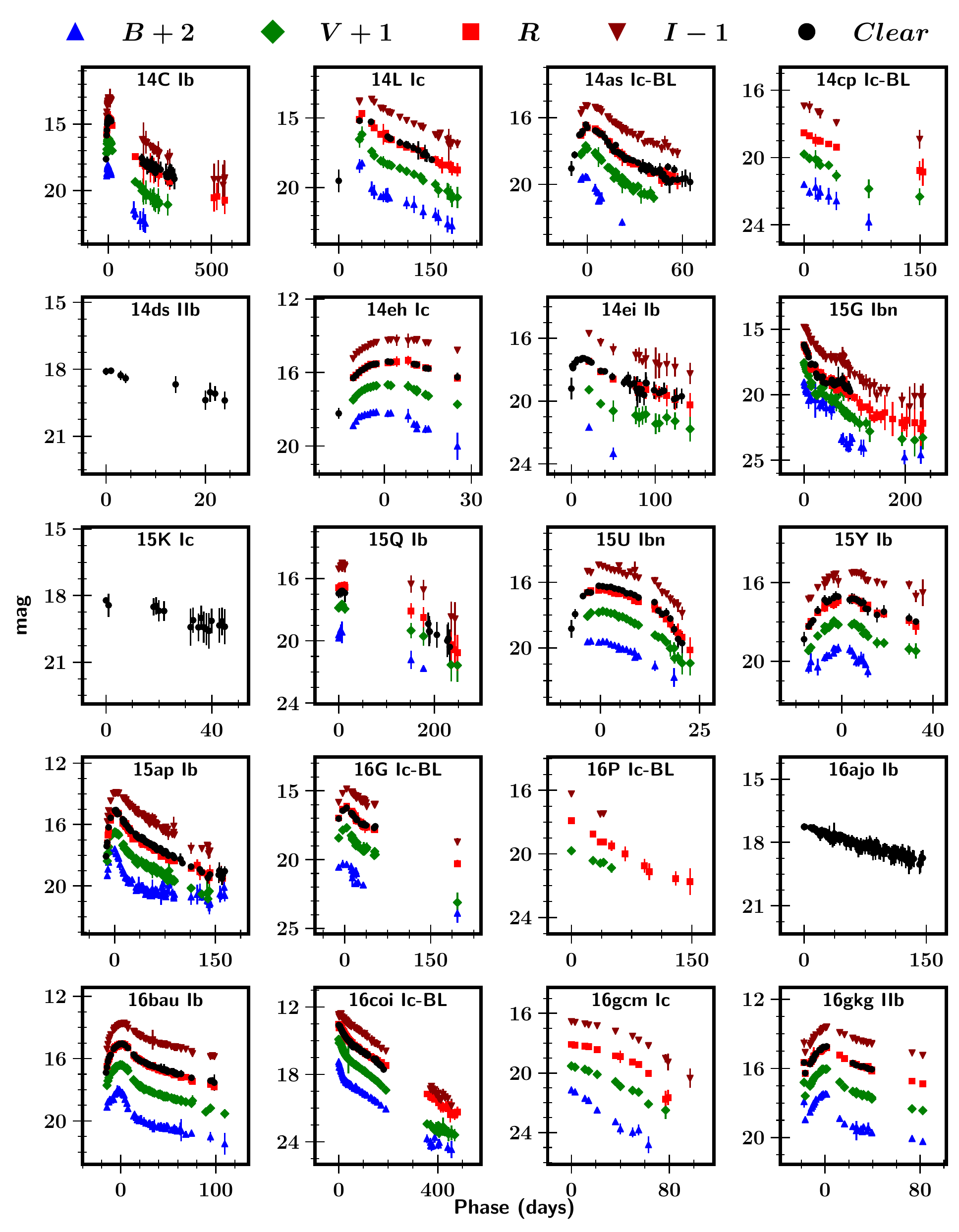}
        \caption{Continued.}
\end{figure*}
\begin{figure*}
\ContinuedFloat
        \includegraphics[width=2.0\columnwidth]{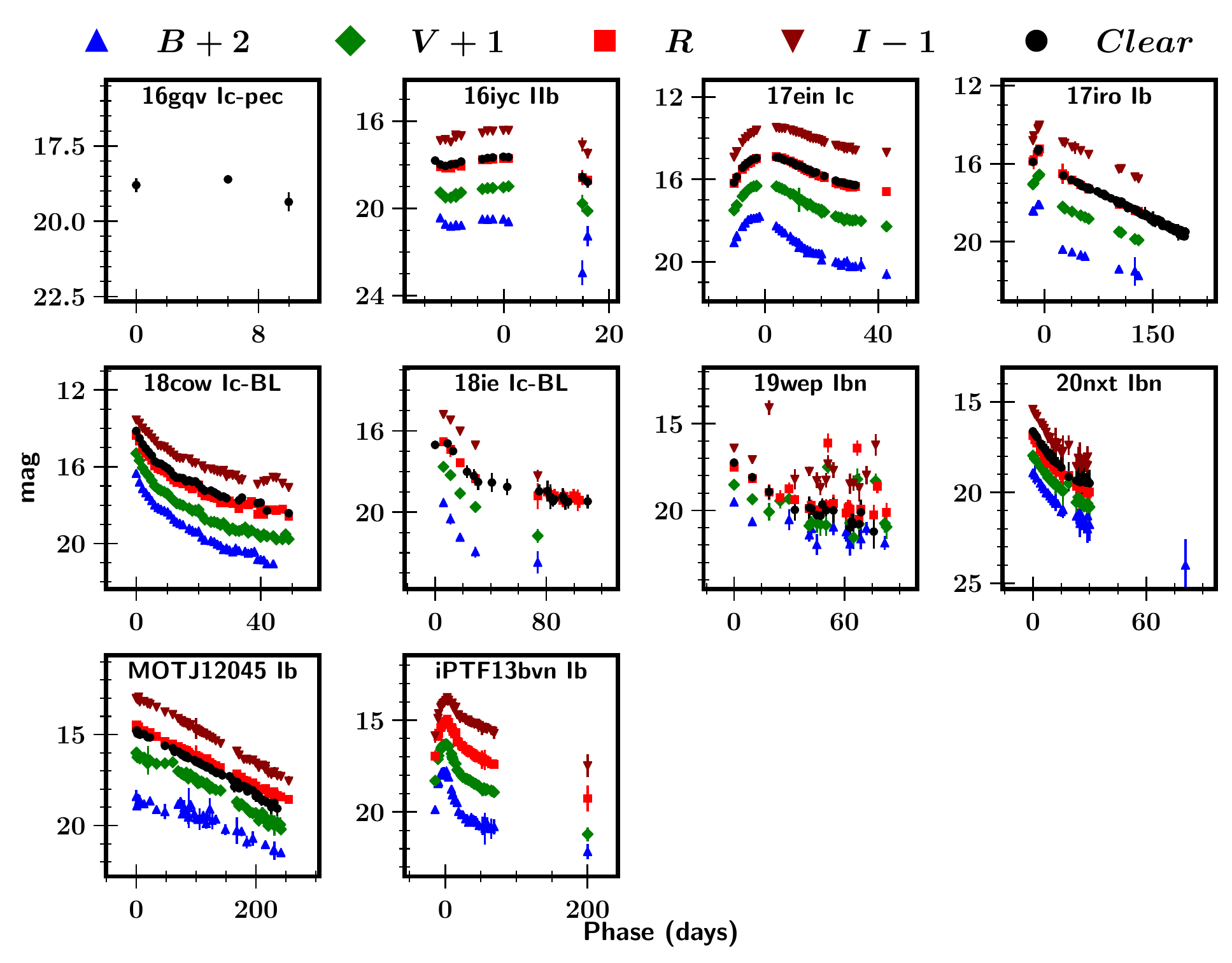}
        \caption{Continued.}
\end{figure*}

%\subsection{Light-curve fitting information}
\begin{figure*}
        \includegraphics[width=0.49\columnwidth]{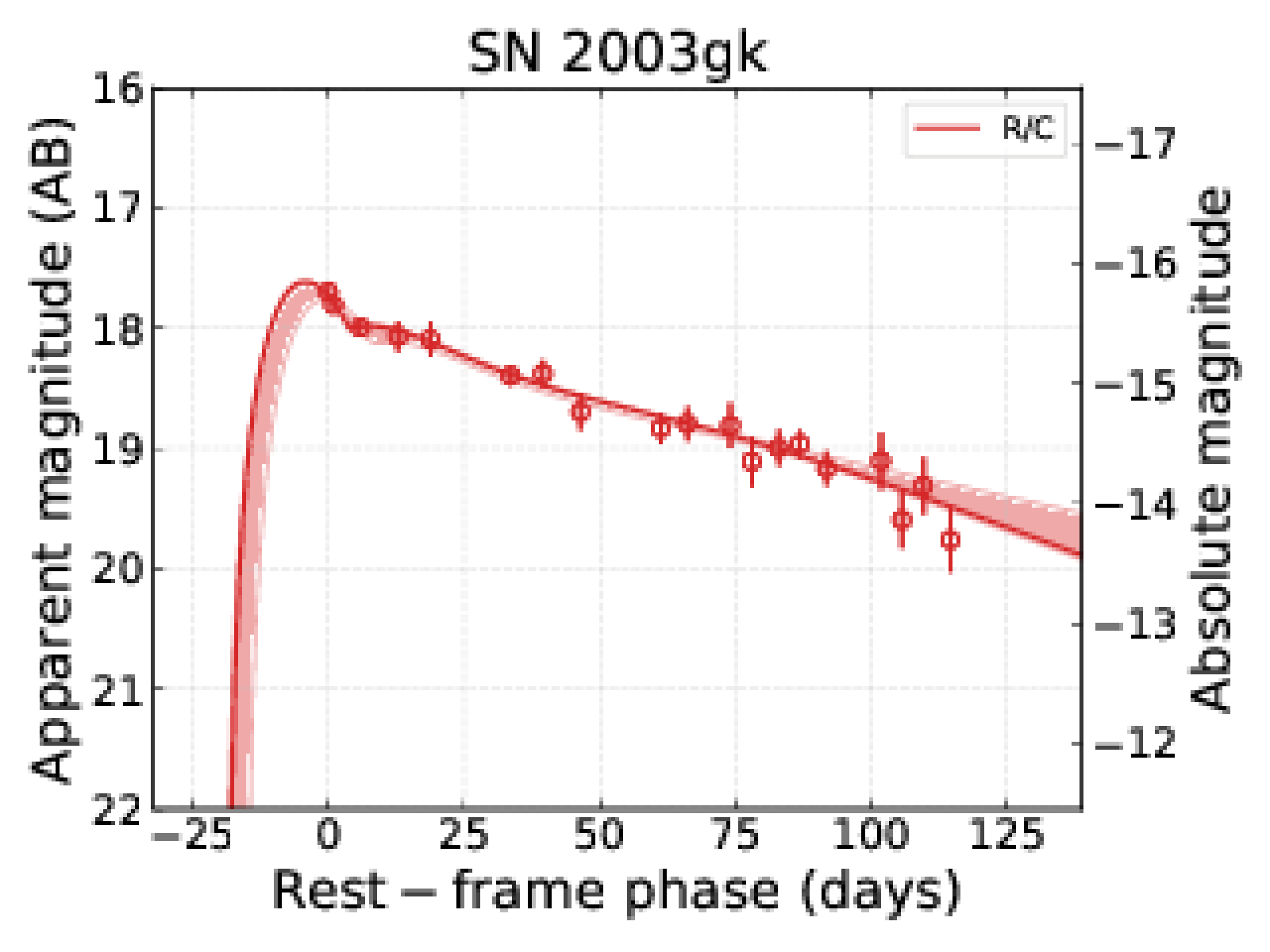}
        \includegraphics[width=0.49\columnwidth]{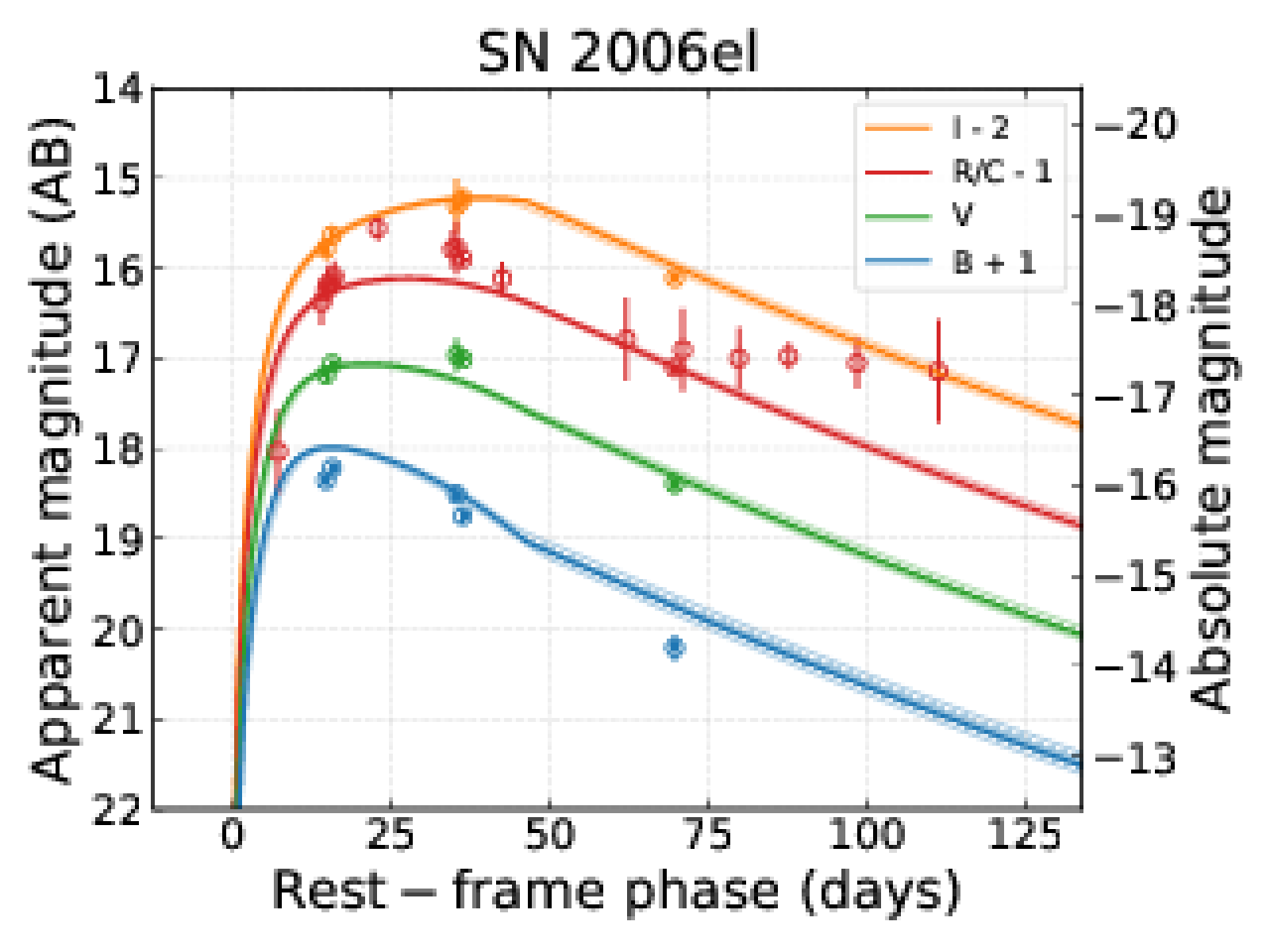}
        \includegraphics[width=0.49\columnwidth]{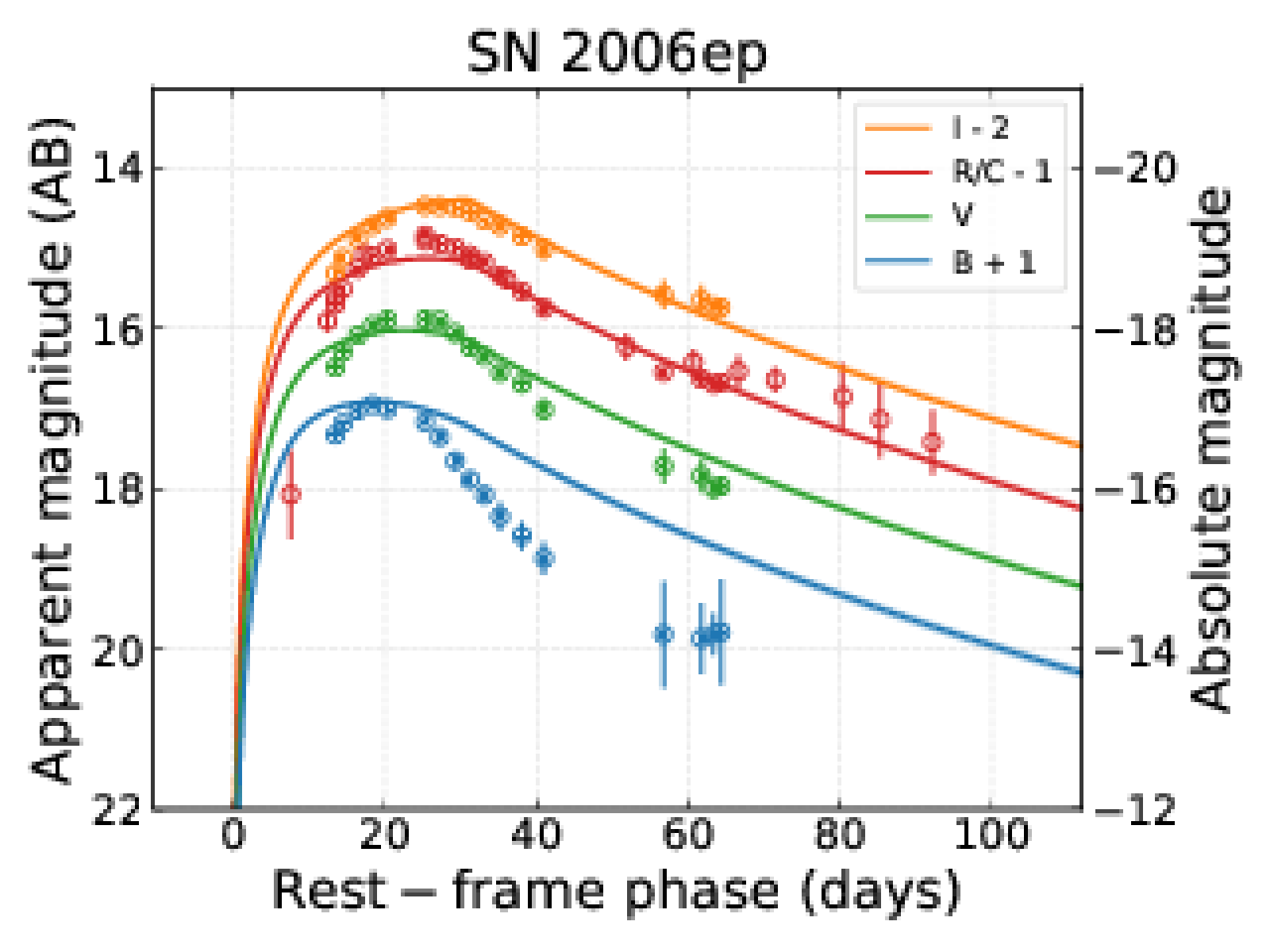}
        \includegraphics[width=0.49\columnwidth]{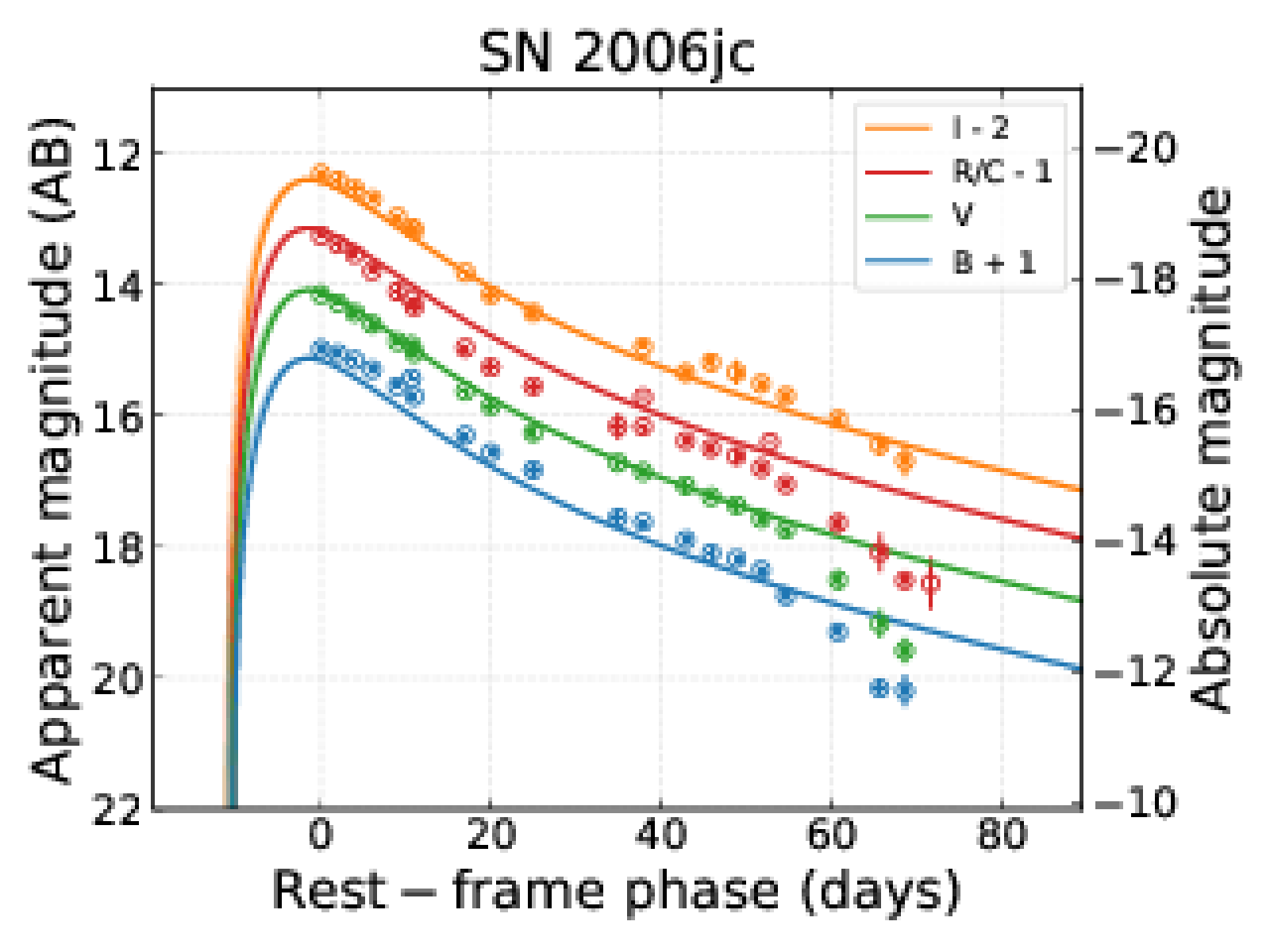}
        \includegraphics[width=0.49\columnwidth]{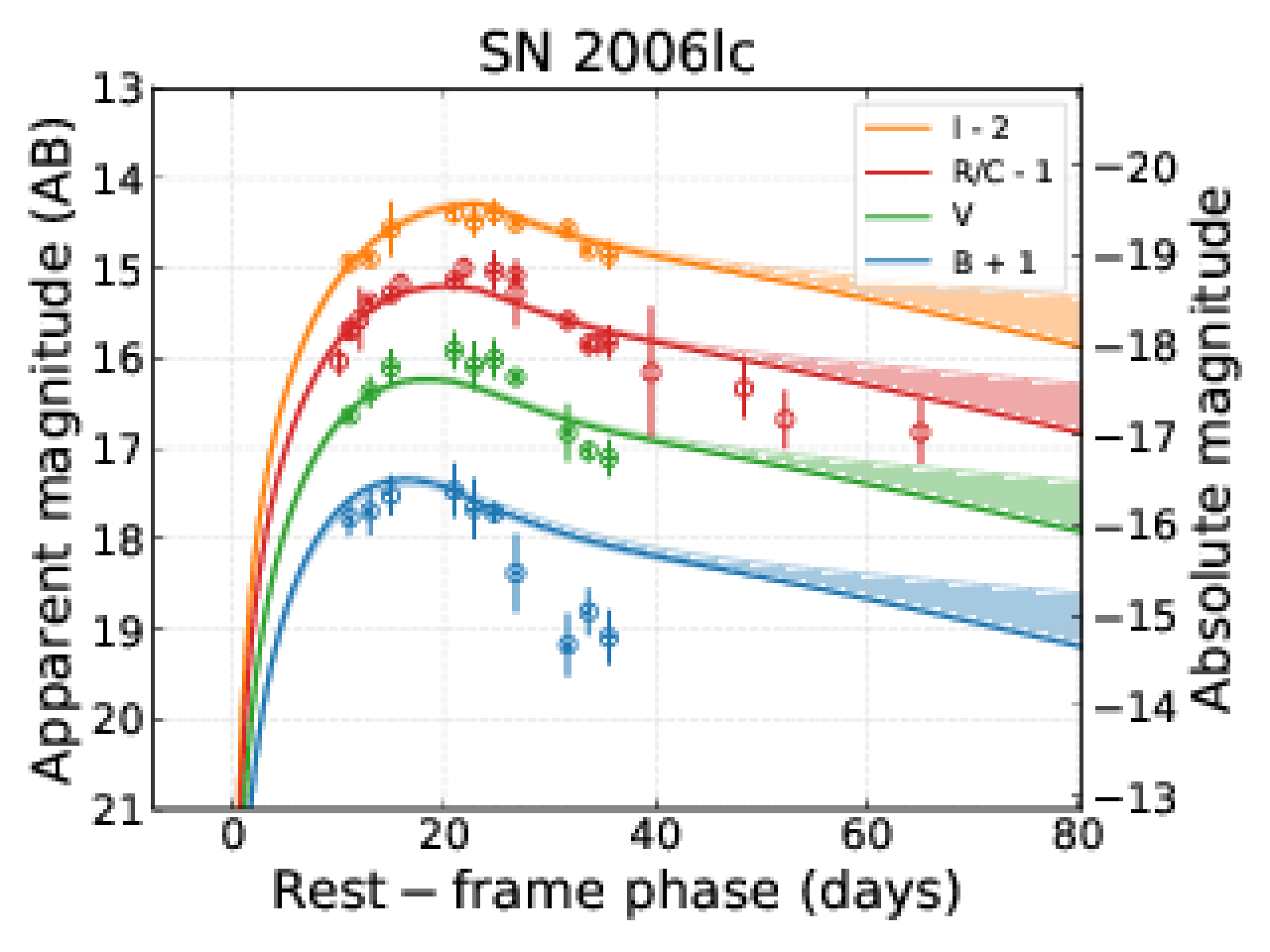}
        \includegraphics[width=0.49\columnwidth]{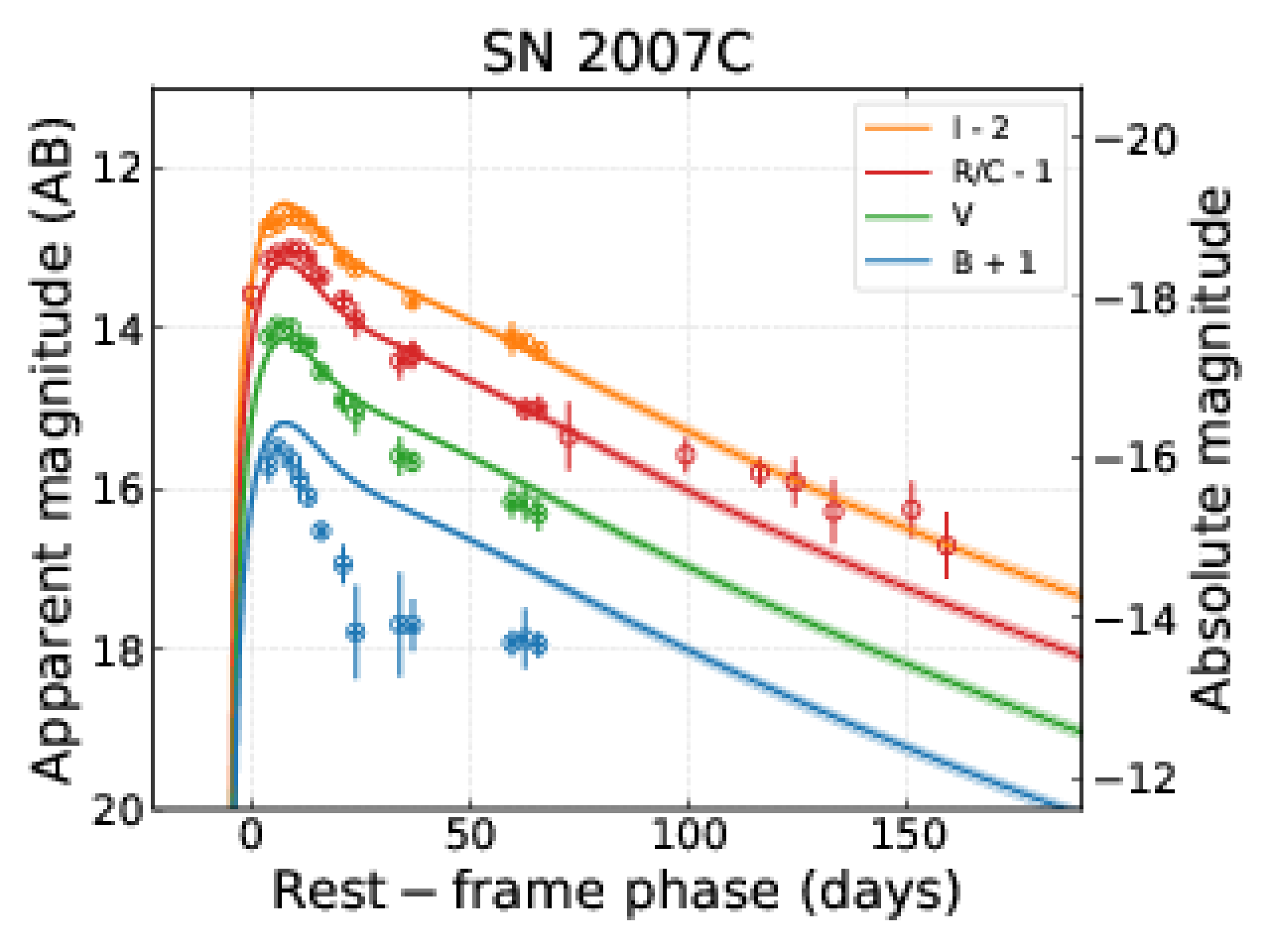}
        \includegraphics[width=0.49\columnwidth]{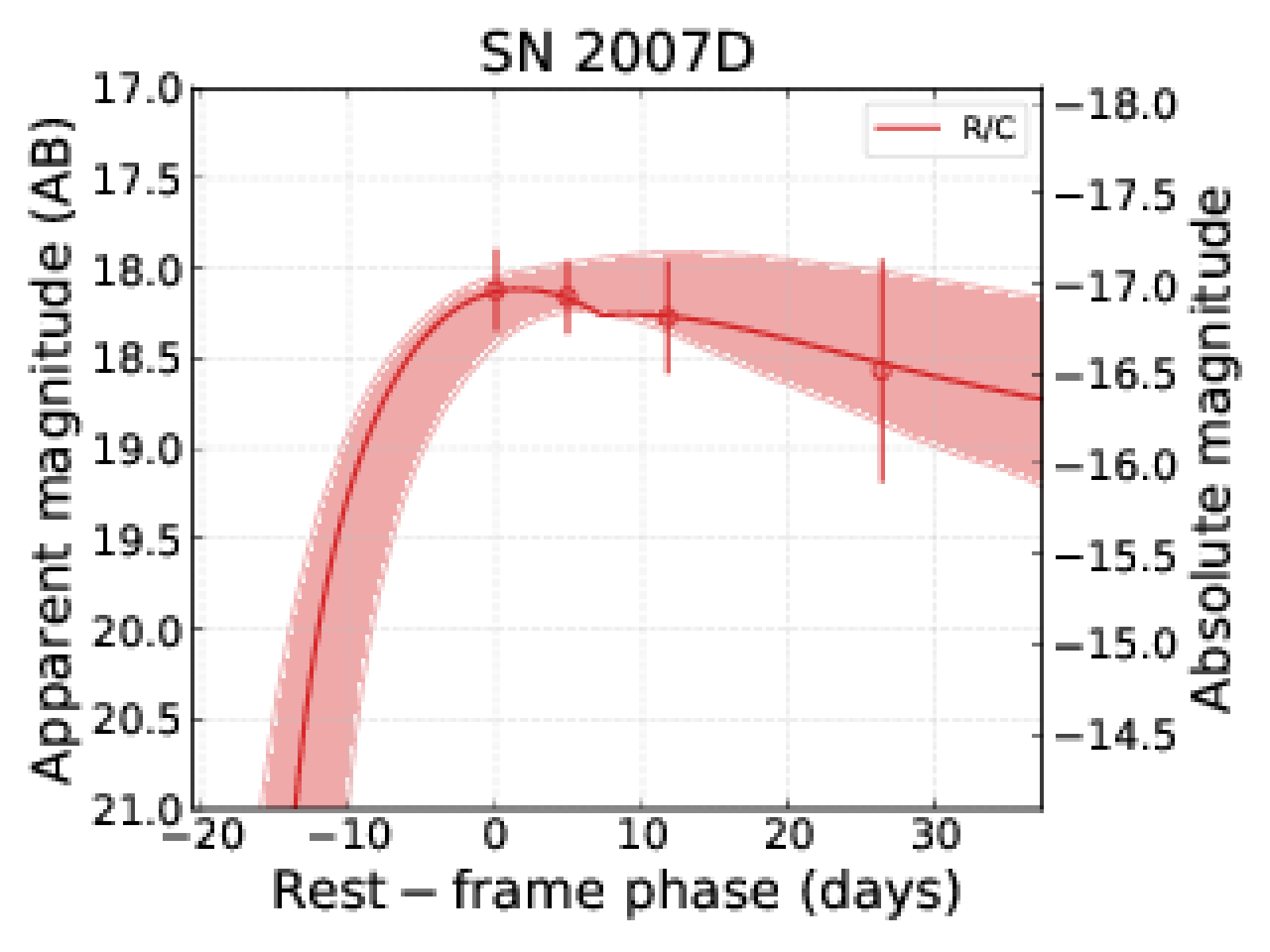}
        \includegraphics[width=0.49\columnwidth]{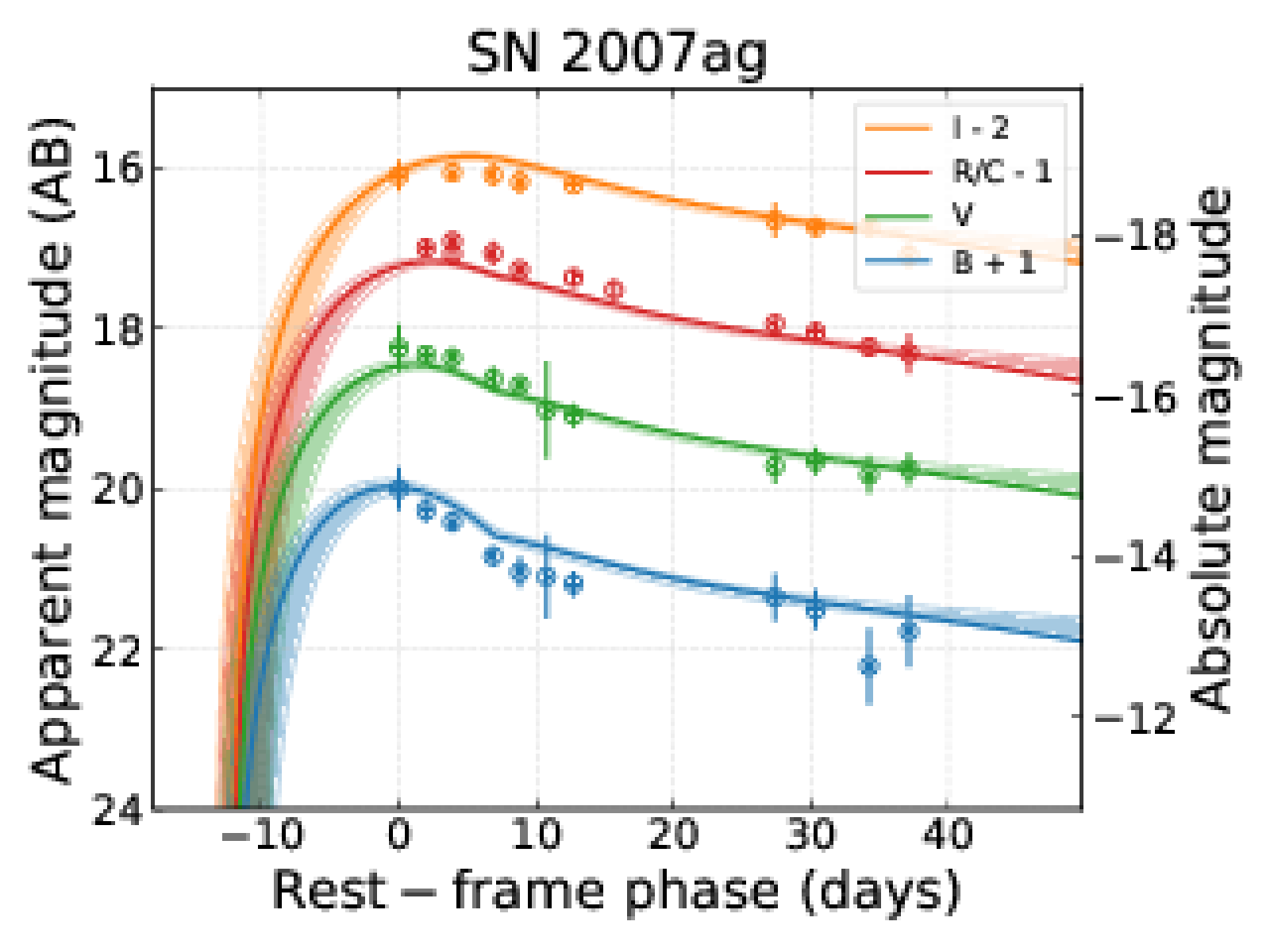}
        \includegraphics[width=0.49\columnwidth]{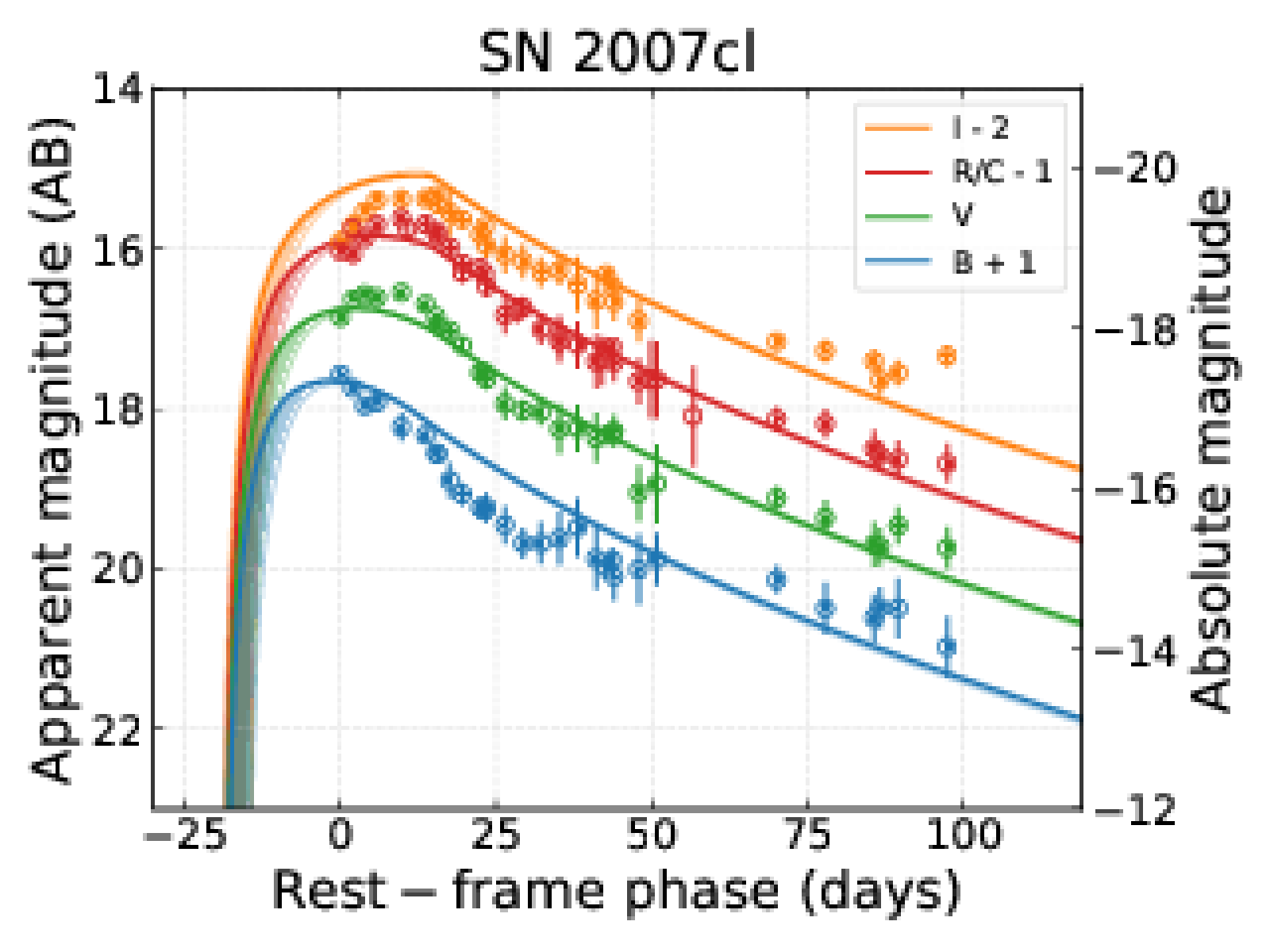}
        \includegraphics[width=0.49\columnwidth]{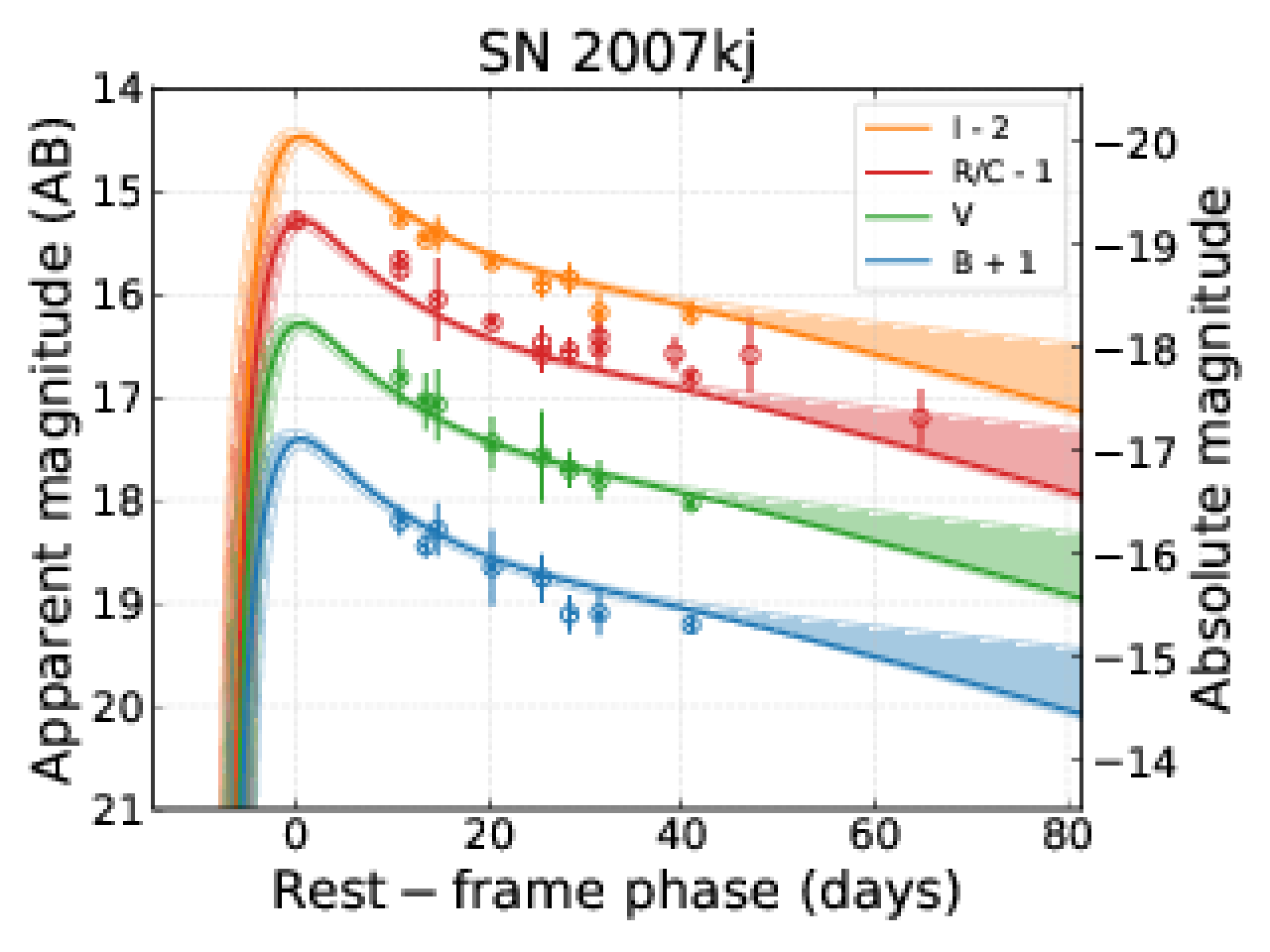}
        \includegraphics[width=0.49\columnwidth]{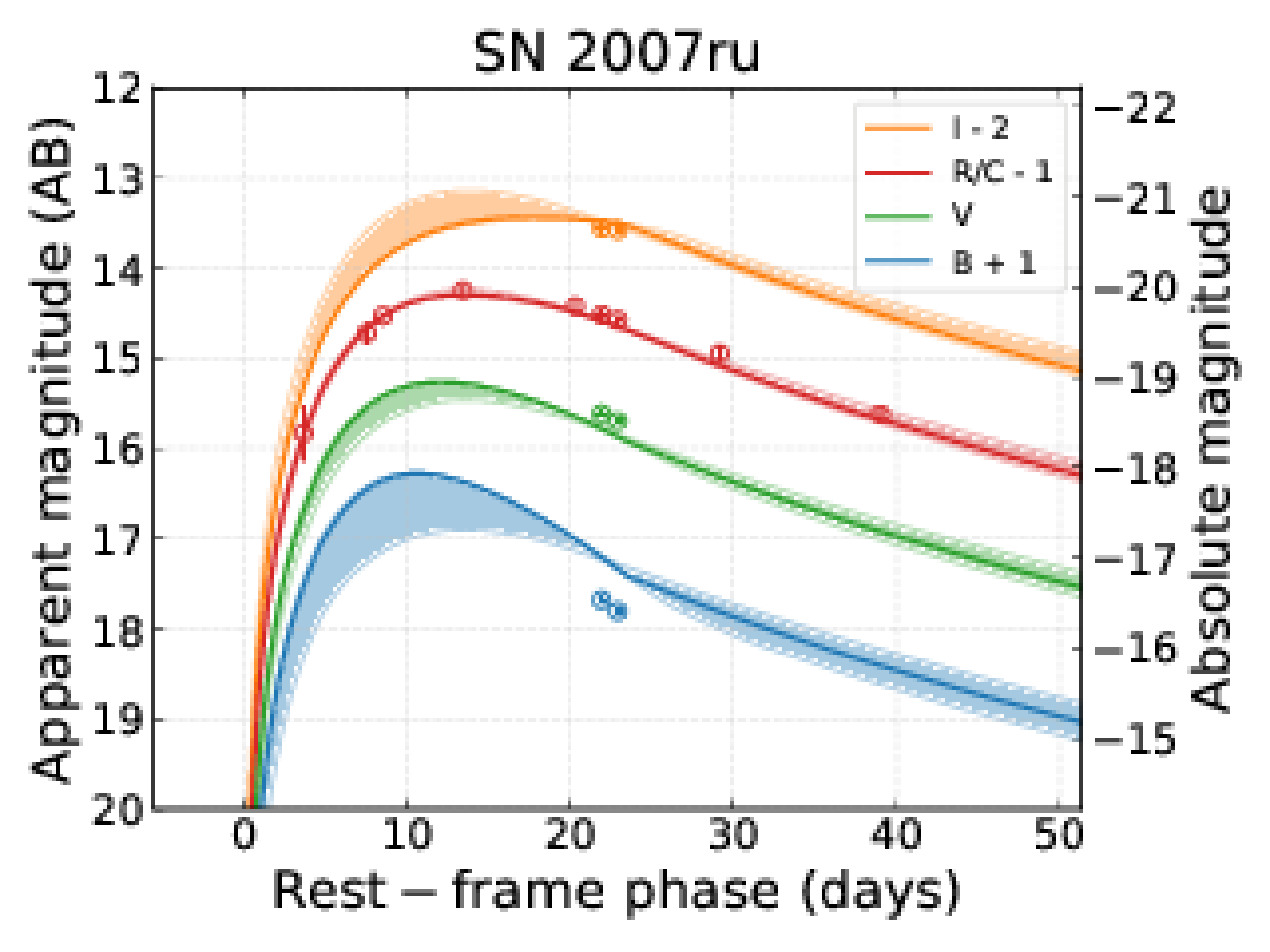}
        \includegraphics[width=0.49\columnwidth]{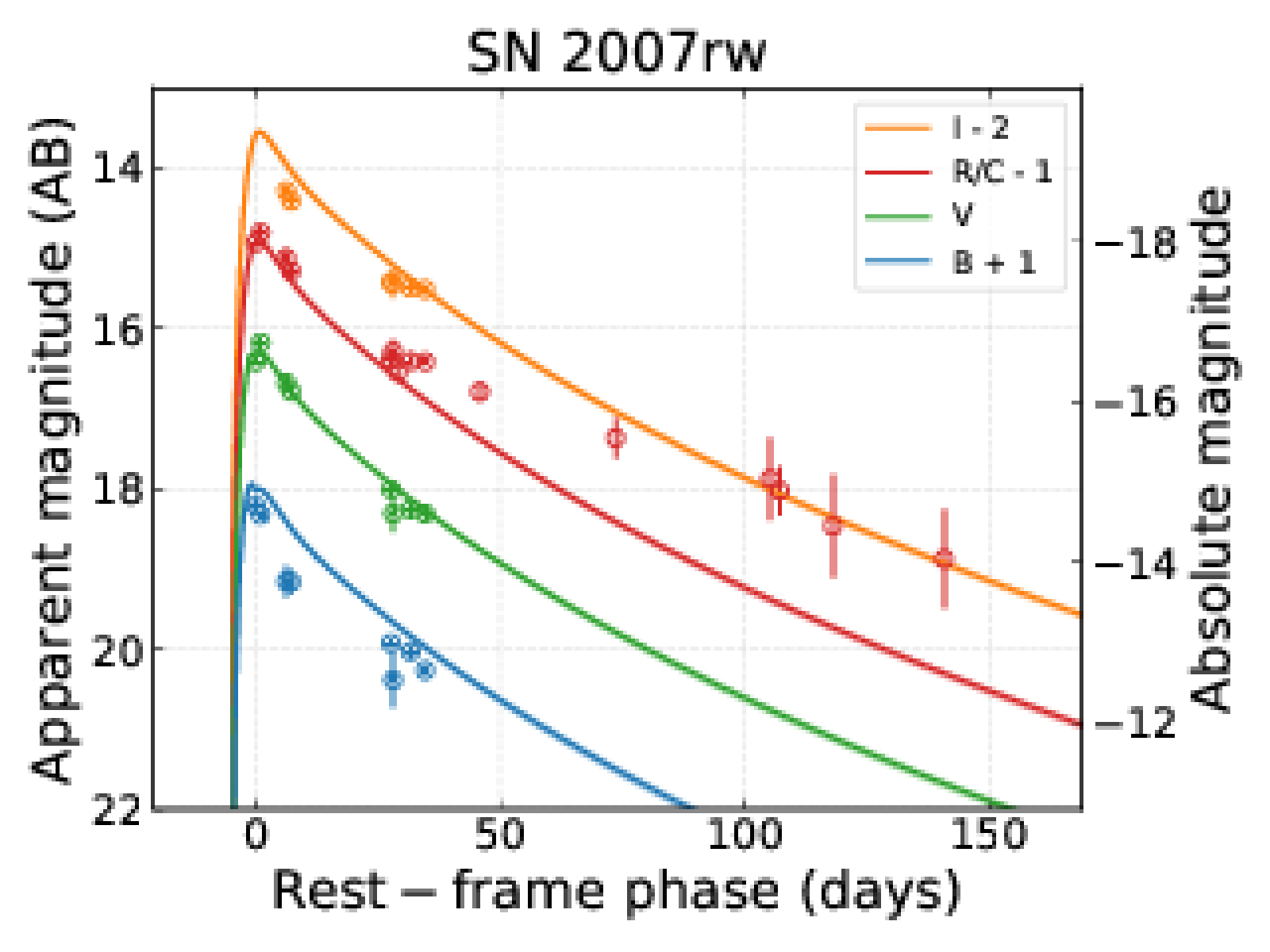}
        \includegraphics[width=0.49\columnwidth]{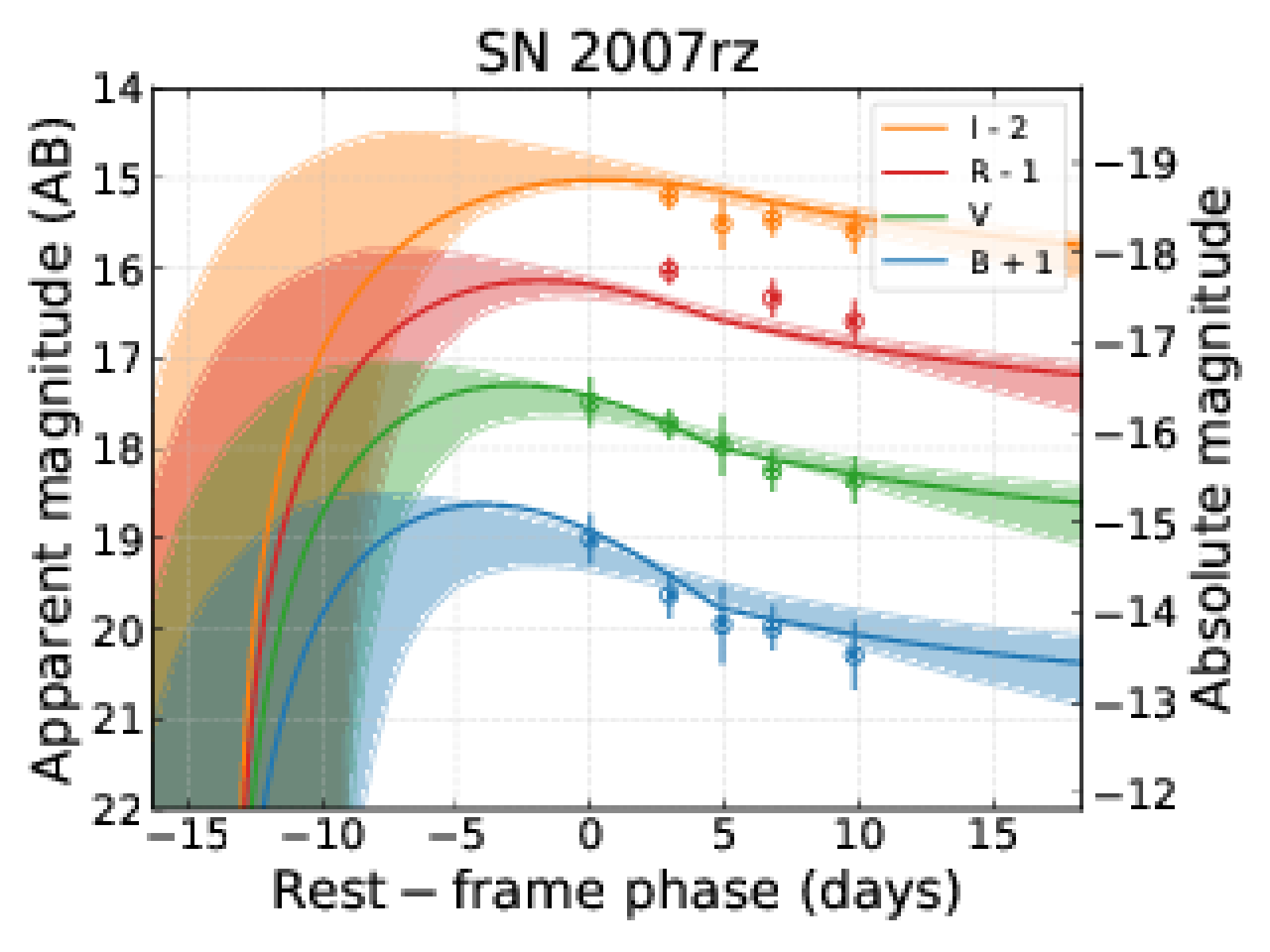}
        \includegraphics[width=0.49\columnwidth]{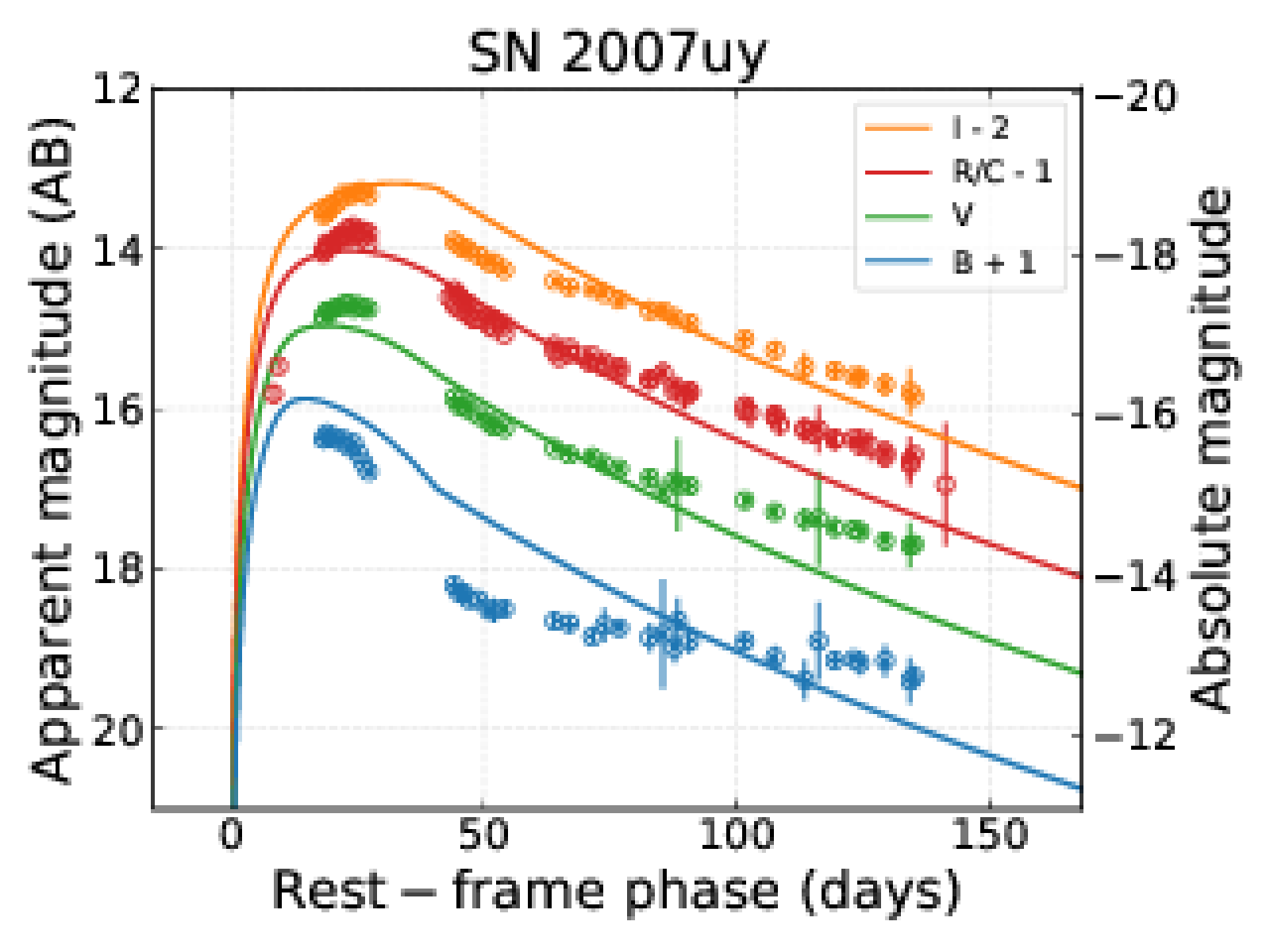}
        \includegraphics[width=0.49\columnwidth]{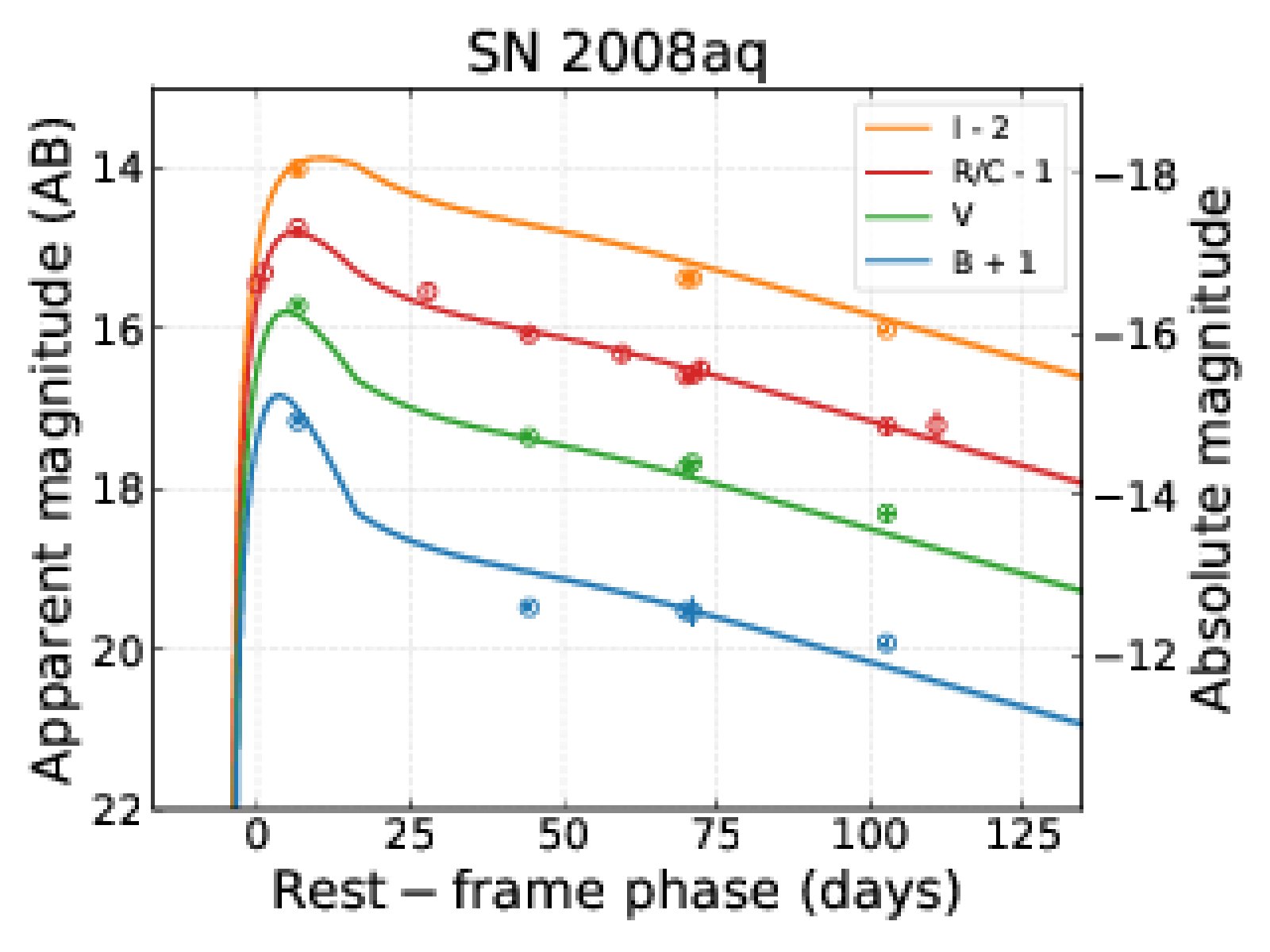}
        \includegraphics[width=0.49\columnwidth]{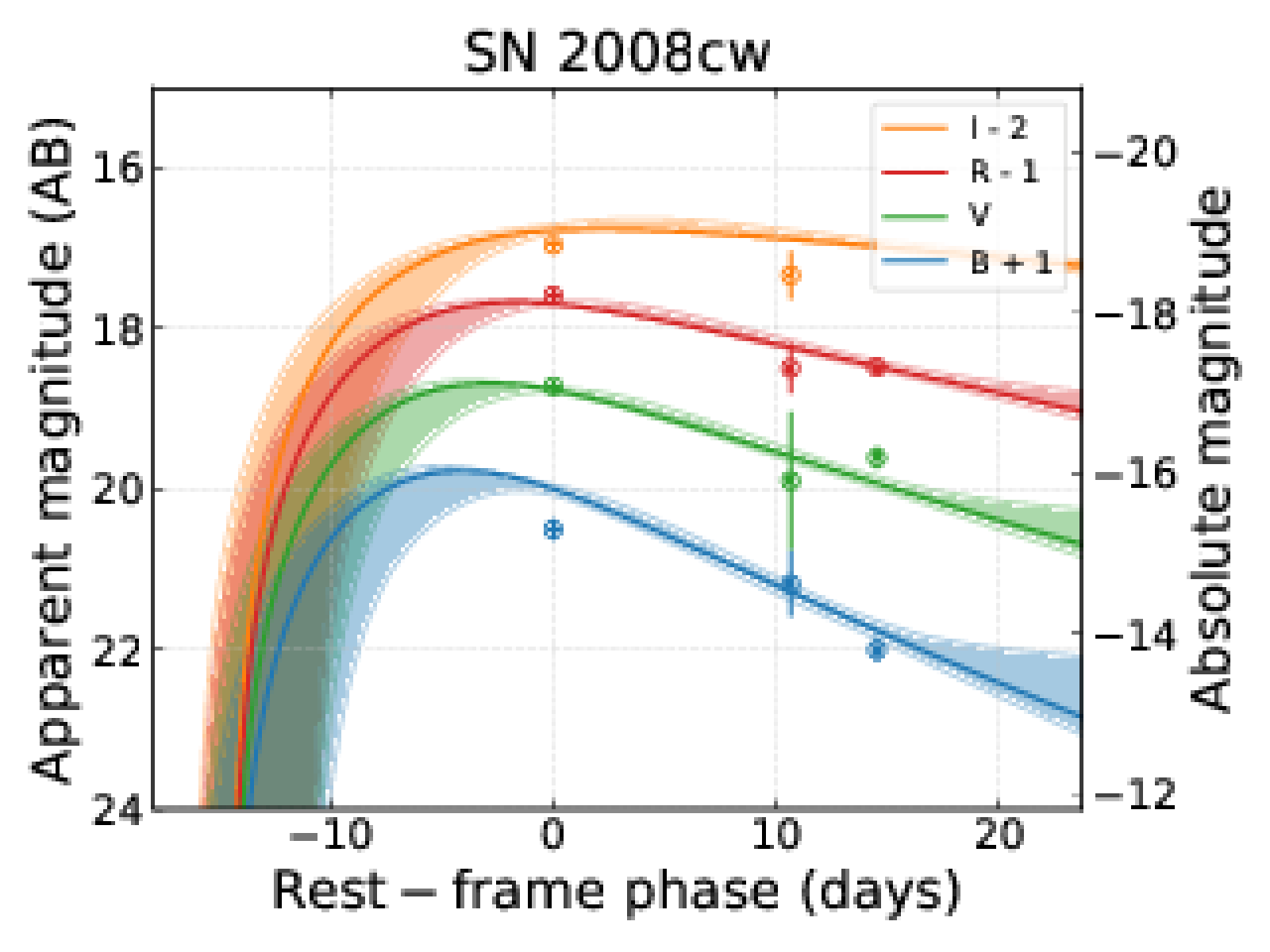}
        \includegraphics[width=0.49\columnwidth]{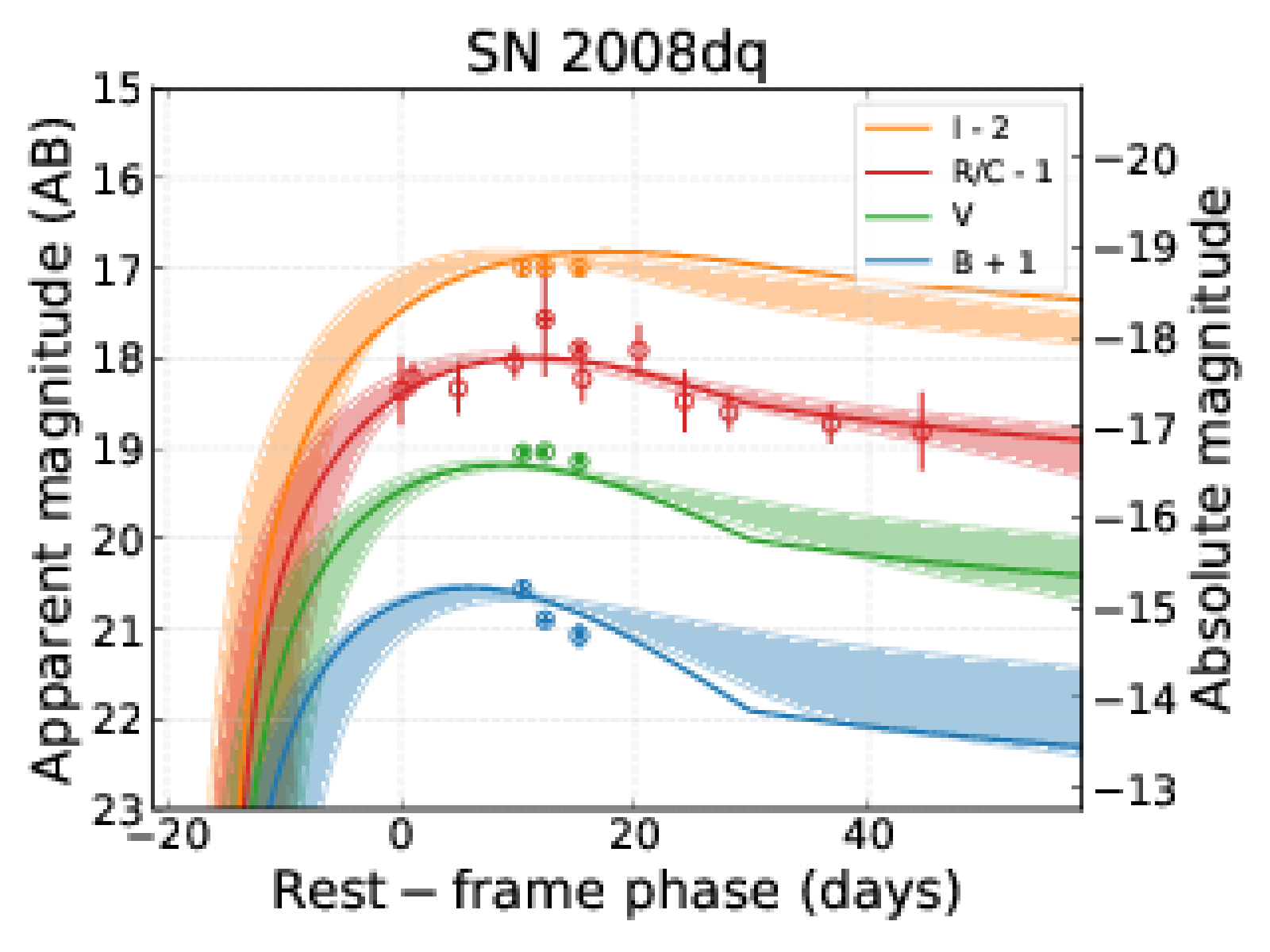}
        \includegraphics[width=0.49\columnwidth]{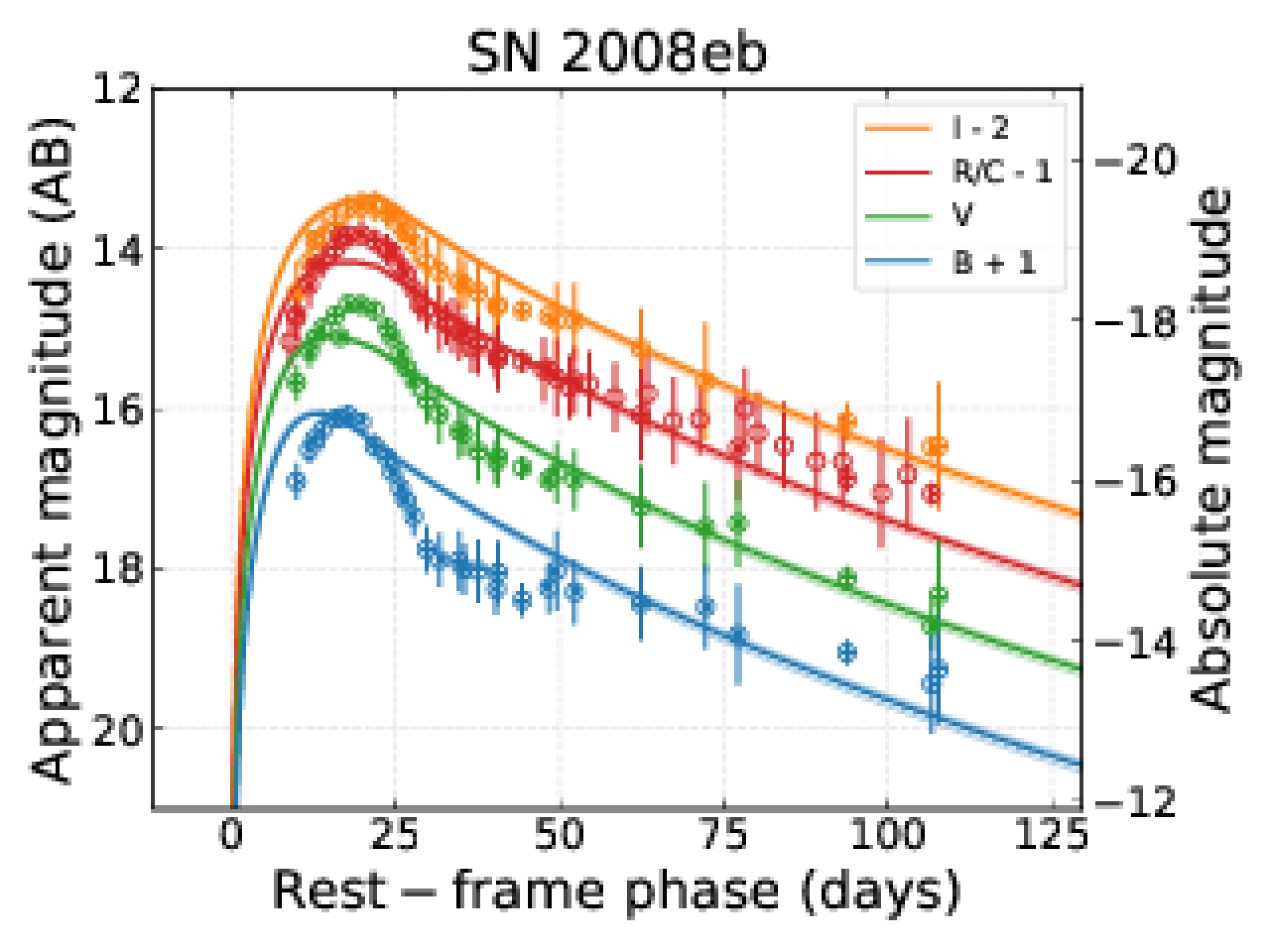}
        \includegraphics[width=0.49\columnwidth]{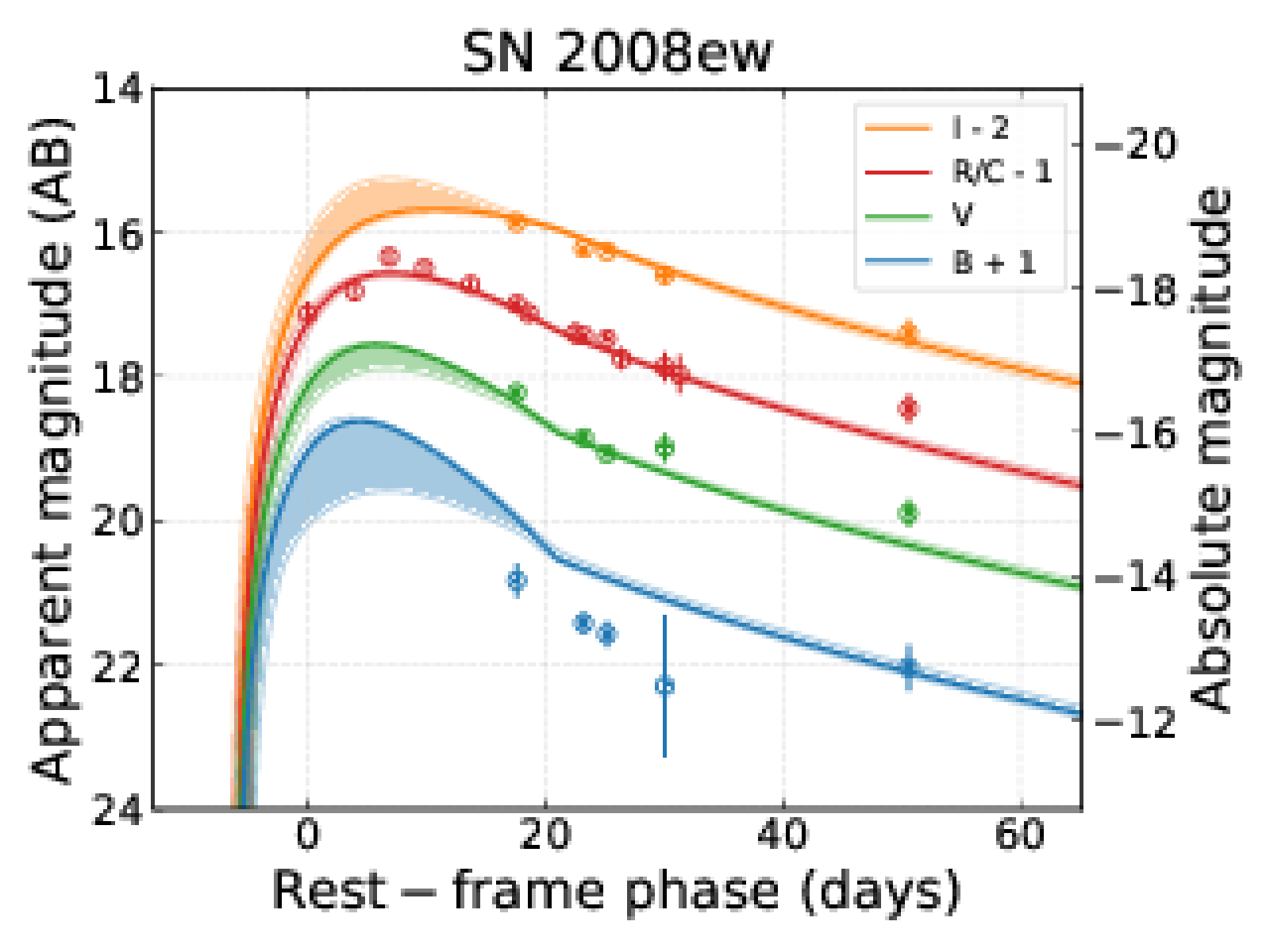}
        \includegraphics[width=0.49\columnwidth]{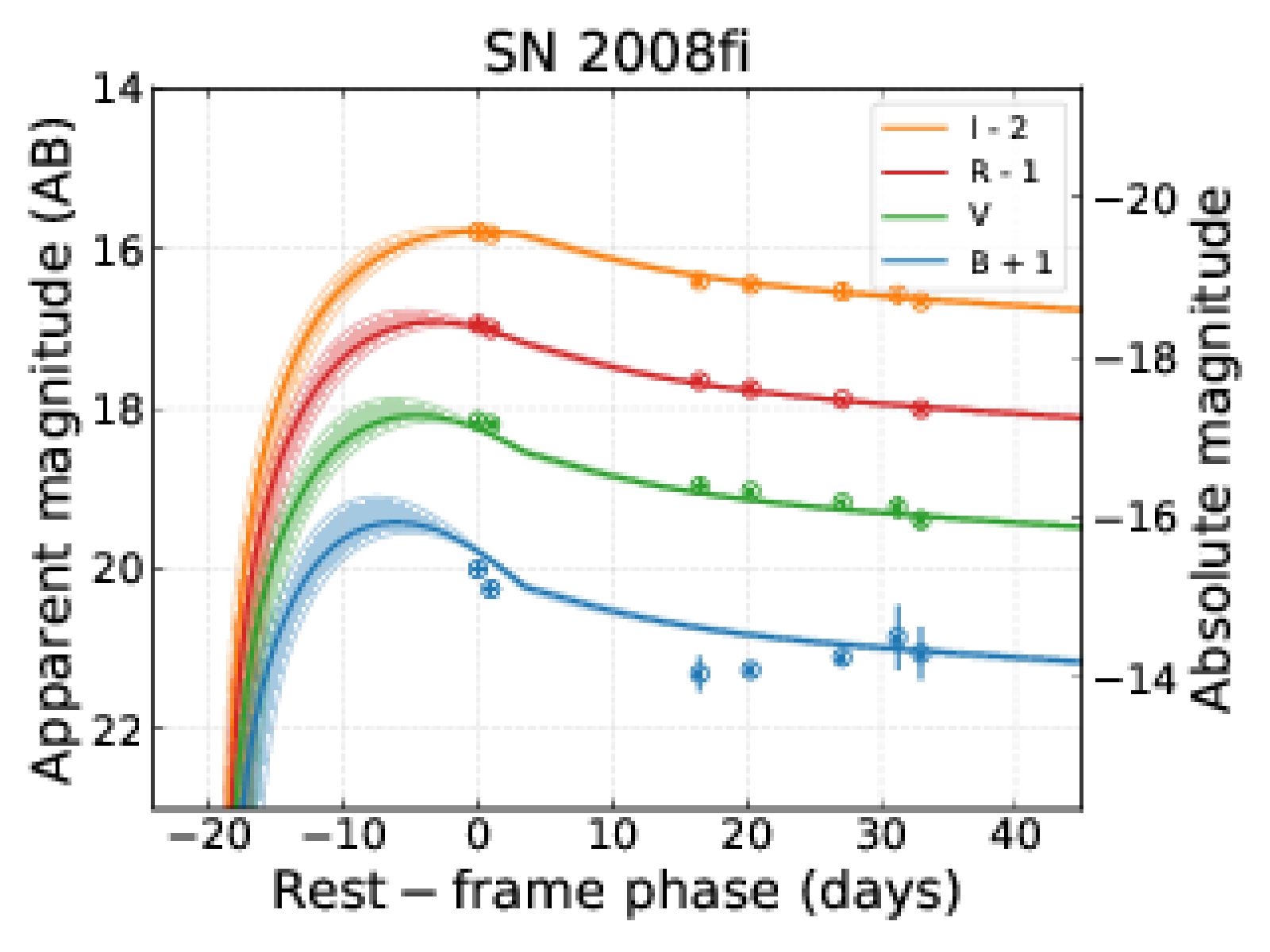}
        \includegraphics[width=0.49\columnwidth]{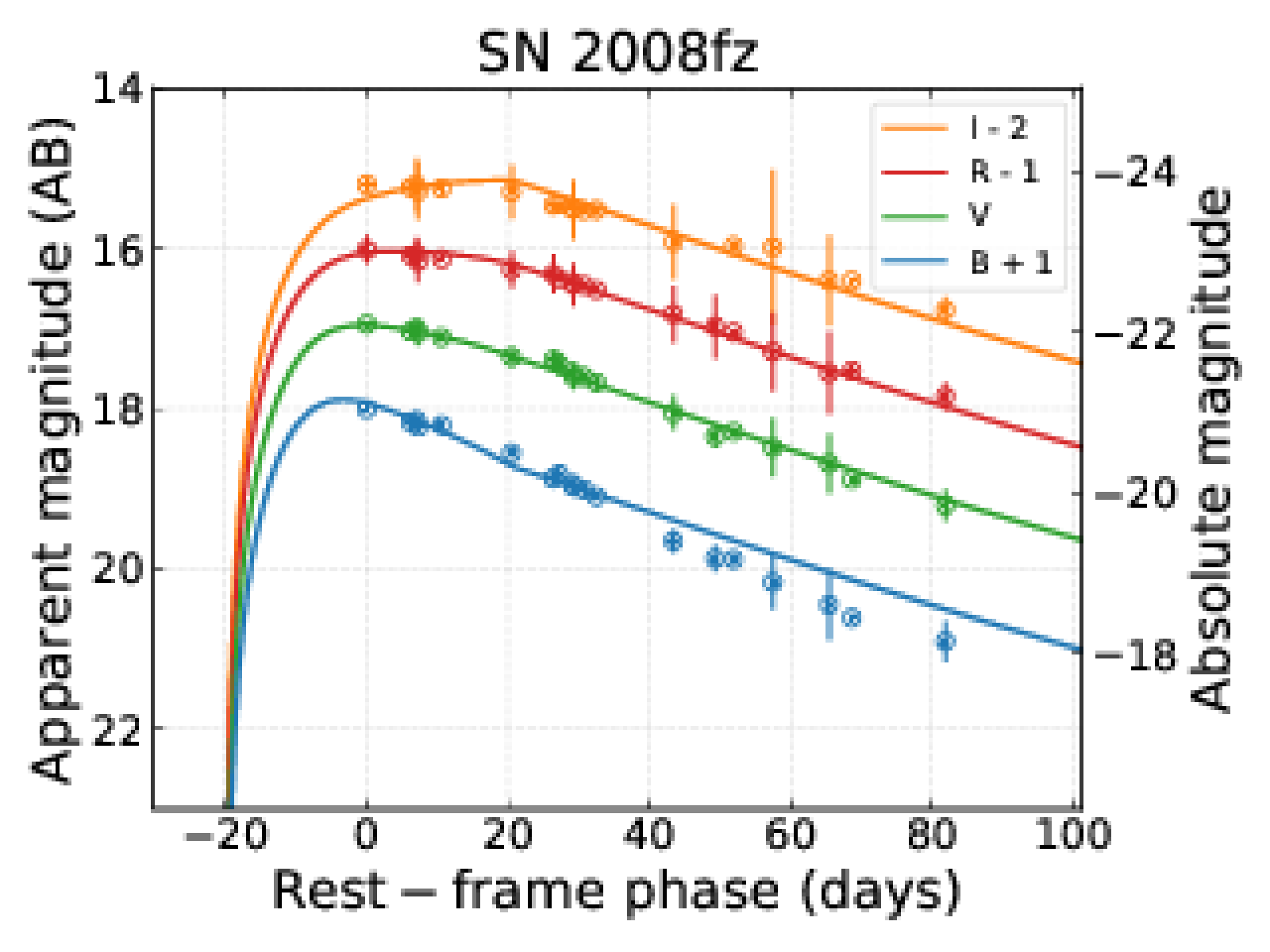}
        \includegraphics[width=0.49\columnwidth]{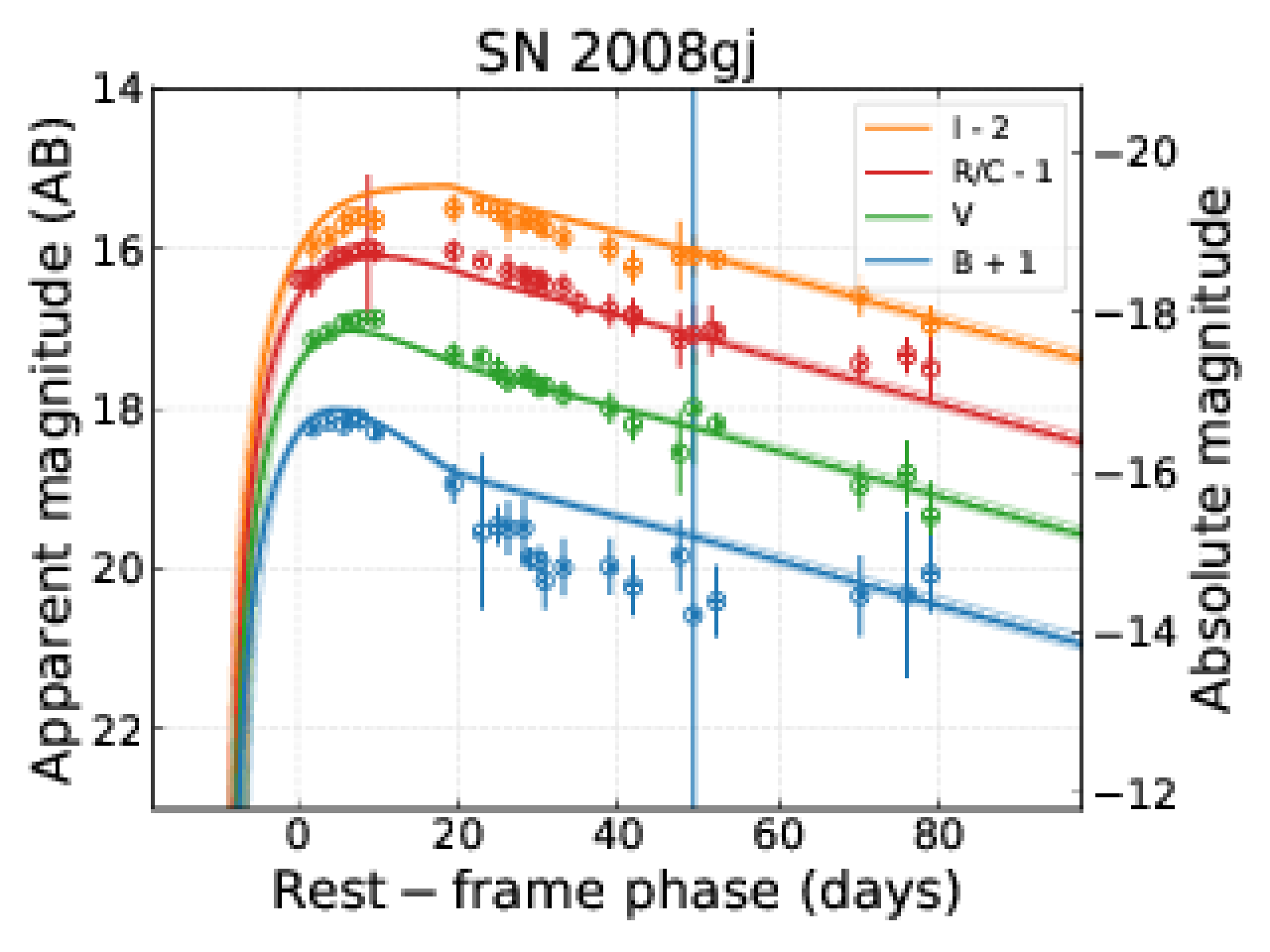}
        \includegraphics[width=0.49\columnwidth]{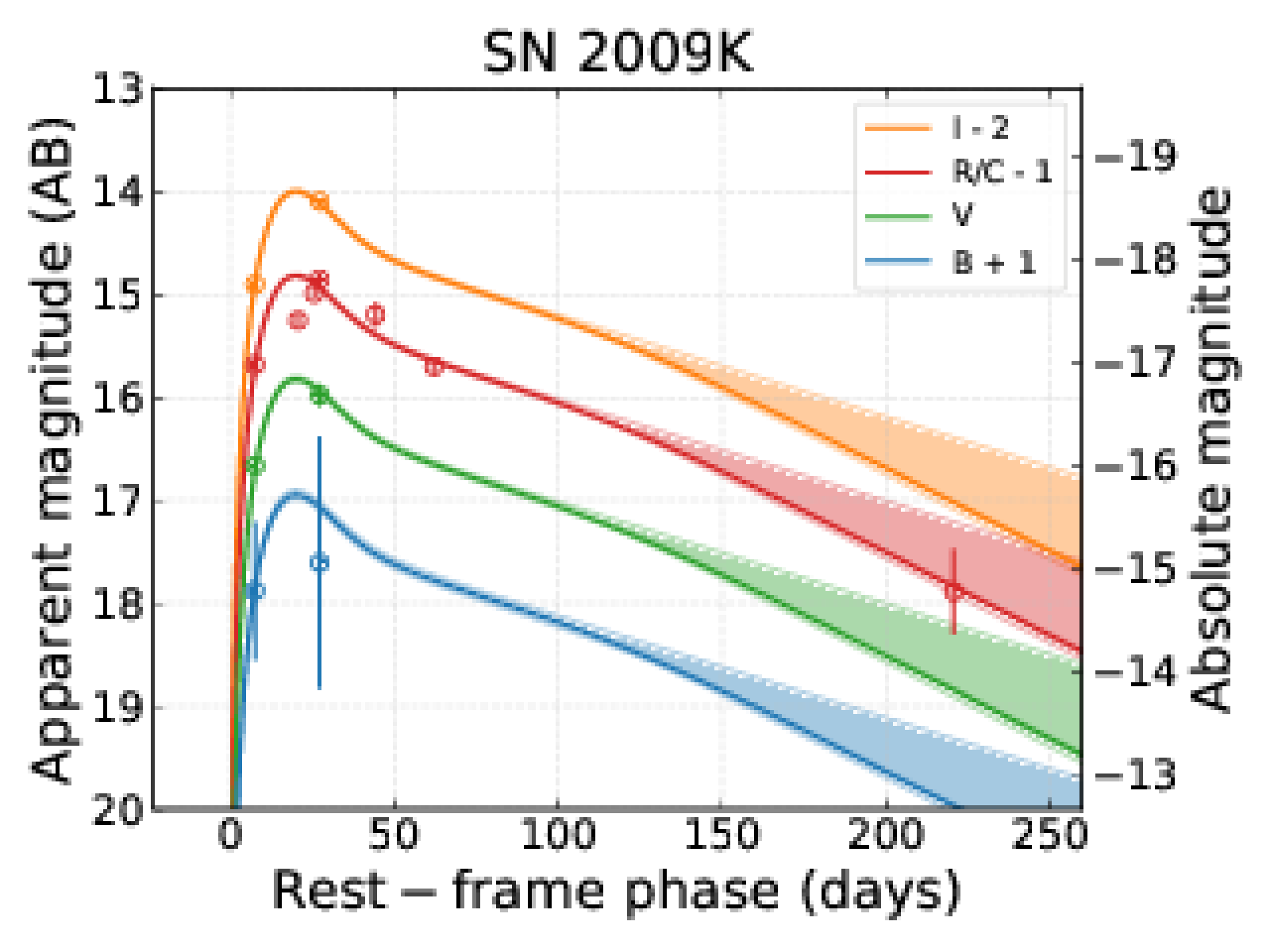}
        \includegraphics[width=0.49\columnwidth]{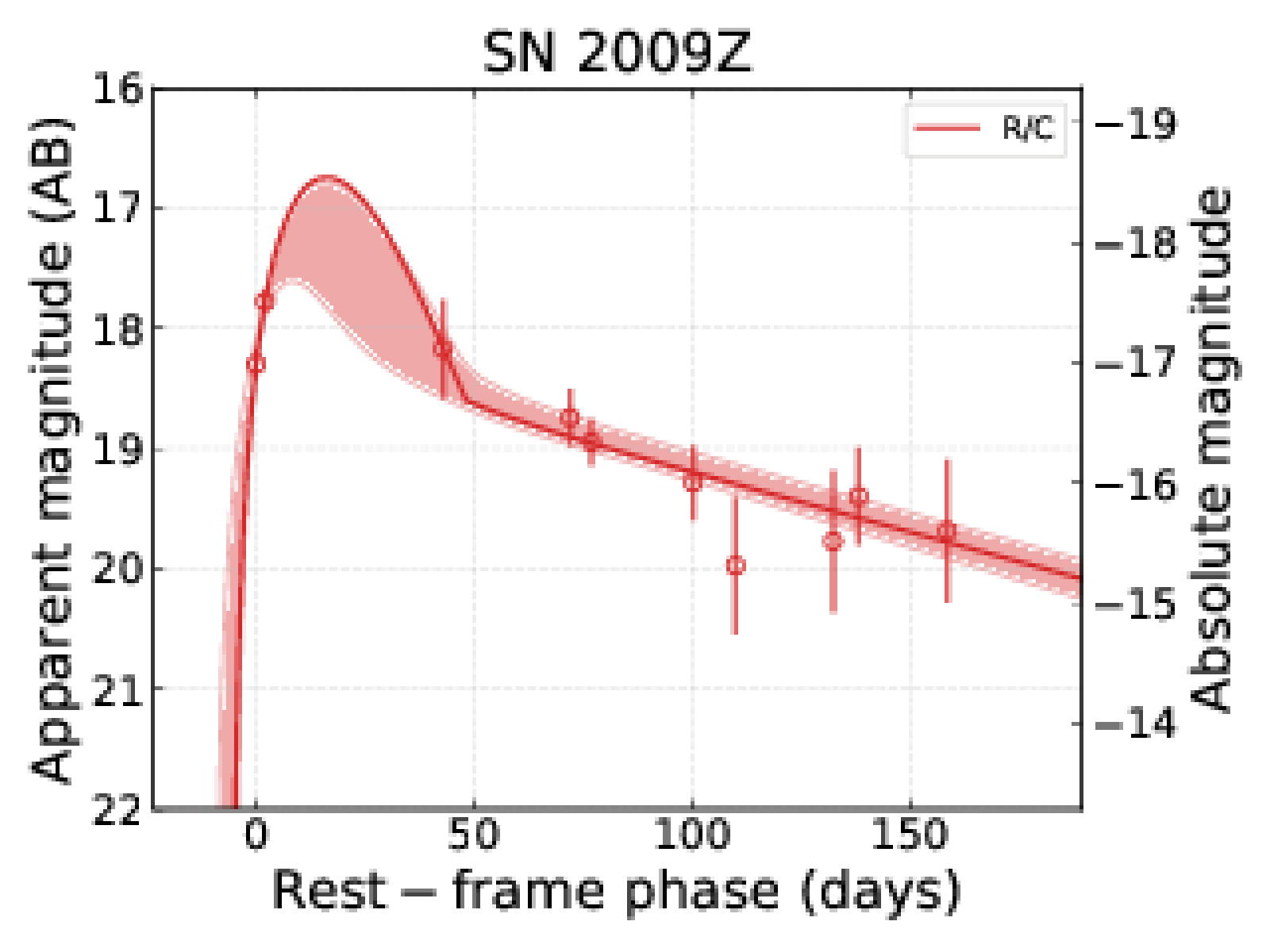}
        \includegraphics[width=0.49\columnwidth]{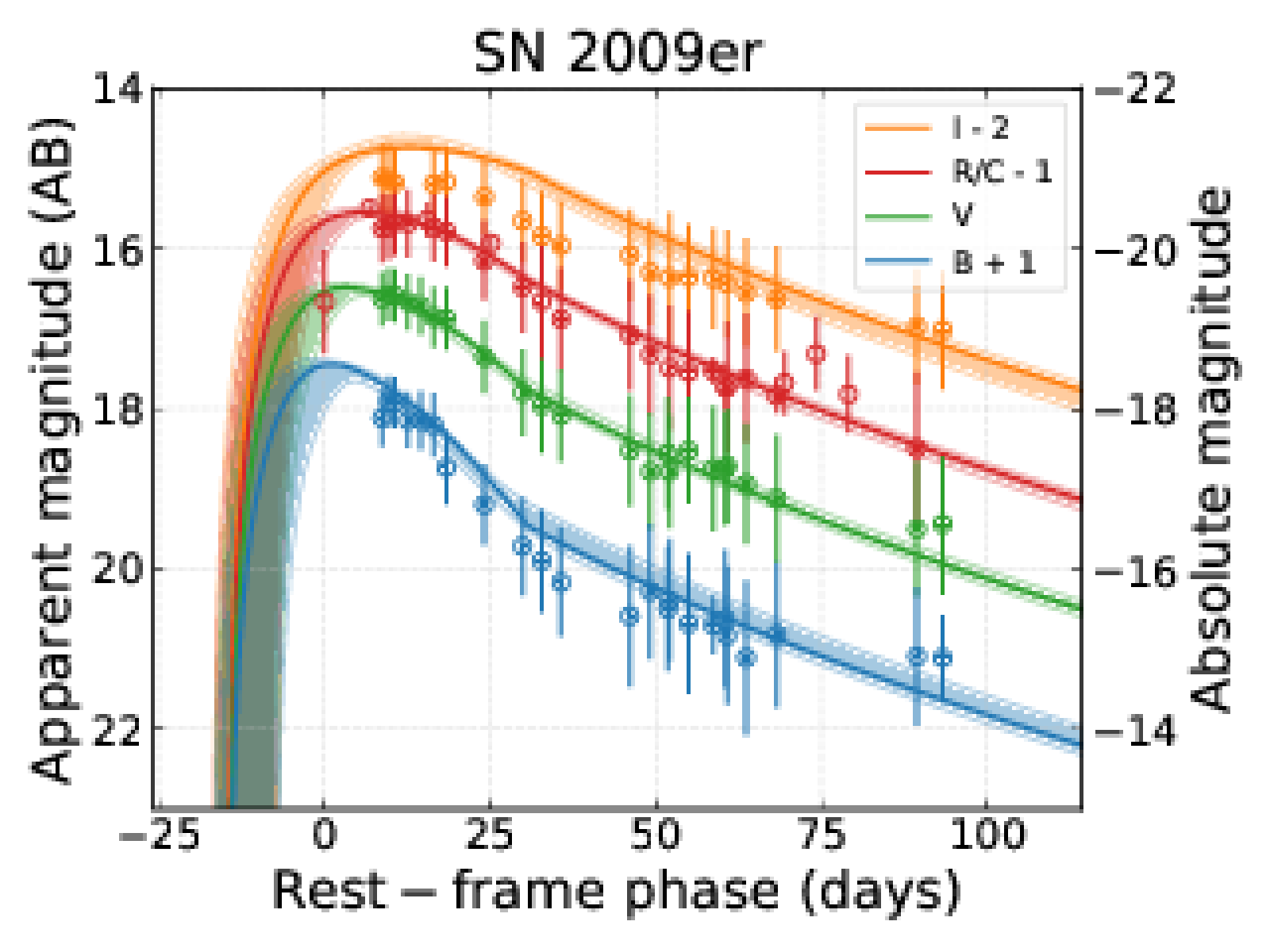}
        \includegraphics[width=0.49\columnwidth]{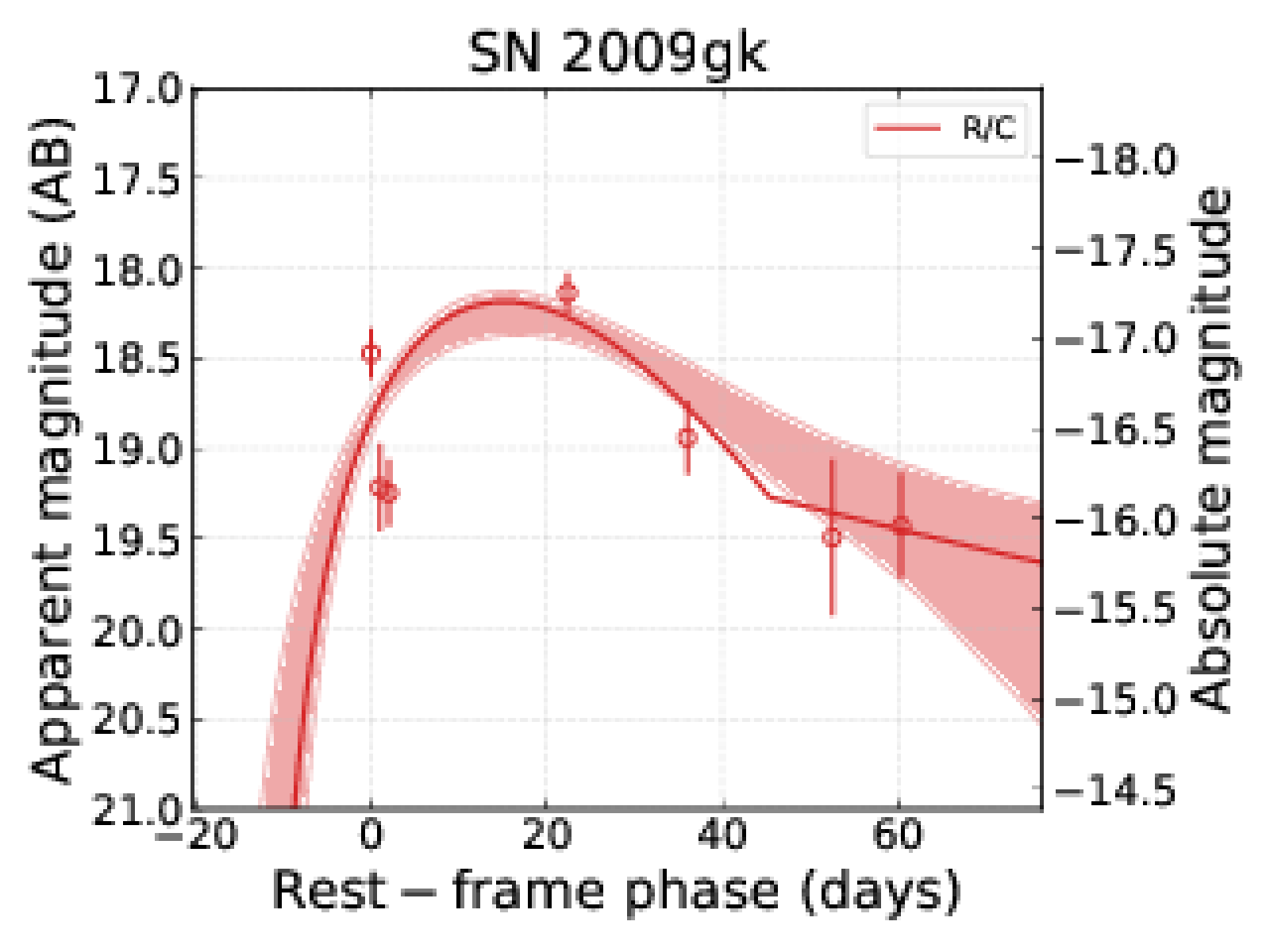}
        \includegraphics[width=0.49\columnwidth]{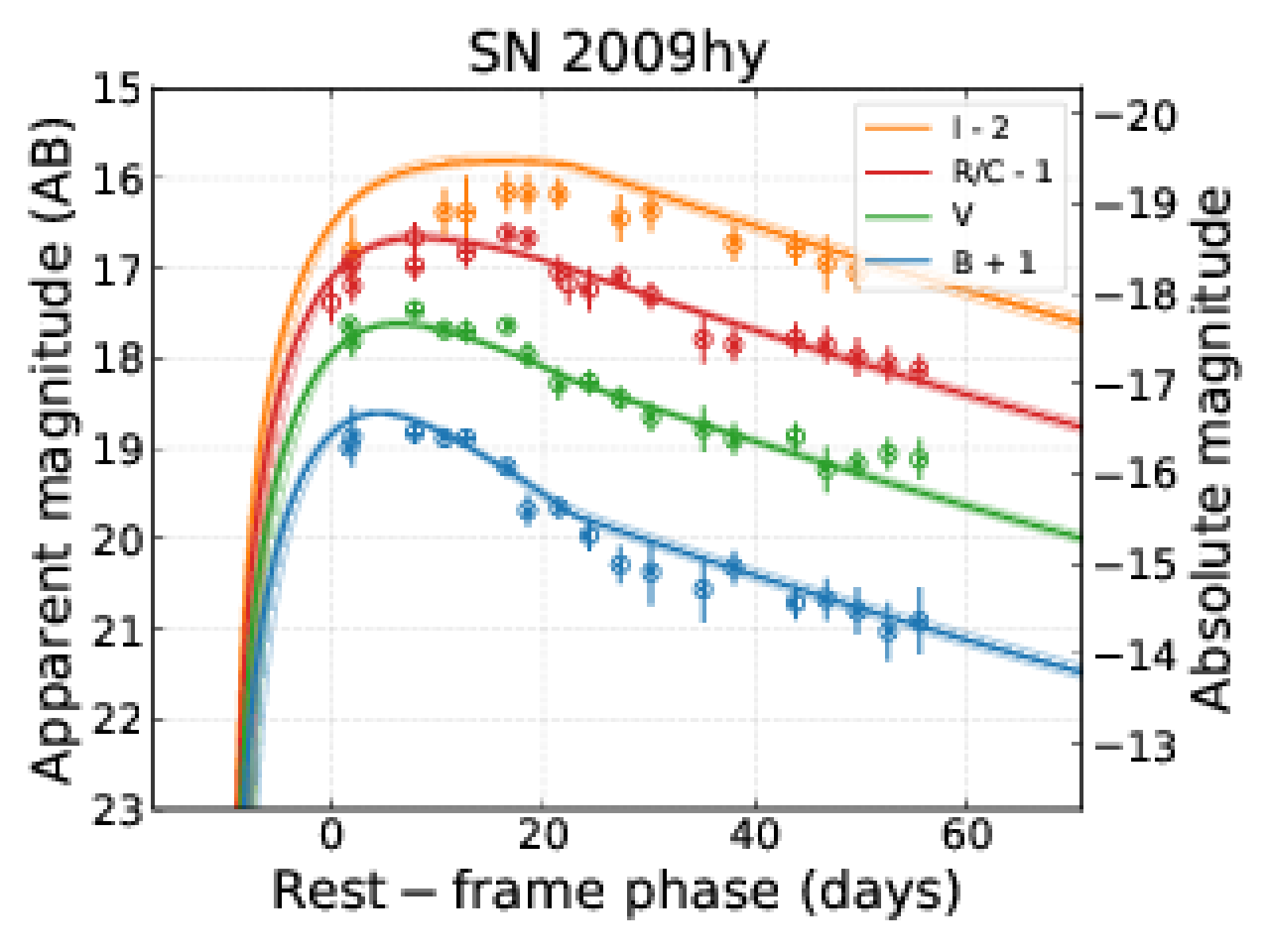}
        \includegraphics[width=0.49\columnwidth]{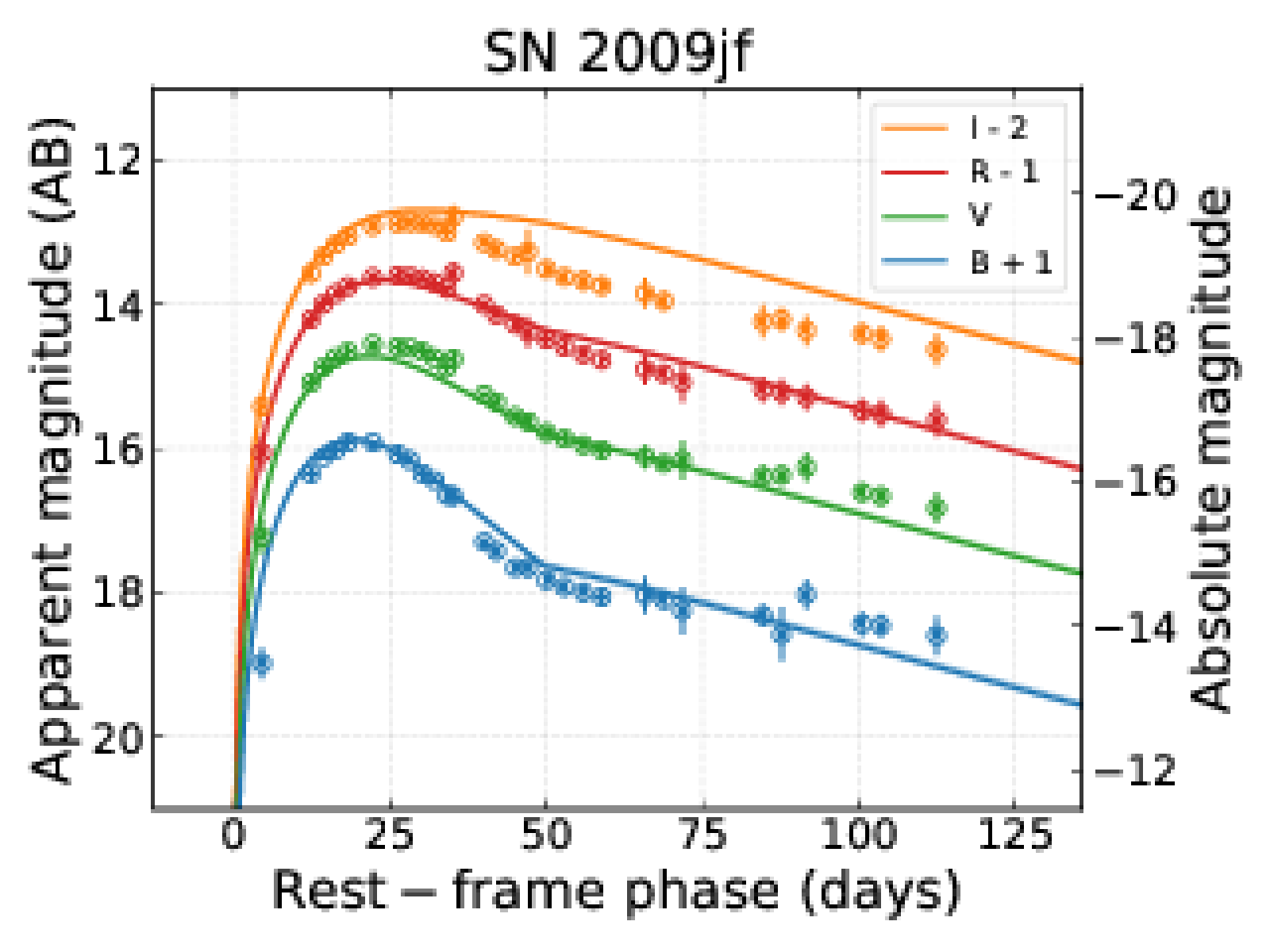}
        \caption{Model fitting of the 69 SNe with $^{56}$Ni model. The shaded regions present the 1$\sigma$ error region of the whole fitting area.}
\label{modelfittinglcallsesne}
\end{figure*}

\begin{figure*}
\ContinuedFloat
        \includegraphics[width=0.49\columnwidth]{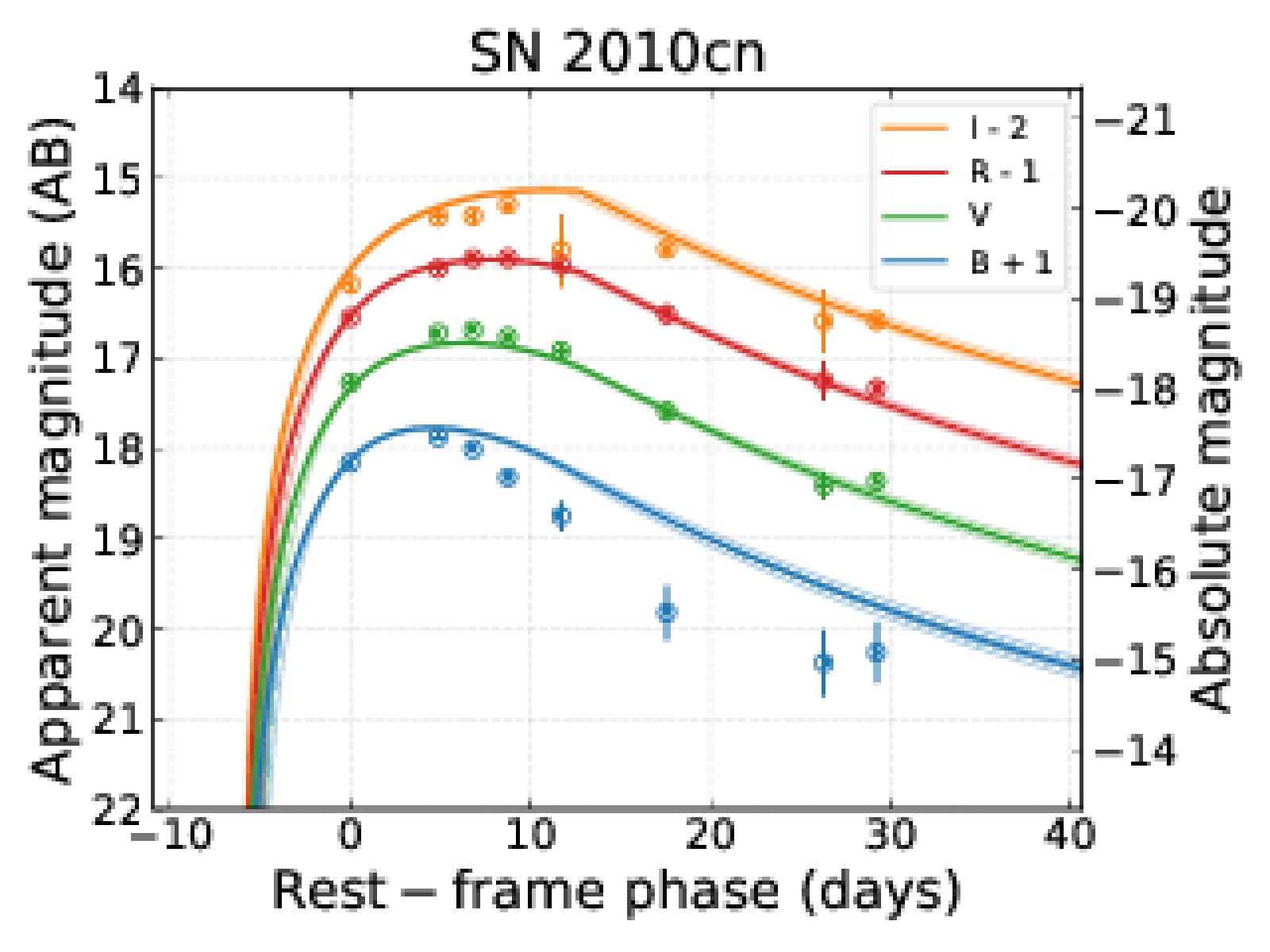}
        \includegraphics[width=0.49\columnwidth]{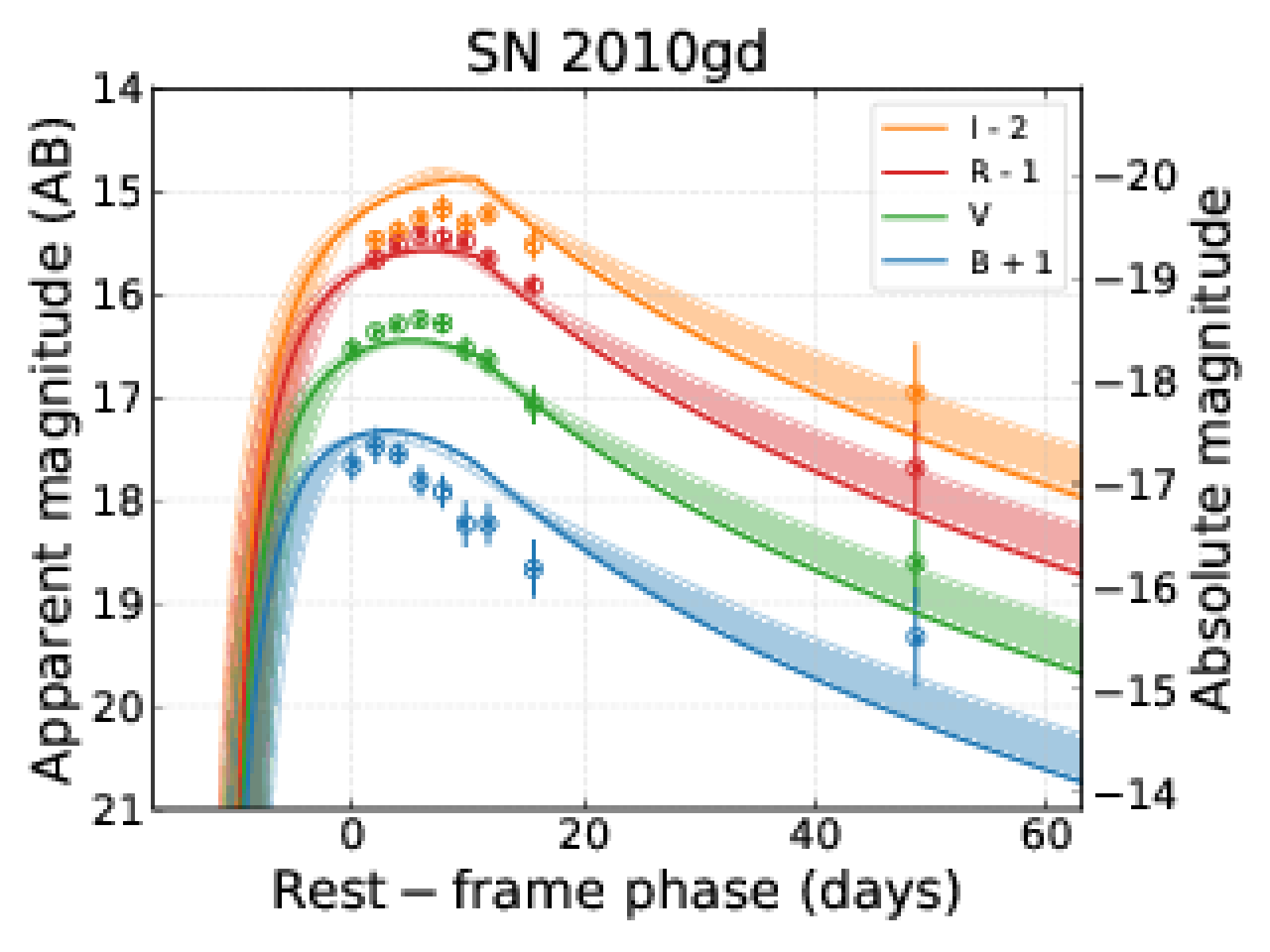}
        \includegraphics[width=0.49\columnwidth]{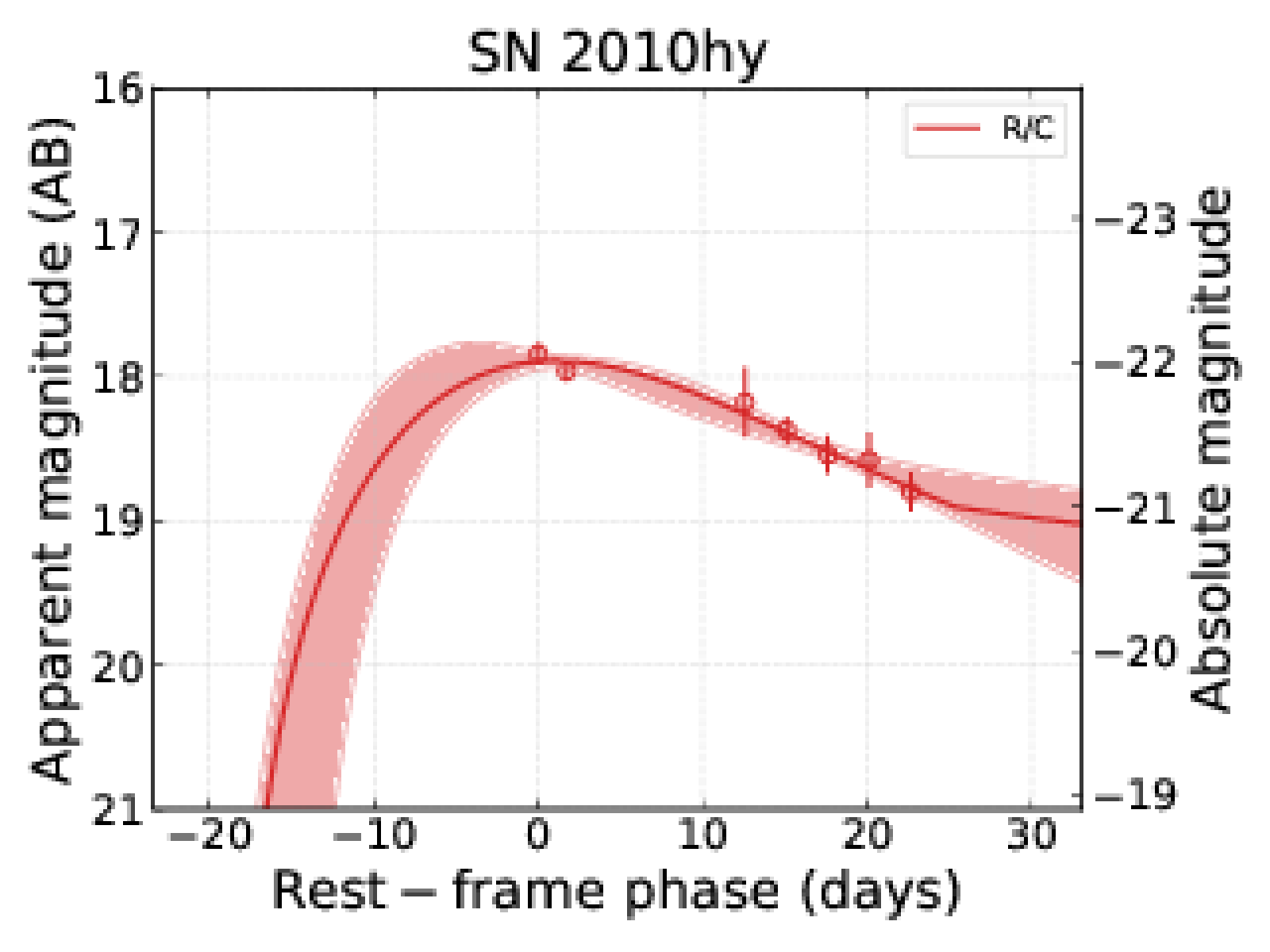}
        \includegraphics[width=0.49\columnwidth]{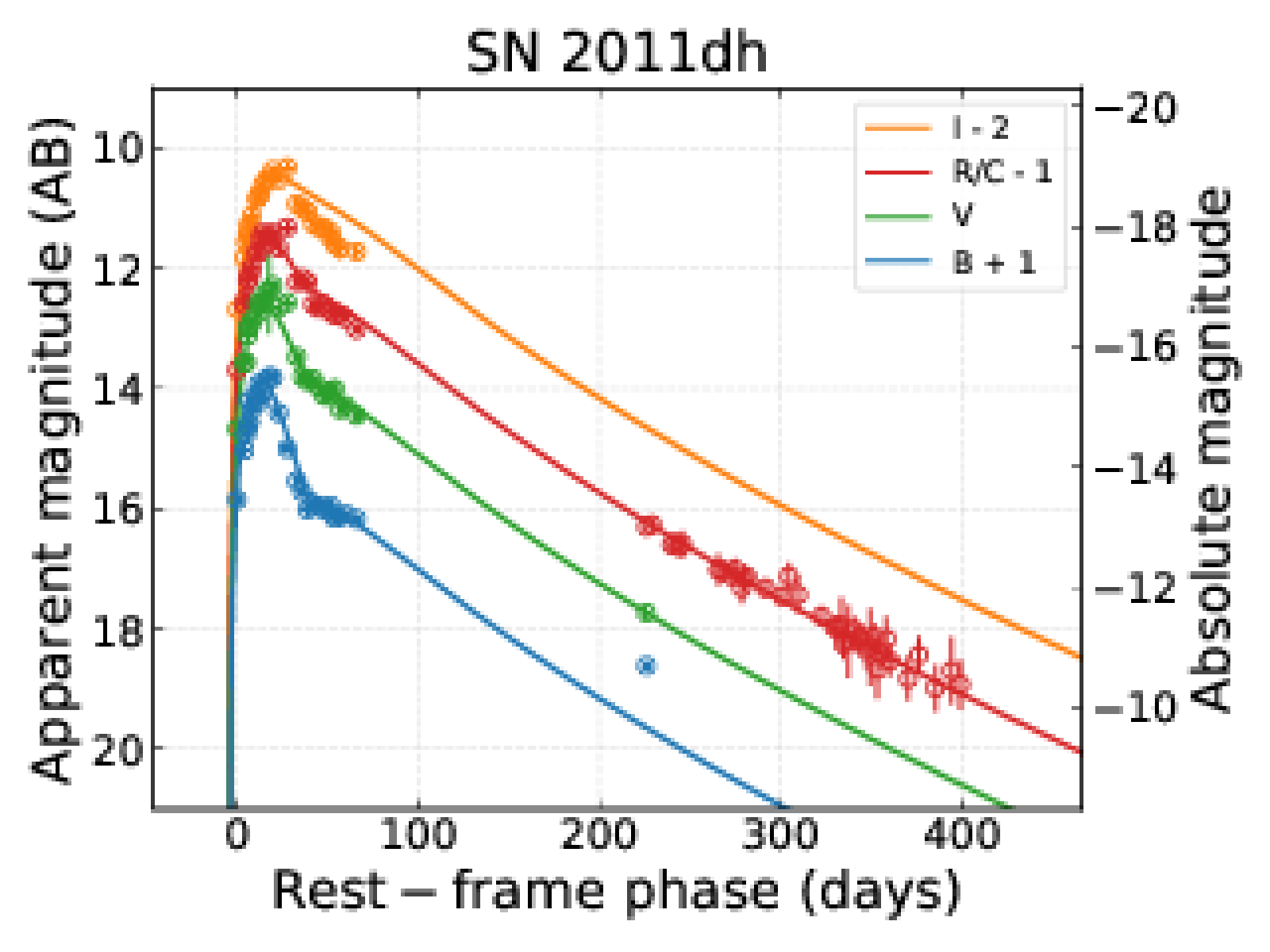}
        \includegraphics[width=0.49\columnwidth]{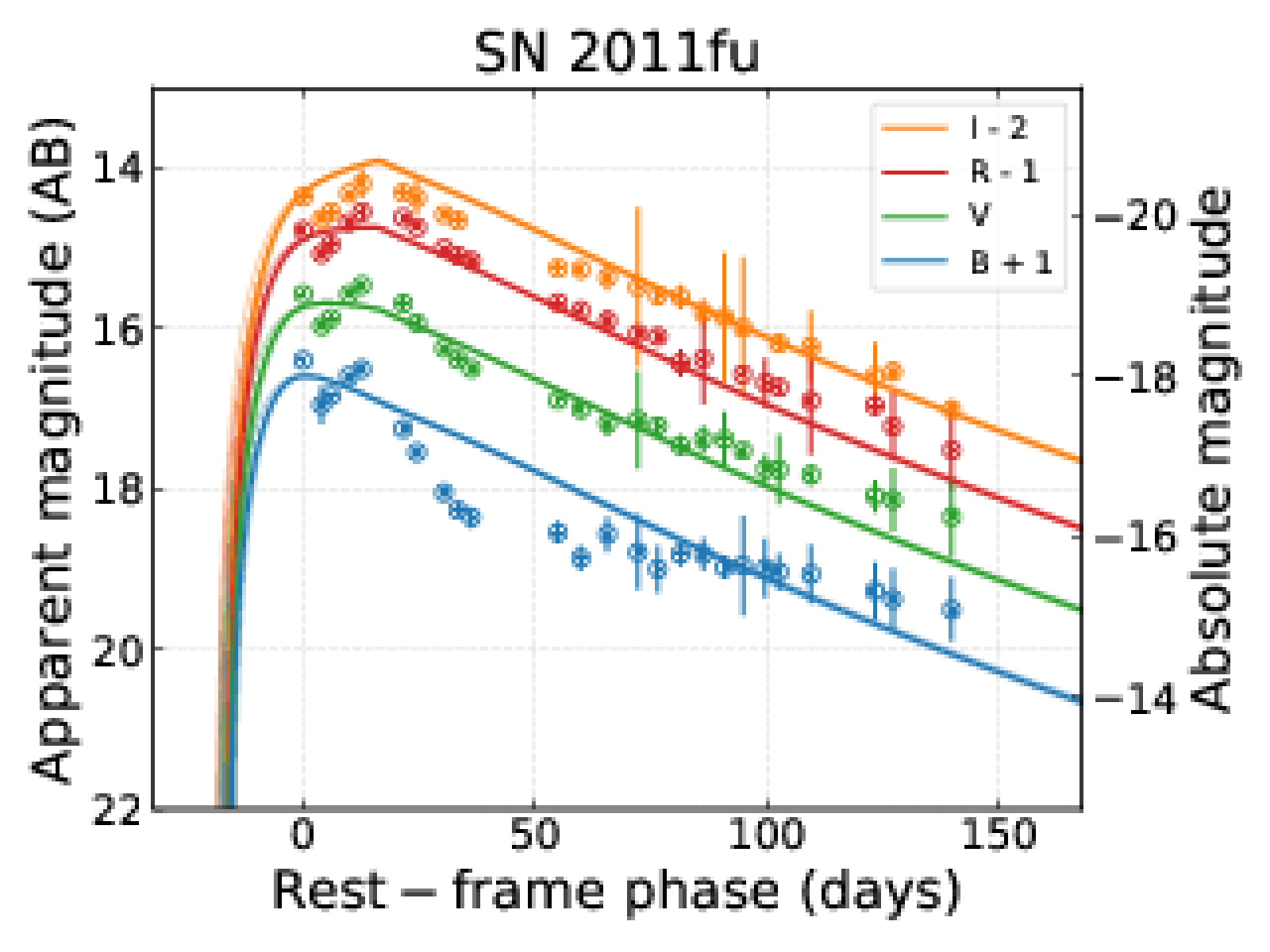}
        \includegraphics[width=0.49\columnwidth]{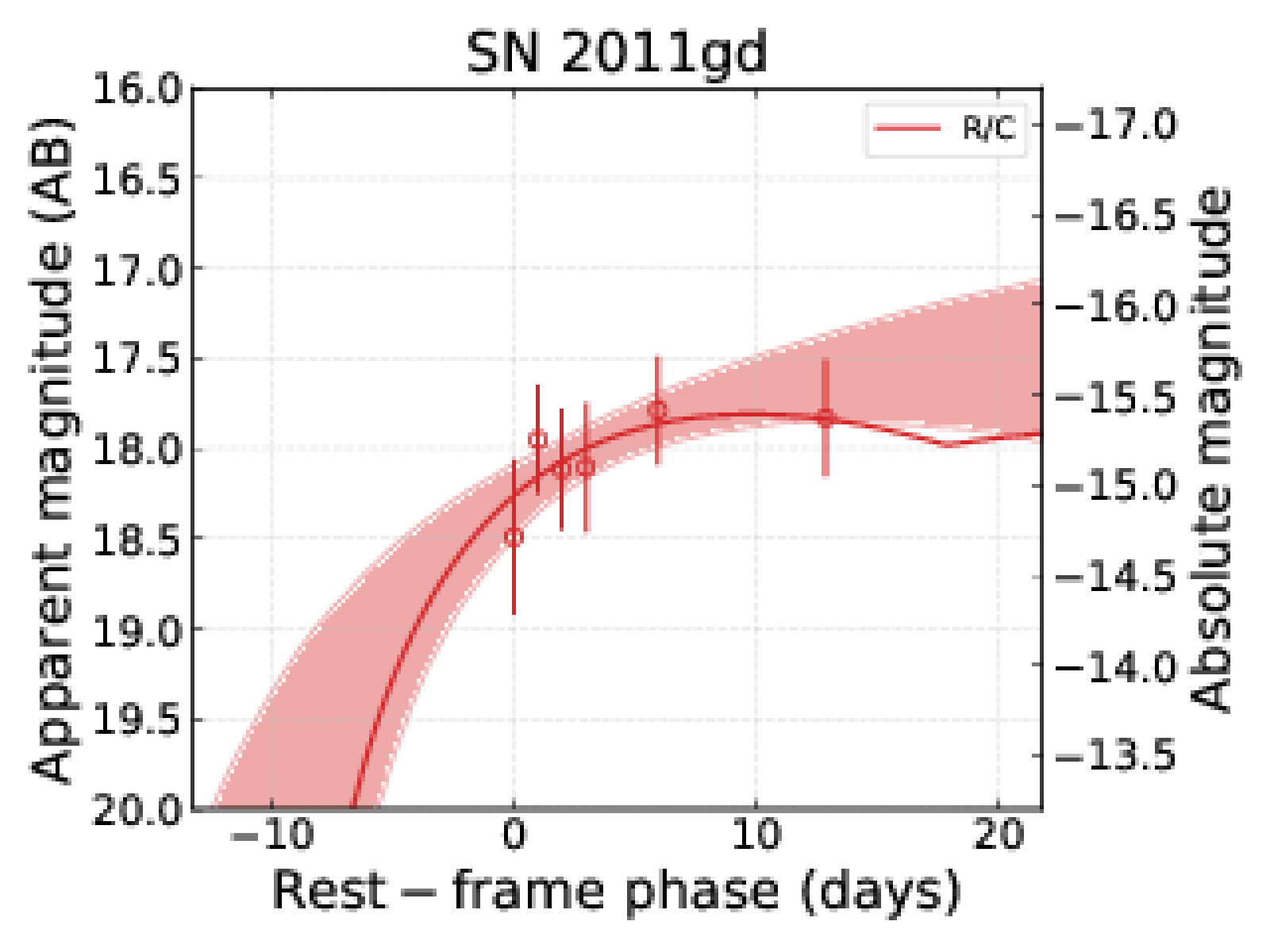}
        \includegraphics[width=0.49\columnwidth]{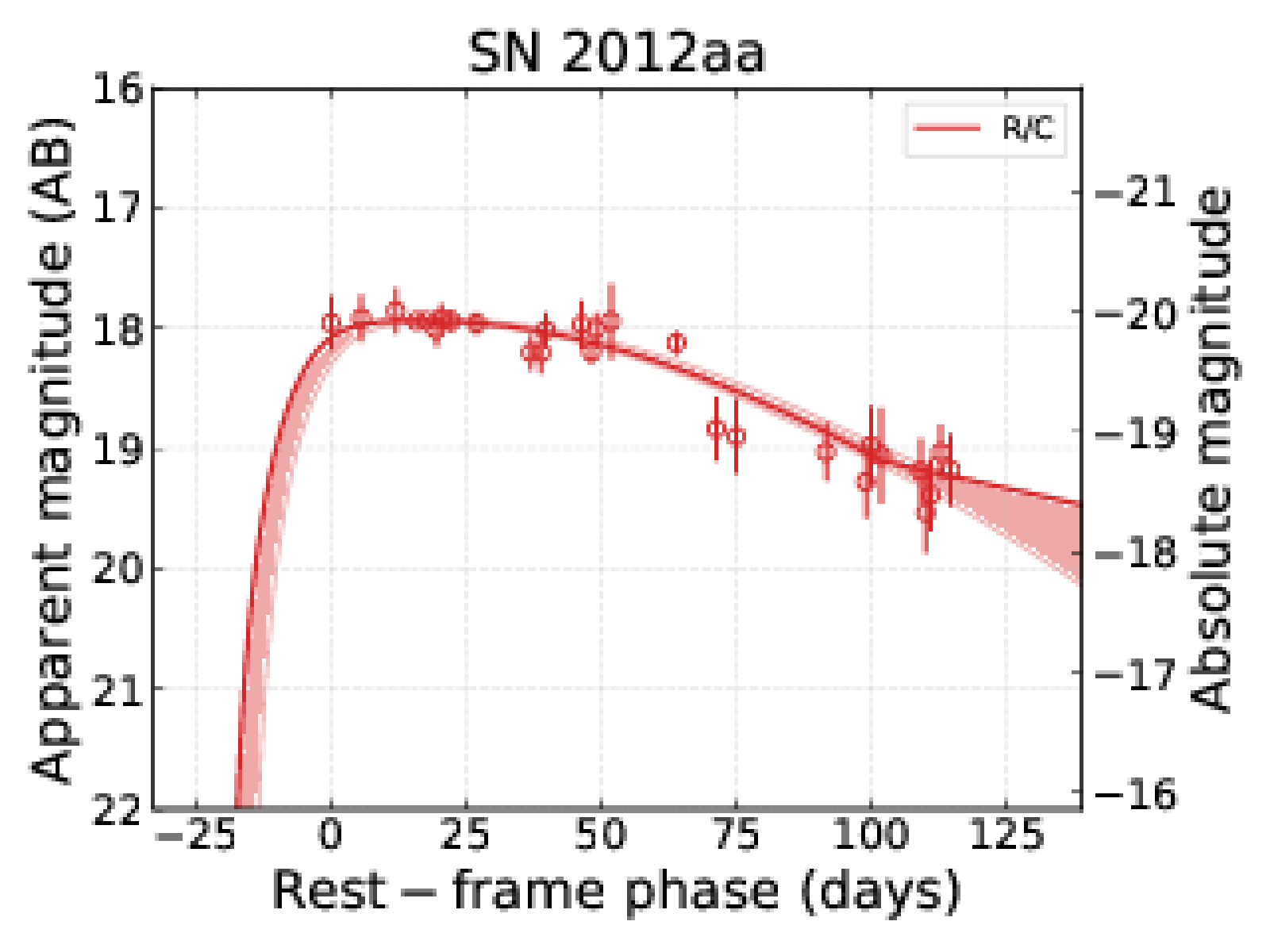}
        \includegraphics[width=0.49\columnwidth]{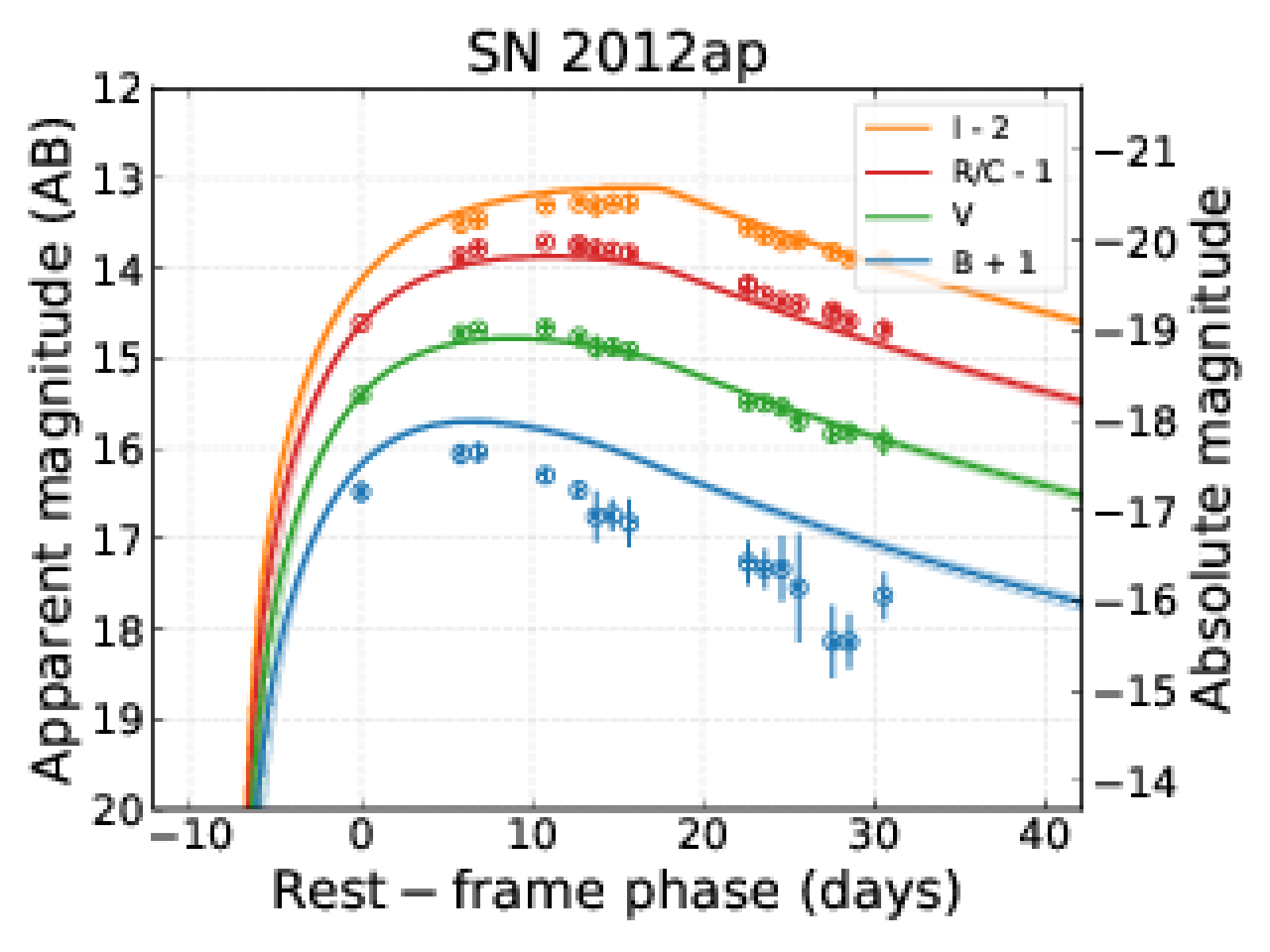}
        \includegraphics[width=0.49\columnwidth]{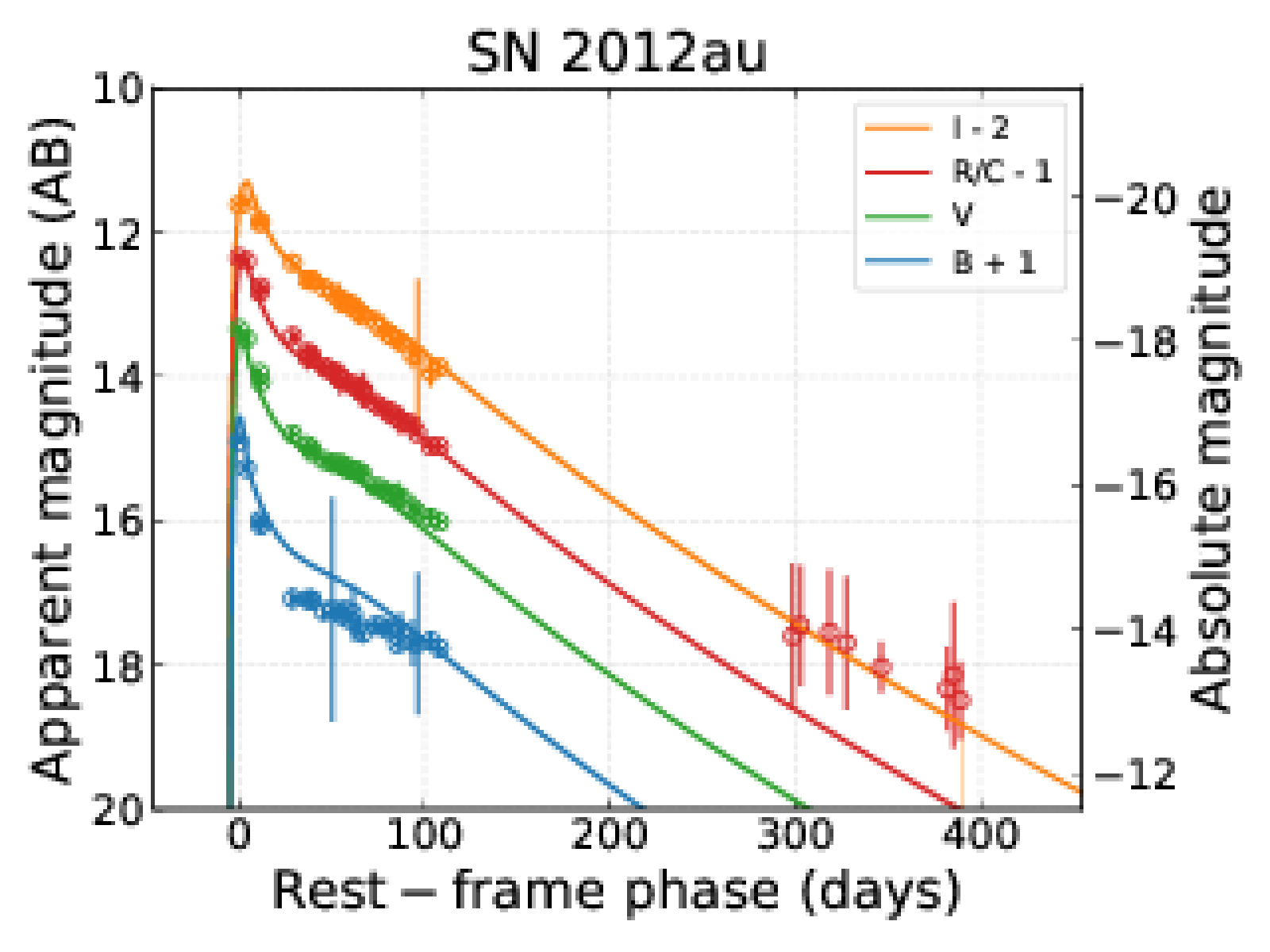}
        \includegraphics[width=0.49\columnwidth]{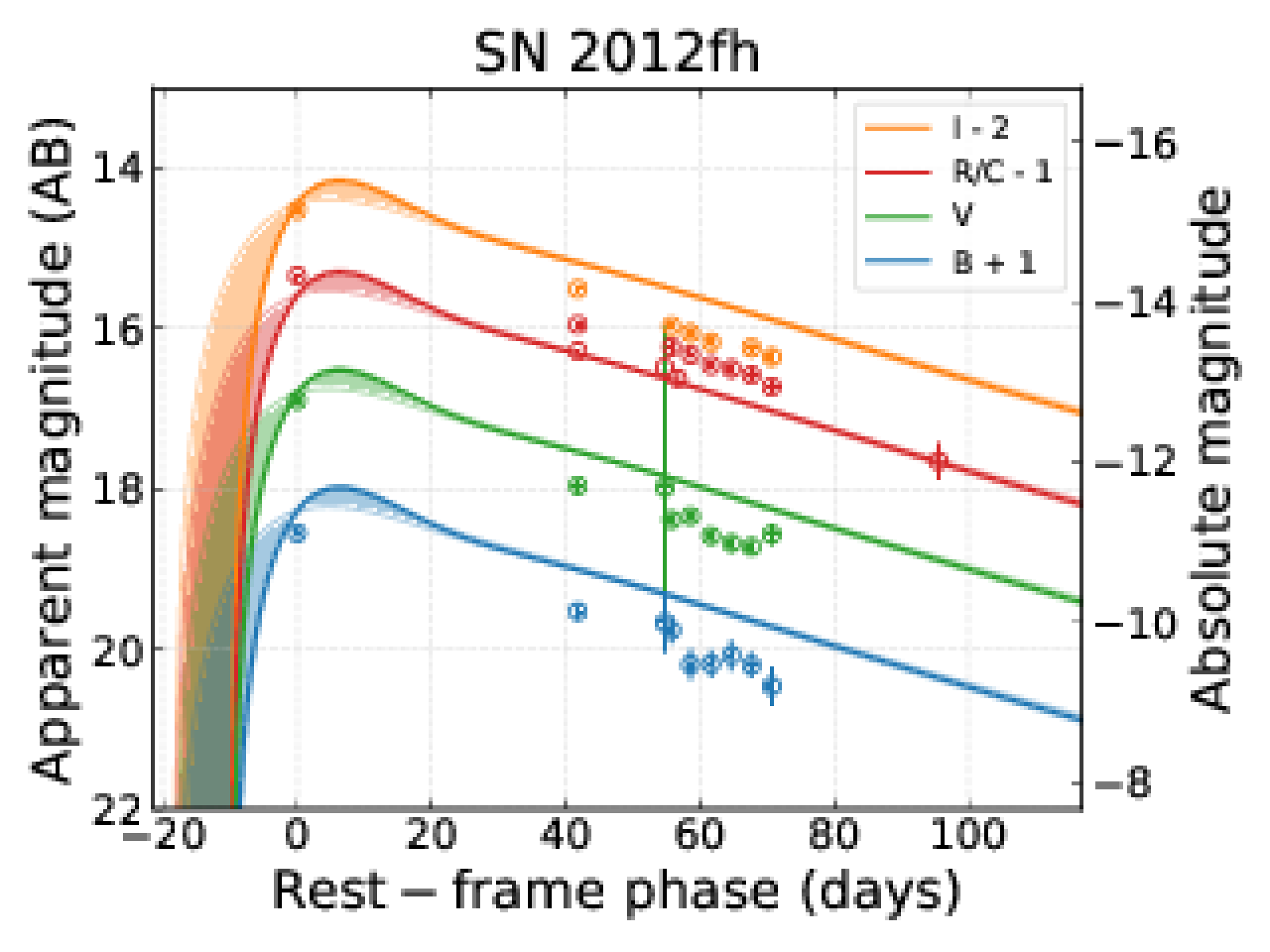}
        \includegraphics[width=0.49\columnwidth]{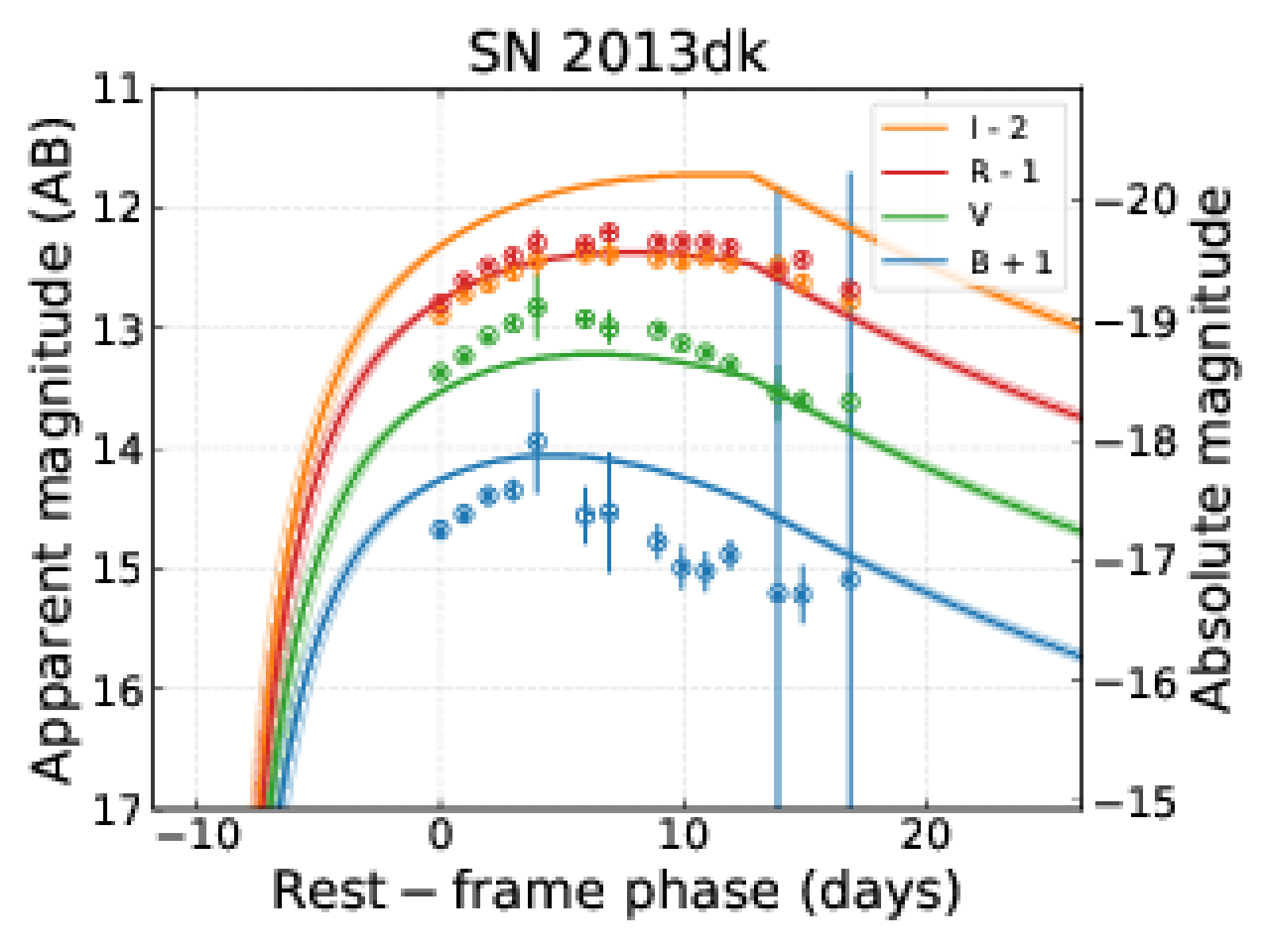}
        \includegraphics[width=0.49\columnwidth]{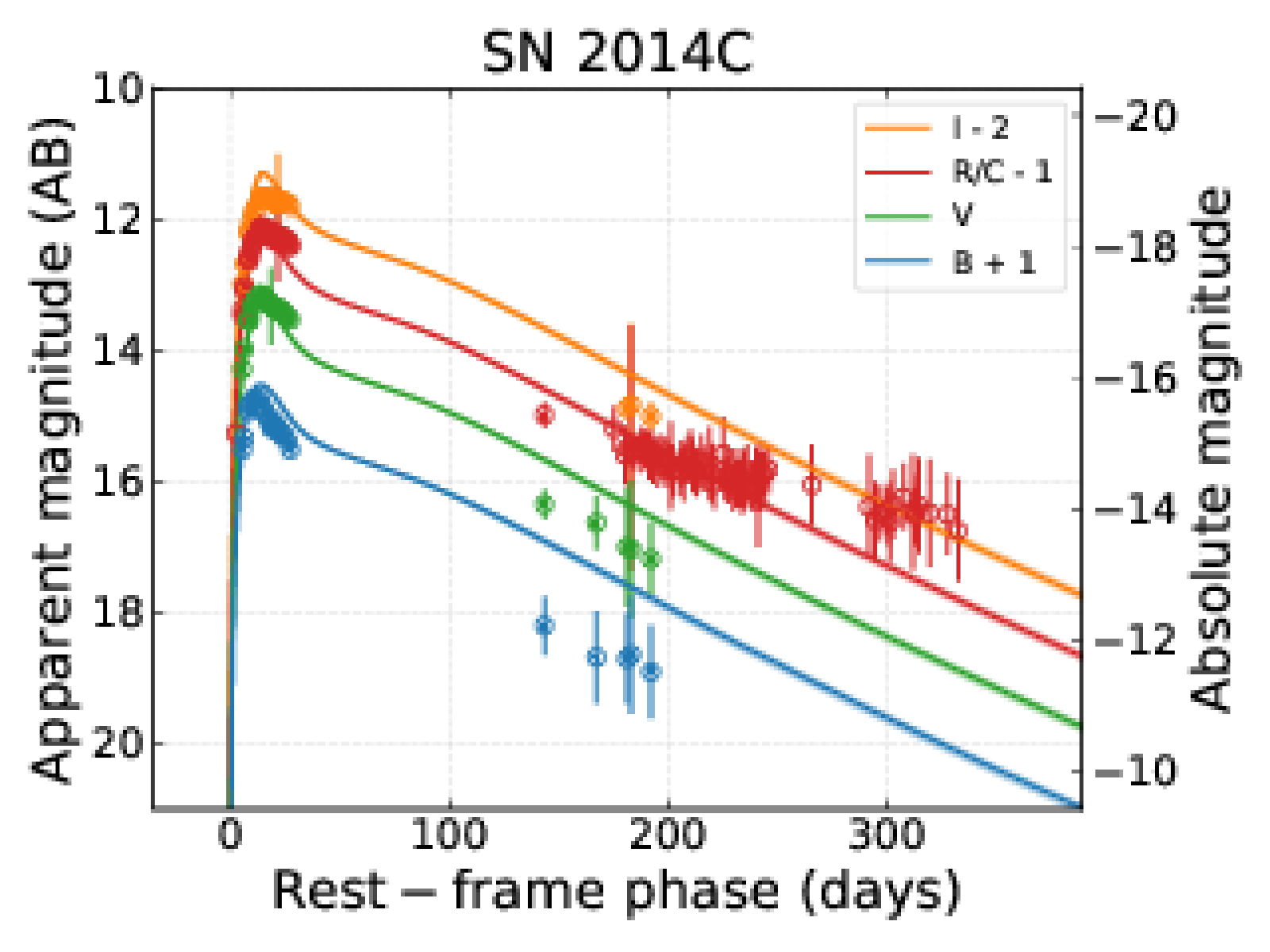}
        \includegraphics[width=0.49\columnwidth]{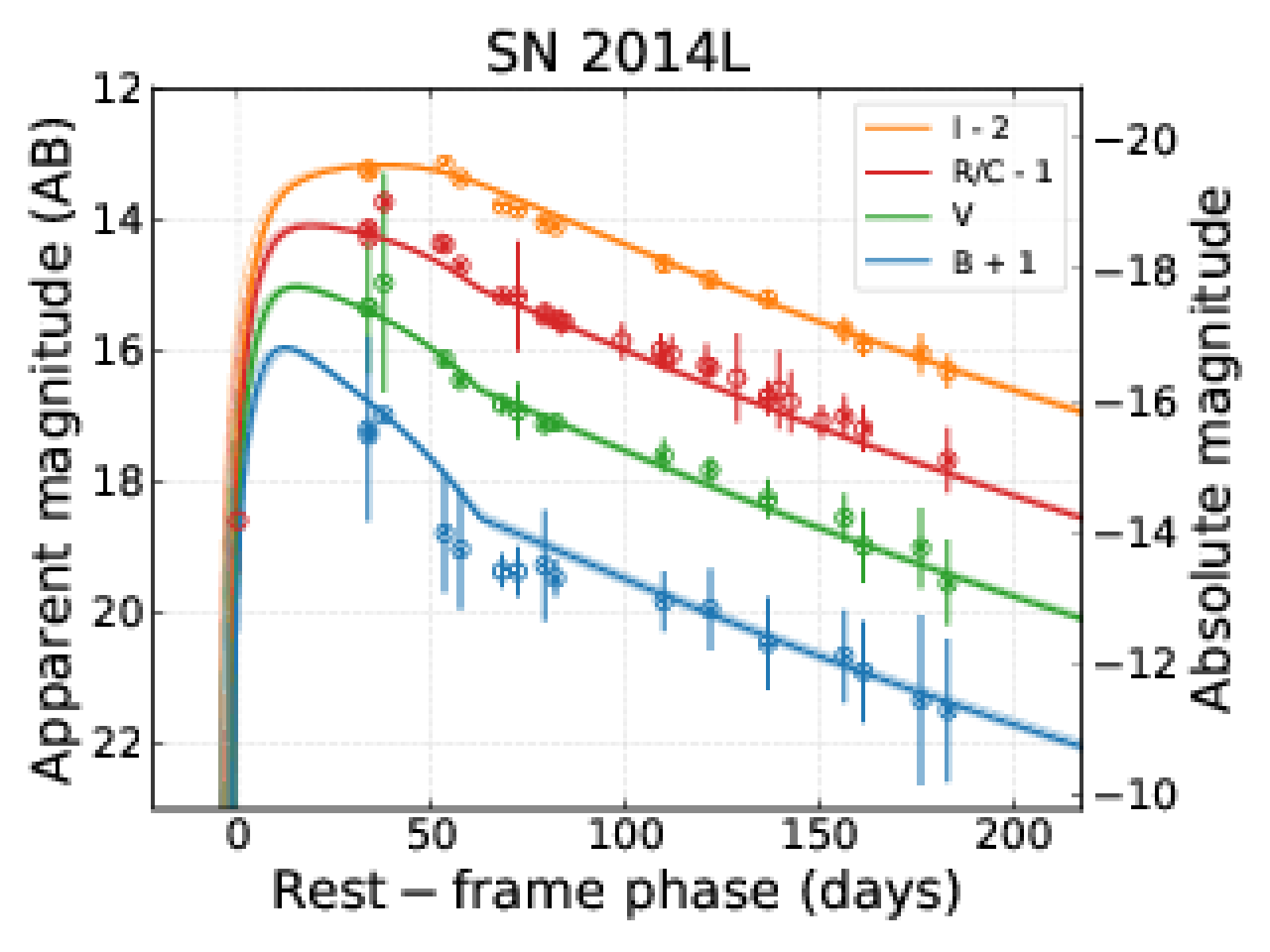}
        \includegraphics[width=0.49\columnwidth]{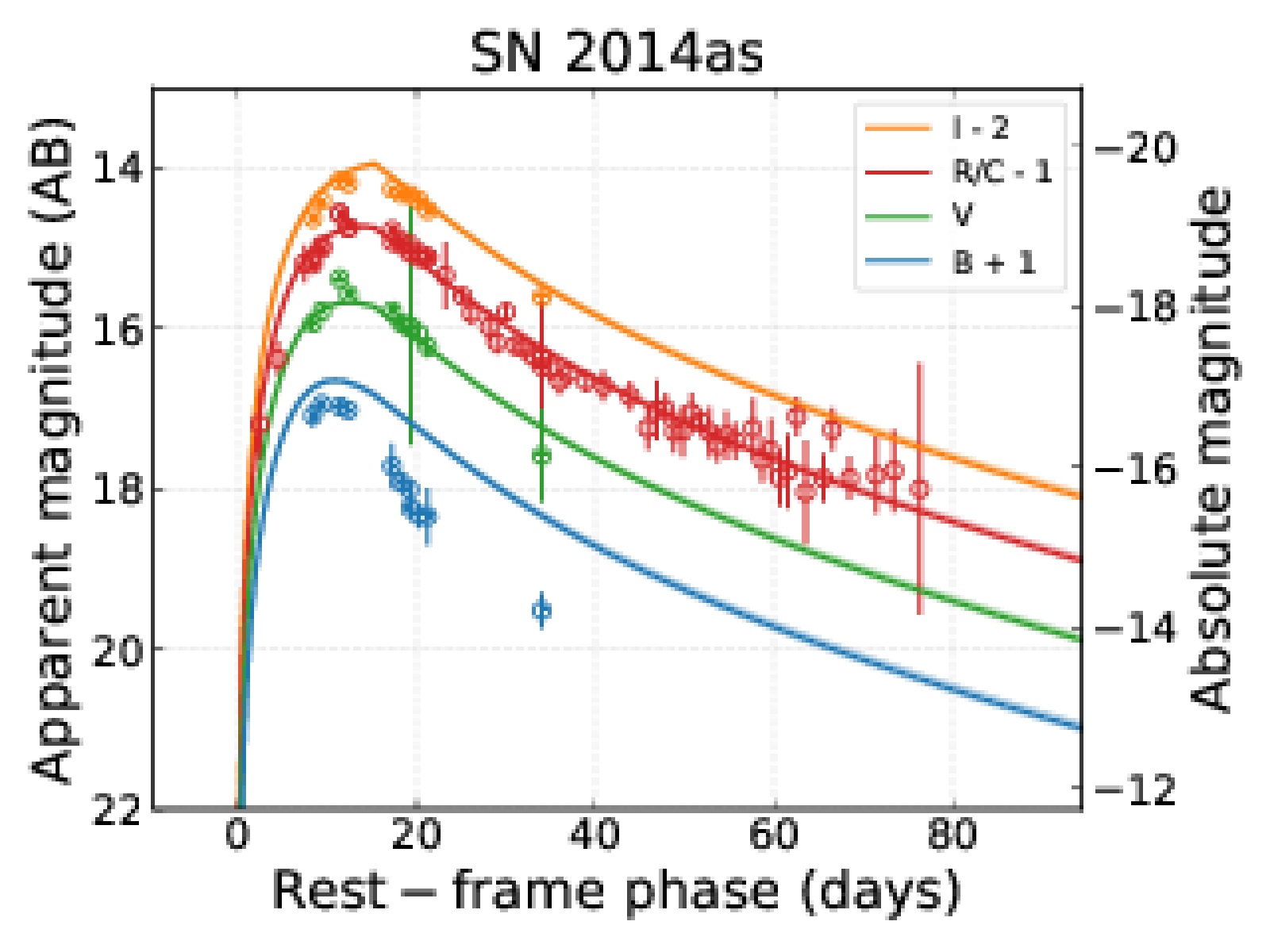}
        \includegraphics[width=0.49\columnwidth]{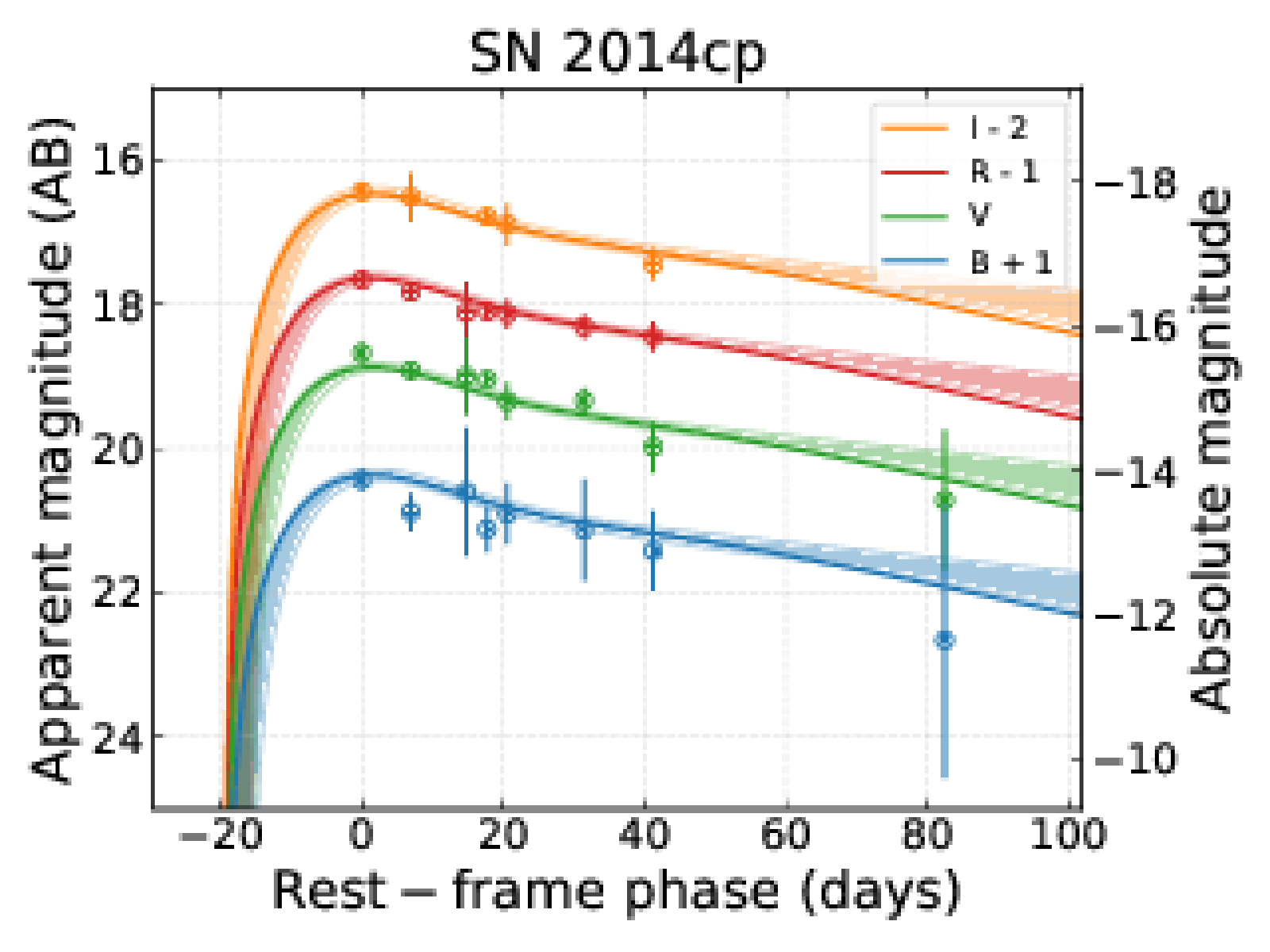}
        \includegraphics[width=0.49\columnwidth]{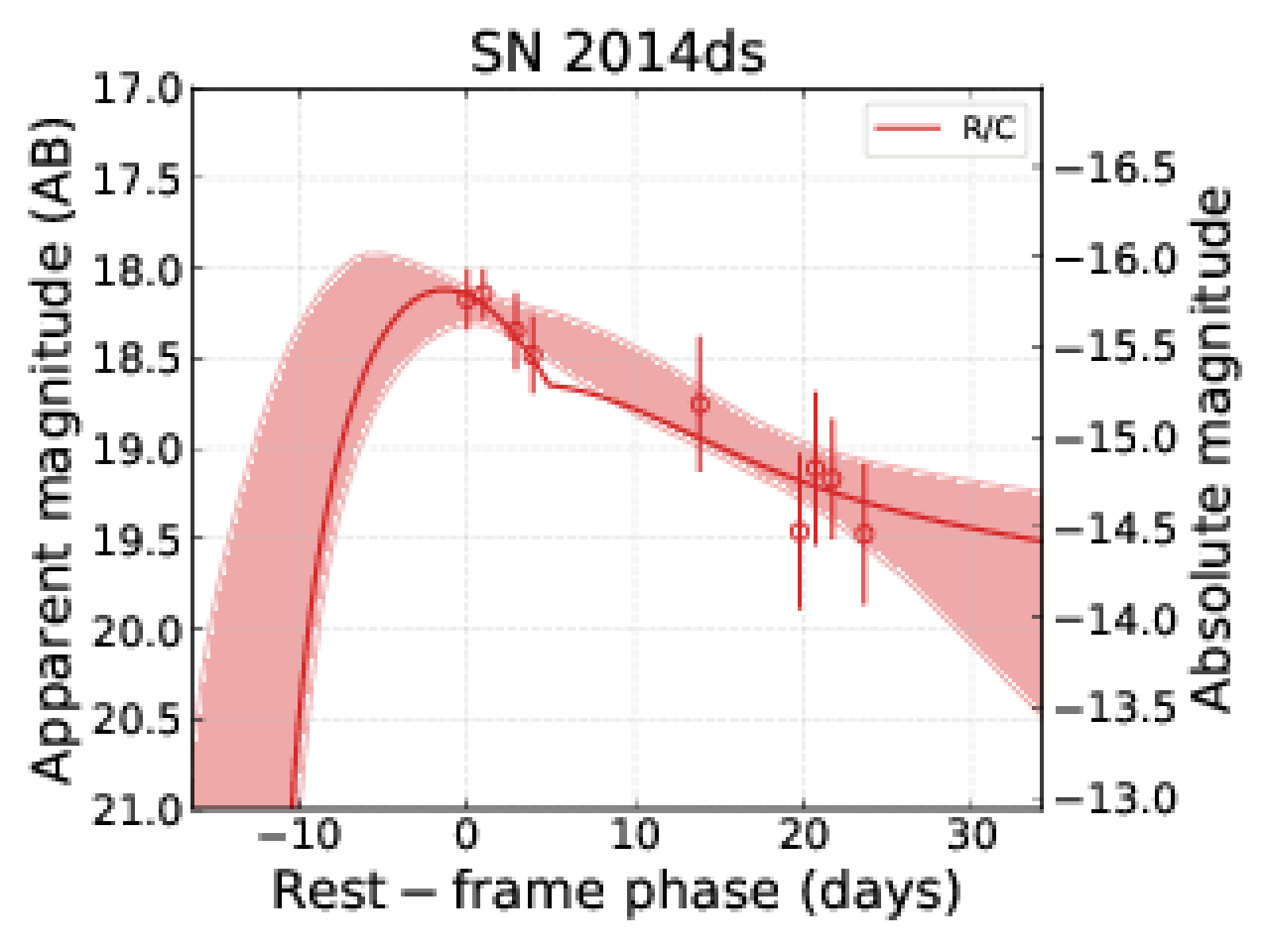}
        \includegraphics[width=0.49\columnwidth]{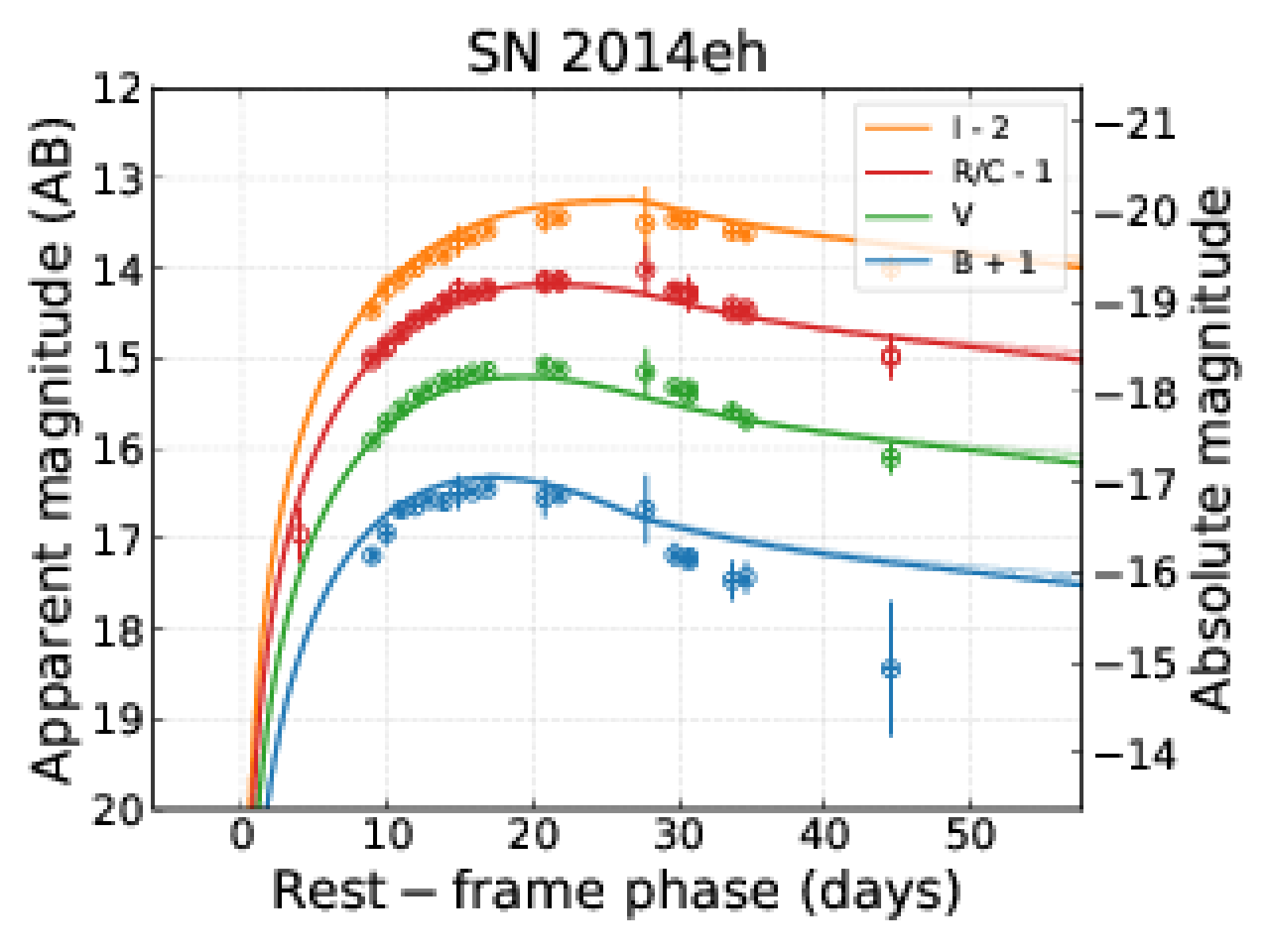}
        \includegraphics[width=0.49\columnwidth]{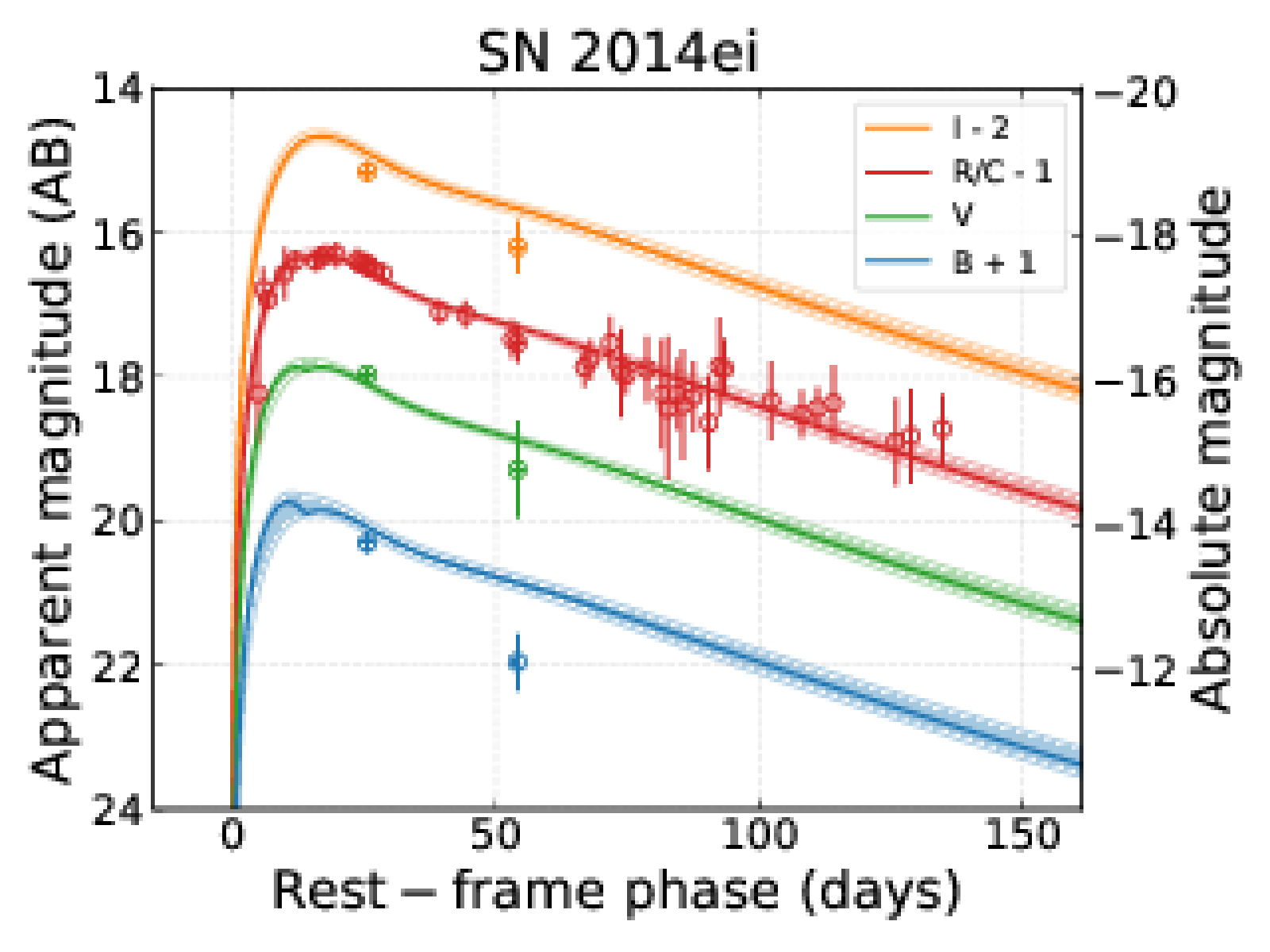}
        \includegraphics[width=0.49\columnwidth]{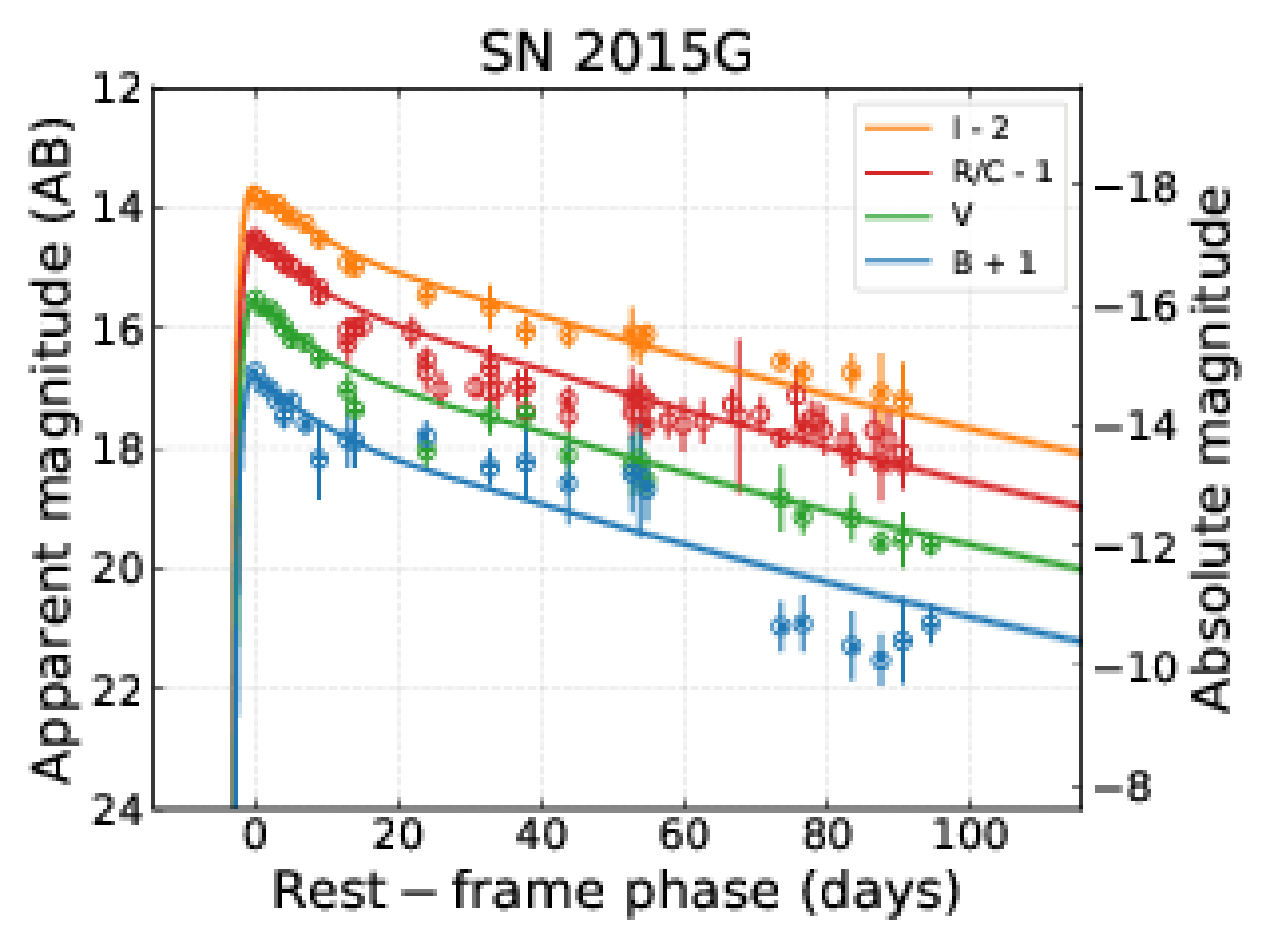}
        \includegraphics[width=0.49\columnwidth]{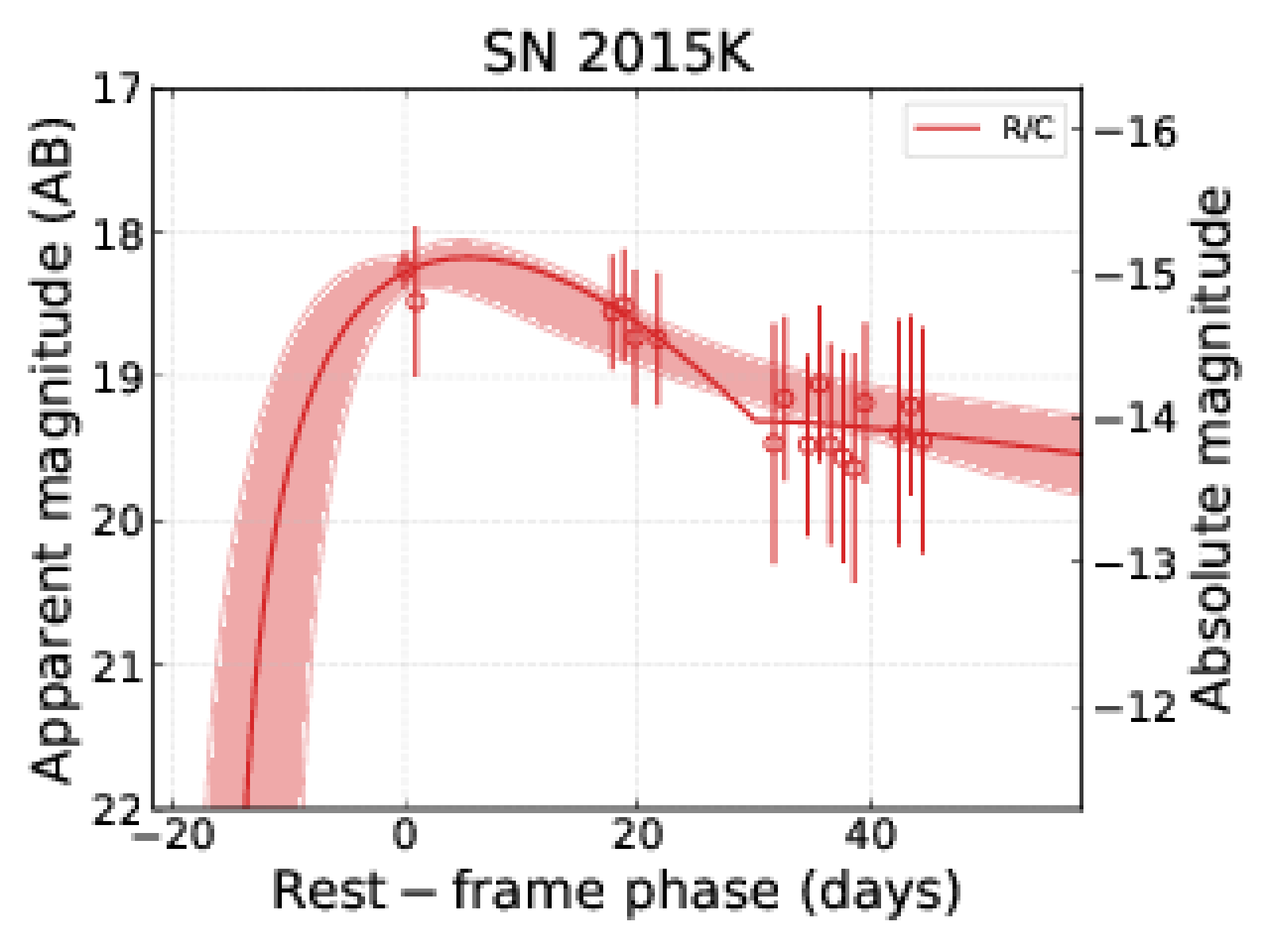}
        \includegraphics[width=0.49\columnwidth]{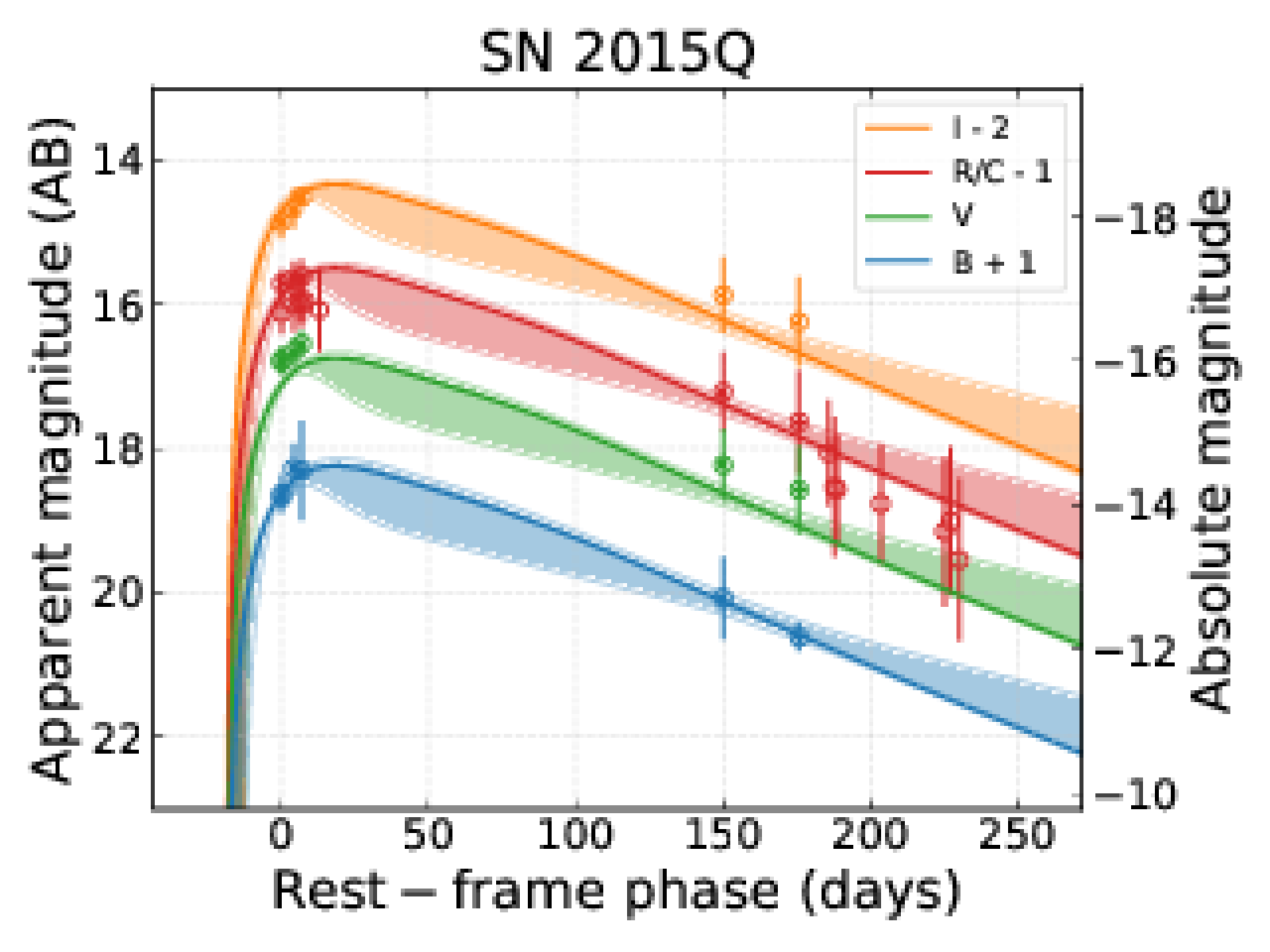}
        \includegraphics[width=0.49\columnwidth]{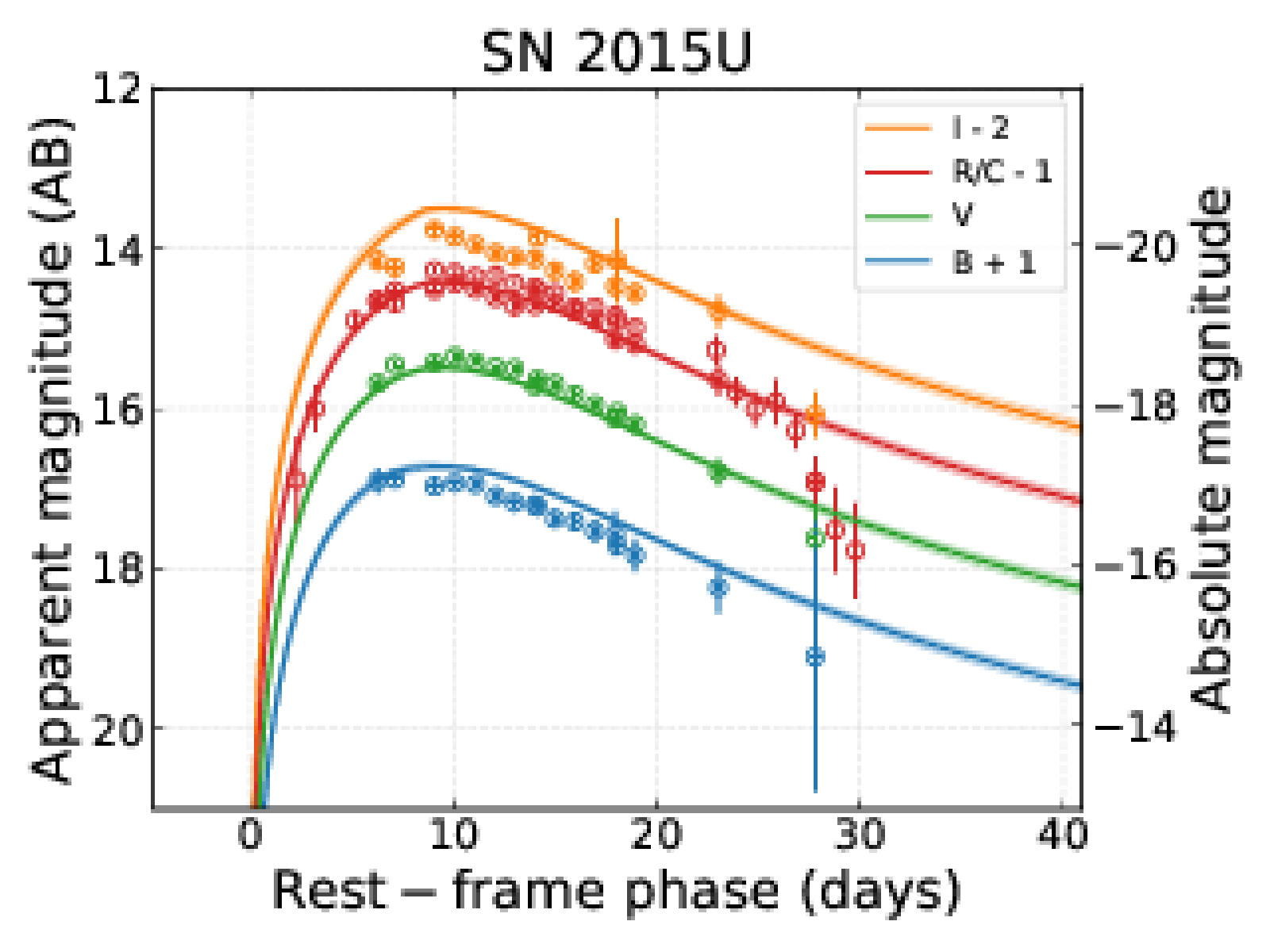}
        \includegraphics[width=0.49\columnwidth]{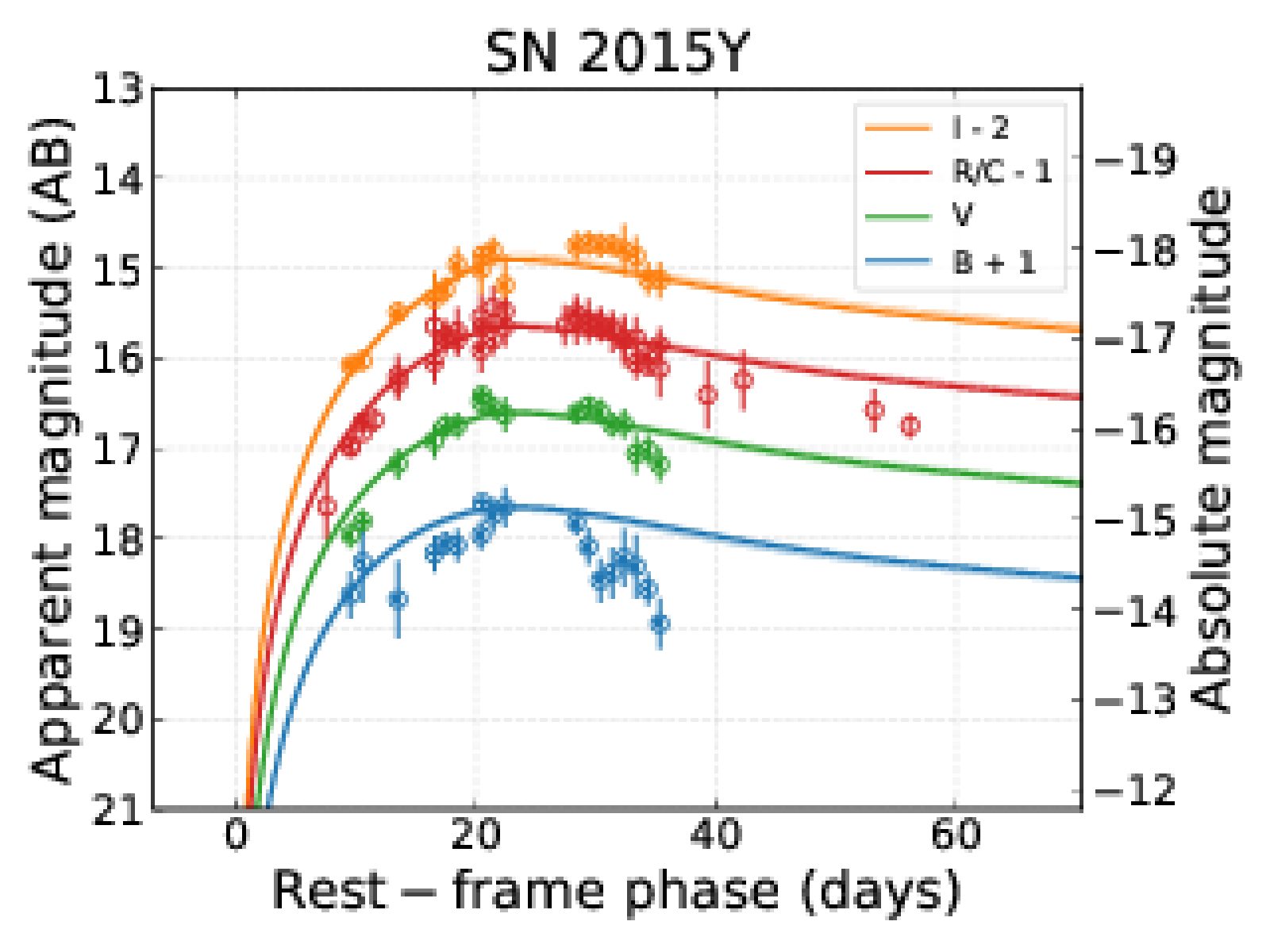}
        \includegraphics[width=0.49\columnwidth]{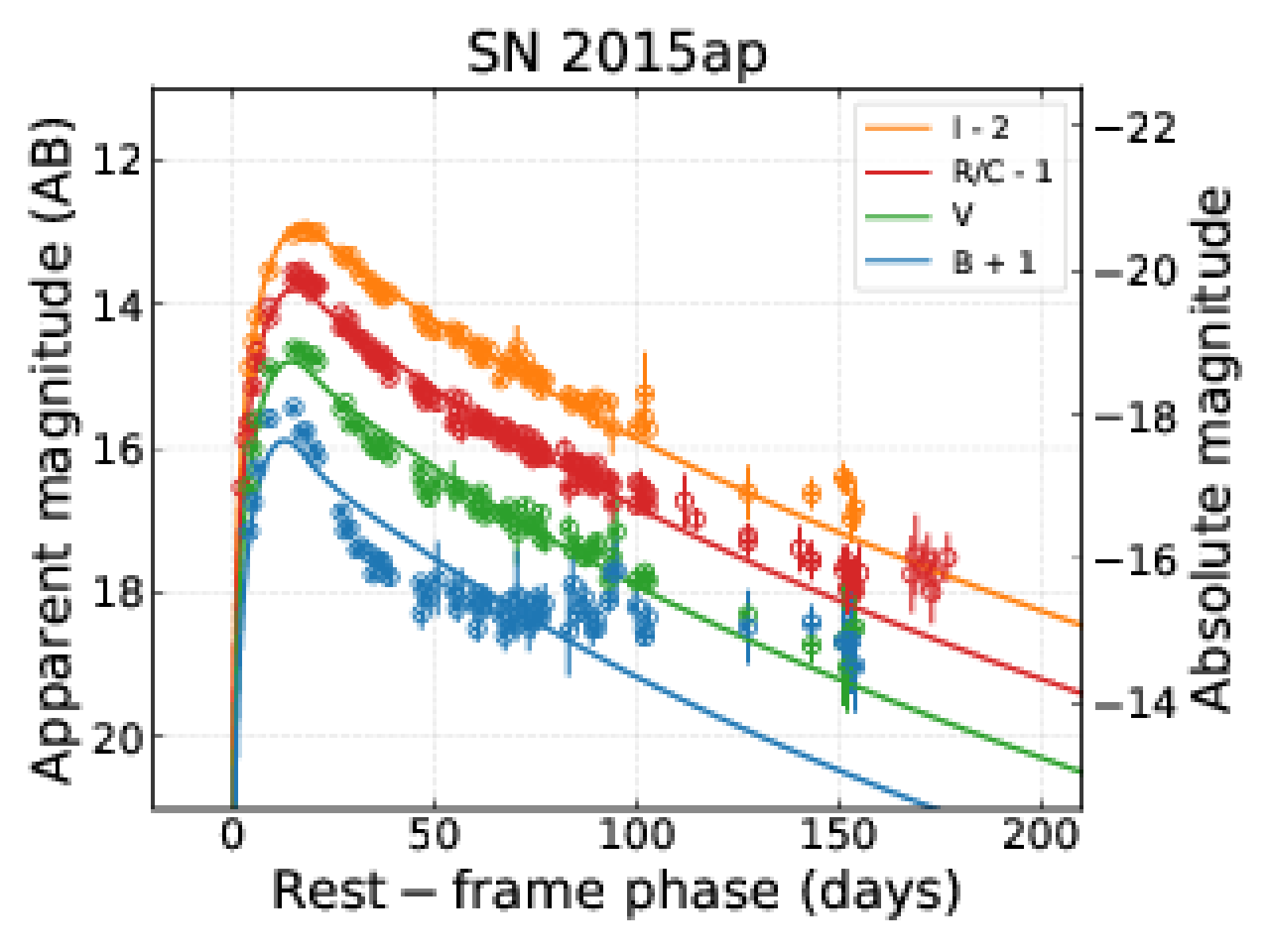}
        \includegraphics[width=0.49\columnwidth]{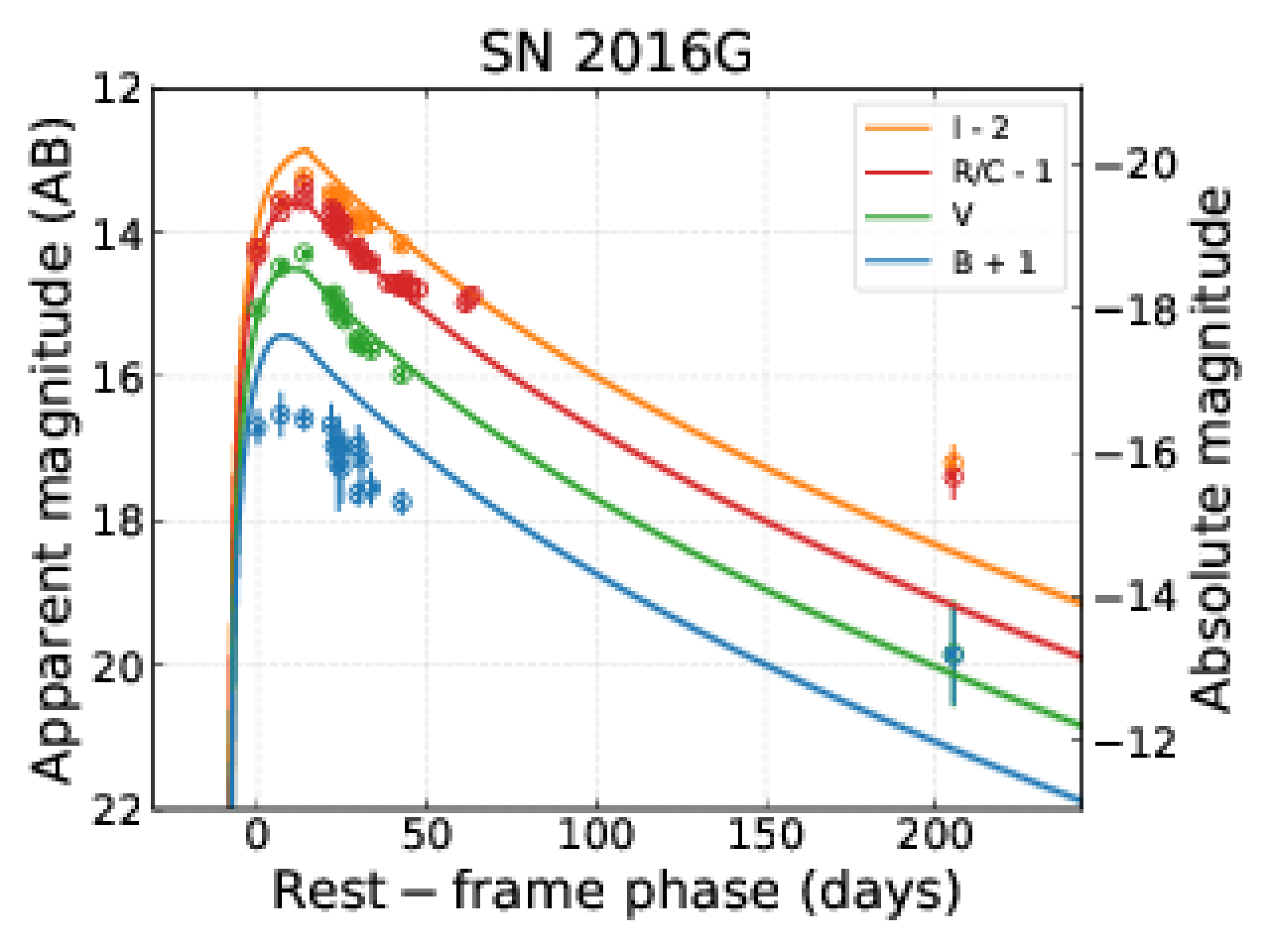}
        \includegraphics[width=0.49\columnwidth]{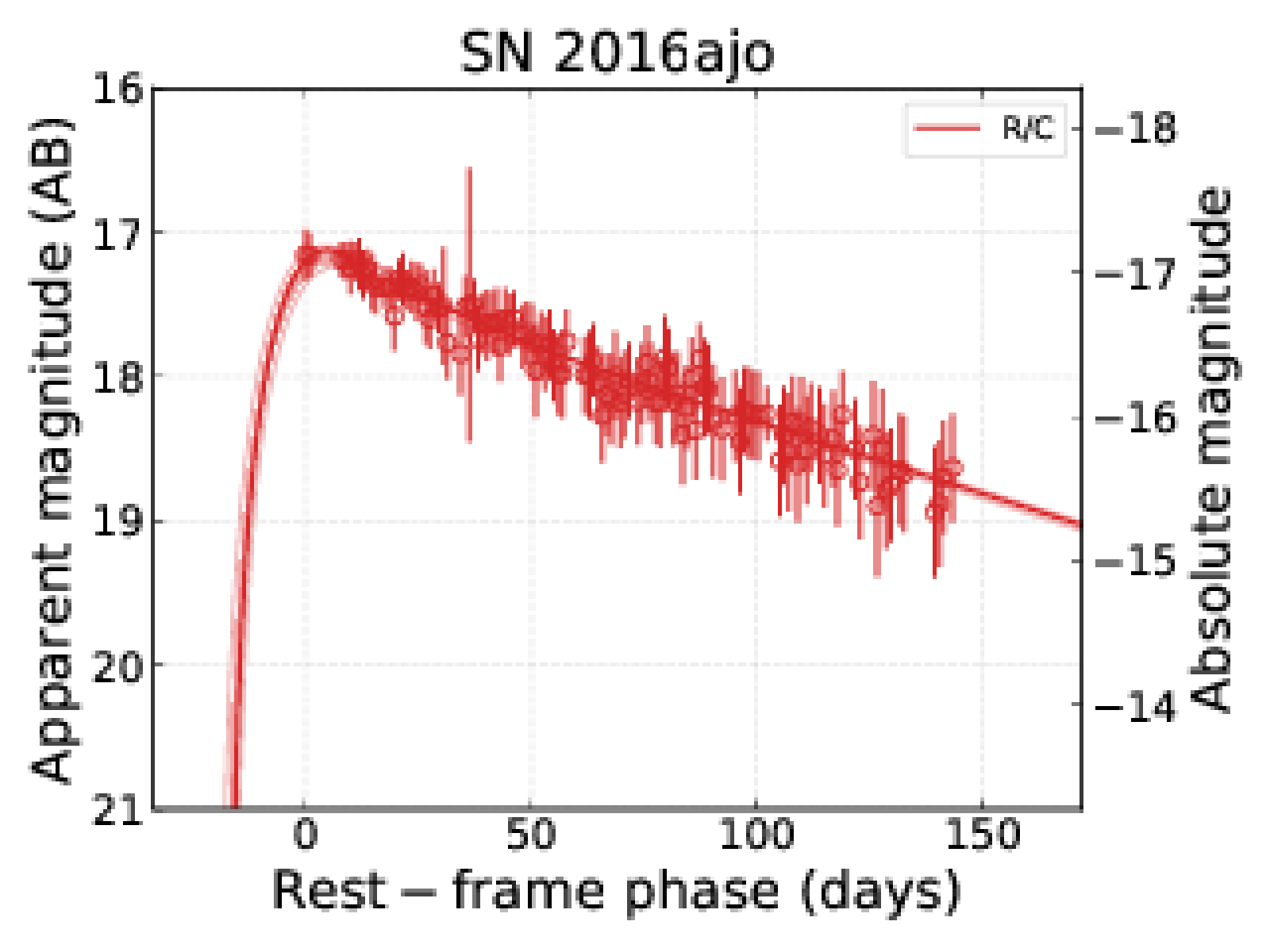}
        \includegraphics[width=0.49\columnwidth]{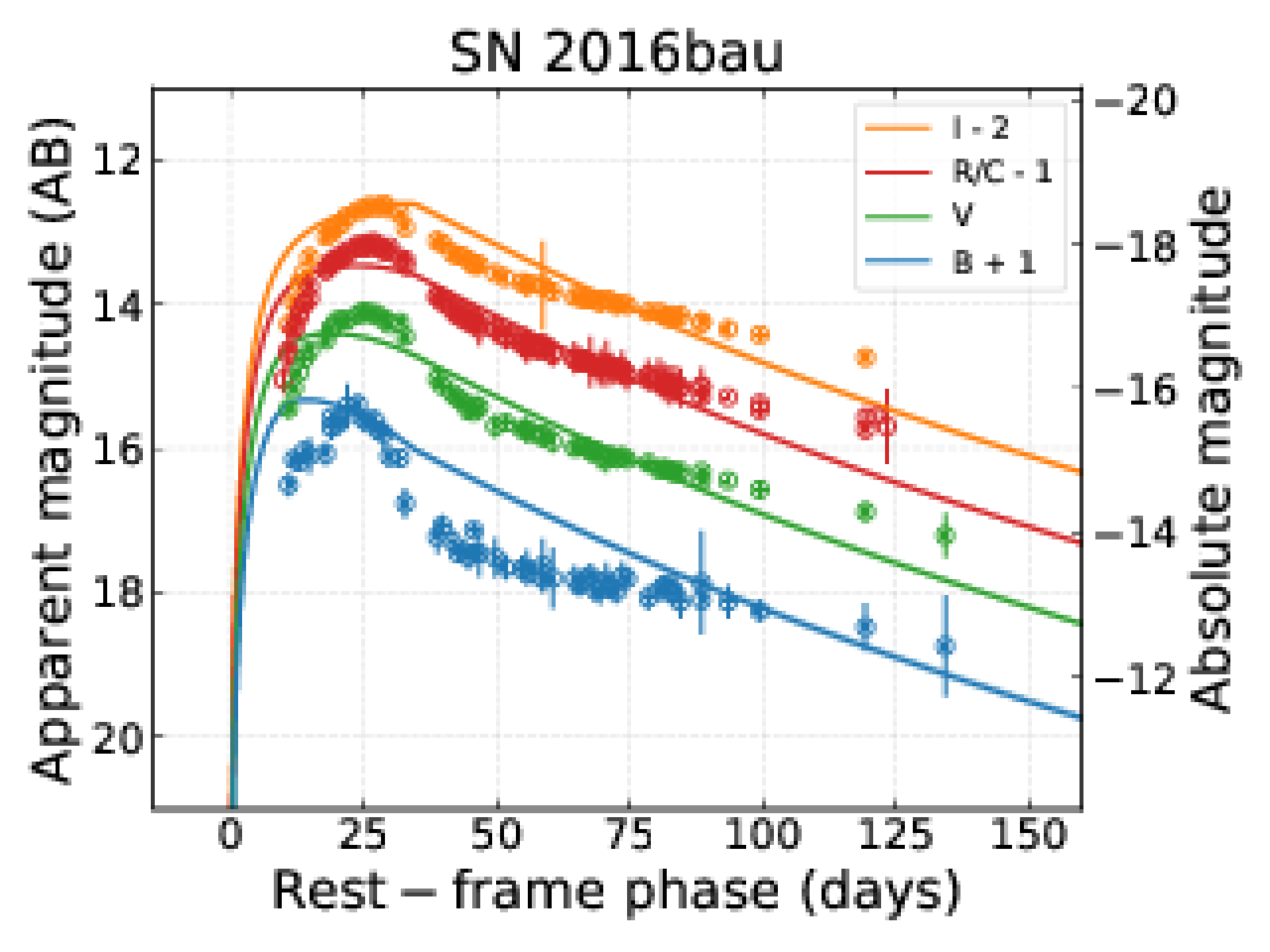}
        \includegraphics[width=0.49\columnwidth]{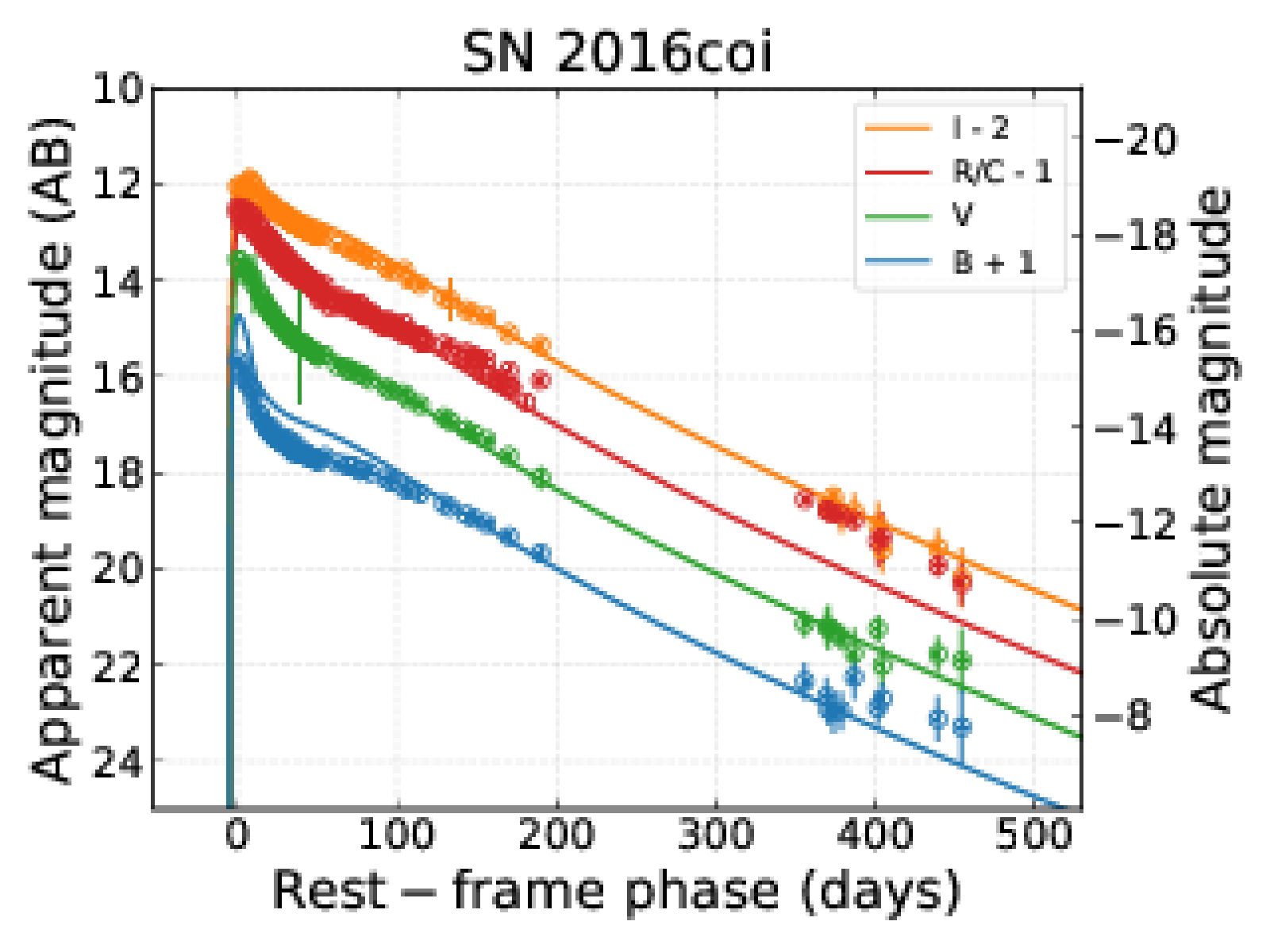}
        \caption{Continued.}
\end{figure*}

\begin{figure*}
\ContinuedFloat
        \includegraphics[width=0.49\columnwidth]{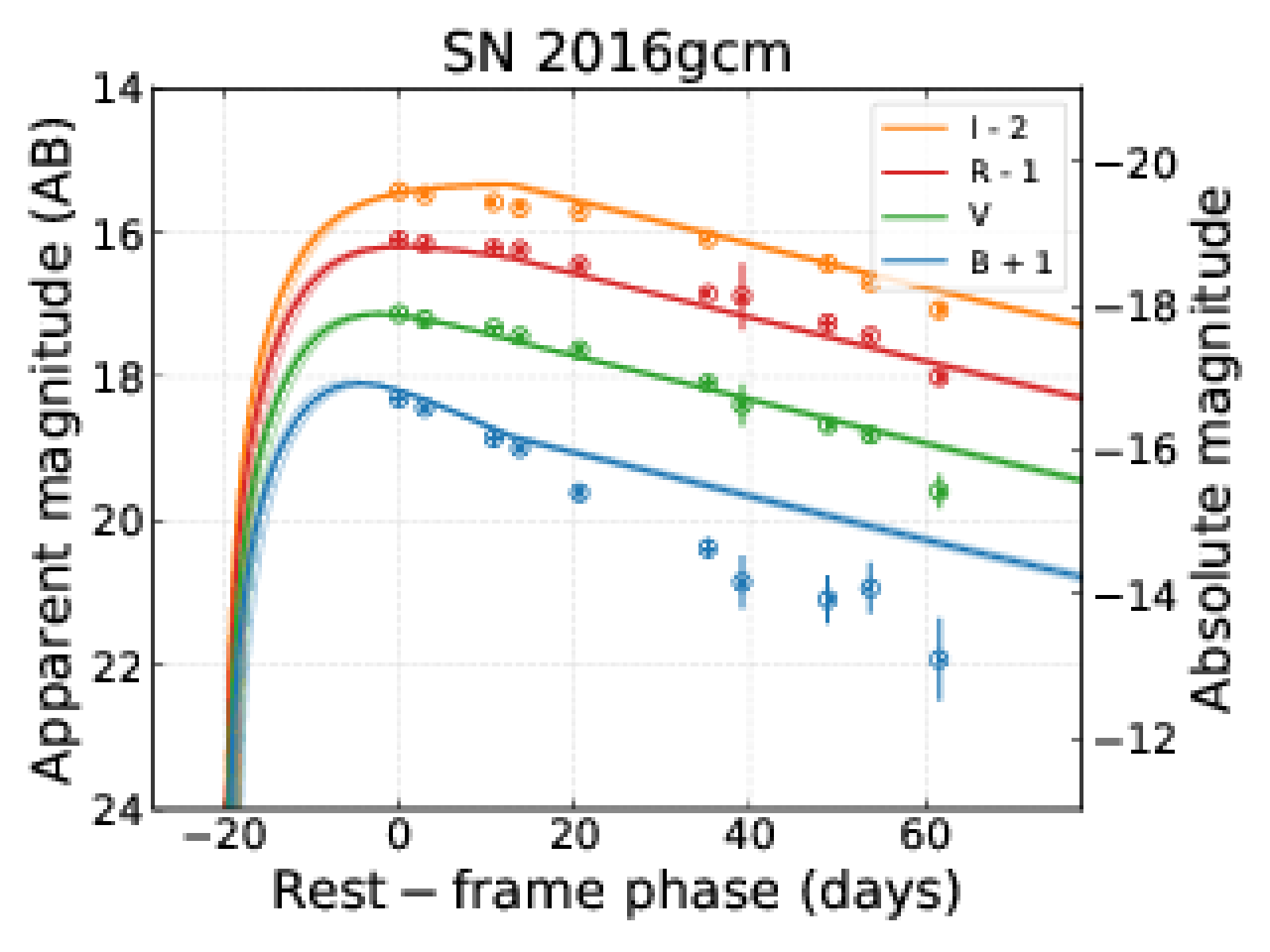}
        \includegraphics[width=0.49\columnwidth]{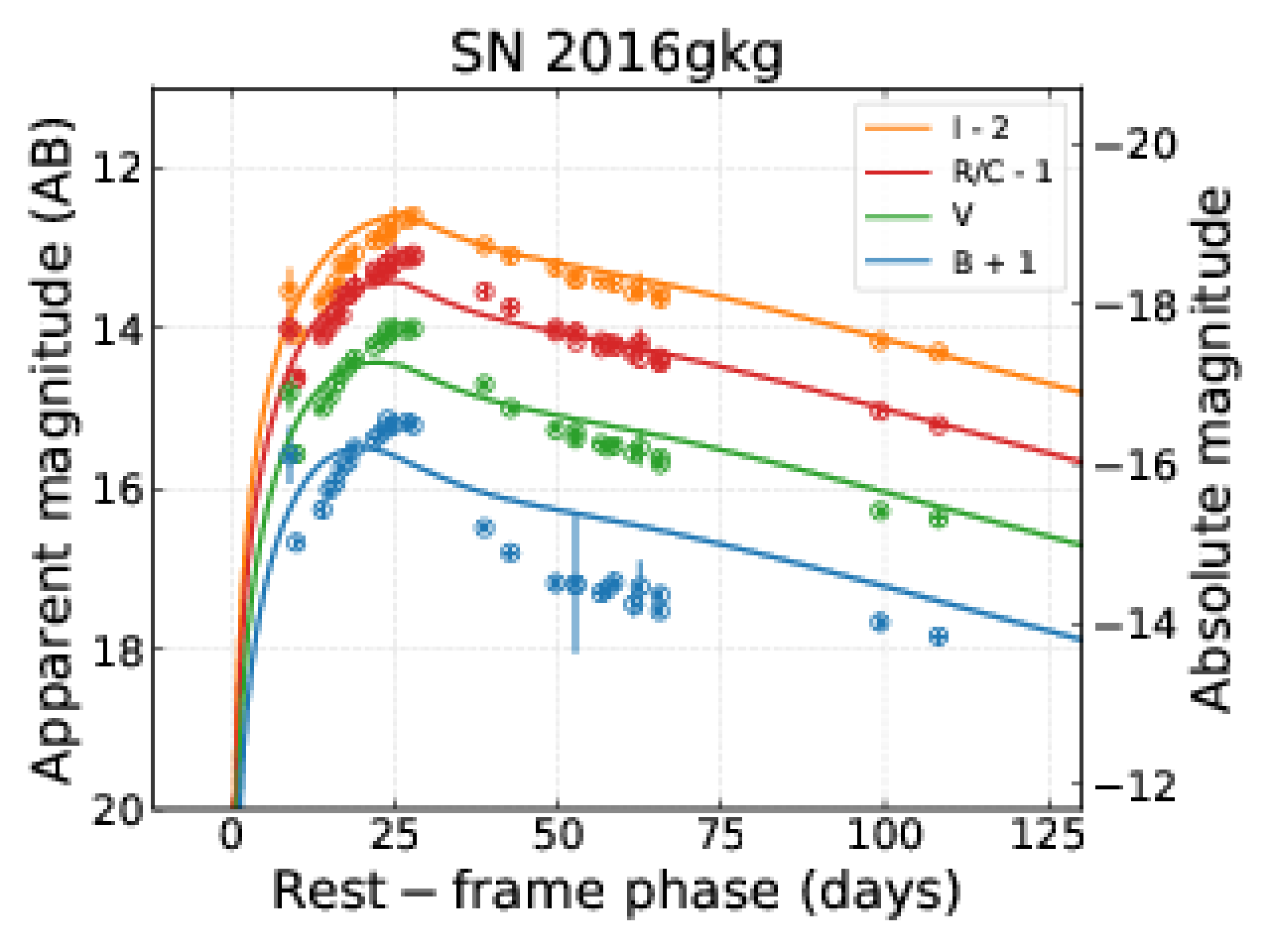}
        \includegraphics[width=0.49\columnwidth]{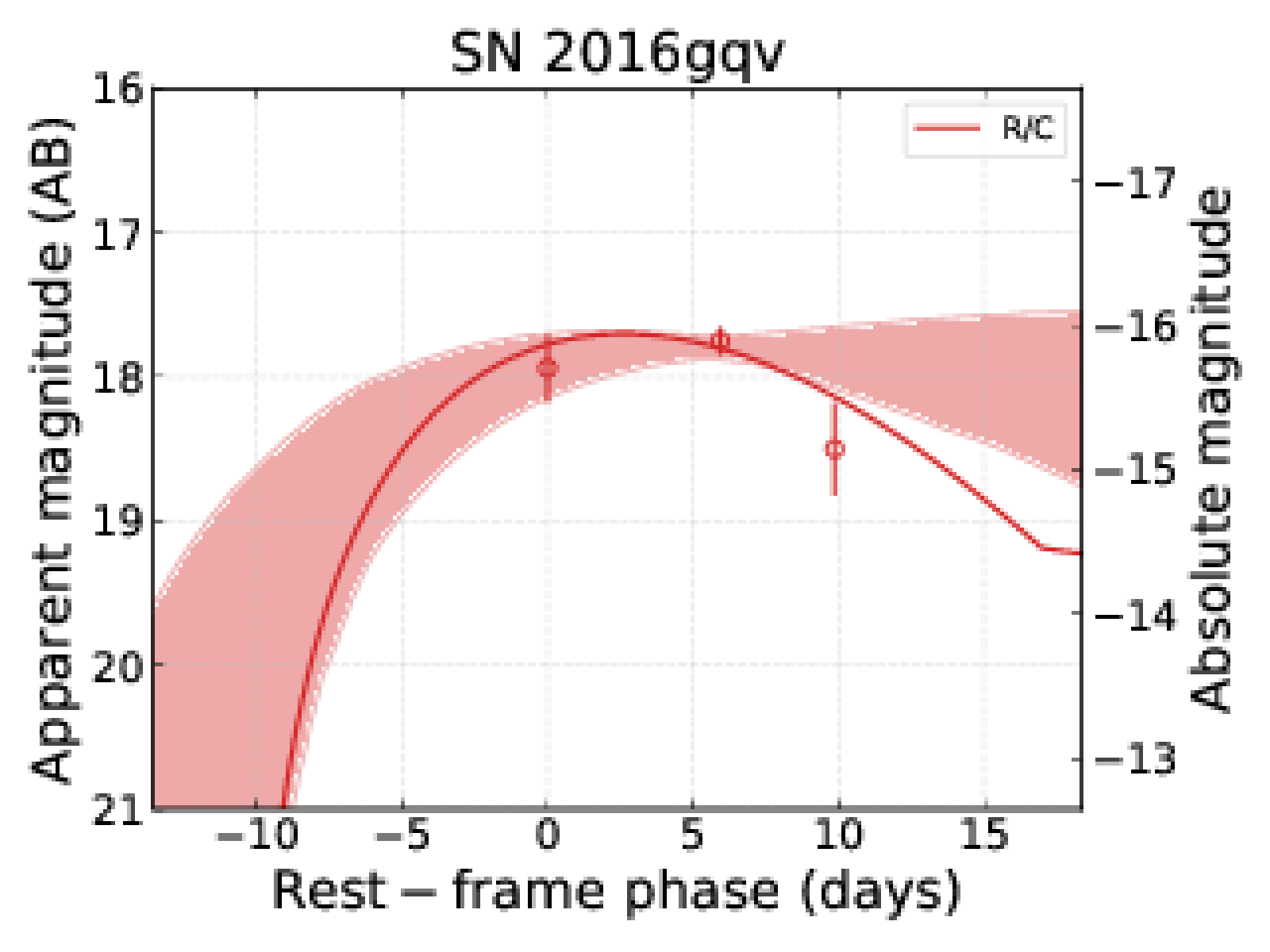}
        \includegraphics[width=0.49\columnwidth]{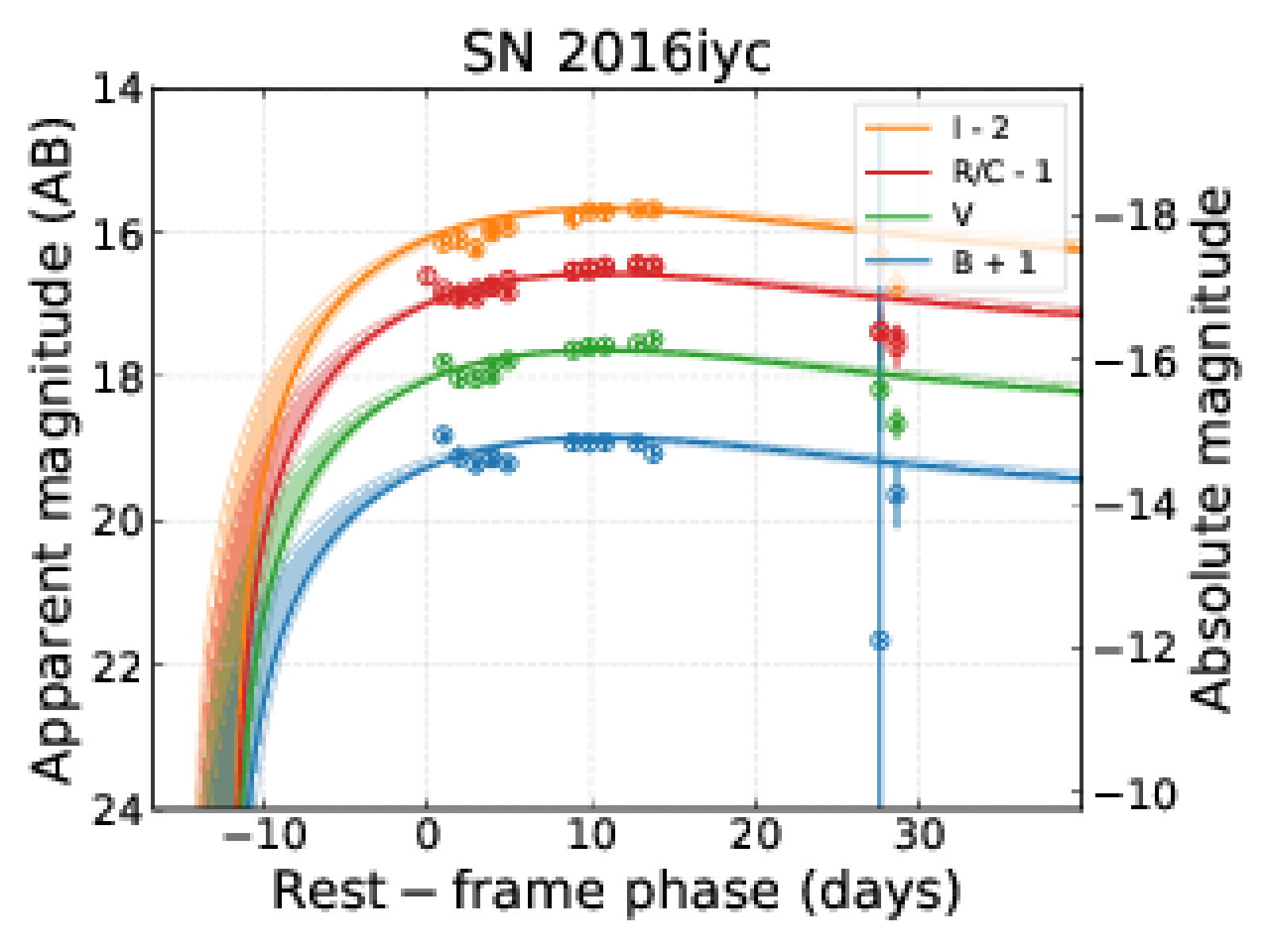}
        \includegraphics[width=0.49\columnwidth]{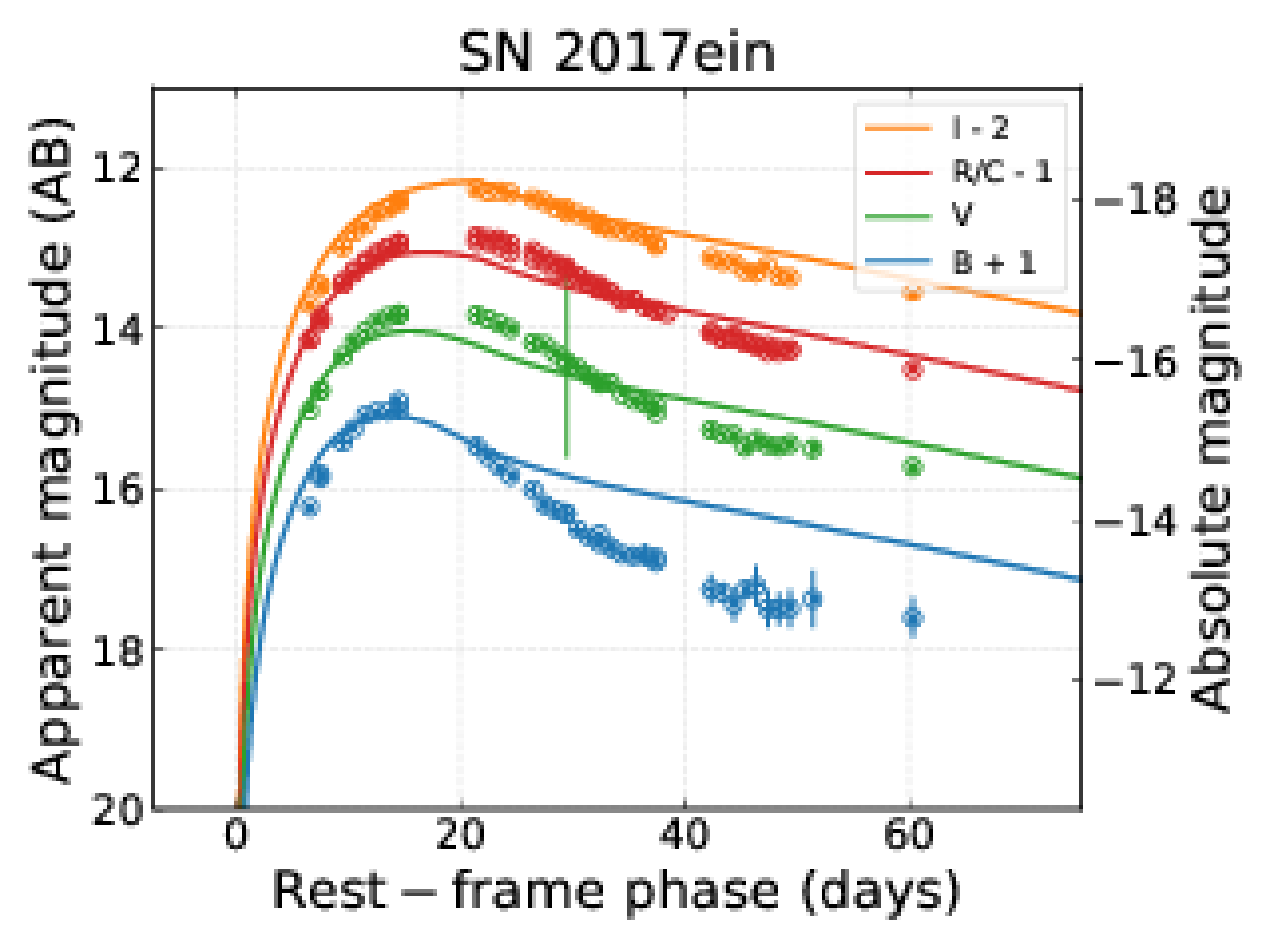}
        \includegraphics[width=0.49\columnwidth]{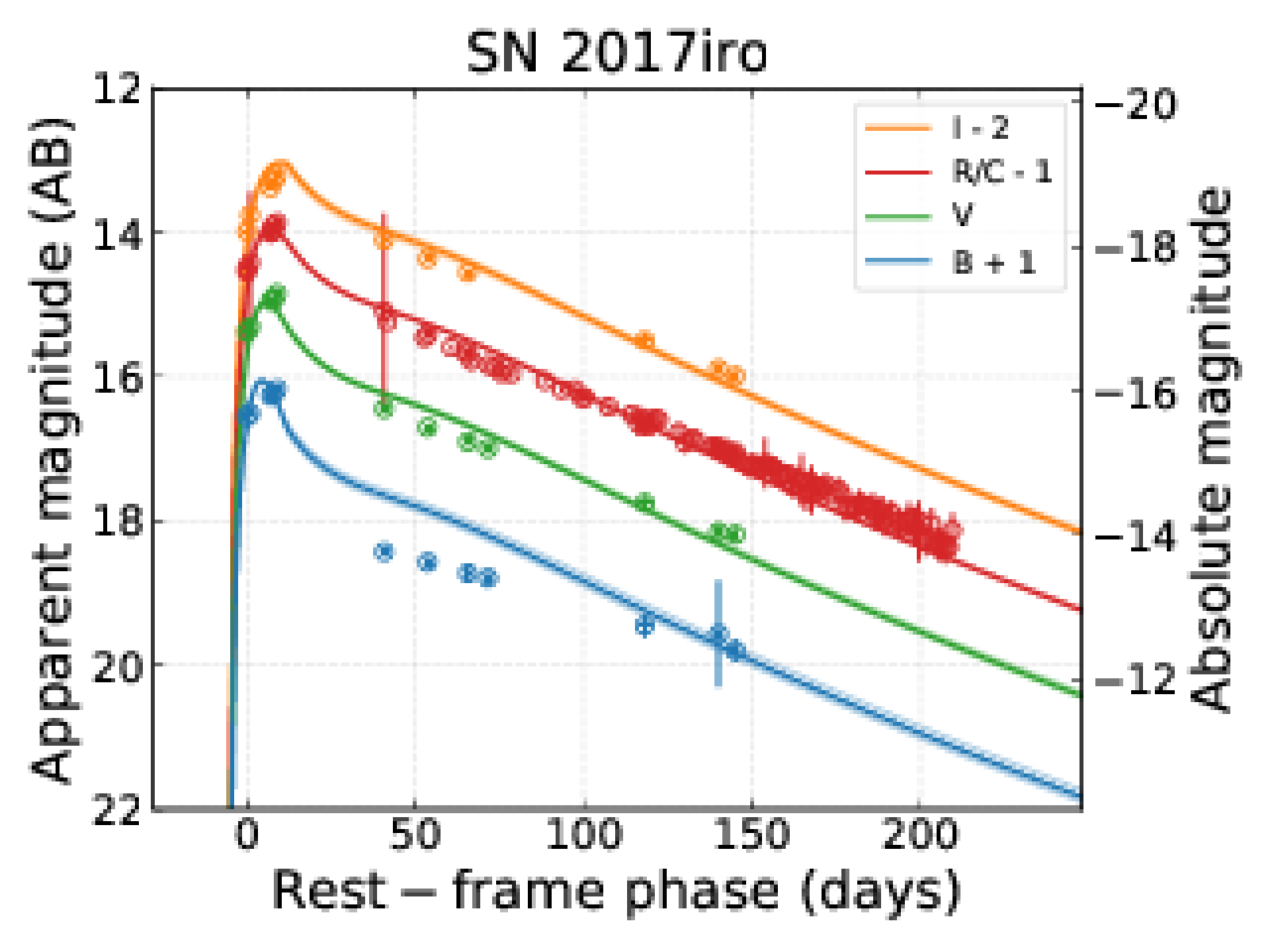}
        \includegraphics[width=0.49\columnwidth]{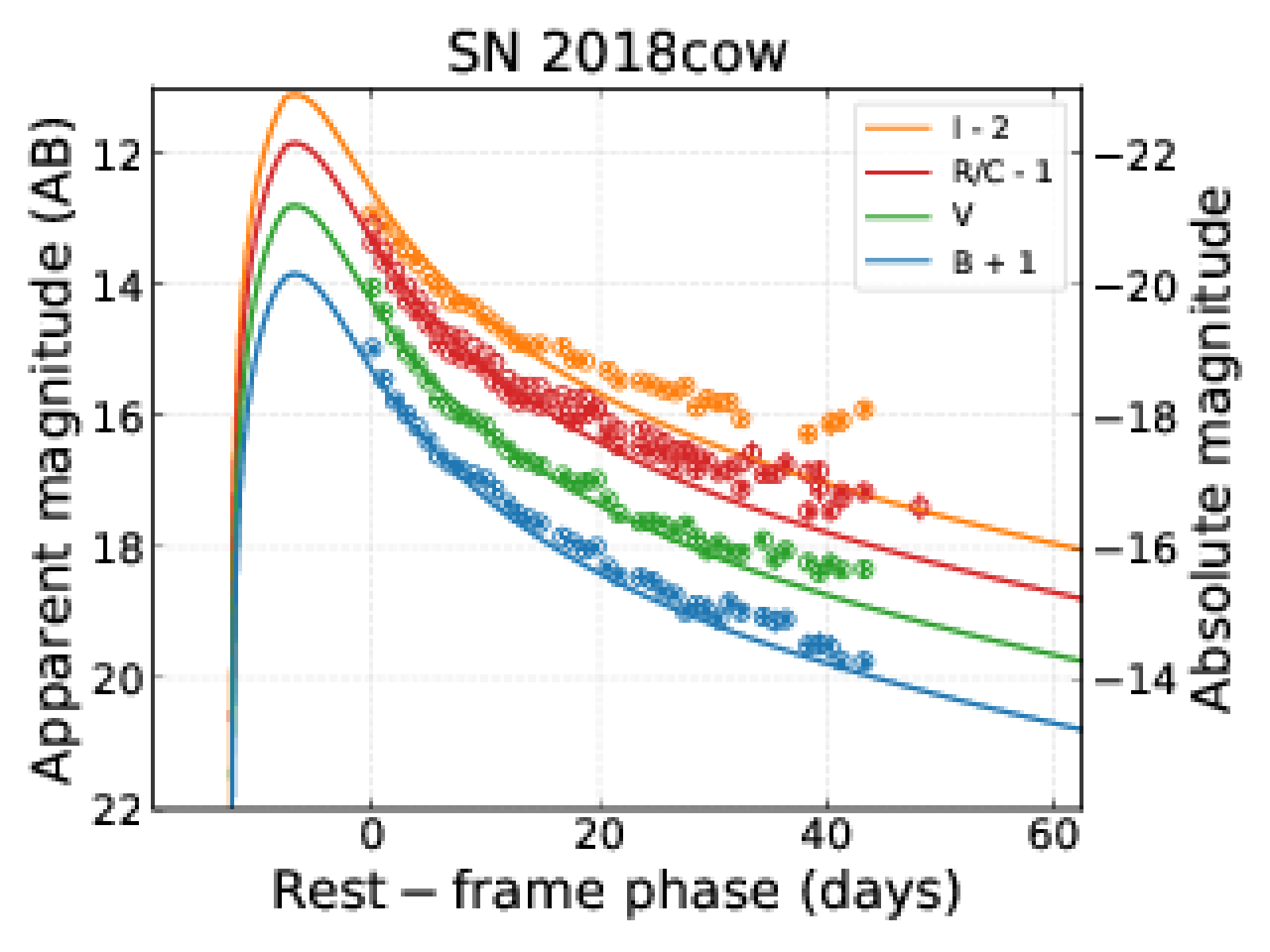}
        \includegraphics[width=0.49\columnwidth]{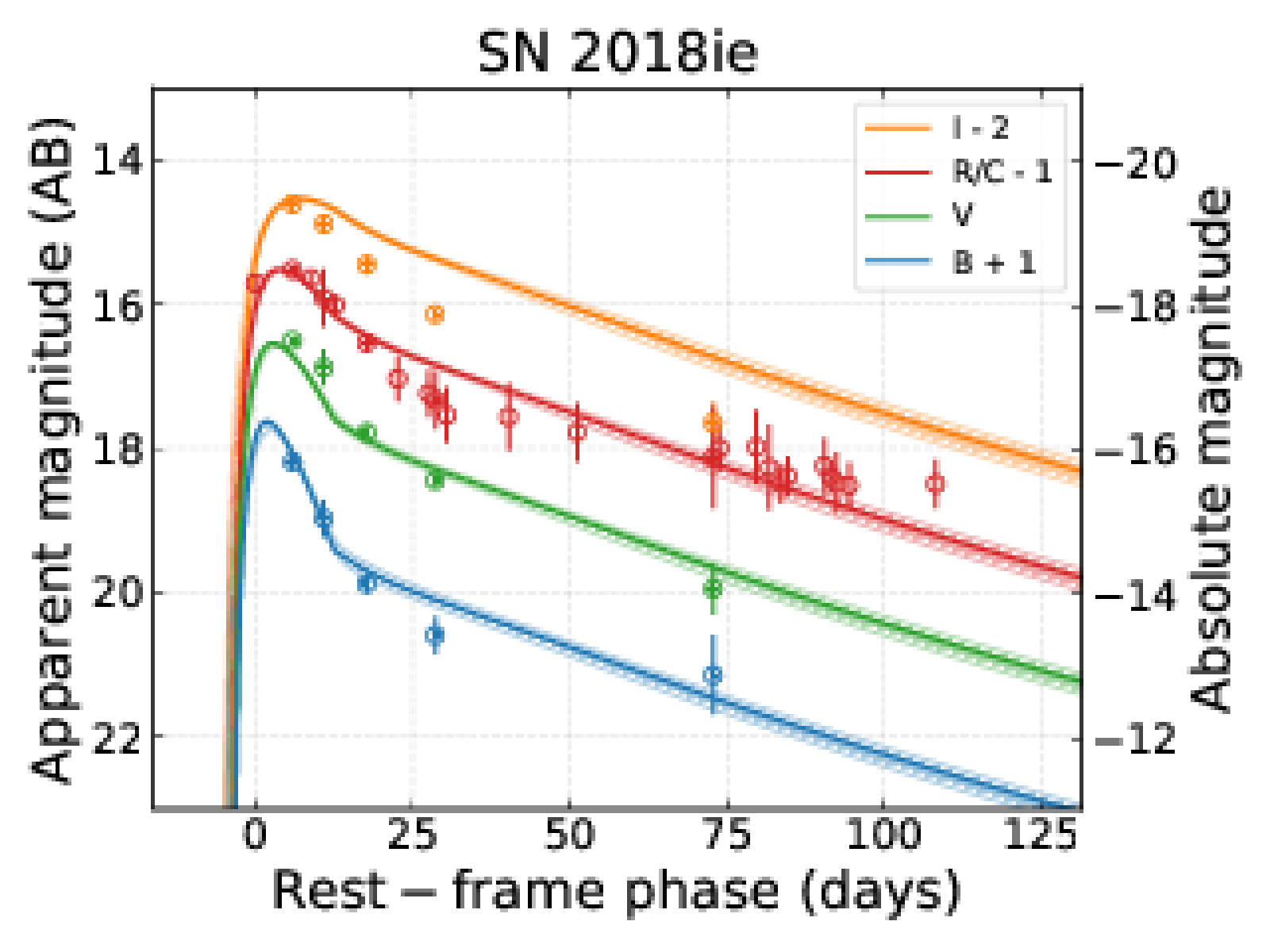}
        \includegraphics[width=0.49\columnwidth]{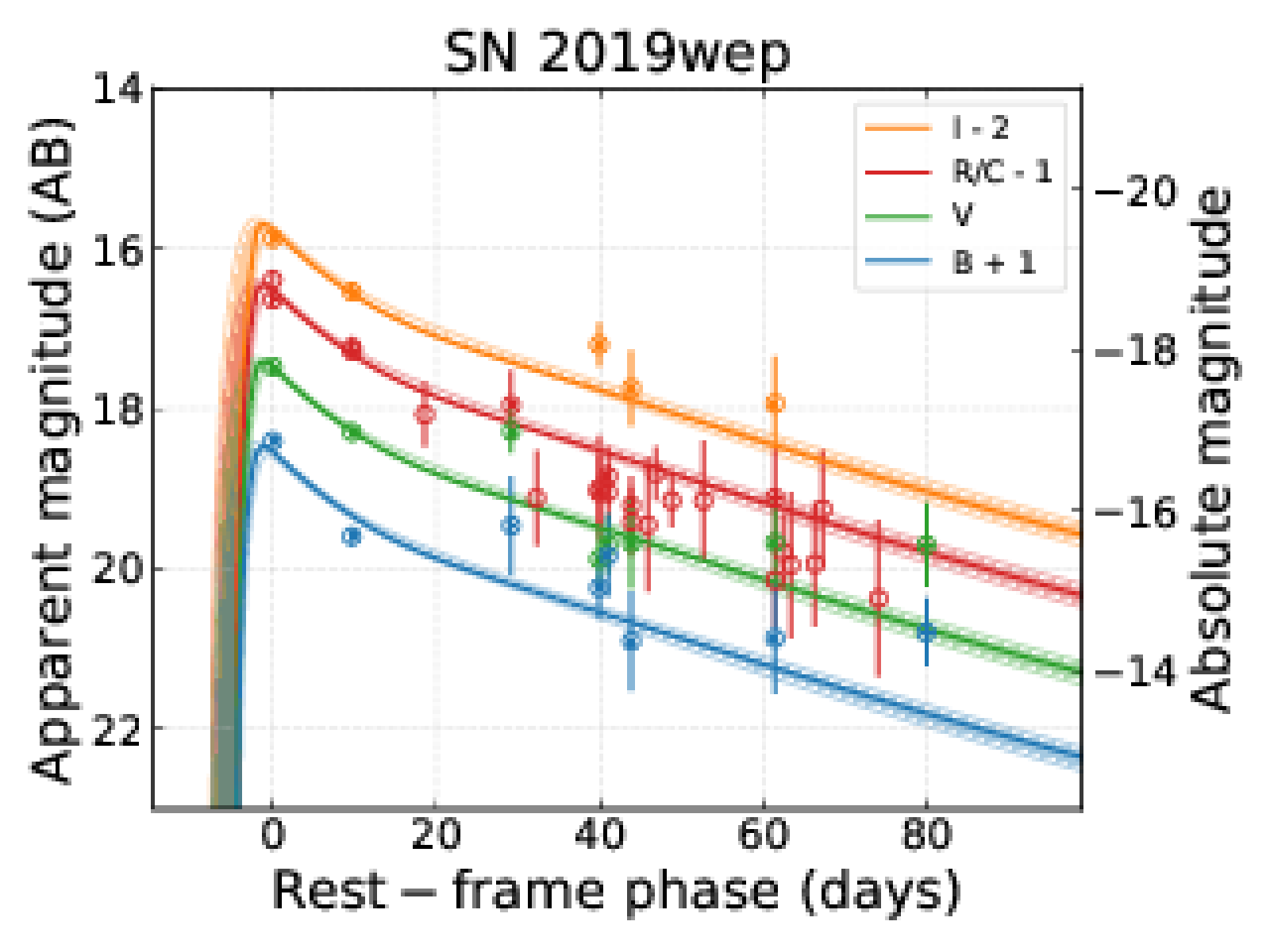}
        \includegraphics[width=0.49\columnwidth]{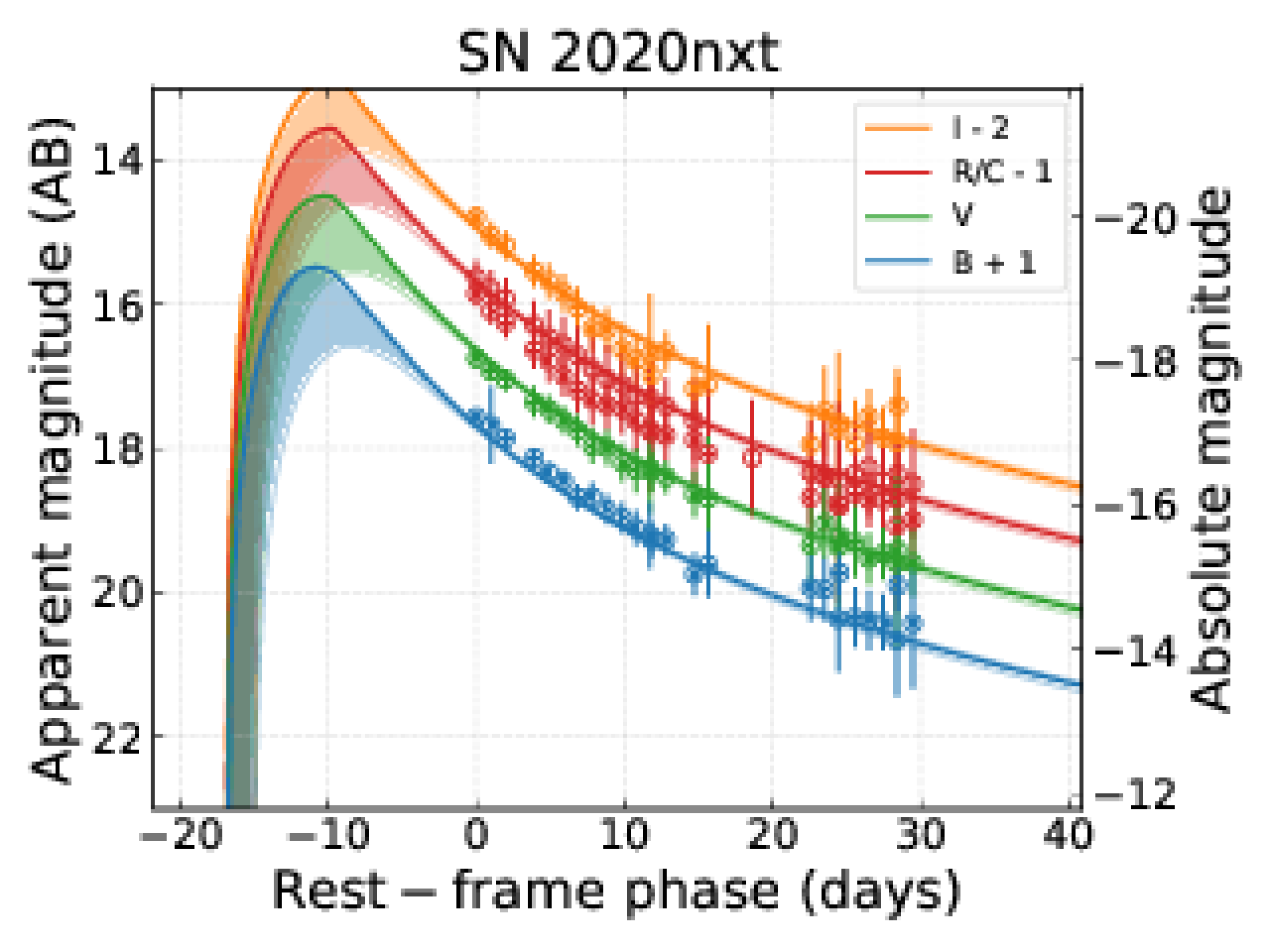}
        \includegraphics[width=0.49\columnwidth]{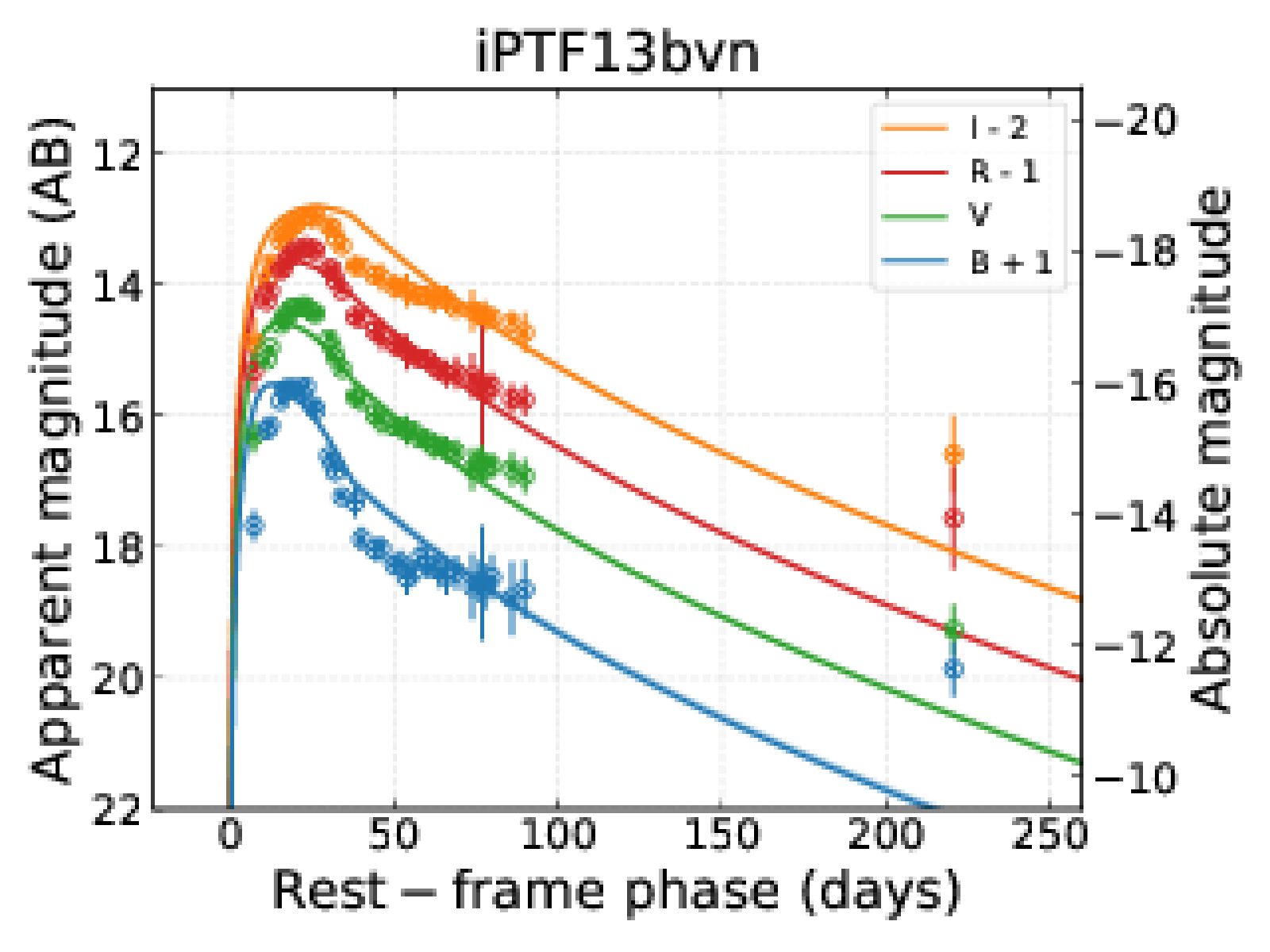}
        \includegraphics[width=0.49\columnwidth]{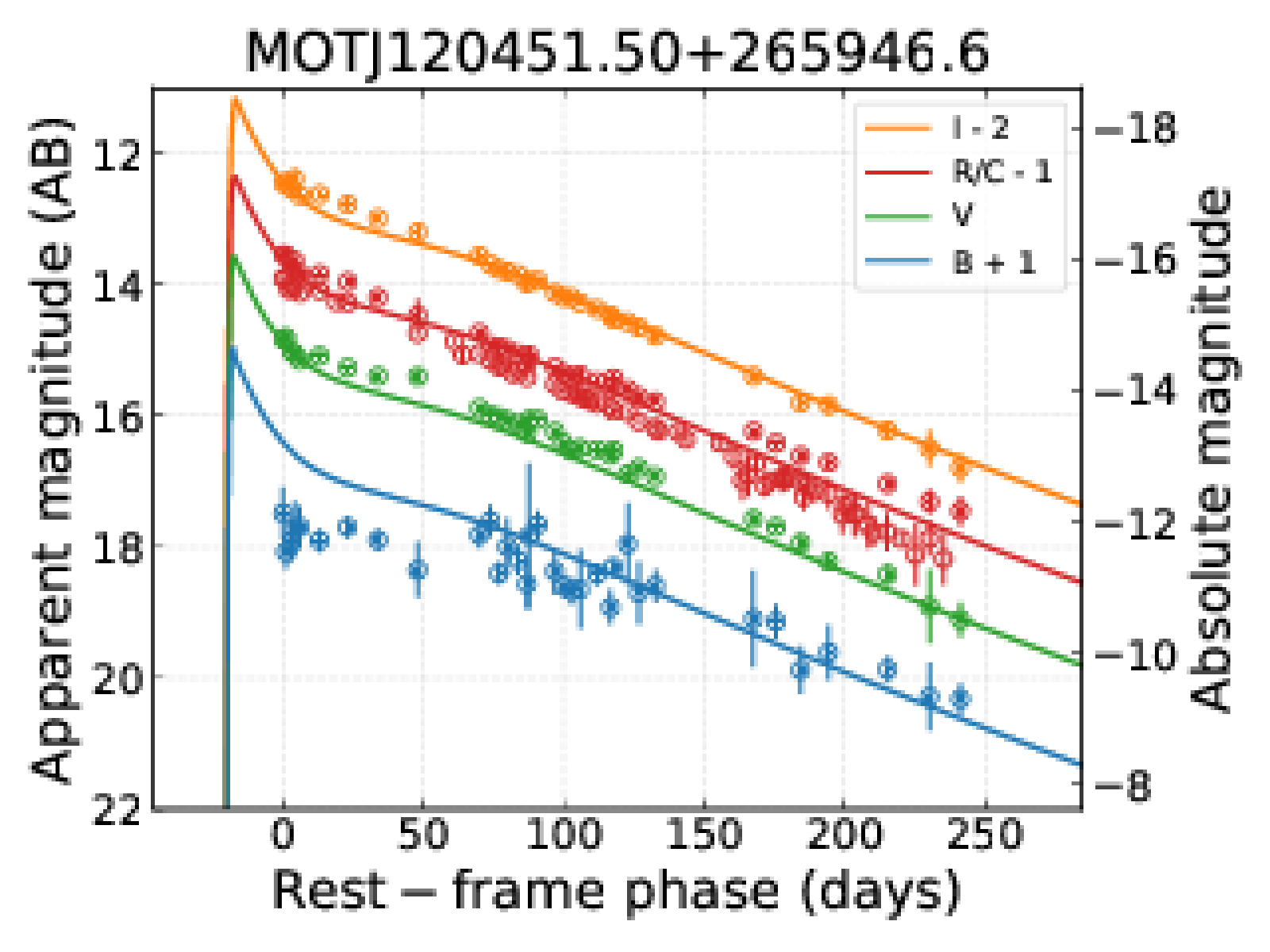}
        \caption{Continued.}
\end{figure*}

\begin{figure*}
        \includegraphics[width=0.49\columnwidth]{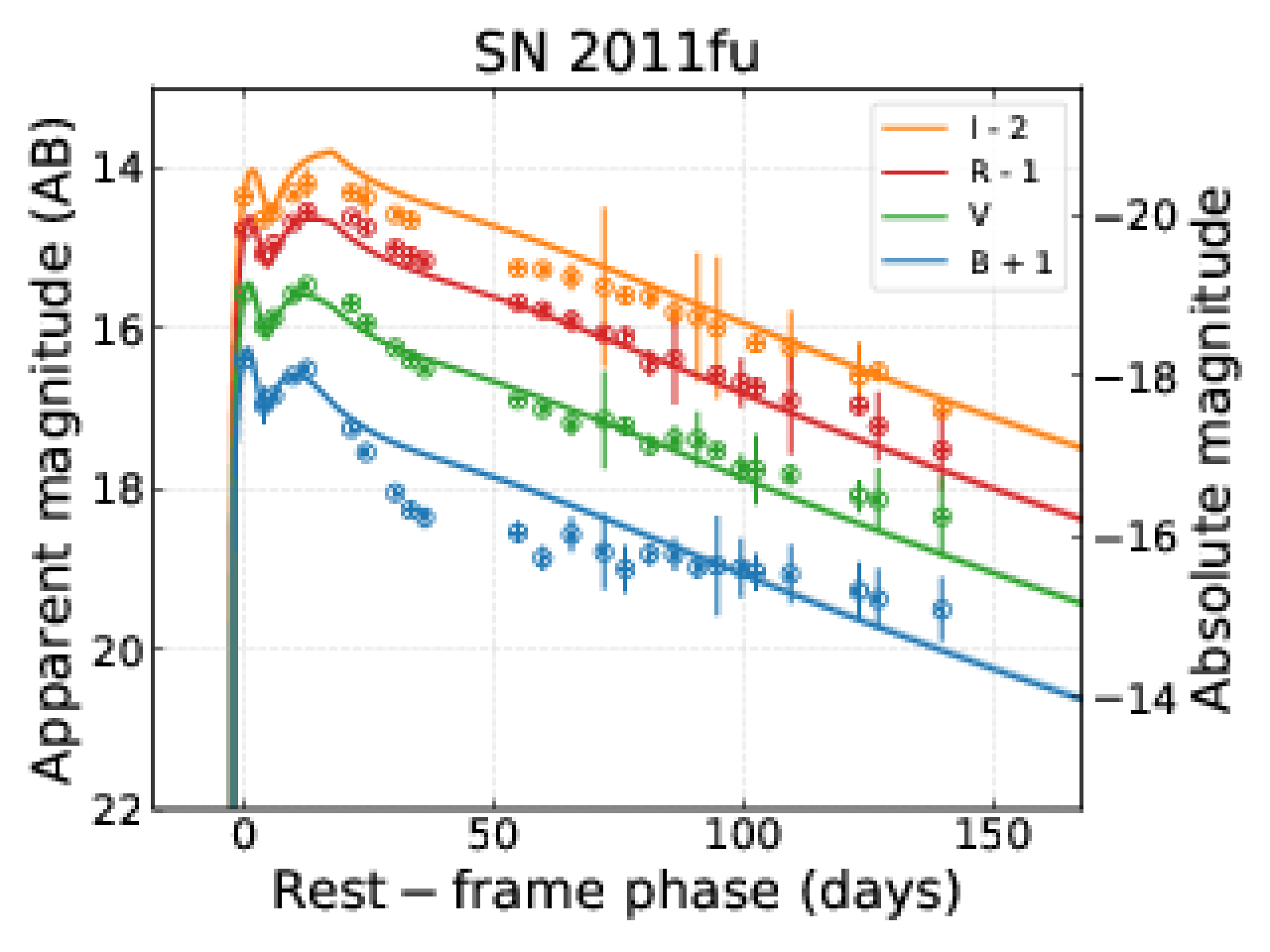}
        \includegraphics[width=0.49\columnwidth]{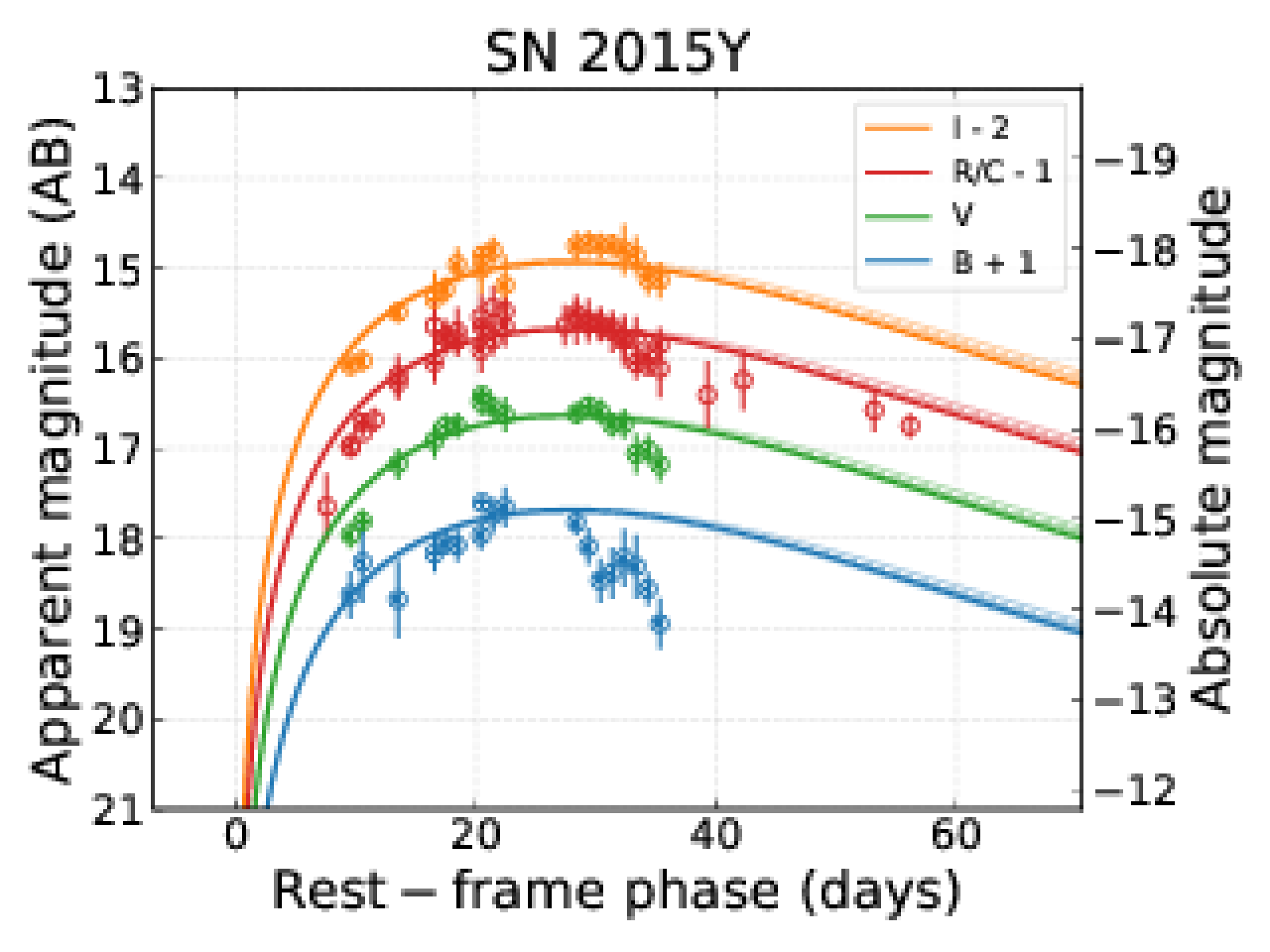}
        \includegraphics[width=0.49\columnwidth]{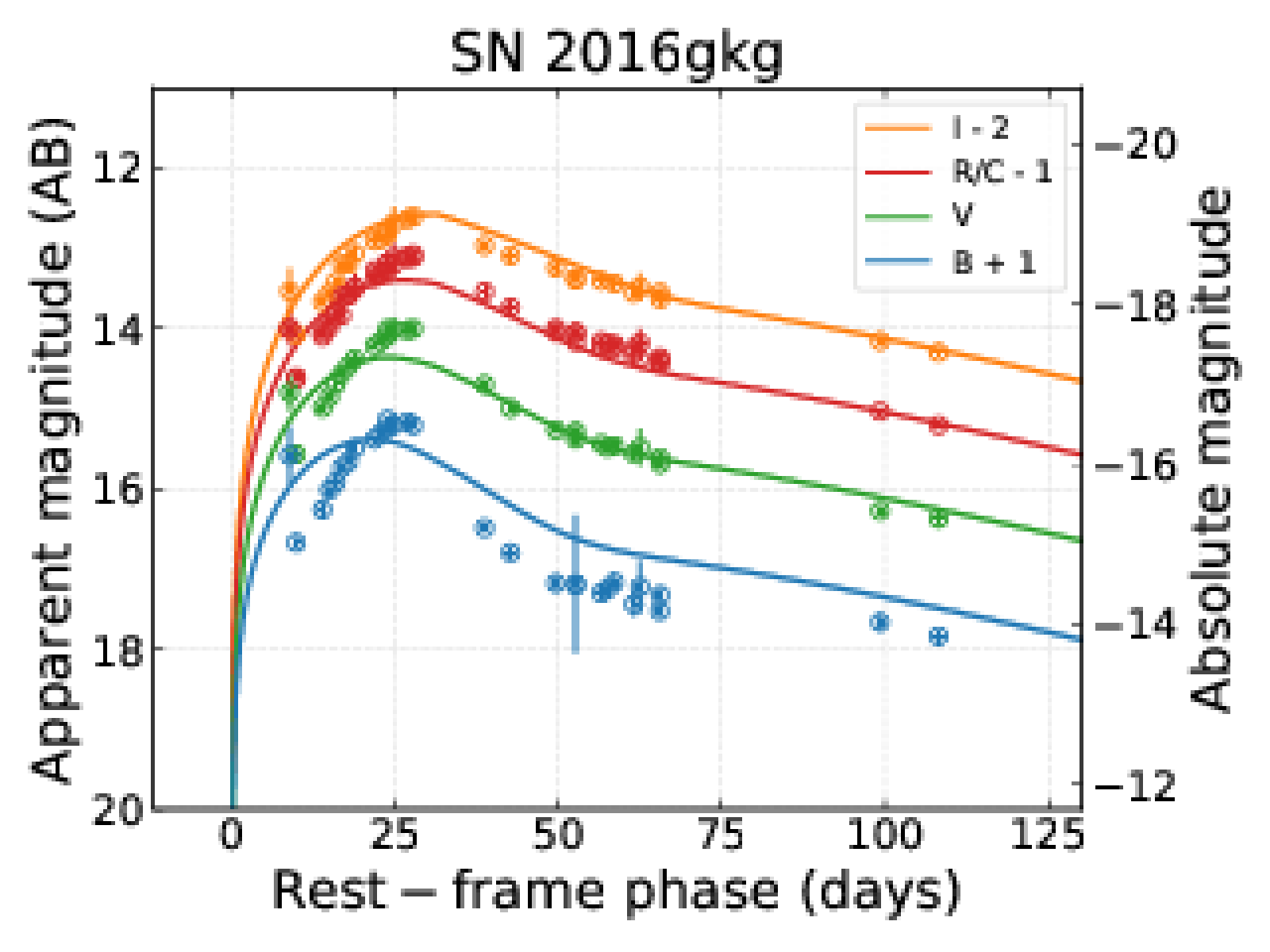}
        \includegraphics[width=0.49\columnwidth]{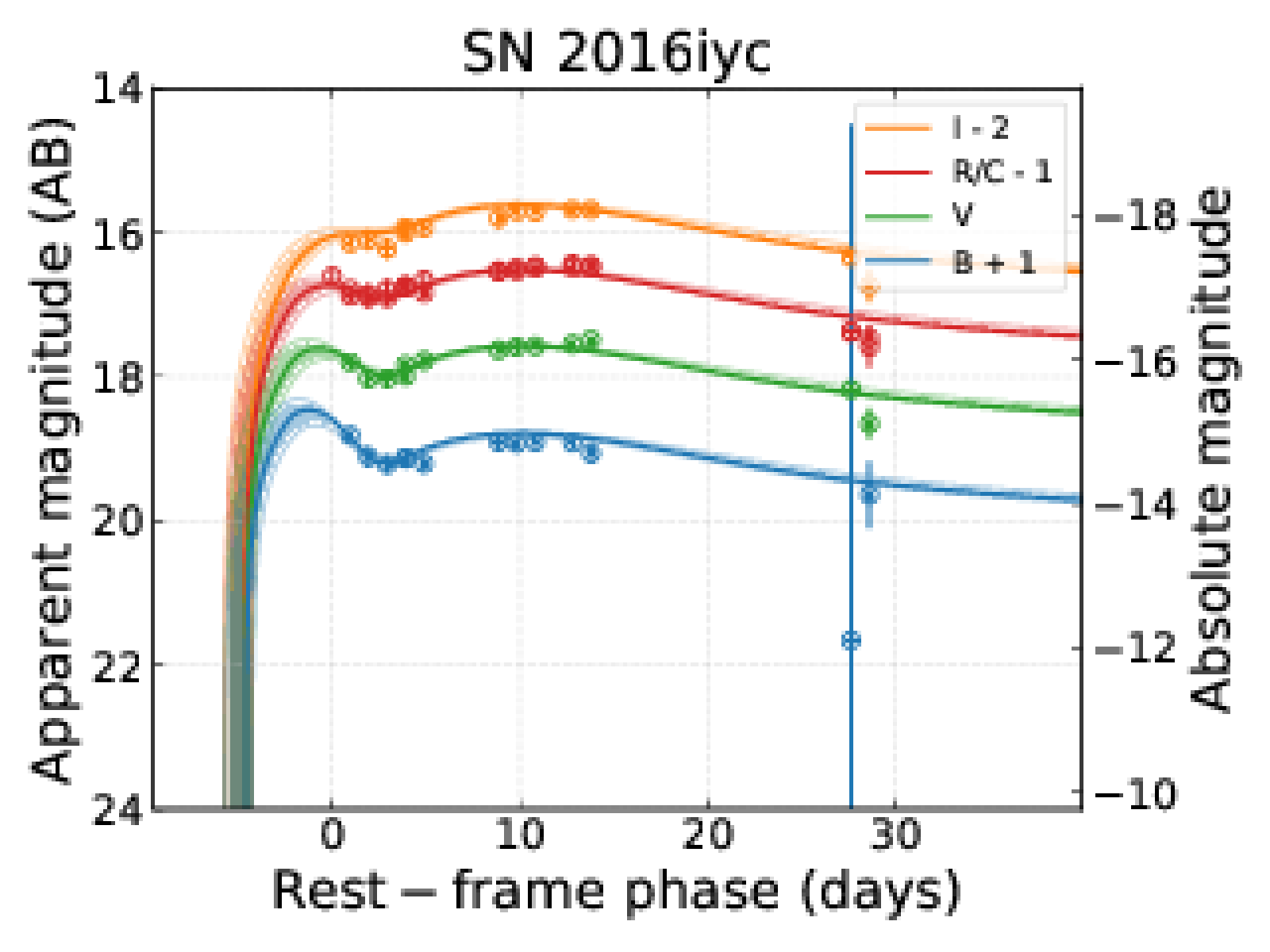}
        \caption{Model fitting of the four SNe with the cooling model}
\label{modelfittinglcof4coolingmodelsesne}
\end{figure*}

\begin{figure*}
        \includegraphics[width=0.49\columnwidth]{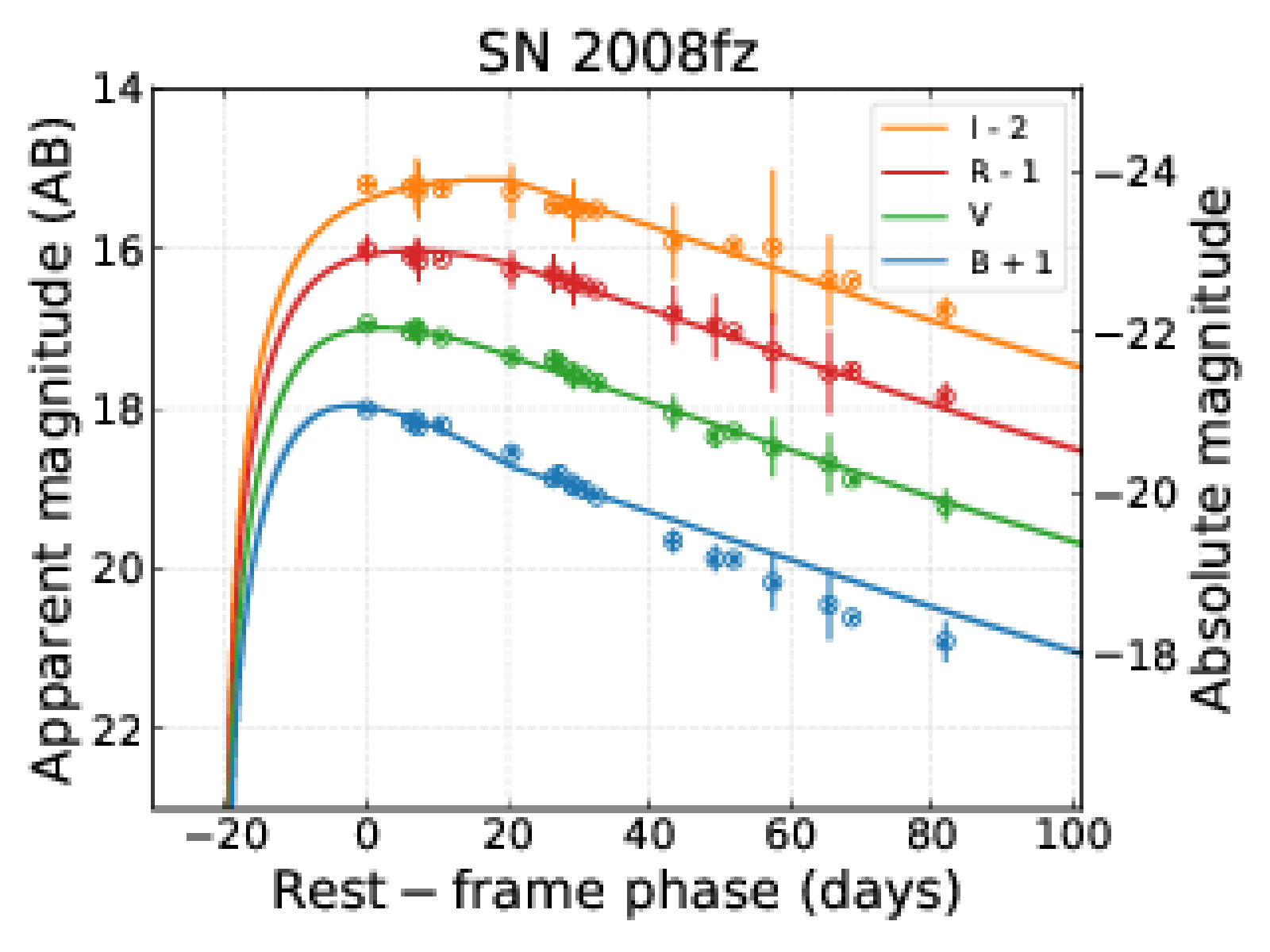}
        \includegraphics[width=0.49\columnwidth]{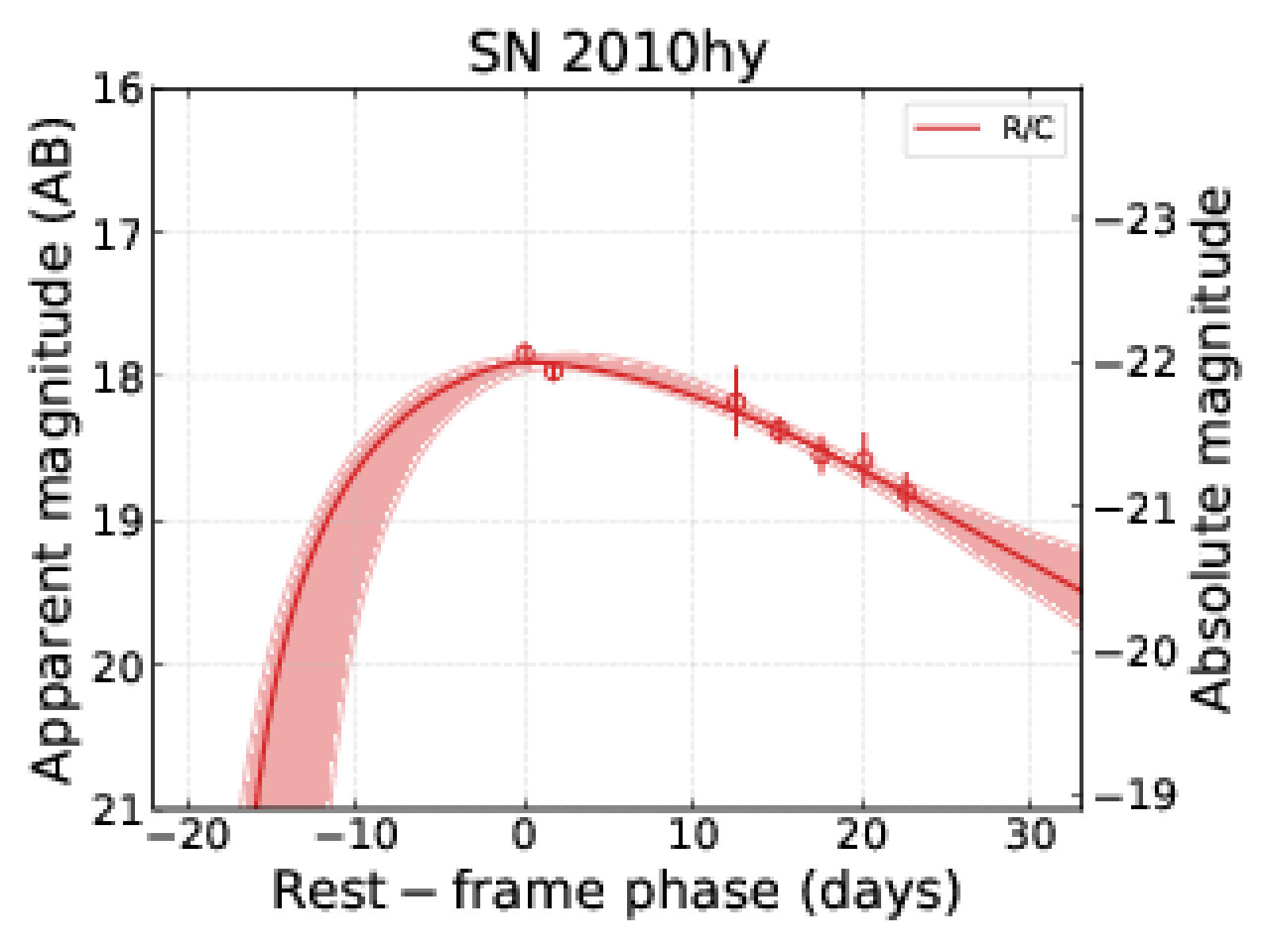}
        \includegraphics[width=0.49\columnwidth]{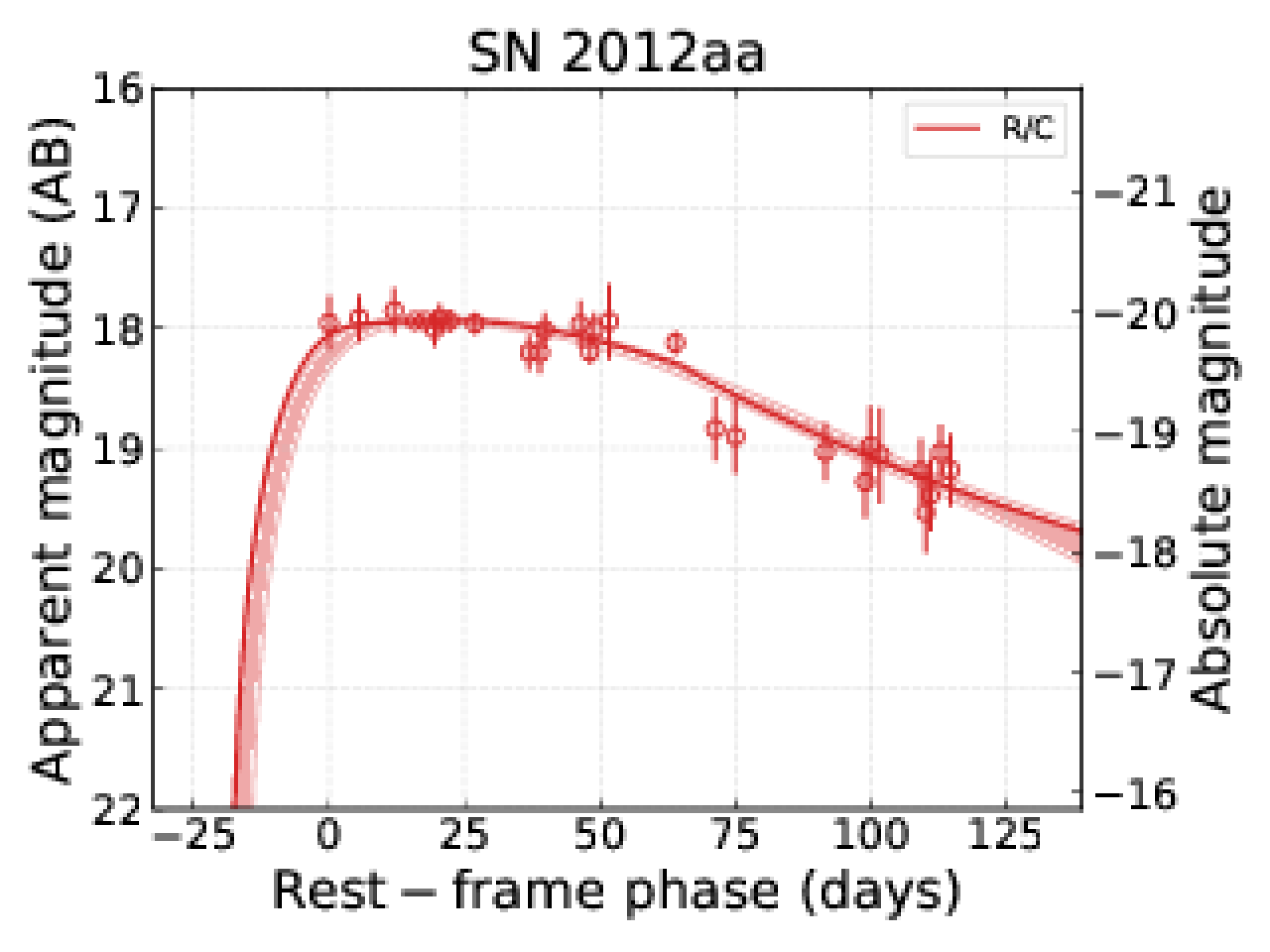}
        \includegraphics[width=0.49\columnwidth]{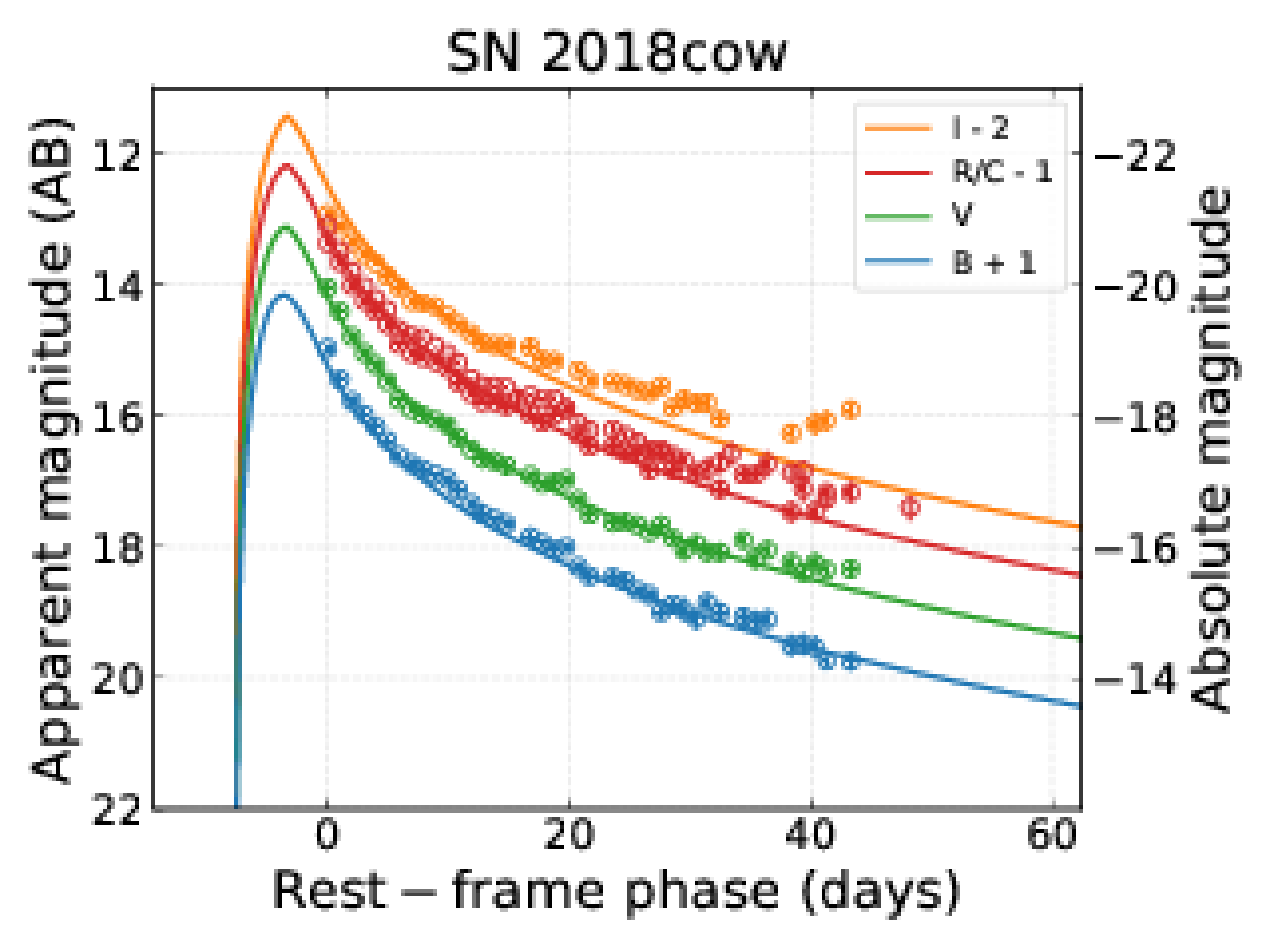}
        \caption{Model fitting of the four SNe with the magnetar model}
\label{modelfittinglcof4magnitarmodelsesne}
\end{figure*}

\begin{landscape}
\begin{table}
\centering\setlength{\tabcolsep}{4pt}      % column separation
\renewcommand{\arraystretch}{1.2}          %row space
\caption{\label{table:LC_Param_Ni}Best-fitting parameters of the $^{56}$Ni model for the SNe.}

\hspace{-110pt}
\begin{tabular}{lcccccccccccc}   % including 68 SNe (except for SN~2009C)
\hline\hline\noalign{\smallskip}
Name   &  $M_{\rm ej}$   &  $v_{9}$             &  $M_{\rm Ni}$   & $\log(\kappa_\gamma, {\rm Ni})$ & $T_{\rm f}$  & $t_{\rm shift}$ &  $\chi^2$/dof \\
       & (M$_{\odot}$)   & ($10^{9}$\,cm\,s$^{-1}$) & (M$_{\odot}$) &   cm$^2$\,g$^{-1}$    & ($10^{3}$\,K) & (days)          &               \\

\noalign{\smallskip}\hline\noalign{\smallskip}
SN~2003gk               & $29.11  ^{+7.11   }_{-7.00  } (25.04 )$ & $4.61   ^{+0.72  }_{-1.10 } (3.99  )$ & $0.17    ^{+0.04    }_{-0.04   } (0.15   )$ & $-0.47   ^{+2.87   }_{-0.43  } (-0.23  )$ & $2.64   ^{+0.19  }_{-0.11 } (2.57  )$ & $-18.66  ^{+2.23  }_{-1.71 } (-17.55 )$ & 0.45    (0.63 ) \\
SN~2006el               & $0.11   ^{+0.06   }_{-0.02  } (0.12  )$ & $0.34   ^{+0.01  }_{-0.01 } (0.34  )$ & $0.22    ^{+0.01    }_{-0.01   } (0.22   )$ & $-1.19   ^{+0.09   }_{-0.17  } (-1.25  )$ & $6.22   ^{+0.34  }_{-0.29 } (6.18  )$ &            ---                        & 6.18    (6.24 ) \\
SN~2006ep               & $0.10   ^{+0.01   }_{-0.01  } (0.10  )$ & $0.36   ^{+0.01  }_{-0.01 } (0.36  )$ & $0.52    ^{+0.01    }_{-0.02   } (0.52   )$ & $-1.51   ^{+0.03   }_{-0.03  } (-1.53  )$ & $9.41   ^{+0.22  }_{-0.22 } (9.48  )$ &            ---                        & 11.38   (11.40) \\
SN~2006jc               & $1.41   ^{+0.52   }_{-0.65  } (1.84  )$ & $2.90   ^{+0.93  }_{-1.28 } (3.67  )$ & $0.15    ^{+0.01    }_{-0.01   } (0.16   )$ & $-1.13   ^{+0.11   }_{-0.19  } (-1.06  )$ & $10.00  ^{+0.01  }_{-0.01 } (10.00 )$ & $-10.48  ^{+0.45  }_{-0.49 } (-10.69 )$ & 104.49  (104.72) \\
SN~2006lc               & $1.20   ^{+0.22   }_{-0.19  } (1.28  )$ & $0.64   ^{+0.04  }_{-0.04 } (0.64  )$ & $0.18    ^{+0.01    }_{-0.01   } (0.18   )$ & $-1.50   ^{+2.55   }_{-1.74  } (0.23   )$ & $7.44   ^{+0.29  }_{-0.28 } (7.36  )$ &            ---                        & 2.75    (2.86 ) \\
SN~2007C                & $4.04   ^{+1.09   }_{-1.49  } (3.65  )$ & $4.03   ^{+1.03  }_{-1.45 } (3.57  )$ & $0.14    ^{+0.01    }_{-0.01   } (0.14   )$ & $-0.50   ^{+0.13   }_{-0.22  } (-0.58  )$ & $9.93   ^{+0.05  }_{-0.10 } (9.98  )$ & $-4.47   ^{+0.43  }_{-0.51 } (-4.56  )$ & 9.94    (9.96 ) \\
SN~2007D                & $14.30  ^{+18.13  }_{-14.61 } (22.76 )$ & $2.49   ^{+1.28  }_{-1.45 } (3.13  )$ & $0.18    ^{+0.20    }_{-0.08   } (0.19   )$ & $3.06    ^{+1.87   }_{-1.88  } (1.26   )$ & $4.82   ^{+3.78  }_{-2.35 } (3.55  )$ & $-15.98  ^{+3.85  }_{-2.57 } (-16.40 )$ &    ---          \\
SN~2007ag               & $2.18   ^{+0.24   }_{-0.21  } (2.03  )$ & $1.17   ^{+0.17  }_{-0.16 } (1.22  )$ & $0.058   ^{+0.004   }_{-0.003  } (0.056  )$ & $-1.27   ^{+2.24   }_{-1.98  } (0.71   )$ & $4.67   ^{+0.09  }_{-0.09 } (4.68  )$ & $-12.15  ^{+1.64  }_{-2.06 } (-11.48 )$ & 2.13    (2.20 ) \\
SN~2007cl               & $0.12   ^{+0.07   }_{-0.03  } (0.15  )$ & $0.46   ^{+0.06  }_{-0.04 } (0.48  )$ & $0.58    ^{+0.06    }_{-0.08   } (0.54   )$ & $-1.37   ^{+0.13   }_{-0.09  } (-1.44  )$ & $8.15   ^{+0.13  }_{-0.13 } (8.17  )$ & $-18.26  ^{+2.42  }_{-1.55 } (-17.34 )$ & 6.96    (7.02 ) \\
SN~2007kj               & $0.91   ^{+0.53   }_{-0.35  } (0.67  )$ & $3.86   ^{+0.85  }_{-1.07 } (3.80  )$ & $0.17    ^{+0.02    }_{-0.01   } (0.17   )$ & $0.19    ^{+1.71   }_{-1.41  } (1.50   )$ & $8.92   ^{+0.32  }_{-0.29 } (8.93  )$ & $-6.31   ^{+1.44  }_{-2.03 } (-6.07  )$ & 2.08    (2.17 ) \\
SN~2007ru               & $0.65   ^{+4.61   }_{-0.18  } (0.75  )$ & $1.42   ^{+2.43  }_{-0.05 } (1.46  )$ & $0.49    ^{+0.03    }_{-0.05   } (0.46   )$ & $-1.25   ^{+0.13   }_{-0.11  } (-1.21  )$ & $6.09   ^{+0.15  }_{-0.71 } (5.92  )$ &            ---                        & 38.59   (39.35) \\
SN~2007rw               & $0.71   ^{+0.05   }_{-0.05  } (0.74  )$ & $4.99   ^{+0.03  }_{-0.07 } (4.96  )$ & $0.040   ^{+0.001   }_{-0.001  } (0.040  )$ & $0.0097  ^{+0.049  }_{-0.050 } (-0.021 )$ & $4.95   ^{+0.03  }_{-0.03 } (4.95  )$ & $-4.80   ^{+0.12  }_{-0.14 } (-4.88  )$ & 33.39   (33.47) \\
SN~2007rz               & $0.78   ^{+2.07   }_{-0.84  } (1.31  )$ & $1.06   ^{+1.38  }_{-2.11 } (3.10  )$ & $0.043   ^{+0.011   }_{-0.011  } (0.047  )$ & $1.40    ^{+2.08   }_{-1.84  } (0.92   )$ & $4.96   ^{+0.28  }_{-0.44 } (4.75  )$ & $-13.29  ^{+5.06  }_{-3.81 } (-14.55 )$ & 1.45    (1.80 ) \\
SN~2007uy               & $0.15   ^{+0.01   }_{-0.01  } (0.15  )$ & $0.33   ^{+0.01  }_{-0.01 } (0.33  )$ & $0.16    ^{+0.01    }_{-0.01   } (0.16   )$ & $-1.57   ^{+0.02   }_{-0.01  } (-1.56  )$ & $6.17   ^{+0.05  }_{-0.05 } (6.17  )$ &            ---                        & 61.42   (61.43) \\
SN~2008aq               & $0.12   ^{+0.03   }_{-0.02  } (0.13  )$ & $0.64   ^{+0.02  }_{-0.02 } (0.63  )$ & $0.032   ^{+0.001   }_{-0.001  } (0.032  )$ & $-0.29   ^{+0.10   }_{-0.10  } (-0.34  )$ & $5.12   ^{+0.04  }_{-0.04 } (5.13  )$ & $-3.84   ^{+0.23  }_{-0.28 } (-3.96  )$ & 34.25   (34.25) \\
SN~2008cw               & $0.41   ^{+0.32   }_{-0.29  } (0.56  )$ & $0.74   ^{+0.19  }_{-0.08 } (0.79  )$ & $0.081   ^{+0.007   }_{-0.008  } (0.080  )$ & $0.44    ^{+1.84   }_{-1.83  } (1.30   )$ & $3.26   ^{+1.38  }_{-1.54 } (3.73  )$ & $-14.45  ^{+2.95  }_{-2.35 } (-13.73 )$ & 7.46    (7.85 ) \\
SN~2008dq               & $2.95   ^{+8.87   }_{-8.27  } (11.12 )$ & $0.52   ^{+1.46  }_{-2.49 } (3.01  )$ & $0.097   ^{+0.011   }_{-0.010  } (0.089  )$ & $0.84    ^{+1.94   }_{-1.94  } (1.15   )$ & $6.15   ^{+0.19  }_{-1.77 } (4.45  )$ & $-14.69  ^{+3.21  }_{-3.81 } (-13.28 )$ & 2.63    (3.16 ) \\
SN~2008eb               & $0.35   ^{+0.07   }_{-0.07  } (0.32  )$ & $0.60   ^{+0.03  }_{-0.03 } (0.59  )$ & $0.22    ^{+0.01    }_{-0.01   } (0.23   )$ & $-1.57   ^{+0.04   }_{-0.01  } (-1.56  )$ & $8.02   ^{+0.22  }_{-0.21 } (8.02  )$ &            ---                        & 3.01    (3.02 ) \\
SN~2008ew               & $0.36   ^{+2.76   }_{-0.14  } (0.36  )$ & $0.73   ^{+2.14  }_{-0.07 } (0.76  )$ & $0.090   ^{+0.006   }_{-0.005  } (0.088  )$ & $-1.54   ^{+0.25   }_{-0.10  } (-1.44  )$ & $4.87   ^{+0.14  }_{-0.11 } (4.84  )$ & $-5.94   ^{+0.74  }_{-0.86 } (-5.70  )$ & 4.91    (5.12 ) \\
SN~2008fi               & $1.82   ^{+0.40   }_{-0.75  } (1.74  )$ & $1.10   ^{+0.08  }_{-0.05 } (1.12  )$ & $0.11    ^{+0.01    }_{-0.01   } (0.11   )$ & $2.47    ^{+1.72   }_{-1.72  } (1.47   )$ & $5.13   ^{+0.06  }_{-0.06 } (5.13  )$ & $-18.83  ^{+1.31  }_{-0.77 } (-18.50 )$ & 8.63    (8.76 ) \\
SN~2008fz               & $2.43   ^{+0.29   }_{-0.28  } (2.46  )$ & $2.59   ^{+0.03  }_{-0.03 } (2.60  )$ & $17.52   ^{+0.18    }_{-0.18   } (17.44  )$ & $-0.73   ^{+0.05   }_{-0.05  } (-0.73  )$ & $7.41   ^{+0.05  }_{-0.05 } (7.41  )$ & $-19.97  ^{+0.13  }_{-0.06 } (-19.92 )$ & 7.10    (7.12 ) \\
SN~2008gj               & $0.40   ^{+0.13   }_{-0.17  } (0.39  )$ & $0.61   ^{+0.05  }_{-0.04 } (0.62  )$ & $0.21    ^{+0.01    }_{-0.01   } (0.20   )$ & $-1.15   ^{+0.32   }_{-0.18  } (-1.11  )$ & $6.77   ^{+0.17  }_{-0.17 } (6.75  )$ & $-8.10   ^{+1.19  }_{-1.00 } (-7.90  )$ & 2.92    (2.93 ) \\
SN~2009K                & $12.37  ^{+3.76   }_{-5.29  } (12.40 )$ & $3.51   ^{+1.06  }_{-1.50 } (3.52  )$ & $0.11    ^{+0.01    }_{-0.01   } (0.11   )$ & $-0.20   ^{+2.10   }_{-1.21  } (0.91   )$ & $8.86   ^{+0.29  }_{-0.26 } (8.83  )$ &            ---                        & 8.47    (8.80 ) \\
SN~2009Z                & $27.09  ^{+20.69  }_{-11.65 } (16.55 )$ & $3.30   ^{+1.12  }_{-1.37 } (3.36  )$ & $0.77    ^{+0.38    }_{-0.26   } (0.44   )$ & $2.68    ^{+1.61   }_{-1.62  } (1.63   )$ & $3.03   ^{+4.18  }_{-0.44 } (2.61  )$ & $-5.68   ^{+1.64  }_{-3.28 } (-6.56  )$ & 0.59    (1.35 ) \\
SN~2009er               & $1.02   ^{+1.33   }_{-0.54  } (0.89  )$ & $1.07   ^{+0.49  }_{-0.26 } (1.27  )$ & $1.27    ^{+0.37    }_{-0.21   } (1.11   )$ & $-1.45   ^{+0.40   }_{-0.22  } (-1.27  )$ & $5.33   ^{+0.53  }_{-0.43 } (5.13  )$ & $-14.51  ^{+4.34  }_{-4.94 } (-11.99 )$ & 0.38    (0.42 ) \\
SN~2009gk               & $14.85  ^{+13.87  }_{-15.34 } (29.67 )$ & $1.23   ^{+1.64  }_{-0.66 } (1.70  )$ & $0.24    ^{+0.14    }_{-0.15   } (0.32   )$ & $-0.99   ^{+2.10   }_{-2.04  } (0.91   )$ & $2.65   ^{+2.21  }_{-1.01 } (3.06  )$ & $-12.22  ^{+2.96  }_{-3.33 } (-13.38 )$ & 23.75   (32.35) \\
SN~2009hy               & $0.47   ^{+0.12   }_{-0.13  } (0.42  )$ & $0.57   ^{+0.05  }_{-0.05 } (0.60  )$ & $0.19    ^{+0.01    }_{-0.01   } (0.19   )$ & $-1.54   ^{+0.17   }_{-0.07  } (-1.48  )$ & $6.04   ^{+0.25  }_{-0.24 } (6.02  )$ & $-9.22   ^{+0.90  }_{-0.79 } (-8.63  )$ & 1.89    (1.92 ) \\
SN~2009jf               & $1.96   ^{+0.07   }_{-0.07  } (1.97  )$ & $0.63   ^{+0.01  }_{-0.01 } (0.63  )$ & $0.23    ^{+0.01    }_{-0.01   } (0.23   )$ & $-1.57   ^{+0.01   }_{-0.01  } (-1.56  )$ & $4.57   ^{+0.05  }_{-0.05 } (4.57  )$ &            ---                        & 7.81    (7.82 ) \\
SN~2010cn               & $0.26   ^{+0.06   }_{-0.07  } (0.22  )$ & $0.95   ^{+0.06  }_{-0.05 } (0.98  )$ & $0.40    ^{+0.02    }_{-0.02   } (0.40   )$ & $-1.50   ^{+0.21   }_{-0.12  } (-1.40  )$ & $8.05   ^{+0.48  }_{-0.46 } (8.04  )$ & $-6.00   ^{+0.52  }_{-0.42 } (-5.71  )$ & 7.43    (7.44 ) \\
SN~2010gd               & $0.11   ^{+0.11   }_{-0.03  } (0.14  )$ & $0.60   ^{+0.17  }_{-0.09 } (0.66  )$ & $0.60    ^{+0.11    }_{-0.12   } (0.49   )$ & $-1.53   ^{+0.29   }_{-0.08  } (-1.46  )$ & $9.86   ^{+0.11  }_{-0.27 } (9.95  )$ & $-10.33  ^{+1.92  }_{-1.98 } (-10.05 )$ & 5.36    (5.56 ) \\
SN~2010hy               & $12.75  ^{+14.45  }_{-9.36  } (12.50 )$ & $4.84   ^{+2.06  }_{-2.16 } (6.97  )$ & $10.91   ^{+1.66    }_{-1.18   } (10.62  )$ & $0.48    ^{+1.76   }_{-1.79  } (1.43   )$ & $6.10   ^{+2.79  }_{-3.36 } (4.67  )$ & $-19.12  ^{+3.13  }_{-1.90 } (-17.35 )$ & 1.91    (15.08) \\
SN~2011dh               & $1.49   ^{+0.05   }_{-0.05  } (1.48  )$ & $0.53   ^{+0.01  }_{-0.01 } (0.53  )$ & $0.077   ^{+0.001   }_{-0.001  } (0.077  )$ & $-1.57   ^{+0.01   }_{-0.01  } (-1.56  )$ & $4.28   ^{+0.03  }_{-0.03 } (4.28  )$ & $-4.10   ^{+0.07  }_{-0.07 } (-4.09  )$ & 8.78    (8.78 ) \\
SN~2011fu               & $0.44   ^{+0.09   }_{-0.08  } (0.47  )$ & $0.64   ^{+0.04  }_{-0.04 } (0.61  )$ & $0.97    ^{+0.05    }_{-0.05   } (1.00   )$ & $-1.12   ^{+0.14   }_{-0.14  } (-1.19  )$ & $8.61   ^{+0.16  }_{-0.16 } (8.58  )$ & $-16.24  ^{+1.52  }_{-1.72 } (-17.01 )$ & 16.12   (16.12) \\
SN~2011gd               & $33.48  ^{+11.73  }_{-16.03 } (33.34 )$ & $2.37   ^{+2.16  }_{-1.18 } (1.49  )$ & $0.14    ^{+0.22    }_{-0.08   } (0.12   )$ & $0.34    ^{+1.91   }_{-1.86  } (1.20   )$ & $5.22   ^{+3.48  }_{-2.75 } (2.78  )$ & $-9.94   ^{+4.45  }_{-3.85 } (-14.18 )$ &    ---          \\
\noalign{\smallskip}\hline\noalign{\smallskip}
\end{tabular}
\smallskip

\end{table}

\clearpage

\begin{table*}
\ContinuedFloat
\centering%
\setlength{\tabcolsep}{4pt}% column separation
\renewcommand{\arraystretch}{1.2}%row space
\caption{\label{table:LC_Param_Ni}...continued.}

\hspace{-110pt}
\begin{tabular}{lcccccccccccc}
\hline\hline\noalign{\smallskip}
Name  & $M_{\rm ej}$    &  $v_{9}$            &  $M_{\rm Ni}$   & log$(\kappa_\gamma, Ni)$ & $T_{\rm f}$   & $t_{\rm shift}$ &  $\chi^2$/dof \\
      & (M$_{\odot}$)   & ($10^{9}$\,cm\,s$^{-1}$) &  (M$_{\odot}$)  &   cm$^2$\,g$^{-1}$   & ($10^{3}$\,K) & (days)       &               \\
\noalign{\smallskip}\hline\noalign{\smallskip}
SN~2012aa               & $1.33   ^{+0.98   }_{-0.73  } (1.30  )$ & $0.84   ^{+0.05  }_{-0.05 } (0.84  )$ & $2.96    ^{+0.11    }_{-0.10   } (2.93   )$ & $3.17    ^{+1.61   }_{-1.64  } (1.63   )$ & $3.19   ^{+0.99  }_{-1.48 } (4.05  )$ & $-19.34  ^{+2.56  }_{-1.54 } (-17.87 )$ & 0.79    (0.86 ) \\
SN~2012ap               & $0.42   ^{+0.08   }_{-0.09  } (0.37  )$ & $0.85   ^{+0.03  }_{-0.03 } (0.86  )$ & $0.74    ^{+0.05    }_{-0.04   } (0.75   )$ & $-1.55   ^{+0.10   }_{-0.05  } (-1.51  )$ & $8.08   ^{+0.22  }_{-0.22 } (8.10  )$ & $-7.08   ^{+0.37  }_{-0.29 } (-6.88  )$ & 6.92    (6.98 ) \\
SN~2012au               & $0.37   ^{+0.02   }_{-0.02  } (0.37  )$ & $2.77   ^{+0.07  }_{-0.06 } (2.76  )$ & $0.13    ^{+0.01    }_{-0.01   } (0.13   )$ & $0.63    ^{+0.04   }_{-0.04  } (0.63   )$ & $5.69   ^{+0.02  }_{-0.02 } (5.68  )$ & $-5.13   ^{+0.13  }_{-0.13 } (-5.15  )$ & 32.89   (32.89) \\
SN~2012fh               & $5.53   ^{+7.02   }_{-6.05  } (11.53 )$ & $2.65   ^{+0.99  }_{-1.42 } (3.63  )$ & $0.0028  ^{+0.0002  }_{-0.0002 } (0.0029 )$ & $-0.89   ^{+0.23   }_{-0.22  } (-0.97  )$ & $5.97   ^{+0.06  }_{-0.06 } (5.97  )$ & $-9.82   ^{+5.35  }_{-3.78 } (-14.68 )$ & 42.34   (46.11) \\
SN~2013dk               & $0.10   ^{+0.01   }_{-0.01  } (0.10  )$ & $0.67   ^{+0.02  }_{-0.02 } (0.68  )$ & $0.98    ^{+0.02    }_{-0.02   } (0.98   )$ & $-1.57   ^{+0.01   }_{-0.01  } (-1.57  )$ & $9.97   ^{+0.02  }_{-0.05 } (9.99  )$ & $-8.12   ^{+0.35  }_{-0.37 } (-8.10  )$ & 33.03   (33.09) \\
SN~2014C                & $1.32   ^{+0.05   }_{-0.06  } (1.31  )$ & $0.85   ^{+0.02  }_{-0.02 } (0.85  )$ & $0.094   ^{+0.001   }_{-0.001  } (0.094  )$ & $-0.67   ^{+0.03   }_{-0.03  } (-0.67  )$ & $7.55   ^{+0.07  }_{-0.08 } (7.55  )$ &            ---                        & 12.07   (12.07) \\
SN~2014L                & $0.24   ^{+0.26   }_{-0.21  } (0.41  )$ & $0.50   ^{+0.02  }_{-0.03 } (0.48  )$ & $0.25    ^{+0.01    }_{-0.01   } (0.25   )$ & $-0.95   ^{+0.31   }_{-0.23  } (-1.20  )$ & $4.21   ^{+0.09  }_{-0.08 } (4.21  )$ & $-1.45   ^{+1.07  }_{-2.25 } (-2.47  )$ & 3.62    (3.63 ) \\
SN~2014as               & $0.32   ^{+0.03   }_{-0.03  } (0.32  )$ & $0.83   ^{+0.02  }_{-0.02 } (0.83  )$ & $0.24    ^{+0.01    }_{-0.01   } (0.24   )$ & $-1.57   ^{+0.01   }_{-0.01  } (-1.56  )$ & $9.15   ^{+0.20  }_{-0.19 } (9.14  )$ &            ---                        & 7.22    (7.24 ) \\
SN~2014cp               & $15.47  ^{+4.56   }_{-5.04  } (11.56 )$ & $4.17   ^{+1.07  }_{-1.50 } (3.50  )$ & $0.029   ^{+0.001   }_{-0.001  } (0.029  )$ & $-0.63   ^{+1.91   }_{-1.83  } (1.19   )$ & $5.88   ^{+0.24  }_{-0.22 } (5.93  )$ & $-19.20  ^{+2.24  }_{-1.36 } (-18.11 )$ & 0.76    (0.81 ) \\
SN~2014ds               & $8.42   ^{+30.48  }_{-6.02  } (7.61  )$ & $4.15   ^{+1.05  }_{-1.23 } (3.47  )$ & $0.067   ^{+0.145   }_{-0.030  } (0.053  )$ & $-0.37   ^{+1.92   }_{-1.87  } (1.19   )$ & $3.06   ^{+4.81  }_{-1.31 } (2.77  )$ & $-11.89  ^{+4.03  }_{-3.03 } (-15.61 )$ & 0.40    (4.45 ) \\
SN~2014eh               & $1.69   ^{+0.11   }_{-0.11  } (1.68  )$ & $0.79   ^{+0.02  }_{-0.02 } (0.78  )$ & $0.31    ^{+0.01    }_{-0.01   } (0.31   )$ & $-1.34   ^{+2.18   }_{-2.15  } (0.80   )$ & $6.77   ^{+0.11  }_{-0.11 } (6.75  )$ &            ---                        & 3.42    (3.43 ) \\
SN~2014ei               & $5.55   ^{+2.07   }_{-1.59  } (6.68  )$ & $2.41   ^{+1.32  }_{-0.91 } (3.07  )$ & $0.084   ^{+0.007   }_{-0.006  } (0.083  )$ & $-0.92   ^{+0.20   }_{-0.22  } (-0.76  )$ & $4.15   ^{+0.17  }_{-0.16 } (4.15  )$ &            ---                        & 0.81    (0.83 ) \\
SN~2015G                & $0.10   ^{+0.02   }_{-0.01  } (0.11  )$ & $4.76   ^{+0.29  }_{-0.55 } (4.61  )$ & $0.015   ^{+0.001   }_{-0.001  } (0.015  )$ & $1.03    ^{+0.08   }_{-0.13  } (0.94   )$ & $8.00   ^{+0.10  }_{-0.09 } (8.00  )$ & $-2.76   ^{+0.25  }_{-0.34 } (-2.93  )$ & 2.20    (2.22 ) \\
SN~2015K                & $30.19  ^{+22.11  }_{-7.42  } (11.59 )$ & $1.95   ^{+1.22  }_{-1.52 } (3.31  )$ & $0.11    ^{+0.104   }_{-0.011  } (0.027  )$ & $3.06    ^{+1.89   }_{-1.92  } (1.21   )$ & $3.46   ^{+4.66  }_{-1.23 } (2.30  )$ & $-15.15  ^{+5.30  }_{-3.51 } (-15.12 )$ & 0.11    (0.45 ) \\
SN~2015Q                & $29.61  ^{+17.37  }_{-11.09 } (21.24 )$ & $2.51   ^{+1.18  }_{-1.28 } (3.26  )$ & $0.080   ^{+0.028   }_{-0.013  } (0.057  )$ & $-1.06   ^{+2.65   }_{-0.72  } (-0.38  )$ & $5.84   ^{+0.26  }_{-0.24 } (5.86  )$ & $-18.36  ^{+3.26  }_{-2.59 } (-16.12 )$ & 0.84    (0.87 ) \\
SN~2015U                & $1.30   ^{+0.26   }_{-0.18  } (1.31  )$ & $2.40   ^{+0.26  }_{-0.17 } (2.43  )$ & $0.24    ^{+0.01    }_{-0.01   } (0.24   )$ & $-1.47   ^{+0.07   }_{-0.06  } (-1.45  )$ & $7.69   ^{+0.13  }_{-0.13 } (7.70  )$ &            ---                        & 6.01    (6.02 ) \\
SN~2015Y                & $1.07   ^{+0.09   }_{-0.10  } (1.07  )$ & $0.23   ^{+0.01  }_{-0.01 } (0.23  )$ & $0.073   ^{+0.002   }_{-0.003  } (0.072  )$ & $3.11    ^{+1.89   }_{-1.89  } (1.21   )$ & $9.73   ^{+0.18  }_{-0.27 } (9.91  )$ &            ---                        & 2.82    (2.82 ) \\
SN~2015ap               & $1.29   ^{+0.04   }_{-0.04  } (1.29  )$ & $1.29   ^{+0.01  }_{-0.01 } (1.29  )$ & $0.42    ^{+0.01    }_{-0.01   } (0.42   )$ & $-1.26   ^{+0.01   }_{-0.01  } (-1.26  )$ & $7.42   ^{+0.04  }_{-0.04 } (7.42  )$ &            ---                        & 11.96   (11.96) \\
SN~2016G                & $0.50   ^{+0.07   }_{-0.10  } (0.45  )$ & $0.63   ^{+0.02  }_{-0.02 } (0.64  )$ & $0.58    ^{+0.01    }_{-0.01   } (0.58   )$ & $-1.55   ^{+0.12   }_{-0.05  } (-1.50  )$ & $9.99   ^{+0.01  }_{-0.02 } (10.00 )$ & $-8.09   ^{+0.46  }_{-0.35 } (-7.90  )$ & 15.81   (15.85) \\
SN~2016ajo              & $21.13  ^{+8.20   }_{-9.21  } (22.79 )$ & $2.14   ^{+1.04  }_{-1.43 } (3.52  )$ & $0.27    ^{+0.14    }_{-0.04   } (0.18   )$ & $2.03    ^{+1.50   }_{-1.52  } (1.78   )$ & $5.28   ^{+3.45  }_{-2.07 } (3.37  )$ & $-17.26  ^{+2.31  }_{-1.11 } (-18.49 )$ & 0.18    (1.60 ) \\
SN~2016bau              & $0.10   ^{+0.01   }_{-0.01  } (0.10  )$ & $0.29   ^{+0.01  }_{-0.01 } (0.29  )$ & $0.12    ^{+0.01    }_{-0.01   } (0.12   )$ & $-1.46   ^{+0.02   }_{-0.04  } (-1.47  )$ & $7.03   ^{+0.11  }_{-0.11 } (7.03  )$ &            ---                        & 15.47   (15.47) \\
SN~2016coi              & $0.10   ^{+0.01   }_{-0.01  } (0.10  )$ & $1.30   ^{+0.01  }_{-0.01 } (1.30  )$ & $0.077   ^{+0.001   }_{-0.001  } (0.077  )$ & $0.53    ^{+0.01   }_{-0.01  } (0.53   )$ & $5.16   ^{+0.01  }_{-0.01 } (5.16  )$ & $-5.23   ^{+0.03  }_{-0.04 } (-5.22  )$ & 180.05  (180.06) \\
SN~2016gcm              & $0.49   ^{+0.12   }_{-0.14  } (0.43  )$ & $0.52   ^{+0.02  }_{-0.02 } (0.54  )$ & $0.27    ^{+0.01    }_{-0.01   } (0.27   )$ & $-1.45   ^{+0.20   }_{-0.12  } (-1.38  )$ & $6.90   ^{+0.17  }_{-0.17 } (6.90  )$ & $-19.81  ^{+0.89  }_{-0.50 } (-19.31 )$ & 9.93    (9.96 ) \\
SN~2016gkg              & $1.05   ^{+0.03   }_{-0.03  } (1.06  )$ & $0.41   ^{+0.01  }_{-0.01 } (0.41  )$ & $0.16    ^{+0.01    }_{-0.01   } (0.16   )$ & $-1.57   ^{+0.01   }_{-0.01  } (-1.57  )$ & $8.16   ^{+0.06  }_{-0.06 } (8.16  )$ &            ---                        & 55.79   (55.80) \\
SN~2016gqv              & $11.50  ^{+17.51  }_{-16.44 } (24.42 )$ & $3.26   ^{+1.31  }_{-1.72 } (3.11  )$ & $0.086   ^{+0.12    }_{-0.07   } (0.11   )$ & $1.94    ^{+1.91   }_{-1.88  } (1.20   )$ & $3.14   ^{+4.80  }_{-1.49 } (2.40  )$ & $-10.54  ^{+4.58  }_{-3.44 } (-15.11 )$ &    ---          \\
SN~2016iyc              & $3.82   ^{+1.32   }_{-1.71  } (3.62  )$ & $0.84   ^{+0.24  }_{-0.43 } (0.77  )$ & $0.058   ^{+0.004   }_{-0.003  } (0.060  )$ & $0.94    ^{+1.90   }_{-1.89  } (1.21   )$ & $7.85   ^{+0.17  }_{-0.27 } (7.88  )$ & $-11.87  ^{+1.11  }_{-1.91 } (-12.55 )$ & 2.75    (2.88 ) \\
SN~2017ein              & $0.39   ^{+0.01   }_{-0.01  } (0.38  )$ & $0.37   ^{+0.01  }_{-0.01 } (0.37  )$ & $0.050   ^{+0.001   }_{-0.001  } (0.050  )$ & $-1.57   ^{+0.01   }_{-0.01  } (-1.57  )$ & $7.28   ^{+0.04  }_{-0.04 } (7.30  )$ &            ---                        & 37.98   (37.99) \\
SN~2017iro              & $0.37   ^{+0.02   }_{-0.02  } (0.37  )$ & $0.90   ^{+0.02  }_{-0.02 } (0.90  )$ & $0.081   ^{+0.001   }_{-0.001  } (0.081  )$ & $-0.48   ^{+0.04   }_{-0.04  } (-0.48  )$ & $6.40   ^{+0.28  }_{-0.27 } (6.35  )$ & $-4.75   ^{+0.14  }_{-0.14 } (-4.78  )$ & 16.01   (16.02) \\
SN~2018cow              & $2.56   ^{+0.03   }_{-0.03  } (2.56  )$ & $10.00  ^{+0.01  }_{-0.01 } (10.00 )$ & $3.57    ^{+0.04    }_{-0.03   } (3.56   )$ & $-1.57   ^{+0.01   }_{-0.01  } (-1.57  )$ & $10.00  ^{+0.01  }_{-0.01 } (10.00 )$ & $-12.43  ^{+0.03  }_{-0.03 } (-12.42 )$ & 241.67  (241.69) \\
SN~2018ie               & $0.15   ^{+0.13   }_{-0.11  } (0.25  )$ & $1.70   ^{+0.12  }_{-0.13 } (1.59  )$ & $0.077   ^{+0.003   }_{-0.003  } (0.079  )$ & $0.057   ^{+0.32   }_{-0.28  } (-0.24  )$ & $4.62   ^{+0.15  }_{-0.13 } (4.63  )$ & $-3.34   ^{+0.63  }_{-0.78 } (-3.97  )$ & 6.63    (6.65 ) \\
SN~2019wep              & $0.17   ^{+0.29   }_{-0.13  } (0.27  )$ & $3.86   ^{+0.83  }_{-1.17 } (3.84  )$ & $0.10    ^{+0.01    }_{-0.01   } (0.11   )$ & $0.67    ^{+0.38   }_{-0.38  } (0.35   )$ & $9.77   ^{+0.17  }_{-0.28 } (9.92  )$ & $-4.56   ^{+1.51  }_{-1.81 } (-5.77  )$ & 1.70    (1.78 ) \\
SN~2020nxt              & $0.65   ^{+0.69   }_{-0.31  } (0.50  )$ & $4.39   ^{+0.62  }_{-1.06 } (4.15  )$ & $1.41    ^{+0.64    }_{-0.43   } (0.94   )$ & $-1.47   ^{+0.32   }_{-0.15  } (-1.37  )$ & $9.92   ^{+0.06  }_{-0.12 } (9.98  )$ & $-16.84  ^{+1.00  }_{-0.82 } (-16.15 )$ & 0.47    (1.36 ) \\
iPTF13bvn               & $0.15   ^{+0.01   }_{-0.01  } (0.15  )$ & $0.37   ^{+0.01  }_{-0.01 } (0.37  )$ & $0.11    ^{+0.01    }_{-0.01   } (0.11   )$ & $-1.57   ^{+0.01   }_{-0.01  } (-1.57  )$ & $5.59   ^{+0.10  }_{-0.10 } (5.57  )$ &            ---                        & 10.14   (10.15) \\
MOTJ120451.50+265946.6  & $0.11   ^{+0.04   }_{-0.02  } (0.12  )$ & $4.57   ^{+0.44  }_{-0.80 } (4.39  )$ & $0.015   ^{+0.001   }_{-0.001  } (0.015  )$ & $1.82    ^{+0.13   }_{-0.19  } (1.70   )$ & $5.69   ^{+0.03  }_{-0.03 } (5.69  )$ & $-20.00  ^{+0.02  }_{-0.01 } (-19.99 )$ & 15.53   (16.03) \\
\noalign{\smallskip}\hline\noalign{\smallskip}
\end{tabular}
\end{table*}
\end{landscape}

\begin{table*}
\centering
\setlength{\tabcolsep}{3pt}% column separation
\renewcommand{\arraystretch}{1.2}%row space
\vspace{2cm}
\caption{\label{table:LC_Param_Cooling}Best-fitting parameters of the cooling plus $^{56}$Ni model for the double-peaked SNe.}
\begin{tabular}{ccccccccccccc}
\hline\hline\noalign{\smallskip}
Name & $M_{\rm e}$ & $R_{\rm e,12}$ & $E_{\rm e,50}$            & $M_{\rm ej}$   &     $v_9$            &  $M_{\rm Ni}$ & log$(\kappa_\gamma)$ &  $T_{\rm f}$   &  $t_{\rm shift}$   & $\chi^2$/dof  \\
   & (M$_{\odot}$) & ($10^{12}$\,cm) & ($10^{50}$\,erg\,s$^{-1}$)  & (M$_{\odot}$)  & ($10^{9}$\,cm\,s$^{-1}$) & (M$_{\odot}$) &   cm$^2$\,g$^{-1}$   & ($10^{3}$\,K)  &     (days)       &               \\
\noalign{\smallskip}\hline\noalign{\smallskip}

               {SN 2011fu} & $0.059^{+0.009}_{-0.008}$ & $103.83^{+135.3}_{-68.4}$ & $0.54^{+0.24}_{-0.23}$ & $0.68^{+0.13}_{-0.13}$ & $1.18^{+0.04}_{-0.04}$ & $0.60^{+0.01}_{-0.01}$ & $-0.66^{+0.09}_{-0.08}$ & $8.05^{+0.13}_{-0.13}$ & $-2.07^{+0.41}_{-0.43}$ & 10.27 \\
                    & (0.054) & (156.23) & (0.37) & (0.78) & (1.18) & (0.61) & (-0.72) & (8.08) & (-2.36) & (10.28) \\\noalign{\smallskip}
               {SN 2015Y} & $19.10^{+8.96}_{-10.15}$ & $386.05^{+1104.94}_{-983.49}$ & $-4.65^{+0.72}_{-0.54}$ & $37.51^{+1.82}_{-1.99}$ & $4.97^{+0.09}_{-0.19}$ & $0.087^{+0.003}_{-0.003}$ & $-1.57^{+0.02}_{-0.01}$ & $9.90^{+0.14}_{-0.20}$ &          ---         & 5.14 \\
                    & (17.03) & (1347.86) & (-4.22) & (37.00) & (4.88) & (0.086) & (-1.56) & (9.80) &   & (5.17)  &  \\\noalign{\smallskip}
               {SN 2016gkg} & $0.21^{+0.01}_{-0.01}$ & $2996.83^{+9.06}_{-19.66}$ & $-1.80^{+0.01}_{-0.01}$ & $1.09^{+0.04}_{-0.04}$ & $0.30^{+0.01}_{-0.01}$ & $0.11^{+0.01}_{-0.01}$ & $-1.57^{+0.01}_{-0.01}$ & $7.62^{+0.07}_{-0.07}$ &           ---        &  41.55 \\
                    & (0.21) & (2987.88) & (-1.80) & (1.09) & (0.30) & (0.11) & (-1.57) & (7.62)  &  &  (41.57) \\\noalign{\smallskip}
               {SN 2016iyc} & $0.011^{+0.002}_{-0.001}$ & $422.40^{+327.06}_{-205.44}$ & $-1.88^{+0.20}_{-0.17}$ & $1.16^{+0.38}_{-0.16}$ & $0.70^{+0.24}_{-0.09}$ & $0.042^{+0.002}_{-0.002}$ & $1.41^{+1.94}_{-1.91}$ & $7.90^{+0.19}_{-0.19}$ & $-4.64^{+0.67}_{-0.76}$ & 1.10 \\
                    & (0.011) & (436.80) & (-1.89) & (1.26) & (0.68) & (0.043) & (1.16) & (7.83) & (-4.98) & (1.12) \\

\noalign{\smallskip}\hline\noalign{\smallskip}
\end{tabular}
\smallskip
\end{table*}

\begin{table*}
\centering
\setlength{\tabcolsep}{4pt}% column separation
\renewcommand{\arraystretch}{1.2}%row space
\vspace{2cm}
\caption{\label{table:LC_Param_Mag}Best-fitting parameters of the magnetar model for the luminous SNe.}
\begin{tabular}{ccccccccccccc}

\hline\hline\noalign{\smallskip}
Name & $M_{\rm ej}$  & $P_{0}$    &   $B_{p,14}$   &        $V_9$           & log$(\kappa_\gamma)$  &   $T_{\rm f}$       & $t_{\rm shift}$ &  $\chi^2$/dof \\
     & (M$_{\odot}$) & (ms)     & ($10^{14}$\,G) & ($10^{9}$\,cm\,s$^{-1}$)    &   cm$^2$\,g$^{-1}$      &   ($10^{3}$\,K)     &     (days)    &               \\
\noalign{\smallskip}\hline\noalign{\smallskip}
SN 2008fz & $1.93^{+0.31}_{-0.37}$$\ (1.86)$ & $2.99^{+0.01}_{-0.01}$$\ (2.99)$ & $1.13^{+0.02}_{-0.02}$$\ (1.14)$ & $2.63^{+0.03}_{-0.03}$$\ (2.63)$ & $-0.25^{+0.10}_{-0.07}$$\ (-0.23)$ & $7.43^{+0.05}_{-0.05}$$\ (7.42)$ & $-19.99^{+0.08}_{-0.04}$$\ (-19.95)$ & 8.01$\ (8.02)$ \\
SN 2010hy & $39.35^{+13.83}_{-24.71}$$\ (31.26)$ & $2.55^{+0.61}_{-1.40}$$\ (3.27)$ & $3.91^{+1.64}_{-1.23}$$\ (3.19)$ & $4.97^{+2.52}_{-2.12}$$\ (5.96)$ & $1.84^{+1.83}_{-1.81}$$\ (1.32)$ & $8.78^{+1.87}_{-2.14}$$\ (7.23)$ & $-17.72^{+3.40}_{-2.46}$$\ (-16.5)$ & -- \\   %   inf$\ (inf)$ 
SN 2012aa & $0.62^{+0.31}_{-0.30}$$\ (0.73)$ & $1.13^{+1.52}_{-0.74}$$\ (1.81)$ & $2.91^{+0.15}_{-0.18}$$\ (2.93)$ & $0.36^{+0.04}_{-0.04}$$\ (0.38)$ & $2.12^{+1.66}_{-1.67}$$\ (1.56)$ & $8.24^{+2.03}_{-3.26}$$\ (6.36)$ & $-18.89^{+2.30}_{-1.34}$$\ (-18.17)$ & 0.76$\ (1.11)$ \\
SN 2018cow & $1.23^{+0.02}_{-0.02}$$\ (1.23)$ & $1.22^{+0.01}_{-0.01}$$\ (1.23)$ & $0.10^{+0.01}_{-0.01}$$\ (0.10)$ & $10.0^{+0.01}_{-0.02}$$\ (9.99)$ & $-2.00^{+0.01}_{-0.01}$$\ (-2.00)$ & $10.0^{+0.01}_{-0.01}$$\ (10.00)$ & $-7.62^{+0.02}_{-0.02}$$\ (-7.62)$ & 191.64$\ (191.72)$ \\
\noalign{\smallskip}\hline\noalign{\smallskip}
\end{tabular}
\smallskip
\end{table*}

\begin{landscape}
\begin{deluxetable}{lcccccc}
 \tabcolsep 0.4mm
 \tablewidth{0pt}
 \tablecaption{Light-Curve Data in the Standard System (only a portion of data is shown here as an example)}
  \tablehead{\colhead{SN} & \colhead{MJD} & \colhead{$B$ (mag)} & \colhead{$V$ (mag)} & \colhead{$R$ (mag)} & \colhead{$I$ (mag)} & \colhead{$Clear$ (mag)}} 
\startdata
                 2003gk &  52821.455 &          ---      &          ---      &          ---      &          ---      &  17.717$\pm$0.061 \\
                 2003gk &  52822.463 &          ---      &          ---      &          ---      &          ---      &  17.828$\pm$0.051 \\
                 2003gk &  52827.480 &          ---      &          ---      &          ---      &          ---      &  18.016$\pm$0.068 \\
                 2003gk &  52834.456 &          ---      &          ---      &          ---      &          ---      &  18.091$\pm$0.124 \\
                 2003gk &  52840.448 &          ---      &          ---      &          ---      &          ---      &  18.114$\pm$0.145 \\
                 2003gk &  52855.405 &          ---      &          ---      &          ---      &          ---      &  18.417$\pm$0.072 \\
                 2003gk &  52861.415 &          ---      &          ---      &          ---      &          ---      &  18.399$\pm$0.113 \\
                 2003gk &  52868.452 &          ---      &          ---      &          ---      &          ---      &  18.718$\pm$0.155 \\
                 2003gk &  52888.281 &          ---      &          ---      &          ---      &          ---      &  18.817$\pm$0.143 \\
                 2003gk &  52896.328 &          ---      &          ---      &          ---      &          ---      &  18.832$\pm$0.188 \\
                 2003gk &  52900.263 &          ---      &          ---      &          ---      &          ---      &  19.131$\pm$0.207 \\
                 2003gk &  52905.289 &          ---      &          ---      &          ---      &          ---      &  19.018$\pm$0.148 \\
                 2003gk &  52909.258 &          ---      &          ---      &          ---      &          ---      &  18.987$\pm$0.112 \\
                 2003gk &  52914.230 &          ---      &          ---      &          ---      &          ---      &  19.187$\pm$0.141 \\
                 2003gk &  52924.305 &          ---      &          ---      &          ---      &          ---      &  19.126$\pm$0.236 \\
                 2003gk &  52928.232 &          ---      &          ---      &          ---      &          ---      &  19.616$\pm$0.238 \\
                 2003gk &  52932.210 &          ---      &          ---      &          ---      &          ---      &  19.334$\pm$0.239 \\
                 2003gk &  52937.237 &          ---      &          ---      &          ---      &          ---      &  19.784$\pm$0.265 \\
                 2003gk &  52821.455 &          ---      &          ---      &          ---      &          ---      &  17.717$\pm$0.061 \\
                 2003gk &  52822.463 &          ---      &          ---      &          ---      &          ---      &  17.828$\pm$0.051 \\
                 2003gk &  52827.480 &          ---      &          ---      &          ---      &          ---      &  18.016$\pm$0.068 \\
                 2003gk &  52834.456 &          ---      &          ---      &          ---      &          ---      &  18.091$\pm$0.124 \\
                 2003gk &  52840.448 &          ---      &          ---      &          ---      &          ---      &  18.114$\pm$0.145 \\
                 2003gk &  52855.405 &          ---      &          ---      &          ---      &          ---      &  18.417$\pm$0.072 \\
                 2003gk &  52861.415 &          ---      &          ---      &          ---      &          ---      &  18.399$\pm$0.113 \\
                 2003gk &  52868.452 &          ---      &          ---      &          ---      &          ---      &  18.718$\pm$0.155 \\
                 2003gk &  52883.306 &          ---      &          ---      &          ---      &          ---      &  18.855$\pm$0.133 \\
                 2003gk &  52888.281 &          ---      &          ---      &          ---      &          ---      &  18.817$\pm$0.143 \\
                 2003gk &  52896.328 &          ---      &          ---      &          ---      &          ---      &  18.832$\pm$0.188 \\
                 2003gk &  52900.263 &          ---      &          ---      &          ---      &          ---      &  19.131$\pm$0.207 \\
                 2003gk &  52905.289 &          ---      &          ---      &          ---      &          ---      &  19.018$\pm$0.148 \\
                 2003gk &  52909.258 &          ---      &          ---      &          ---      &          ---      &  18.987$\pm$0.112 \\
                 2003gk &  52914.230 &          ---      &          ---      &          ---      &          ---      &  19.187$\pm$0.141 \\
                 2003gk &  52924.305 &          ---      &          ---      &          ---      &          ---      &  19.126$\pm$0.236 \\
                 2003gk &  52928.232 &          ---      &          ---      &          ---      &          ---      &  19.616$\pm$0.238 \\
                 2003gk &  52932.210 &          ---      &          ---      &          ---      &          ---      &  19.334$\pm$0.239 \\
                 2003gk &  52937.237 &          ---      &          ---      &          ---      &          ---      &  19.784$\pm$0.265 \\
                 2003gk &  52883.306 &          ---      &          ---      &          ---      &          ---      &  18.855$\pm$0.133 \\
                 2006el &  53965.324 &          ---      &          ---      &          ---      &          ---      &  19.808$\pm$0.464 \\
                 2006el &  53972.281 &          ---      &          ---      &          ---      &          ---      &  18.153$\pm$0.250 \\
                 2006el &  53973.007 &  18.920$\pm$0.069 &  18.335$\pm$0.050 &  18.022$\pm$0.095 &  17.846$\pm$0.121 &          ---      \\
                 2006el &  53973.304 &          ---      &          ---      &          ---      &          ---      &  17.952$\pm$0.215 \\
                 2006el &  53973.977 &  18.782$\pm$0.057 &  18.213$\pm$0.048 &  17.922$\pm$0.067 &  17.695$\pm$0.091 &          ---      \\
                 2006el &  53974.304 &          ---      &          ---      &          ---      &          ---      &  17.865$\pm$0.175 \\
                 2006el &  53981.255 &          ---      &          ---      &          ---      &          ---      &  17.328$\pm$0.140 \\
                 2006el &  53993.265 &          ---      &          ---      &          ---      &          ---      &  17.563$\pm$0.215 \\
                 2006el &  53993.957 &  19.078$\pm$0.093 &  18.118$\pm$0.181 &  17.549$\pm$0.296 &  17.367$\pm$0.301 &          ---      \\
                 2006el &  53994.925 &  19.316$\pm$0.076 &  18.163$\pm$0.059 &  17.668$\pm$0.071 &  17.287$\pm$0.084 &          ---      \\
                 2006el &  54001.269 &          ---      &          ---      &          ---      &          ---      &  17.888$\pm$0.183 \\
                 2006el &  54021.156 &          ---      &          ---      &          ---      &          ---      &  18.561$\pm$0.462 \\
                 2006el &  54028.887 &  20.778$\pm$0.143 &  19.548$\pm$0.079 &  18.877$\pm$0.063 &  18.171$\pm$0.056 &          ---      \\
                 2006el &  54030.193 &          ---      &          ---      &          ---      &          ---      &  18.679$\pm$0.456 \\
                 2006el &  54039.155 &          ---      &          ---      &          ---      &          ---      &  18.772$\pm$0.341 \\
                 2006el &  54047.126 &          ---      &          ---      &          ---      &          ---      &  18.751$\pm$0.152 \\
                 2006el &  54058.109 &          ---      &          ---      &          ---      &          ---      &  18.828$\pm$0.284 \\
                 2006el &  54071.111 &          ---      &          ---      &          ---      &          ---      &  18.917$\pm$0.579 \\
\enddata
%\tablenotetext{a}{In units of \kms\,d$^{-1}$.}
\label{lightcurvedatainstandardsystem}
\end{deluxetable}
\end{landscape}

\begin{landscape}
\begin{deluxetable}{lccccccc}
 \tabcolsep 0.4mm
 \tablewidth{0pt}
 \tablecaption{Light Curve Data in Natural System (only a portion of data is shown here as example)}
  \tablehead{\colhead{SN} & \colhead{MJD} & \colhead{$B$ (mag)} & \colhead{$V$ (mag)} & \colhead{$R$ (mag)} & \colhead{$I$ (mag)} & \colhead{$Clear$ (mag)} & \colhead{System} }
\startdata
                 2003gk &  52821.455 &          ---      &          ---      &          ---      &          ---      &  17.717$\pm$0.061 &    kait2 \\
                 2003gk &  52822.463 &          ---      &          ---      &          ---      &          ---      &  17.828$\pm$0.051 &    kait2 \\
                 2003gk &  52827.480 &          ---      &          ---      &          ---      &          ---      &  18.016$\pm$0.068 &    kait2 \\
                 2003gk &  52834.456 &          ---      &          ---      &          ---      &          ---      &  18.091$\pm$0.124 &    kait2 \\
                 2003gk &  52840.448 &          ---      &          ---      &          ---      &          ---      &  18.114$\pm$0.145 &    kait2 \\
                 2003gk &  52855.405 &          ---      &          ---      &          ---      &          ---      &  18.417$\pm$0.072 &    kait2 \\
                 2003gk &  52861.415 &          ---      &          ---      &          ---      &          ---      &  18.399$\pm$0.113 &    kait2 \\
                 2003gk &  52868.452 &          ---      &          ---      &          ---      &          ---      &  18.718$\pm$0.155 &    kait2 \\
                 2003gk &  52883.306 &          ---      &          ---      &          ---      &          ---      &  18.855$\pm$0.133 &    kait2 \\
                 2003gk &  52888.281 &          ---      &          ---      &          ---      &          ---      &  18.817$\pm$0.143 &    kait2 \\
                 2003gk &  52896.328 &          ---      &          ---      &          ---      &          ---      &  18.832$\pm$0.188 &    kait2 \\
                 2003gk &  52900.263 &          ---      &          ---      &          ---      &          ---      &  19.131$\pm$0.207 &    kait2 \\
                 2003gk &  52905.289 &          ---      &          ---      &          ---      &          ---      &  19.018$\pm$0.148 &    kait2 \\
                 2003gk &  52909.258 &          ---      &          ---      &          ---      &          ---      &  18.987$\pm$0.112 &    kait2 \\
                 2003gk &  52914.230 &          ---      &          ---      &          ---      &          ---      &  19.187$\pm$0.141 &    kait2 \\
                 2003gk &  52924.305 &          ---      &          ---      &          ---      &          ---      &  19.126$\pm$0.236 &    kait2 \\
                 2003gk &  52928.232 &          ---      &          ---      &          ---      &          ---      &  19.616$\pm$0.238 &    kait2 \\
                 2003gk &  52932.210 &          ---      &          ---      &          ---      &          ---      &  19.334$\pm$0.239 &    kait2 \\
                 2003gk &  52937.237 &          ---      &          ---      &          ---      &          ---      &  19.784$\pm$0.265 &    kait2 \\
                 2003gk &  52821.455 &          ---      &          ---      &          ---      &          ---      &  17.717$\pm$0.061 &    kait2 \\
                 2003gk &  52822.463 &          ---      &          ---      &          ---      &          ---      &  17.828$\pm$0.051 &    kait2 \\
                 2003gk &  52827.480 &          ---      &          ---      &          ---      &          ---      &  18.016$\pm$0.068 &    kait2 \\
                 2003gk &  52834.456 &          ---      &          ---      &          ---      &          ---      &  18.091$\pm$0.124 &    kait2 \\
                 2003gk &  52840.448 &          ---      &          ---      &          ---      &          ---      &  18.114$\pm$0.145 &    kait2 \\
                 2003gk &  52855.405 &          ---      &          ---      &          ---      &          ---      &  18.417$\pm$0.072 &    kait2 \\
                 2003gk &  52861.415 &          ---      &          ---      &          ---      &          ---      &  18.399$\pm$0.113 &    kait2 \\
                 2003gk &  52868.452 &          ---      &          ---      &          ---      &          ---      &  18.718$\pm$0.155 &    kait2 \\
                 2003gk &  52883.306 &          ---      &          ---      &          ---      &          ---      &  18.855$\pm$0.133 &    kait2 \\
                 2003gk &  52888.281 &          ---      &          ---      &          ---      &          ---      &  18.817$\pm$0.143 &    kait2 \\
                 2003gk &  52896.328 &          ---      &          ---      &          ---      &          ---      &  18.832$\pm$0.188 &    kait2 \\
                 2003gk &  52900.263 &          ---      &          ---      &          ---      &          ---      &  19.131$\pm$0.207 &    kait2 \\
                 2003gk &  52905.289 &          ---      &          ---      &          ---      &          ---      &  19.018$\pm$0.148 &    kait2 \\
                 2003gk &  52909.258 &          ---      &          ---      &          ---      &          ---      &  18.987$\pm$0.112 &    kait2 \\
                 2003gk &  52914.230 &          ---      &          ---      &          ---      &          ---      &  19.187$\pm$0.141 &    kait2 \\
                 2003gk &  52924.305 &          ---      &          ---      &          ---      &          ---      &  19.126$\pm$0.236 &    kait2 \\
                 2003gk &  52928.232 &          ---      &          ---      &          ---      &          ---      &  19.616$\pm$0.238 &    kait2 \\
                 2003gk &  52932.210 &          ---      &          ---      &          ---      &          ---      &  19.334$\pm$0.239 &    kait2 \\
                 2003gk &  52937.237 &          ---      &          ---      &          ---      &          ---      &  19.784$\pm$0.265 &    kait2 \\
                 2006el &  53965.324 &          ---      &          ---      &          ---      &          ---      &  19.808$\pm$0.464 &    kait3 \\
                 2006el &  53972.281 &          ---      &          ---      &          ---      &          ---      &  18.153$\pm$0.250 &    kait3 \\
                 2006el &  53973.007 &  18.866$\pm$0.068 &  18.366$\pm$0.050 &  18.050$\pm$0.086 &  17.825$\pm$0.126 &          ---      &  nickel1 \\
                 2006el &  53973.304 &          ---      &          ---      &          ---      &          ---      &  17.952$\pm$0.215 &    kait3 \\
                 2006el &  53973.977 &  18.730$\pm$0.056 &  18.243$\pm$0.048 &  17.948$\pm$0.061 &  17.672$\pm$0.095 &          ---      &  nickel1 \\
                 2006el &  53974.304 &          ---      &          ---      &          ---      &          ---      &  17.865$\pm$0.175 &    kait3 \\
                 2006el &  53981.255 &          ---      &          ---      &          ---      &          ---      &  17.328$\pm$0.140 &    kait3 \\
                 2006el &  53993.265 &          ---      &          ---      &          ---      &          ---      &  17.563$\pm$0.215 &    kait3 \\
                 2006el &  53993.957 &  18.990$\pm$0.089 &  18.169$\pm$0.180 &  17.600$\pm$0.269 &  17.334$\pm$0.314 &          ---      &  nickel1 \\
                 2006el &  53994.925 &  19.210$\pm$0.074 &  18.224$\pm$0.058 &  17.712$\pm$0.064 &  17.248$\pm$0.087 &          ---      &  nickel1 \\
                 2006el &  54001.269 &          ---      &          ---      &          ---      &          ---      &  17.888$\pm$0.183 &    kait3 \\
                 2006el &  54021.156 &          ---      &          ---      &          ---      &          ---      &  18.561$\pm$0.462 &    kait3 \\
                 2006el &  54028.887 &  20.665$\pm$0.141 &  19.613$\pm$0.078 &  18.937$\pm$0.056 &  18.110$\pm$0.057 &          ---      &  nickel1 \\
                 2006el &  54030.193 &          ---      &          ---      &          ---      &          ---      &  18.679$\pm$0.456 &    kait3 \\
                 2006el &  54039.155 &          ---      &          ---      &          ---      &          ---      &  18.772$\pm$0.341 &    kait3 \\
                 2006el &  54047.126 &          ---      &          ---      &          ---      &          ---      &  18.751$\pm$0.152 &    kait3 \\
                 2006el &  54058.109 &          ---      &          ---      &          ---      &          ---      &  18.828$\pm$0.284 &    kait3 \\
                 2006el &  54071.111 &          ---      &          ---      &          ---      &          ---      &  18.917$\pm$0.579 &    kait3 \\
\enddata
%\tablenotetext{a}{In units of \kms\,d$^{-1}$.}
\label{lightcurvedatainnaturalsystem}
\end{deluxetable}
\end{landscape}

% Don't change these lines
\bsp    % typesetting comment
\label{lastpage}
\end{document}